\documentclass[final,3p,times,nopreprintline]{elsarticle}
\usepackage{amsmath,amssymb,amsfonts,dsfont}
\usepackage{graphicx,color}
\usepackage{slashed}
\usepackage{booktabs,tabulary}
\usepackage{placeins}
\usepackage{hyperref}
\usepackage{enumerate}
\usepackage{ulem}

\newcommand{\Fpi}{F_\pi}
\newcommand{\mpi}{M_{\pi}}
\newcommand{\mpii}{M_{\pi^0}}

\newcommand{\Order}{\mathcal{O}}

\newcommand{\diff}{\text{d}}
\newcommand{\eps}{\epsilon}
\newcommand{\eV}{\,\text{eV}}
\newcommand{\keV}{\,\text{keV}}
\newcommand{\MeV}{\,\text{MeV}}
\newcommand{\GeV}{\,\text{GeV}}
\newcommand{\TeV}{\,\text{TeV}}
\newcommand{\beq}{\begin{equation}}
\newcommand{\eeq}{\end{equation}}

\newcommand{\SU}{S\!U}

\renewcommand{\vec}[1]{\mathbf{#1}}

\newcommand{\Amp}{\mathcal{A}}
\newcommand{\F}{\mathcal{F}}
\newcommand{\G}{\mathcal{G}}
\newcommand{\M}{\mathcal{M}}
\newcommand{\N}{\mathcal{N}}

\newcommand{\unity}{\mathds{1}}
\renewcommand{\Re}{\text{Re}\,}
\renewcommand{\Im}{\text{Im}\,}
\newcommand{\BR}{\mathcal{B}}
\newcommand{\disc}{\text{disc}\,}
\newcommand{\ep}{\eta^{(\prime)}}

\newcommand{\epgg}{\eta\to\pi^0\gamma\gamma}
\newcommand{\ggpe}{\gamma\gamma\to\pi^0\eta}
\newcommand{\meta}{M_{\eta}}
\newcommand{\metap}{M_{\eta'}}
\newcommand{\mpc}{M_\pi}
\newcommand{\setap}{s_{\eta'}}
\newcommand{\sth}{s_{\text{th}}}
\newcommand{\Qetap}{Q_{\eta'}}
\newcommand{\TM}{\mathcal{T\!M}}

\newcommand{\new}[1]{{\color{red} #1}}

\newcommand{\eolp}{\,.}
\newcommand{\eolc}{\,,}

\newcommand{\mpip}{M_{\pi^+}}

\newcommand{\mpipm}{M_{\pi^\pm}}
\newcommand{\mpiz}{M_{\pi^0}}

\newcommand{\mKp}{M_{K^+}}
\newcommand{\mKpm}{M_{K^\pm}}
\newcommand{\mKz}{M_{K^0}}

\renewcommand{\L}{\mathcal{L}}
\renewcommand{\O}{\mathcal{O}}
\newcommand{\A}{\mathcal{A}}
\renewcommand{\P}{\mathcal{P}}

\newcommand{\toright}[1]{\hspace*{\fill}{\footnotesize{#1}}}

\def\XXint#1#2#3{{\setbox0=\hbox{$#1{#2#3}{\int}$}
     \vcenter{\hbox{$#2#3$}}\kern-0.5\wd0}}

\allowdisplaybreaks[1]

\begin{document}

\renewcommand{\theequation}{\arabic{equation}}

\numberwithin{equation}{section}
 
\setcounter{tocdepth}{3}

\begin{frontmatter}

\title{\toright{\textnormal{JLAB-THY-20-3219}}\\
Precision tests of fundamental physics with $\eta$ and $\eta^\prime$ mesons
}

\author[NC]{Liping Gan}
\author[Bonn]{Bastian Kubis}
\author[Bloomington1,Bloomington2,JLab]{Emilie Passemar}
\author[Toronto]{Sean Tulin}

\address[NC]{Department of Physics and Physical Oceanography, University of North Carolina Wilmington, Wilmington, North Carolina 28403, USA}
\address[Bonn]{Helmholtz-Institut f\"ur Strahlen- und Kernphysik (Theorie) and
   Bethe Center for Theoretical Physics, Universit\"at Bonn, 53115 Bonn, Germany}
\address[Bloomington1]{Department of Physics, Indiana University, Bloomington, Indiana 47405, USA}
\address[Bloomington2]{Center for Exploration of Energy and Matter, Indiana University, Bloomington, Indiana 47408, USA}
\address[JLab]{Theory Center, Thomas Jefferson National Accelerator Facility, Newport News, Virginia 23606, USA}
\address[Toronto]{Department of Physics and Astronomy, York University, Toronto, Ontario M3J 1P3, Canada}

\begin{abstract} 
Decays of the neutral and long-lived $\eta$ and $\eta'$ mesons provide a unique, flavor-conserving laboratory to test 
low-energy Quantum Chromodynamics and search for new physics beyond the Standard Model.  
They have drawn world-wide attention in recent years 
and have inspired broad experimental programs in different high-intensity-frontier centers. 
New experimental data will offer critical inputs to precisely determine the light quark mass ratios, 
$\eta$--$\eta'$ mixing parameters, and hadronic contributions to the anomalous magnetic moment of the muon. 
At the same time, it will provide a sensitive probe to test potential new physics.
This includes searches for hidden photons, light Higgs scalars, and axion-like particles that are complementary to worldwide efforts to detect new light particles below the GeV mass scale, as well as tests of discrete symmetry violation.
In this review, we give an update on theoretical developments, discuss the experimental opportunities, 
and identify future research needed in this field.
\end{abstract}


\end{frontmatter}

\tableofcontents

\section{Introduction}

Quantum Chromodynamics (QCD) is the theory of quarks and gluons, and yet these particles cannot be directly observed as asymptotic states due to confinement. 
All observable particles are color-singlet hadrons and lattice field theory remains the only approach for calculating their spectrum and properties from first principles in QCD. 
Nevertheless, a parallel approach based on symmetries has largely shaped our understanding of QCD at low energies.
These arguments have led to the celebrated framework of chiral perturbation theory ($\chi$PT) as a predictive effective theory for the lowest-lying states of QCD.

The properties of the $\eta$ and $\eta^{\prime}$ mesons, in particular, relate to many different threads within this theme and 
therefore provide a testing ground of the fundamental symmetries of QCD in the Standard Model and beyond.
The symmetry structure of QCD at low energies can be investigated based on its Lagrangian, which, 
neglecting heavy quarks, is given by
\beq \label{eq:QCDLagrangian}
\mathcal{L}_{\rm QCD} = \sum_{q=u,d,s} \bar{q} \left(i  \gamma^\mu  D_\mu - m_q \right) q -\frac{1}{4} G_{\mu \nu}^a G^{a \mu \nu} \, ,
\eeq
where $D_\mu = \partial_\mu + i g_s \lambda^a G_\mu^a/2$ is the covariant derivative, $G_\mu^a$ is the gluon field with strong coupling $g_s$, and $\lambda^a$ are the Gell-Mann matrices.
$G_{\mu \nu}^a=\partial_\mu G_\nu^a-\partial_\nu G_\mu^a - g_s f^{abc} G_\mu^b G_\nu^c$ denotes the corresponding field strength tensor, where $f^{abc}$ are the $\SU(3)$ structure constants.
In the limit where the quark masses are set to zero, $\mathcal{L}_{\rm QCD}$ is invariant under global chiral $U(3)_L \times U(3)_R$ transformations acting on the left- and 
right-handed quark fields.
In nature, however, chiral symmetry is broken in several ways.
The condensation of quark--anti-quark pairs in the QCD vacuum spontaneously breaks $U(3)_L \times U(3)_R$ down to $\SU(3)_V \times U(1)_B$, where $\SU(3)_V$ ($\SU(2)_V$) and $U(1)_B$ are flavor (isospin) and baryon number symmetries, respectively.
Each broken generator results in a massless Nambu--Goldstone boson, which form $\SU(3)_V$ multiplets: the octet of pseudoscalar mesons ($\pi^0$, $\pi^{\pm}$, $K^{\pm}$, $K^0$, $\bar{K}^0$, and $\eta_8$) and, in principle, a ninth pseudoscalar meson ($\eta_0$) that is a singlet for the broken axial $U(1)_A$ symmetry.
The quark masses, which are small compared to the chiral symmetry breaking scale $\sim 1\GeV$, break chiral symmetry explicitly and thereby generate masses for the Nambu--Goldstone bosons, following the mechanism discovered by Gell-Mann, Oakes, and Renner~\cite{GellMann:1968rz}.
Furthermore, the unequal quark masses break the $\SU(3)_V$ flavor symmetry (and, to a lesser extent, isospin). 
This breaks the degeneracy of the pseudoscalar octet and leads to mixing such that the $\eta$ and $\eta^\prime$ mesons are admixtures of the octet $\eta_8$ and singlet $\eta_0$ (plus a small isospin-violating mixing with the $\pi^0$).

The first predictions for $\eta,\eta^\prime$ mesons date back to work in the 1950s by Okun~\cite{Okun:1958}, Yamaguchi~\cite{1958PThPh..19..622Y,1959PThPS..11...37Y}, and Iketa et al.~\cite{Ikeda:1959zz}, based on the Sakata model for strongly-interacting particles~\cite{Sakata:1956hs}.
This framework treated the nucleon $N$ and hyperon $\Lambda$ as the constituents making up mesons and predicted the existence of isospin-singlet $N \bar{N}$ and $\Lambda \bar{\Lambda}$ states to go along with pions and kaons.
The Eightfold Way of Gell-Mann~\cite{GellMann:1961ky} and Ne'eman~\cite{Neeman:1961jhl} also predicted the $\eta$ meson as the last missing piece of the pseudoscalar octet. 
In 1961, Pevsner et al.~\cite{Pevsner:1961pa} discovered this new particle as a three-pion resonance. 
While Gell-Mann did not anticipate a substantial three-pion channel (being forbidden by isospin), Bastien et al.~\cite{Bastien:1962zz} soon connected it to Gell-Mann's work.
On the other hand, the $\eta^\prime$ meson, discovered in 1964~\cite{Kalbfleisch:1964zz,Goldberg:1964zza}, appeared to be an unsuitable candidate for the missing ninth pseudoscalar meson from the spontaneously broken $U(1)_A$ since it was much heavier than the octet~\cite{Glashow:1967vr,Weinberg:1975ui}.

The $U(1)_A$ Noether current has a nonvanishing divergence known as the
Adler--Bell--Jackiw or axial anomaly~\cite{Adler:1969gk,Bell:1969ts}.  The 
$U(1)_A$ symmetry is explicitly broken by the quantum fluctuations of 
the quarks coupling to the gauge fields. It has a purely quantum-mechanical origin, representing
one of the most profound symmetry breaking phenomena in nature. 
The anomaly, associated with the coupling of quarks to the electromagnetic field, is primarily responsible for the two-photon decays of $\pi^0$, $\eta$, and $\eta^{\prime}$.  
The second type of anomaly associated with quarks coupling to the gluon fields prevents the $\eta^{\prime}$ from being a Goldstone boson due to the spontaneous breaking of $U(1)_A$; consequently the $\eta^{\prime}$ acquires a nonvanishing mass in the chiral limit~\cite{tHooft:1976rip}. 
It is important to note that in a limit where the number of colors, $N_c$, is large,  the $\eta^{\prime}$ does become a Goldstone boson~\cite{Witten:1979vv}.  
In this limit the axial anomaly, which is proportional to $1/N_c$, vanishes.  
Lastly, $U(1)_B$ baryon number symmetry is also broken explicitly by the axial anomaly associated with electroweak gauge fields, but this effect is thought to be negligible except in the very early Universe~\cite{Kuzmin:1985mm}.

The underlying symmetries of QCD impose strong constraints on the properties of the light mesons. The low-energy theorems based on the chiral invariance of the underlying QCD Lagrangian are highly developed and well tested in the domain of pionic and kaonic reactions~\cite{Holstein:2001bt}, and it is an obvious next step to attempt to extend this success into the $\eta$ sector. The theoretical developments and experimental measurements for $\eta$ and $\eta^{\prime}$ will play an important role in this effort.

The dominant hadronic decay modes of the $\eta$ meson, $\eta\to\pi^+\pi^-\pi^0$ and $\eta\to3\pi^0$, break isospin symmetry. They constitute one of the relatively rare isospin-breaking hadronic observables in which electromagnetic effects are strongly suppressed~\cite{Bell:1996mi,Sutherland:1966zz}, offering clean experimental access to the light quark mass difference $m_u-m_d$; to establish this link and extract quark mass information with precision, however, sophisticated theoretical methods are required.  
The same holds, in principle, for the three-pion decays of the $\eta'$, although here neither experimental nor theoretical precision are at a competitive level yet.  
The dominant hadronic $\eta'$ decay modes, $\eta'\to\eta\pi^+\pi^-$ and $\eta'\to\eta\pi^0\pi^0$, probe the low-energy dynamics of $S$-wave interactions both in the $\pi\pi$ and the $\pi\eta$ systems.  Similarly, the diphoton decay $\eta\to\pi^0\gamma\gamma$ is sensitive to the subtle interplay between scalar dynamics and (dominant) vector-meson exchanges, in a process for which the leading interaction is of polarizability-type and suppressed in the low-energy expansion of $\chi$PT.

A significant number of $\eta$ and $\eta'$ decays are anomalous or of odd intrinsic parity, most prominently the electromagnetic decays into two photons.  While the (by far dominant) two-photon decay of the $\pi^0$, determining the latter's lifetime, is given to very high accuracy by the Wess--Zumino--Witten anomaly~\cite{Wess:1971yu,Witten:1983tw}, those of $\eta$ and $\eta'$ depend on the subtle mixing of the original $\SU(3)$ octet and singlet states---accurate measurements of these decays therefore allow one to pin down the corresponding mixing angles and decay constants with improved precision.  The transition form factors (TFFs) that parameterize the dependence of all pseudoscalar-to-two-photon couplings on the photons' virtualities are similarly of utmost current interest, as they determine the size of important hadronic quantum corrections in the anomalous magnetic moment of the muon, $(g-2)_\mu$.  
The long-standing discrepancy of now about $4.2$ standard deviations between the Standard Model calculation~\cite{Aoyama:2020ynm} and the combined Brookhaven~\cite{Bennett:2006fi} and Fermilab~\cite{Abi:2021gix,Albahri:2021ixb} measurements of this quantity is widely regarded as one of the most striking hints at the possibility of physics beyond the Standard Model (BSM), which might be substantiated further by updated and improved experimental results at Fermilab~\cite{Grange:2015fou} and J-PARC~\cite{Saito:2012zz} soon.
Several other anomalous processes, such as $\eta'\to2(\pi^+\pi^-)$, $\ep\to\pi^+\pi^-\gamma^{(*)}$, and $\eta'\to\omega\gamma^{(*)}$, are intimately related to the transition form factors in the context of dispersion theory, which allows for a data-driven reconstruction of these form factors given that their hadronic discontinuities are dominated by these channels. 
Finally, the rare dilepton decays $\pi^0,\,\eta,\,\eta'\to\ell^+\ell^-$ are fully calculable within the Standard Model based on the pseudoscalar transition form factors, and deviations from such predictions might hint at physics beyond it. 

The study of $\eta$ and $\eta^\prime$ mesons also provides a unique probe for BSM physics.
Both have relatively small decay widths---many strong and electromagnetic decays are forbidden at leading order due to $P$, $C$, $CP$, $G$-parity, or angular momentum conservation~\cite{Nefkens:2002sa}---which enhances the relative importance of new physics signals.
Traditionally, BSM searches in $\eta,\eta^\prime$ decays have been motivated in the context of flavor-conserving tests of discrete symmetries, namely $C$, $P$, $T$, and combinations thereof~\cite{Nefkens:1995dk,Nefkens:2002sa,Jarlskog:2002zz,Kupsc:2011ah}.
Given its quantum numbers, $J^{PC} = 0^{-+}$, the $\eta$ meson was recognized long ago as a testing ground for discrete symmetry violation predicted for strong or electromagnetic forces~\cite{Kobzarev,Prentki:1965tt,Lee:1965zza,Lee:1965hi,Bernstein:1965hj} similar to those found for weak interactions~\cite{Wu:1957my,Christenson:1964fg}.
While these speculations have not been borne out experimentally thus far, symmetry-violating $\eta,\eta^\prime$ decays remain of interest for BSM searches since the Standard Model rates, sourced by the weak interaction, are negligible.
These tests mostly fall into two cases: {\it (i)} $P,CP$-violating decays, of which $\ep \to 2\pi$ is the most famous example, and {\it (ii)} $C,CP$-violating decays.
Case {\it (i)} has a more solid theoretical footing---there exist connections with underlying models or effective operators---but most signals are already excluded at an observable level by null electric dipole moment (EDM) searches.
To date, the only exception that is demonstrably EDM-safe is ``second generation'' $CP$ violation---$CP$-odd four-fermion operators with strange quarks and muons---that can be tested in $\eta \to \mu^+ \mu^-$ polarization asymmetries~\cite{Sanchez-Puertas:2018tnp,Sanchez-Puertas:2019qwm}.
On the other hand, for case {\it (ii)}, the theoretical framework for $C,CP$-violating decays is largely unexplored and the implications of experiments in terms of BSM physics remains unknown.

Alternatively, $\eta$ and $\eta^\prime$ mesons are a laboratory to search for new weakly-coupled light particles below the GeV scale.
These include dark photons and other hidden gauge bosons, light Higgs scalars, and axion-like particles.
Such states are predicted in connection with dark matter and other BSM frameworks and have become one of the most active areas of research in the phenomenology community over the past decade~\cite{Essig:2013lka,Alekhin:2015byh,Alexander:2016aln,Battaglieri:2017aum}.
In meson decays, new light particles are produced on-shell; when they subsequently decay to Standard Model particles, they appear as resonances in invariant mass distributions~\cite{Nelson:1989fx,Fayet:2007ua,Reece:2009un}.
This signature was considered long ago as a possible discovery channel for the (light) Standard Model Higgs boson $H$, e.g., as a dilepton resonance via $\eta \to \pi^0 H \to \pi^0 \ell^+ \ell^-$~\cite{Ellis:1975ap}.
In the BSM context, the landscape of models and couplings leads to many possible signals in $\eta,\eta^\prime$ decays, in some cases mimicking decays that \new{are} highly suppressed in the Standard Model.
For example, the doubly-radiative decay $\eta\to B\gamma\to\pi^0\gamma\gamma$ offers an experimental opportunity to search for a sub-GeV leptophobic $B$ boson that couples predominantly to quarks and arises from a new $U(1)_B$ baryon number gauge symmetry~\cite{Nelson:1989fx,Tulin:2014tya}. 
A newly developed JLab Eta Factory experiment~\cite{JEF-PAC42} will search for a $B$ boson in this $\eta$ decay channel and has projected sensitivity to the baryonic fine structure constant as low as $10^{-7}$ in the $B$ boson mass region of $0.14$--$0.55\GeV$.
On the other hand, axion-like particles can yield exotic four- and five-body $\ep$ decays that have not been studied previously and have no direct experimental constraints.
Suffice to say, there are many opportunities to explore the rich physics of light BSM particles in $\eta,\eta^\prime$ decays.
On the experimental front, many possible channels can be searched for, while on the theoretical front, the predictions and connections with existing constraints need to be mapped out more quantitatively.

\begin{table}[t!]
\centering
\renewcommand{\arraystretch}{1.3}
\begin{tabular}{llll} 
\toprule
Channel     & Expt.\ branching ratio   & Discussion & Sect.\ \\ 
\midrule
$\eta \to 2\gamma$     & $39.41(20)\% $     & chiral anomaly, $\eta$--$\eta^{\prime}$ mixing  & \ref{sec:eta-2g} \\
$\eta \to 3\pi^0$       & $32.68(23)\% $    &  $m_u-m_d$ & \ref{sec:eta-3pi}\\ 
$\eta \to \pi^0 \gamma\gamma$& $2.56(22)\times 10^{-4}$ & $\chi$PT at $\Order(p^6)$, leptophobic $B$ boson, & \ref{sec:eta-pi0gg}, \ref{sec:BSM-vector},\\[-1mm]
& & \qquad  light Higgs scalars & \quad \ref{sec:scalars} \\
$\eta \to \pi^0\pi^0\gamma\gamma$& $<1.2\times 10^{-3}$ & $\chi$PT, axion-like particles (ALPs) & \ref{sec:axions}\\
$\eta \to 4\gamma$      & $<2.8\times 10^{-4}$     & $<10^{-11}$\cite{Bratkovskaya:1995hk} &  \\ 
$\eta \to \pi^+\pi^-\pi^0$& $22.92(28)\% $    &  $m_u-m_d$, $C/CP$ violation,  & \ref{sec:eta-3pi}, \ref{sec:BSMCV},   \\ [-1mm]
 & &  \qquad light Higgs scalars & \quad\ref{sec:scalars} \\
$\eta \to \pi^+\pi^-\gamma$  &  $4.22(8)\%$  &  chiral anomaly, theory input for singly-virtual TFF   & \ref{sec:eta-pipigamma}, \\[-1mm]
& & \qquad and $(g-2)_\mu$, $P/CP$ violation & \quad \ref{sec:eta-2pi} \\
$\eta \to \pi^+\pi^-\gamma\gamma$& $<2.1\times 10^{-3}$ & $\chi$PT, ALPs & \ref{sec:axions}\\
$\eta \to e^+e^-\gamma$ & $6.9(4)\times 10^{-3}$  & theory input for $(g-2)_\mu$,  & \ref{sec:DRetaTFF}, \\[-1mm]
& &  \qquad dark photon, protophobic $X$ boson & \quad \ref{sec:BSM-vector} \\
$\eta \to \mu^+\mu^-\gamma$ & $3.1(4)\times 10^{-4}$  & theory input for $(g-2)_\mu$, dark photon & \ref{sec:DRetaTFF}, \ref{sec:BSM-vector}\\ 
$\eta \to e^+e^-$ & $<7\times 10^{-7}$  & theory input for $(g-2)_\mu$, BSM weak decays & \ref{sec:P-ll}, \ref{sec:weak} \\ 
$\eta \to \mu^+\mu^-$ & $5.8(8)\times 10^{-6}$  & theory input for $(g-2)_\mu$,  BSM weak decays, & \ref{sec:P-ll}, \ref{sec:weak},\\[-1mm] 
& & \qquad $P/CP$ violation& \quad  \ref{sec:eta-2pi} \\
$\eta \to \pi^0\pi^0 \ell^+\ell^-$   &  & $C/CP$ violation, ALPs  & \ref{sec:BSMCV}, \ref{sec:axions} \\
$\eta \to \pi^+\pi^-e^+e^-$   & $2.68(11)\times 10^{-4}$ & theory input for doubly-virtual TFF and $(g-2)_\mu$,   & \ref{sec:eta-pipill}, \\[-1mm]
& & \qquad $P/CP$ violation, ALPs & \quad \ref{sec:eta-2pi}, \ref{sec:axions} \\
$\eta \to \pi^+\pi^-\mu^+\mu^-$   & $<3.6\times 10^{-4}$ & theory input for doubly-virtual TFF and $(g-2)_\mu$, & \ref{sec:eta-pipill}, \\[-1mm]
& & \qquad $P/CP$ violation,  ALPs & \quad \ref{sec:eta-2pi}, \ref{sec:axions} \\
$\eta \to e^+e^-e^+e^-$ & $2.40(22)\times 10^{-5}$  & theory input for $(g-2)_\mu$ & \ref{sec:doubleDalitz}\\ 
$\eta \to e^+e^-\mu^+\mu^-$ & $<1.6\times 10^{-4}$  & theory input for $(g-2)_\mu$  & \ref{sec:doubleDalitz}\\ 
$\eta \to \mu^+\mu^-\mu^+\mu^-$ & $<3.6\times 10^{-4}$  & theory input for $(g-2)_\mu$ & \ref{sec:doubleDalitz}\\ 
$\eta \to \pi^+\pi^- \pi^0 \gamma$ & $<5\times 10^{-4}$  & direct emission only & \ref{sec:eta-omegag} \\  
$\eta\to\pi^\pm e^\mp \nu_e$ & $< 1.7\times 10^{-4}$ & second-class current & \ref{sec:weak} \\
$\eta \to \pi^+ \pi^-$   &  $<4.4\times 10^{-6}$~\cite{Babusci:2020jwb}  & $P/CP$ violation & \ref{sec:eta-2pi}\\
$\eta \to 2\pi^0$   &  $<3.5\times 10^{-4}$    & $P/CP$ violation & \ref{sec:eta-2pi}\\ 
$\eta \to 4\pi^0$       &  $<6.9\times 10^{-7}$    & $P/CP$ violation & \ref{sec:etap-4pi}, \ref{sec:eta-2pi}\\ 
\bottomrule
\end{tabular}
\renewcommand{\arraystretch}{1.0}
\caption{Summary of $\eta$ meson decays.
Experimental information on the branching ratios is taken from the Particle Data Group (PDG) review~\cite{Tanabashi:2018oca} unless otherwise indicated. 
The total $\eta$ width is $\Gamma_{\eta}=1.31(5)\keV$~\cite{Tanabashi:2018oca}.
\label{tab:eta}}
\end{table}

\begin{table}[t!]
\centering
\renewcommand{\arraystretch}{1.3}
\begin{tabular}{llll} 
\toprule
Channel     & Expt.\ branching ratio   & Discussion & Sect. \\ 
\midrule
$\eta^\prime \to \eta\pi^+\pi^-$ &  $42.6(7)\%$  & large-$N_c$ $\chi$PT, light Higgs scalars & \ref{sec:etap-etapipi}, \ref{sec:scalars}  \\
$\eta^\prime \to \pi^+\pi^-\gamma$  &  $28.9(5)\%$  &  chiral anomaly, theory input for singly-virtual TFF  & \ref{sec:eta-pipigamma},  \\[-1mm]
& & \qquad and $(g-2)_\mu$, $P/CP$ violation & \quad \ref{sec:eta-2pi} \\
$\eta^\prime \to \eta\pi^0\pi^0$ &  $22.8(8)\%$  & large-$N_c$ $\chi$PT & \ref{sec:etap-etapipi}\\
$\eta^\prime \to \omega\gamma$  &  $2.489(76)\%$~\cite{Ablikim:2019wop}  &  theory input for singly-virtual TFF and $(g-2)_\mu$ & \ref{sec:eta-omegag}\\
$\eta^\prime \to \omega e^+e^-$& $2.0(4)\times 10^{-4} $    &  theory input for doubly-virtual TFF and $(g-2)_\mu$ & \ref{sec:eta-omegag}\\  
$\eta^\prime \to 2\gamma$ &  $2.331(37)\%$~\cite{Ablikim:2019wop}  & chiral anomaly, $\eta$--$\eta^{\prime}$ mixing & \ref{sec:eta-2g} \\  
$\eta^\prime \to 3\pi^0$ &  $2.54(18)\%~^{(\star)}$      & $m_u-m_d$ & \ref{sec:etap-3pi}\\   
$\eta^\prime \to \mu^+\mu^-\gamma$ & $1.09(27)\times10^{-4}$ & theory input for $(g-2)_\mu$, dark photon & \ref{sec:DRetaTFF}, \ref{sec:BSM-vector}\\  
$\eta^\prime \to e^+e^-\gamma$ & $4.73(30)\times 10^{-4}$  & theory input for $(g-2)_\mu$, dark photon & \ref{sec:DRetaTFF}, \ref{sec:BSM-vector}\\  
$\eta^\prime \to \pi^+\pi^-\mu^+\mu^-$& $<2.9\times 10^{-5} $    &  theory input for doubly-virtual TFF and $(g-2)_\mu$, & \ref{sec:eta-pipill}, \ref{sec:eta-2pi},\\[-1mm]
& & \qquad $P/CP$ violation,  dark photon, ALPs & \quad \ref{sec:BSM-vector}, \ref{sec:axions}\\
$\eta^\prime \to \pi^+\pi^-e^+e^-$& $2.4\big({}^{+1.3}_{-1.0}\big)\times 10^{-3} $  &  theory input for doubly-virtual TFF and $(g-2)_\mu$, & \ref{sec:eta-pipill}, \ref{sec:eta-2pi},\\[-1mm]
& & \qquad $P/CP$ violation, dark photon, ALPs & \quad \ref{sec:BSM-vector}, \ref{sec:axions}\\
$\eta^\prime \to \pi^0\pi^0\ell^+\ell^-$&     &  $C/CP$ violation, ALPs  &  \ref{sec:BSMCV}, \ref{sec:axions}\\
$\eta^\prime \to \pi^+\pi^-\pi^0$ &  $3.61(17)\times 10^{-3}$  & $m_u-m_d$, $C/CP$ violation, & \ref{sec:etap-3pi}, \ref{sec:BSMCV}, \\   [-1mm]
 & &  \qquad light Higgs scalars & \quad\ref{sec:scalars} \\
$\eta^\prime \to 2(\pi^+\pi^-)$ &  $8.4(9)\times 10^{-5}$      & theory input for doubly-virtual TFF and $(g-2)_\mu$ & \ref{sec:etap-4pi}\\   
$\eta^\prime \to \pi^+\pi^-2\pi^0$ &  $1.8(4)\times 10^{-4}$  & & \ref{sec:etap-4pi}\\  
$\eta^\prime \to 2(\pi^+\pi^-)\pi^0$ &  $<1.8\times 10^{-3}$  & ALPs & \ref{sec:axions}\\  
$\eta'\to K^\pm\pi^\mp$ & $<4\times 10^{-5}$ & weak interactions & \ref{sec:weak} \\
$\eta'\to\pi^\pm e^\mp \nu_e$ & $< 2.1\times 10^{-4}$ & second-class current & \ref{sec:weak} \\
$\eta^\prime \to \pi^0 \gamma\gamma$& $3.20(24)\times 10^{-3}$ & vector and scalar dynamics, $B$ boson,  & \ref{sec:etap-Pgg}, \ref{sec:BSM-vector}, \\[-1mm]  
& & \qquad light Higgs scalars & \quad \ref{sec:scalars} \\
$\eta^\prime \to \eta \gamma\gamma$& $8.3(3.5)\times10^{-5}$~\cite{Ablikim:2019wsb} & vector and scalar dynamics, $B$ boson, & \ref{sec:etap-Pgg}, \ref{sec:BSM-vector},\\[-1mm]
& & \qquad light Higgs scalars & \quad \ref{sec:scalars} \\
$\eta^\prime \to 4\pi^0$ &  $<4.94\times 10^{-5}$~\cite{Ablikim:2019msz}      & ($S$-wave) $P/CP$ violation & \ref{sec:etap-4pi}\\   
$\eta^\prime \to e^+e^-$ & $<5.6\times 10^{-9}$  & theory input for $(g-2)_\mu$, BSM weak decays & \ref{sec:P-ll}, \ref{sec:weak} \\
$\eta^\prime \to \mu^+\mu^-$ &   & theory input for $(g-2)_\mu$, BSM weak decays & \ref{sec:P-ll}, \ref{sec:weak}\\
$\eta^\prime \to \ell^+\ell^-\ell^+\ell^-$ &   & theory input for $(g-2)_\mu$ & \ref{sec:doubleDalitz}\\  
$\eta^\prime \to \pi^+\pi^- \pi^0 \gamma$ &   & $B$ boson & \ref{sec:BSM-vector} \\  
$\eta^\prime \to \pi^+ \pi^-$   & $<1.8\times 10^{-5}$       & $P/CP$ violation & \ref{sec:eta-2pi}\\
$\eta^\prime \to 2\pi^0$   &  $<4\times 10^{-4}$    & $P/CP$ violation & \ref{sec:eta-2pi} \\ 
\bottomrule
\end{tabular}
\renewcommand{\arraystretch}{1.0}
\caption{Summary of $\eta^\prime$ meson decays. Experimental information on the branching ratios is taken from the PDG review~\cite{Tanabashi:2018oca} unless otherwise indicated. 
We remark that for $\BR(\eta'\to3\pi^0)$ marked with ${}^{(\star)}$ above, there is significant tension between the
PDG \textit{fit} and \textit{average}; see the discussion in Sect.~\ref{sec:etap-3pi}.
Also, in this review, we take the PDG \textit{fit} value for the total $\eta^\prime$ width $\Gamma_{\eta^{\prime}}=196(9)\MeV$, which differs somewhat from the PDG \textit{average} that is dominated by the COSY measurement $\Gamma_{\eta^{\prime}}=226(17)(14)\keV$~\cite{Czerwinski:2010my}.
\label{tab:etaprime}}
\end{table}

On the experimental side, there have been broad programs on $\eta^{(\prime)}$ mesons at the high-intensity facilities worldwide, using different production mechanisms and experimental techniques with complementary energies. 
Among the $e^+e^-$ collider facilities, the KLOE experiment at the DA$\Phi$NE $\phi$-factory is running at the $\phi(1020)$ mass peak~\cite{Giovannella:2016gqx}, while the BESIII experiment at the Beijing charm factory is running near the $J/\psi(3097)$ mass peak~\cite{Fang:2015vva}.
The $\eta^{(\prime)}$ mesons are produced through the radiative decays of $\phi\to \eta^{(\prime)}\gamma$ or $J/\psi\to\eta^{(\prime )}\gamma$, respectively, and are tagged by detecting the mono-energetic decay photons. KLOE collected $\sim 1.0\times 10^8$ $\eta$ and $\sim 0.5\times 10^6$ $\eta^{\prime}$ mesons in its initial program, which was completed in 2006.  The upgraded KLOE-II experiment, which finished data taking in March 2018, increased this data set by about a factor of three~\cite{Kang:2019meo}.  BESIII  collected a sample of $\sim 1\times 10^{10}$ $J/\psi$ events from 2009 to 2019, which yielded $\sim 1\times 10^7$ $\eta$ and $\sim 5\times 10^7$ $\eta^{\prime}$ mesons, respectively~\cite{Ablikim:2019hff}. The WASA-at-COSY experiment at the cooler synchrotron COSY storage ring~\cite{Maier:1997zj} produced $\eta$ or $\eta^{\prime}$ through hadronic interactions in collisions of a proton or deuteron beam with a hydrogen or deuterium pellet target near the threshold. The production of mesons was tagged by detecting the forward boosted protons or helium ions.
A large $\eta$ data sample was produced through the reactions of $pd\rightarrow \eta \,^3\text{He}$ ($\sim 3\times 10^7$ $\eta$~\cite{Adlarson:2014aks,Husken:2016edi,Husken:2019dou}) and $pp\rightarrow \eta pp$ ($\sim 5\times 10^8$ $\eta$~\cite{Goswami:2015wdd}).
At the MAMI electron accelerator facility, the Crystal Ball/TAPS experiment~\cite{Unverzagt:2014hya} uses a real-photon beam derived from the production of bremsstrahlung radiation off a $1.6\GeV$ electron beam passing through a thin radiator and tagged with the Glasgow tagging spectrometer. The $\eta$ and $\eta^{\prime}$ mesons were produced from the $\gamma p\to \ep p$ reaction near threshold. About $2.5\times 10^8$ $\eta$ were collected~\cite{Unverzagt:2009vm}. The physics program was extended to the $\eta^{\prime}$ in 2014, with $6\times 10^6$ $\eta'$ events
taken in 10~weeks of beam time~\cite{Steffen:2017jem}.
 
These experiments described above have one thing in common: the $\eta^{(\prime)}$s are produced at relatively low energies in the laboratory frame, where the energies of the decay particles from $\eta^{(\prime)}$s are relatively close to the detection thresholds.
Several ongoing and new $\eta^{(\prime)}$ programs developed at Jefferson Lab (JLab) will produce $\eta^{(\prime)}$s  using an electron or photon beam with energies up to $\sim 12\GeV$. The energies of the decay particles will be much higher than the detection thresholds, which will offer experimental data sets with significantly different systematics compared to the experiments at other facilities. 
The upcoming JLab Eta Factory experiment (JEF)~\cite{JEF-PAC42} aims to measure various $\eta$ and $\eta^{\prime}$ decays with emphasis on rare neutral modes,  with about a factor of two orders of magnitude improvement in background suppression compared to all other existing or planned experiments in the world.  A data sample at the level of $\sim 10^9$ for tagged $\eta$ and $\eta^{\prime}$ will be expected from JEF during the JLab $12\GeV$ era.  

Looking forward, a new proposal ``Rare Eta Decays with a TPC for Optical Photons'' (REDTOP)~\cite{Gonzalez:2017fku, Gatto:2016rae} has been under development. About $\sim 2\times 10^{12}$ $\eta$ mesons per year (and could be a factor of ten increase in phase II) will be produced by a $\sim 1.8\GeV$ proton beam. It would be a super $\eta$ machine on the horizon. 
These high-quality, high-statistics data from worldwide facilities will offer a fruitful, exciting age for $\eta$ and $\eta^{\prime}$ physics.

This report aims to update, extend, and complement several other reviews that have focused on the physics of $\eta$ and
$\eta'$ mesons in the past.  
We believe it is a timely task to summarize the progress made beyond the comprehensive documentation of the state of the art in the conference proceedings of the Uppsala workshop in 2001~\cite{Bijnens:2002zy}.  
Herein we largely concentrate on $\eta$ and $\eta'$ decays, and disregard their interaction with nucleons, nuclei, and nuclear matter that constitutes a large part of Ref.~\cite{Bass:2018xmz}.  
We also acknowledge a review specifically of the BESIII physics program with $\eta$ and $\eta'$ mesons~\cite{Fang:2017qgz}.
Lastly, for completeness, we provide an overview of the most important $\eta$ and $\eta'$ decay modes, their most recently published branching ratios~\cite{Tanabashi:2018oca}, as well as their main points of theoretical interest, in Tables~\ref{tab:eta} and~\ref{tab:etaprime}.
Most of these are discussed in detail in the course of this review.
We except $C$-violating decays that will be listed separately
in Sect.~\ref{sec:BSMfundsym}, Table~\ref{tab:CVdecays}.

This article is structured as follows.  
In Sect.~\ref{sec:experiments}, we review recent, current, and planned experimental activities on $\eta$ and $\eta'$ physics at various particle physics laboratories around the world.  
Sections~\ref{sec:ChPT} and \ref{sec:DispTheory} present the main theoretical tools for model-independent analyses in this realm: chiral perturbation theory for mesons, with an eye on its large-$N_c$ extension that is essential for the inclusion of the $\eta'$, and dispersion theory.
Next, we discuss $\eta$ and $\eta^\prime$ decays within the Standard Model: hadronic decays (Sect.~\ref{sec:hadronic}), anomalous decays (Sect.~\ref{sec:radiative}), and $\eta\to\pi^0 \gamma\gamma$ and related decays (Sect.~\ref{sec:eta-pi0gg}). 
Weak interaction decays in the Standard Model (as well as similar processes from BSM physics) are given in Sect.~\ref{sec:weak}.
Our main discussion on BSM physics in $\eta$ and $\eta^\prime$ decays follows, including tests of discrete spacetime symmetries and lepton flavor violation (Sect.~\ref{sec:BSMfundsym}) and searches for new weakly-coupled hidden particles in the MeV--GeV mass range (Sect.~\ref{sec:BSM-lightparticles}).
We close with a summary of $\eta$ and $\eta^{\prime}$ physics and an outlook for the future research in this field.

\section{Experimental activities}\label{sec:experiments}
Using $\eta^{(\prime)}$ decays as tests of the Standard Model and as probes for BSM physics  has drawn attention from the physics communities since the 1990s.  In recent decades, there have been intensive experimental activities with different production mechanisms and different experimental detection techniques in complementary energy regions. These activities, including both fixed-target and collider experiments, will be highlighted below. More details about some decay channels will be described in the later sections.

\subsection{Precision experiments at JLab}

The high-duty-factor, high-precision, continuous electron beam of $6\GeV$ (and recently upgraded to $12\GeV$) at Jefferson Lab (JLab) offers a great opportunity for $\pi^0$, $\eta$, and $\eta^{\prime}$ physics at the frontier of precision measurements with photo- or electroproduction on fixed targets.

\subsubsection{Primakoff experimental program at JLab \label{exp-pri} }

\begin{figure}
\centering
\includegraphics*[width=0.3\linewidth]{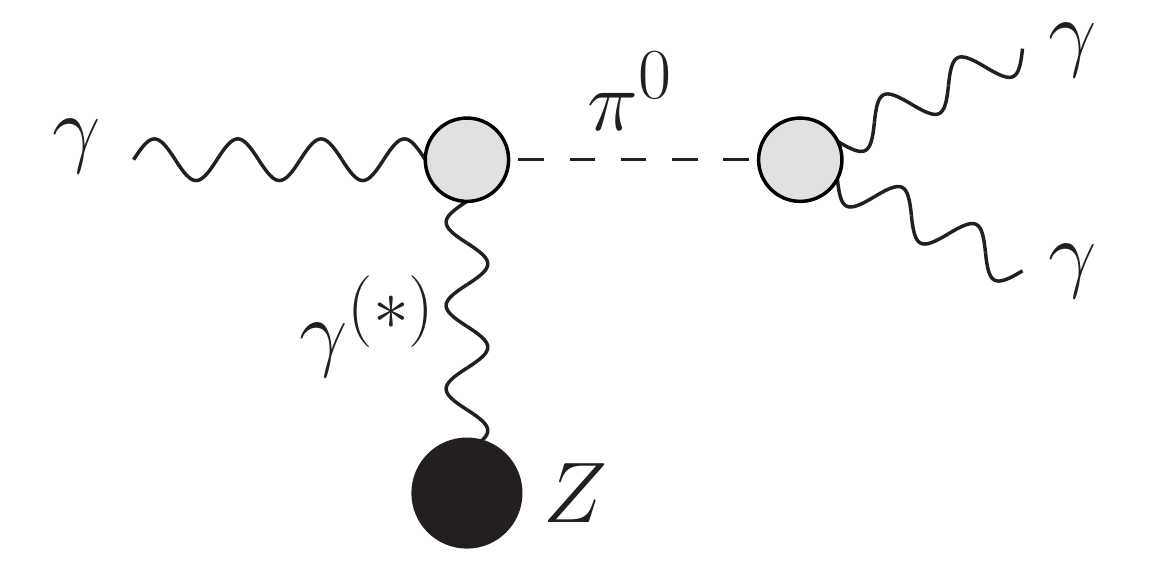}
\caption{Schematic representation of $\pi^0$ Coulomb photoproduction off a nucleus of charge $Z$ (Primakoff effect).}
\label{fig:Primakoff}
\end{figure}

A comprehensive Primakoff experimental program was developed by the PrimEx collaboration in the past decades, and it served as one of the key physics programs driving the JLab $12\GeV$ upgrade~\cite{white-paper,CDR-12GeV}. 
This program includes high-precision measurements of the two-photon decay widths $\Gamma(P\to\gamma\gamma)$ and the transition form factors $F_{P\gamma^*\gamma^*}(-Q^2,0)$ (see Sects.~\ref{sec:eta-2g}, \ref{sec:pi0-eta-etapTFF}, and \ref{sec:DRetaTFF}) at four-momentum transfers $Q^2 = 0.001 \ldots 0.5\GeV^2$, where $P$ represents $\pi^0$, $\eta$, and $\eta^{\prime}$~\cite{Gan:2014pna}.  The main experimental approach is to use high-energy photo- [for $\Gamma(P\to\gamma\gamma)$] or electroproduction [for $F_{P\gamma^*\gamma^*}(-Q^2,0)$] of mesons in the Coulomb field of a nucleus (as shown in Fig.~\ref{fig:Primakoff} for the case of the $\pi^0$). 
This process, known as the Primakoff effect~\cite{Primakoff:1951pj}, has distinguishable characteristics: (1) it is a coherent process; (2) the cross section is peaked at extremely forward angles ($\sim {M_P^2}/{2E^2}$); and (3) it has strong beam energy dependence [$\sigma_P\sim Z^2\log(E)$]. These characteristics serve as natural kinematic filters for a clear experimental signature to extract the Primakoff amplitude from the measured total differential cross sections.
Two experiments (PrimEx-I and PrimEx-II) on the $\pi^0$ radiative decay width were carried out during the JLab $6\GeV$ era. The result of PrimEx-I was published in 2011~\cite{Larin:2010kq} and the result of PrimEx-II was published recently in 2020~\cite{Larin:2020bhc}. The weighted average of PrimEx I and II results is $\Gamma (\pi^0\to\gamma\gamma)=7.802 (052)_{\rm stat}(105)_{\rm syst}\eV$~\cite{Larin:2020bhc}. This result with its 1.50\% total uncertainty represents the most accurate measurement of this fundamental quantity to date. It confirms the chiral anomaly and is $2\sigma$ below the theoretical calculations~\cite{Goity:2002nn,Ananthanarayan:2002kj,Kampf:2009tk,Ioffe:2007eg} based on higher-order corrections to the anomaly. Measurements of the $\eta$ and $\eta^{\prime}$ radiative decay widths and the transition form factors of $\pi^0$, $\eta$, and $\eta^{\prime}$ will be performed during the JLab $12\GeV$ era.  Two data sets for a measurement of $\Gamma(\eta\to\gamma\gamma)$ on $^4$He were collected in spring 2019 and recently in fall 2021.
 The data analysis is in progress. The results from this program will provide insight into our understanding of the chiral anomaly and chiral symmetry breaking.

\subsubsection{\label{sect-JEF} The JLab Eta Factory experimental program with GlueX} 

The approved JLab Eta Factory experiment~\cite{Gan:2009zzd, JEF-PAC42, Gan:2015nyc, JEF-PAC45} will measure various $\eta$ and $\eta^{\prime}$ decays with the GlueX apparatus~\cite{Adhikari:2020cvz, Lawrence:2009zz} and an upgraded forward calorimeter to test QCD and to search for BSM physics. 

The baseline GlueX apparatus~\cite{Adhikari:2020cvz} includes: a high-energy photon tagger for a photon beam up to $12\GeV$ energy, a central drift chamber and forward drift chambers inside a solenoid magnet ($\sim 2\,\text{T}$) for charged-particle tracking, a scintillating hodoscope start counter surrounding the physics target and a forward time-of-flight wall for charged-particle identification,  a barrel calorimeter and a forward calorimeter (FCAL) for electromagnetic particle detection, and a  pair spectrometer for photon flux monitoring.  In addition, two upgrade projects are currently in progress. One is a high-granularity, high-resolution, and radiation-resistant lead tungstate crystal (PbWO$_4$) calorimeter insert ($\sim 1\times 1\,\text{m}^2$ in size) to replace the conventional Pb-glass detector modules in the central region of the existing FCAL (FCAL-II), in order to reach the projected experimental sensitivity for the rare $\eta$ and $\eta^{\prime}$ decays. The mass production of the  PbWO$_4$ detector modules has been in progress and the detector installation  will be expected in 2023.
The second is a new muon detector, required by the charged-pion polarizability experiment (E12-13-008)~\cite{Rory-pp}, which will be located downstream of the FCAL to offer a capability for muon detection. This detector will be ready for experiments in 2022.

The JEF experiment has unique features: (1)~highly boosted $\eta$ and $\eta^{\prime}$ are simultaneously produced by 
a $\sim 12\GeV$ tagged photon beam through the two-body reaction $\gamma p\rightarrow \eta^{(\prime)} p$, hence the detection efficiencies for $\eta^{(\prime)}$ are not sensitive to the detector thresholds; (2)~all $\eta^{(\prime)}$ events are tagged by measuring the recoil protons to reject noncoplanar backgrounds; and (3)~the 
electromagnetic particles from the $\eta^{(\prime)}$ decays ($\gamma$, or $e^+$ and $e^-$ in some cases) will be  measured by an upgraded FCAL-II, a state-of-the-art, high-resolution, high-granularity, hybrid of  PbWO$_4$ crystal and Pb-glass calorimeter with significantly suppressed shower overlaps and improved energy and position resolutions.  The combination of all these experimental techniques makes the JEF experiment a unique factory for $\eta$ and $\eta^{\prime}$ with unprecedented low backgrounds, particularly 
for the rare neutral decay modes. It offers up to two orders of magnitude reduction in backgrounds compared to all other existing or planned experiments in the world. 
The JEF experiment can be run in parallel to the other experiments in Hall~D, such as the GlueX meson spectroscopy experiment~\cite{Meyer:2006az} or any future experiments using a hydrogen target. This offers the possibility to continuously accumulate data throughout the JLab $12\GeV$ era.
The data for nonrare 
$\eta$ and $\eta^{\prime}$ decays
has been collected in parallel to the GlueX experiment using the baseline GlueX apparatus. The 
recently completed phase~I GlueX run at low luminosity accumulated about
$2\times 10^7$ tagged $\eta$ and tagged $\eta^{\prime}$ events (with detected recoil protons). The production rates have been further increased to about $6.5\times 10^7$ for the  tagged $\eta$ and $4.9\times 10^7$ for $\eta^{\prime}$  with every 100 days of beam time at the high-luminosity run since fall 2019.  About $\sim 10^9$ tagged $\eta^{(\prime)}$ events will be collected in the coming years.
Within the first 100 days of running with the future  upgraded FCAL-II, the upper limits of many rare $\eta^{(\prime)}$ decay branching ratios (for both neutral and charged modes) will be improved by up to two orders of magnitude, and sufficient sensitivity will be achieved for the first time to map the Dalitz distribution of $\eta \to\pi^0\gamma\gamma$ to probe the scalar meson dynamics in $\chi$PT.

\subsubsection{Potential opportunities with CLAS and CLAS12}
The CEBAF large acceptance spectrometer (CLAS)~\cite{Mecking:2003zu} was used in Hall~B at JLab to study photo- and electro-induced nuclear and hadronic reactions in the years 1998--2012. There is a large data sample for $\eta$ and $\eta^{\prime}$ in the existing CLAS data set, which a group of CLAS collaborators has been actively mining for the decays to $e^+e^-\gamma$, $\pi^+\pi^-\gamma$, and $\pi^+\pi^-\pi^0$~\cite{Kunkel:2016upp}. Currently the CLAS detector has been replaced by a new spectrometer CLAS12~\cite{Burkert:2008rj} for use with a $12\GeV$ electron beam.  More data for $\eta$ and $\eta^{\prime}$ mesons is expected in the future CLAS12 running period. 

\subsection{Other fixed-target experiments}
There are several fixed-target experimental programs across the world. The A2 collaboration at the Mainz Microtron (MAMI) has been playing one of the leading roles in studying $\eta^{(\prime)}$ decays involving electromagnetic particles  ($\gamma$, $e^+$, and $e^-$) in the final state.  A high-precision tagged photon beam was used to produce $\eta^{(\prime)}$ near the production threshold through the $\gamma p\rightarrow \eta^{(\prime)} p$ reaction. The electromagnetic particles from $\eta^{(\prime)}$ decays were detected by a nonmagnetic detector that combines two electromagnetic calorimeters: a spherical Crystal Ball (with 672 NaI crystals) covering $\sim 93\%$ of $4\pi$ solid angle and a forward TAPS detector (consisting of 384 BaF$_2$ initially and the central 18 BaF$_2$ modules being replaced by 72 PbWO$_4$ crystals for the $\eta^{\prime}$ run in 2014) covering the small polar angles from $1^\circ$ to $20^\circ$~\cite{Prakhov:2008ff,McNicoll:2010qk,Adlarson:2019nwa}. About $\sim 10^8$ $\eta$ events were collected during three run periods~\cite{Kashevarov:2017kqb} from 2007 to 2014.

The WASA-at-COSY collaboration completed a program for the hadronic production of $\eta$ on an internal pellet target (frozen hydrogen or deuterium) from 2006 to 2014 at the cooler synchrotron COSY storage ring~\cite{Maier:1997zj,Husken:2019dou} in J\"ulich, Germany. 
A large $\eta$ data sample was produced in the proton--nucleus fusion reaction $pd\rightarrow \eta \,^3\text{He}$ ($\sim 3\times 10^7$ $\eta$~\cite{Husken:2016edi,Husken:2019dou}) or $pp\rightarrow \eta pp$ ($\sim 5\times 10^8$ $\eta$~\cite{Goswami:2015wdd}) with a proton beam energy slightly above the threshold; the $\eta$ events were tagged by measuring the recoil nucleus in the forward WASA detector ($3^\circ$--$18^\circ$)~\cite{Adam:2004ch}. The central detector system ($20^\circ$--$169^\circ$), consisting of a superconducting solenoid magnet, a drift chamber, a plastic scintillator barrel, and a CsI(Na) crystal electromagnetic calorimeter, was used for identification and reconstruction of mesons. 

In addition, investigations of $\eta^{(\prime)}$ were also performed in high-energy fixed-target experiments, such as the NA60 experiment~\cite{Arnaldi:2016pzu} using a $400\GeV$ proton beam at the Super-Proton-Synchrotron (SPS) in the CERN North Area, and the LHCb experiment~\cite{Aaij:2016jaa} at the Large Hadron Collider (LHC).

A new proposal ``Rare Eta Decays with a TPC for Optical Photons'' (REDTOP)~\cite{Gonzalez:2017fku, Gatto:2016rae} has been under development recently. 
It was originally proposed to utilize the booster system at Fermilab and the delivery ring accelerator complex for the Muon $g-2$ and Mu2e experiments to deliver a $1.8 \GeV$ proton beam for $\eta$ (a $3.5 \GeV$ beam  for $\eta^{\prime}$) production; an alternative option to host it at CERN~\cite{Beacham:2019nyx} is currently under consideration.  
A proton beam with energy of $\sim 1.8\GeV$ and intensity of $\sim 10^{11}$ per second was proposed to 
produce $\eta$ mesons on a target system consisting of ten Be foils.  
Due to the nature of hadronic production, a high production rate is expected at the level of  $\sim 2\times 10^{12}$ $\eta$ mesons per year; the background level, however, will be high as well with an estimated ratio of 1:200 for $\eta$ production relative to the inelastic proton scattering background.
The main detector proposed for REDTOP includes several subsystems  with a $0.6\,\text{T}$ solenoid magnet:
 an Optical Time Projection Chamber (OTPC), a Dual-Readout Integrally Active Nonsegmented Option (ADRIANO) calorimeter, a muon polarimeter (and an optional photon polarimeter). Due to the Cherenkov detection mechanism of the OTPC, this apparatus will be optimal for fast particles from $\eta$ decays and ``blind'' to slower hadronic background. An upgraded version of t-REDTOP~\cite{t-REDTOP} suggests a phase~II REDTOP at the PIP-II facility at Fermilab.  The $\eta$ mesons will be produced via $p d\to\eta \,{}^3\text{He}$ with a $\sim 900\MeV$ proton beam and will be tagged by detecting the recoil ${}^3\text{He}$.  The $\eta$ production rate will be increased by a factor of ten compared to the initial design for phase~I and the background will be largely reduced with a tagging technique. The design for some of the subdetector systems will be modified in order to withstand the high luminosity condition. 

\subsection{Collider experiments}
DA$\Phi$NE in Frascati, Italy and BEPC-II in Beijing, China are $e^+e^-$ collider facilities operating with symmetric and relatively low $e^+$ and $e^-$ beam energies with designed luminosity of $\sim 10^{33}\text{cm}^{-2}\text{s}^{-1}$. 
The KLOE-II collaboration at DA$\Phi$NE and the BESIII collaboration at BEPC-II produce $\eta^{(\prime)}$ mesons through vector-meson radiative decays $V\to\eta^{(\prime)}\gamma$, where $V$ is $\phi(1020)$ for KLOE-II and $J/\psi(3097)$ for BESIII, respectively. The spectrometers at both facilities are magnetic detectors and have the capability to access decay channels with hadronic and electromagnetic particles in the final state. KLOE-II completed its data taking campaign in 2018. Together with the earlier KLOE data, a total of $3\times 10^8$ $\eta$ events were collected~\cite{Kang:2019meo}. 
Due to a larger center-of-mass energy compared to the KLOE experiments, BESIII can access both $\eta$ and $\eta^{\prime}$. A total of $\sim 1\times 10^{10}$ $J/\psi(3097)$ events were accumulated by BESIII~\cite{Ablikim:2019hff} so far, corresponding to about $1\times 10^7$ for $\eta$ and $5\times 10^7$  for $\eta^{\prime}$ from the $J/\psi$ radiative decays. Another $e^+e^-$ symmetric collider, VEPP-2000~\cite{Solodov:2016klh}, with a smaller designed luminosity of $\sim 10^{32}\text{cm}^{-2}\text{s}^{-1}$ at the maximum center-of-mass energy of $2\GeV$ has been operating since 2010. A magnetic detector (CMD-3~\cite{Khazin:2008zz}) and a nonmagnetic detector (SND~\cite{Shatunov:2000zc}), installed in two interaction regions of VEPP-2000, have already collected an integrated luminosity of $\sim 60\,\text{pb}^{-1}$.  A goal of $\sim 1\,\text{fb}^{-1}$ is expected in the coming decade. 

Two previous $B$-factory experiments, the Belle experiment~\cite{Abashian:2000cg} at the KEKB collider in KEK and the BaBar experiment~\cite{Aubert:2001tu} at the PEP-II collider in SLAC, collected $\sim 1.5\,\text{ab}^{-1}$ of integrated luminosity on $\Upsilon$ resonances. Both colliders have asymmetric positron and electron beam energies ($4.0\GeV$ and $7.0\GeV$ for KEKB, $3.1\GeV$ and $9\GeV$ for PEP-II, respectively) to boost the produced particles in the lab frame, offering opportunities to study the $\eta^{(\prime)}$ transition form factors at $Q^2$ up to $100\GeV^2$.  After a decade of running, PEP-II was turned off in 2008.  The KEKB collider and the Belle detector, on the other hand,  were upgraded to SuperKEKB and Belle-II.  About one order of magnitude more statistics will be expected from Belle-II in the coming years~\cite{Konno:2015zna}. 

In addition, the light mesons can also be produced directly  at the $e^+e^-$ collider facilities by the two-photon fusion reaction $e^+e^-\rightarrow \gamma^*\gamma^*e^+e^-\rightarrow \eta^{(\prime)} e^+e^-$ or $e^+e^-$ annihilation.

\section{Chiral perturbation theory}\label{sec:ChPT}
As $\eta$ and $\eta'$ are hadrons, the description of their decays necessarily involves strong interactions. 
However, because the strong coupling constant is large at low energies, a perturbative description of QCD in terms of quarks and gluons is no longer valid.
Instead, an effective-theory description of QCD based on chiral symmetry, known as chiral perturbation theory ($\chi$PT), has emerged as a successful and powerful tool to describe the dynamics of QCD at low energies.
The theory is predictive up to a handful of low-energy constants fitted from observables, although in the case of $\eta$ and $\eta'$ decays, it often needs to be supplemented with dispersion theory in order to reach sufficient precision. 
Lattice QCD is the only truly {\it ab initio} nonperturbative method to solve QCD at low energies, but it is not yet able to describe the complex decay dynamics of $\eta$ and $\eta'$ mesons comprehensively. 
We will therefore point towards important lattice results where relevant, but not introduce the methods themselves.

In the remainder of this section, we offer a brief summary of $\chi$PT and its large-$N_c$ extension that is particularly relevant for $\eta$ and $\eta'$ physics. 
The interested reader is referred the pioneering works of Weinberg~\cite{Weinberg:1978kz} and Gasser and Leutwyler~\cite{Gasser:1983yg,Gasser:1984gg}, as well as to various reviews~\cite{Bijnens:1994qh,Ecker:1994gg,Pich:1995bw,Scherer:2002tk,Gasser:2003cg,Bernard:2006gx,Scherer:2012xha} for a broader survey of the subject. 
Several applications will follow in later sections. 

\subsection{Effective Lagrangian}

The low-energy hadronic world is very difficult to describe theoretically in terms of partonic degrees of freedom due to the nonperturbative nature of the strong interaction.
However, the spectrum of the theory is rather simple at low energies, containing
only the octet of light pseudoscalar mesons: pions, kaons, and the $\eta$. 
Collectively, these are the pseudo-Goldstone bosons of the spontaneously broken chiral symmetry $\SU(3)_L \times \SU(3)_R \to \SU(3)_V$. 
Experimentally, we also observe that pseudoscalar mesons interact weakly at very low energies, both
among themselves and with nucleons. 
Chiral perturbation theory offers a framework to reintroduce a perturbative expansion of a different kind for the strong interactions at low energies, provided a change of fundamental 
degrees of freedom is performed. 

This amounts to replacing the QCD Lagrangian~\eqref{eq:QCDLagrangian},
expressed in terms of the light-quark and gluon fields, 
with the effective Lagrangian of chiral perturbation theory 
\begin{equation} \label{eq:L-ChPT}
\mathcal{L}_{\chi\text{PT}} = \sum_{d \ge 1} \mathcal{L}_{2d}  \,, 
\end{equation}
where $\mathcal{L}_{2d}$ represents terms in the Lagrangian of $\Order(p^{2d})$ in the low-energy power counting scheme and $p$ refers to a generic (small) 4-momentum.
Due to the Gell-Mann--Oakes--Renner relation~\cite{GellMann:1968rz} (further discussed in Sect.~\ref{sec:eta3pi-lightquarkmasses}), pseudoscalar meson masses scale as $M_{\pi/K/\eta}^2 \propto m_q$ and therefore light quark masses are counted as $m_q = \Order(p^2)$.
The degrees of freedom of the theory are no longer quarks and gluons but the pseudo-Goldstone bosons themselves ($\pi,K,\eta$).
The expansion parameter is no longer the strong coupling constant $\alpha_s \equiv g^2_s/4\pi$ but 
the ratio of $p$ to the typical hadronic scale $\Lambda = 4\pi F_\pi\sim 1\GeV$, which is small provided we 
stay at low energies, $p \ll \Lambda$. 

The leading-order Lagrangian is written as
\begin{equation}
\mathcal{L}_2 =
\frac{F_0^2}{4}\langle D_\mu U D^\mu U^\dagger+\chi U^\dagger +  \chi^\dagger U  \rangle \,,
\label{eq:Lp2}
\end{equation}
where
\begin{equation} \label{eq:defs}
U = \exp\left(i \frac{\sqrt{2}\Phi}{F_0}\right)\,, \qquad 
D_\mu U = \partial_\mu U -i (v_\mu + a_\mu) U + i U (v_\mu - a_\mu)\,, \qquad
\chi = 2B_0 (s+ip)\,. 
\end{equation}
$F_0$ is the decay constant of the pseudo-Goldstone bosons in the chiral limit ($m_u = m_d = m_s = 0$); throughout this review, we work with the convention $F_\pi = 92.28(9)\MeV$~\cite{Tanabashi:2018oca}. $\langle\ldots\rangle$ denotes the trace in flavor space. $s$, $p$, $v_\mu$, and $a_\mu$ represent scalar, pseudoscalar, vector, and axial-vector external sources, and 
$B_0$ is a low-energy constant related to the quark condensate according to $\langle 0 | \bar q^i q^j| 0 \rangle = -F_0^2 B_0 \,\delta^{ij}$. In the absence of external fields, i.e., when $v_\mu = a_\mu = p = 0$, the scalar source is set to
$s=\M = \text{diag}(m_u, m_d, m_s)$, with the quark masses $m_q$ encoding the explicit chiral symmetry breaking. The pseudo-Goldstone bosons are collected in the matrix
\begin{equation}
\label{eq:goldstones}
\Phi=\left(
\begin{array}{ccc}
\frac{\pi^0}{\sqrt{2}} + \frac{\eta_8}{\sqrt{6}}  & \pi^+ & K^+\\[1ex]
\pi^- & -\frac{\pi^0}{\sqrt{2}} +\frac{\eta_8}{\sqrt{6}} & K^0\\[1ex]
K^- & \bar{K}^0 & -\frac{2\eta_8}{\sqrt{6}} 
\end{array}
\right)\,,
\end{equation}
where $\eta_8$ denotes the octet component of the $\eta$ meson.

In a mass-independent regularization scheme, quantum corrections of this theory are suppressed by two powers in the low-momentum
expansion per loop order.
The theory is renormalizable and unitary order by order in the expansion, which means that each loop order requires
the introduction of new counterterms to absorb ultraviolet infinities.
At next-to-leading order (NLO) or $\mathcal{O} (p^4)$ the Lagrangian is given as~\cite{Gasser:1984gg}
\begin{align}
\mathcal{L}_{4} &=  L_1 \langle D_\mu U^\dagger  D^\mu U \rangle^2 + L_2 \langle D_\mu U^\dagger  D_\nu U \rangle 
\langle D^\mu U^\dagger  D^\nu U \rangle  + L_3 \langle D_\mu U^\dagger  D^\mu U  D_\nu U^\dagger  D^\nu U \rangle \nonumber \\
& + L_4  \langle D_\mu U^\dagger  D^\mu U \rangle \langle \chi^\dagger U +  \chi U^\dagger \rangle +  
L_5  \langle D_\mu U^\dagger  D^\mu U \left( \chi^\dagger U +  \chi U^\dagger \right) \rangle + L_6 
\langle \chi^\dagger U +  \chi U^\dagger \rangle^2 \nonumber \\
& + L_7 \langle \chi^\dagger U -  \chi U^\dagger \rangle^2 + 
L_8 \langle \chi^\dagger U \chi^\dagger U + \chi U^\dagger \chi U^\dagger \rangle -i L_9 
\langle F_R^{\mu \nu} D_\mu U  D_\nu U^\dagger + F_L^{\mu \nu} D_\mu U^\dagger  D_\nu U  \rangle  \nonumber \\
& + L_{10} \langle U^\dagger F_R^{\mu \nu} UF_{L \mu \nu} \rangle\,,
\label{eq:LchiPTp4}
\end{align}
with the additional definitions
\begin{align}
F_L^{\mu \nu} &\equiv \partial^\mu l^\nu -  \partial^\nu l^\mu -i [l^\mu, l^\nu]\,, &
 l_\mu &\equiv v_\mu - a_\mu\,, 
\nonumber \\ 
F_R^{\mu \nu} &\equiv \partial^\mu r^\nu -  \partial^\nu r^\mu -i [r^\mu, r^\nu]\,, 
& r_\mu &\equiv v_\mu + a_\mu\, .
\end{align} 
The $\beta$-functions determining the infinite parts and the scale dependence of the low-energy
constants (LECs) $L_i$ are known~\cite{Gasser:1984gg}.
Today many quantities and processes of high physical interest have been computed up to two loops or $\mathcal{O} (p^6)$ in $\chi$PT in the meson sector; see, e.g., Ref.~\cite{Bijnens:2006zp} for a review.
The main obstacle in reaching a high degree of precision in the theory predictions is the determination of the LECs. 
$\mathcal{L}_{2}$ in Eq.~\eqref{eq:Lp2} contains 2 such constants ($F_0$ and $B_0$),
$\mathcal{L}_{4}$ in Eq.~\eqref{eq:LchiPTp4} already has 10 of these unknowns, the $L_i$. 
At $\mathcal{O} (p^6)$ there are 90 LECs (in the three-flavor theory)~\cite{Bijnens:1999sh,Bijnens:1999hw}. 
Fortunately only a limited number of LECs appears for any given physical process. These LECs are not predicted by the symmetries of the theory alone; they would 
be if QCD was dynamically solvable. Instead they need to be (1) determined from measurements, (2) estimated with models such as resonance exchanges or large $N_c$~\cite{Ecker:1988te}, or (3) computed on the lattice~\cite{Aoki:2019cca}.

The chiral Lagrangians discussed so far, Eqs.~\eqref{eq:Lp2} and \eqref{eq:LchiPTp4}, display 
an accidental symmetry that is not present in QCD in general: in the absence of external
currents (except the quark masses), it is invariant under $\phi \leftrightarrow -\phi$.
This misses interaction terms of odd intrinsic parity such as $K\bar K\to 3\pi$ or, 
when coupled to electroweak currents, $\pi^0\to\gamma\gamma$.  
This is related to the fact that a generating functional built on $\L_2+\L_4$ 
is invariant under chiral transformations, while that of QCD is affected by anomalies.
An effective action to reproduce the chiral anomaly was first given by Wess and 
Zumino~\cite{Wess:1971yu}, and is often given in the form derived by 
Witten~\cite{Witten:1983tw}:
\begin{align}
S[U,l,r]_{\text{WZW}} &= - \frac{iN_c}{240\pi^2} \int_{M^5} \diff^5x \eps^{ijklm} 
\big\langle \Sigma^L_i \Sigma^L_j \Sigma^L_k \Sigma^L_l \Sigma^L_m \big\rangle
- \frac{iN_c}{48\pi^2} \int \diff^4 x \eps^{\mu\nu\alpha\beta}
\Big( W(U,l,r)_{\mu\nu\alpha\beta} - W(1,l,r)_{\mu\nu\alpha\beta} \Big) \,, \notag\\
W(U,l,r)_{\mu\nu\alpha\beta} &= \Big\langle U l_\mu l_\nu l_\alpha U^\dagger r_\beta
+ \frac{1}{4}U l_\mu U^\dagger r_\nu U l_\alpha U^\dagger r_\beta 
+ i U \partial_\mu l_\nu l_\alpha U^\dagger r_\beta
+ i \partial_\mu r_\nu U l_\alpha U^\dagger r_\beta 
- i \Sigma^L_\mu l_\nu U^\dagger r_\alpha U l_\beta 
+ \Sigma^L_\mu U^\dagger \partial_\nu r_\alpha U l_\beta \notag\\
&\quad -\Sigma^L_\mu \Sigma^L_\nu U^\dagger r_\alpha U l_\beta 
+ \Sigma^L_\mu l_\nu \partial_\alpha l_\beta 
+ \Sigma^L_\mu \partial_\nu l_\alpha l_\beta
- i \Sigma^L_\mu l_\nu l_\alpha l_\beta 
+ \frac{1}{2}\Sigma^L_\mu l_\nu \Sigma^L_\alpha l_\beta
-i \Sigma^L_\mu \Sigma^L_\nu \Sigma^L_\alpha l_\beta \Big\rangle - (L \leftrightarrow R) \,,
\label{eq:WZW-Lagr}
\end{align}
where $\Sigma^L_\mu = U^\dagger \partial_\mu U$, $\Sigma^R_\mu = U \partial_\mu U^\dagger$, 
and $(L\leftrightarrow R)$ denotes the combined replacements
$U\leftrightarrow U^\dagger$, $l_\mu \leftrightarrow r_\mu$, and 
$\Sigma^L_\mu\leftrightarrow\Sigma^R_\mu$.
Notably, the first part independent of external currents cannot be written in terms of 
a local Lagrangian in four dimensions, but is formulated on a five-dimensional manifold
$M^5$ whose boundary is four-dimensional Minkowski space.  
A further striking property of the anomaly is that it is free of unknown low-energy constants,
and entirely fixed in terms of known parameters such as $N_c$ and $F_0$. 
The odd-intrinsic-parity meson Lagrangian of $\Order(p^6)$ that is necessary to 
renormalize one-loop calculations based on Eq.~\eqref{eq:WZW-Lagr} has 
also been worked out~\cite{Ebertshauser:2001nj,Bijnens:2001bb}.
Several examples of anomalous $\eta$ and $\eta'$ decays, whose leading-order amplitudes
are determined by the Wess--Zumino--Witten action~\eqref{eq:WZW-Lagr},
are discussed in Sect.~\ref{sec:radiative}.

While $\chi$PT confined to its $\SU(2)$ subsector---relying on the expansion in $m_u$ and $m_d$ only---is truly a high-precision theory, chiral series involving strange quarks are often much more problematic.
This is particularly due to the much larger masses of kaons and $\eta$ compared to the pions.
This is of immediate concern for all decays of the $\eta$ meson:
as the energy increases, strong final-state interactions become more and more important, such as in $\eta \to 3 \pi$.
$\chi$PT alone is not able to describe such interactions with good accuracy and it needs
to be complemented with explicit resonances or dispersion theory to better account for large higher-order corrections. 
In particular, dispersion theory allows one to extrapolate chiral amplitudes to higher energies based on experimental data for phase shifts, etc.  

\subsection{Large-$N_c$ $\chi$PT and $\eta$--$\eta'$ mixing angle}\label{sec:large-Nc}
Large-$N_c$ chiral perturbation theory allows for the explicit inclusion of the $\eta'$ meson in an effective-Lagrangian framework~\cite{Rosenzweig:1979ay,DiVecchia:1980yfw,Nath:1979ik,Witten:1980sp,Kawarabayashi:1980dp}.
When $N_c \to \infty$, the $U(1)_A$ anomaly is absent and the pseudoscalar singlet $\eta_0$ becomes the 
ninth Goldstone boson associated with the spontaneous symmetry breaking of $U(3)_L \times U(3)_R \to U(3)_V$~\cite{Gasser:1984gg}. 
Both chiral and large-$N_c$ corrections are treated perturbatively. The effective Lagrangian is thus organized as
\begin{equation}
\L_{\text{eff}} = \L^{(0)} + \L^{(1)} + \L^{(2)} + \ldots \,,
\end{equation}
where the contributions of order $1$, $\delta$, $\delta^2$, $\ldots$ follow the ordering of the series with the following counting rule:
\begin{equation}
\partial_\mu = \O \left( \sqrt{\delta} \right), \quad m_q=\O \left(\delta\right), \quad 1/N_c = \O \left(\delta\right).
\end{equation}
At leading order (LO), Eq.~\eqref{eq:Lp2} is amended according to 
\begin{equation}
\label{L0}
{\cal L}^{(0)}=
\frac{F^2}{4}\langle\partial_\mu U^\dagger\partial^\mu U +  \chi U^\dagger + \chi^\dagger U \rangle -
\frac{1}{2}M_0^2\eta_0^2 \,,
\end{equation}
where we note the large-$N_c$ scaling $B_0\sim\Order(N^0_c)$, 
$F \sim\Order(\sqrt{N_c})$ is the pion decay constant in the chiral limit,\footnote{Note that at leading order in the combined $\delta$ counting, $F$ coincides with both octet and singlet decay constants.} and
$M_0^2 \sim\Order(N_f/N_c)$ is the $U(1)_A$ anomaly contribution to the $\eta_0$ mass ($N_f=3$ refers to the number of relevant flavors), 
which vanishes in the large-$N_c$ limit~\cite{Witten:1979vv}.\footnote{In Ref.~\cite{Girlanda:2001pc}, alternative scenarios such as $N_{f,c}\to\infty$, $N_f/N_c=\text{fixed}$ have been investigated.  The scaling of $M_0^2$ with $N_f$ is used to argue why the mass of the $\eta'$ is rather large, despite the large-$N_c$ limit suggesting otherwise.}
$U=\exp{(i\sqrt{2}\Phi/F)}$ is now enlarged to take into account the $\eta_0$: 
\begin{equation}
\label{eq:goldstones-prime}
\Phi=\left(
\begin{array}{ccc}
\frac{\pi^0}{\sqrt{2}} + \frac{\eta_8}{\sqrt{6}} + \frac{\eta_0}{\sqrt{3}} & \pi^+ & K^+\\[1ex]
\pi^- & -\frac{\pi^0}{\sqrt{2}} +\frac{\eta_8}{\sqrt{6}} + \frac{\eta_0}{\sqrt{3}}& K^0\\[1ex]
K^- & \bar{K}^0 & -\frac{2\eta_8}{\sqrt{6}} + \frac{\eta_0}{\sqrt{3}}
\end{array}
\right) \,.
\end{equation}
The octet--singlet eigenstates $(\eta_8,\eta_0)$ are related to the physical mass eigenstates $(\eta,\eta^\prime)$ by
\begin{equation}
\label{LOmixing}
\left(
\begin{array}{c}
\eta_8\\
\eta_0
\end{array}
\right)=
\left(
\begin{array}{cc}
\cos\theta_P & \sin\theta_P\\
-\sin\theta_P  & \cos\theta_P
\end{array}
\right)
\left(
\begin{array}{c}
\eta\\
\eta^\prime
\end{array}
\right) \,,
\end{equation}
where $\theta_P$ is the $\eta$--$\eta^\prime$ mixing angle in the octet--singlet basis at this order.
Higher-order corrections to the Lagrangian~\eqref{L0} have been worked out 
systematically~\cite{HerreraSiklody:1996pm,HerreraSiklody:1997kd,Kaiser:2000gs}.

Modern phenomenological analyses of $\eta$--$\eta^\prime$ mixing are largely based on 
large-$N_c$ $\chi$PT~\cite{Leutwyler:1997yr,Feldmann:1998vh,Feldmann:1998sh,Feldmann:1999uf,Escribano:2005qq},
which allows for a common treatment of the $\eta'$ together with the pseudo-Goldstone octet of the lightest
pseudoscalars.  
Since the single-angle mixing scheme in Eq.~\eqref{LOmixing} is only valid at leading order~\cite{Schechter:1992iz,Kiselev:1992ms},
the modern treatment is based on a careful analysis
of the coupling of the flavor-neutral mesons to axial currents.
The meson decay constants are defined in terms of the matrix elements of the 
axial-vector currents $A_\mu^a = \bar q \gamma_\mu \gamma_5 \frac{\lambda^a}{2} q$, using the standard Gell-Mann matrices
$\lambda^a$ and $\lambda^0 = \sqrt{2/3}\unity$ for the singlet current, according to
\beq
\langle 0 | A_\mu^a (0) | P(p) \rangle = i F_P^a p_\mu  \,. \label{eq:def-decay-const}
\eeq
One can therefore define singlet and octet decay constants of $\eta$ and $\eta'$, which can be written in terms
of two basic decay constants $F_8$, $F_0$ and two mixing angle $\theta_8$, $\theta_0$ as
\beq
\left( \begin{array}{cc} F_\eta^8 & F_\eta^0 \\ F_{\eta'}^8 & F_{\eta'}^0 \end{array} \right)
= \left(\begin{array}{cc} F_8\cos\theta_8 & -F_0\sin\theta_0 \\ F_8\sin\theta_8 & F_0\cos\theta_0 \end{array} \right) \,.
\label{eq:mixing-08}
\eeq
Both decay constants and mixing angles can equivalently be translated into a flavor (nonstrange vs.\ strange) basis
instead of the singlet--octet basis~\cite{Feldmann:1998vh,Feldmann:1999uf,Escribano:2005qq}: one defines nonstrange ($F_P^q$) and strange ($F_P^s$) decay constants in analogy to Eq.~\eqref{eq:def-decay-const}, employing
the axial-vector currents $(\bar u \gamma_\mu \gamma_5 u + \bar d \gamma_\mu \gamma_5 d )/\sqrt{2}$
and $\bar s \gamma_\mu \gamma_5 s$, respectively.  These are then written in terms of two decay constants $F_q$
and $F_s$ and two \textit{different} mixing angles $\phi_q$, $\phi_s$ according to
\beq
\left( \begin{array}{cc} F_\eta^q & F_\eta^s \\ F_{\eta'}^q & F_{\eta'}^s \end{array} \right)
= \left(\begin{array}{cc} F_q\cos\phi_q & -F_s\sin\phi_s \\ F_q\sin\phi_q & F_s\cos\phi_s \end{array} \right) \,.
\label{eq:mixing-qs}
\eeq
While the two schemes Eqs.~\eqref{eq:mixing-08} and \eqref{eq:mixing-qs} are in principle equivalent, it turns out 
phenomenologically that they behave somewhat differently: while the difference of the octet--singlet mixing angles
is an $\SU(3)$-breaking effect, $F_8 F_0 \sin(\theta_8-\theta_0) = -2\sqrt{2}/3(F_K^2-F_\pi^2)$, 
the difference $\phi_q-\phi_s$ is suppressed in the large-$N_c$ counting and turns out to be compatible with zero
in most extraction of $\eta$--$\eta'$ mixing from experimental data~\cite{Escribano:2005qq,Escribano:2015yup}.  
Neglecting these $1/N_c$ corrections and using one single flavor-mixing angle $\phi \approx \phi_q \approx \phi_s$, 
the relation between the two schemes simplifies to
\beq
F_8^2 = \frac{1}{3}F_q^2 + \frac{2}{3}F_s^2 \,, \quad 
F_0^2 = \frac{2}{3}F_q^2 + \frac{1}{3}F_s^2 \,, \quad
\theta_8 = \phi - \arctan \Bigg(\frac{\sqrt{2}F_s}{F_q}\Bigg) \,, \quad
\theta_0 = \phi - \arctan \Bigg(\frac{\sqrt{2}F_q}{F_s}\Bigg) \,. 
\eeq
Next-to-next-to-leading-order corrections, which include meson-loop effects, 
have been considered in Refs.~\cite{Guo:2015xva,Bickert:2016fgy} (cf.\ also already a much earlier analysis in the context of mixing corrections to $\pi^0\to\gamma\gamma$~\cite{Goity:2002nn}); loop corrections in a scheme that does not rely
on the large-$N_c$ expansion were considered earlier in Ref.~\cite{Borasoy:2003yb}.  As a conclusion from these studies, we caution that chiral higher-order corrections may affect the extracted mixing angles significantly~\cite{Goity:2002nn} or at least increase the associated uncertainties by a lot~\cite{Bickert:2016fgy}.
Below, we only discuss the scheme advocated, e.g., in Refs.~\cite{Feldmann:1998vh,Feldmann:1999uf,Escribano:2005qq},
which neglects quark-mass effects, but seems to be phenomenologically very successful.

In Refs.~\cite{Feldmann:1998vh,Feldmann:1998sh,Feldmann:1999uf}, a large variety of scattering and decay processes
involving $\eta$ and $\eta'$ mesons have been analyzed simultaneously to extract the two decay constants and
mixing angles, with the result
\begin{align}
F_8 &= 1.26(4)F_\pi \,, & F_0 &= 1.17(3)F_\pi \,, &
\theta_8 &= -21.2(1.6)^\circ \,, & \theta_0 &= -9.2(1.7)^\circ \,; 
\notag\\
F_q &= 1.07(2)F_\pi \,, & F_s &= 1.34(6)F_\pi \,, &
\phi &= 39.3(1.0)^\circ \,.
\label{eq:mixing-Feldmann}
\end{align}
The two-photon decays play a central role in this analysis.
Ref.~\cite{Escribano:2005qq} partially updates the experimental data base of their mixing analysis,
but is not quite as comprehensive in the list of reactions analyzed; the authors obtain
\begin{align}
F_8 &= 1.51(5)F_\pi \,, & F_0 &= 1.29(4)F_\pi \,, &
\theta_8 &= -23.8(1.4)^\circ \,, & \theta_0 &= -2.4(1.9)^\circ \,; 
\notag\\
F_q &= 1.09(3)F_\pi \,, & F_s &= 1.66(6)F_\pi \,, &
\phi_q &= 39.9(1.3)^\circ \,, & \phi_s &= 41.4(1.4)^\circ .
\end{align}
Of particular interest in the context of this review is the extraction of Ref.~\cite{Escribano:2015yup} (compare also the extensive discussion in Ref.~\cite{Sanchez-Puertas:2017sih}) based exclusively 
on the transition form factors of $\eta$ and $\eta'$, see Sects.~\ref{sec:pi0-eta-etapTFF} and \ref{sec:DRetaTFF}, which combines information on the 
two-photon decay widths with high-energy space-like data on $\gamma\gamma^*\to\ep$, but avoids potential theoretical bias when 
combining these with other, completely unrelated experimental input.  The authors arrive at
\begin{align}
F_8 &= 1.27(2)F_\pi \,, & F_0 &= 1.14(5)F_\pi \,, &
\theta_8 &= -21.2(1.9)^\circ \,, & \theta_0 &= -6.9(2.4)^\circ \,; 
\notag\\
F_q &= 1.03(4)F_\pi \,, & F_s &= 1.36(4)F_\pi \,, &
\phi_q &= 39.6(2.3)^\circ \,, & \phi_s &= 40.8(1.8)^\circ ,
\end{align}
which is much closer to Eq.~\eqref{eq:mixing-Feldmann}.  
A combined study of diphoton decays of the lightest pseudoscalars as well as vector-meson conversion
decays in resonance chiral theory~\cite{Chen:2012vw} finds results similarly compatible with Eq.~\eqref{eq:mixing-Feldmann}.

Finally, simulations of the $\eta$--$\eta'$ system in lattice QCD~\cite{Ottnad:2012fv,Michael:2013gka,Ottnad:2017bjt} are becoming 
extremely competitive in accuracy with phenomenological extractions, with the latest results on the flavor-mixing scheme 
translating into~\cite{Ottnad:2017bjt}
\beq
F_q = 0.960(37)(46)F_\pi \,,  \qquad  F_s = 1.143(23)(05)F_K = 1.363(27)(06)F_\pi  \,,  \qquad
\phi = 38.8(2.2)(2.4)^\circ \,, 
\eeq
where the second errors refer to uncertainties induced by chiral extrapolations to the physical point.

\section{Dispersion theory}\label{sec:DispTheory}
Dispersive techniques are powerful, model-independent methods based on the fundamental principles of analyticity (the mathematical manifestation of causality) and unitarity (a consequence of probability conservation). 
By exploiting nonperturbative relations between amplitudes, they allow for a resummation of rescattering effects between final-state particles, in contrast to a strictly perturbative $\chi$PT expansion in which such effects would be treated order-by-order only. 
Dispersion theory, coupled with $\chi$PT, therefore allows one to extend the $\chi$PT effective description of strong dynamics from low energy to an intermediate-energy range where resonances start to appear.

\subsection{Analyticity}

\begin{figure}
  \begin{center}
    \includegraphics[width=0.3\linewidth]{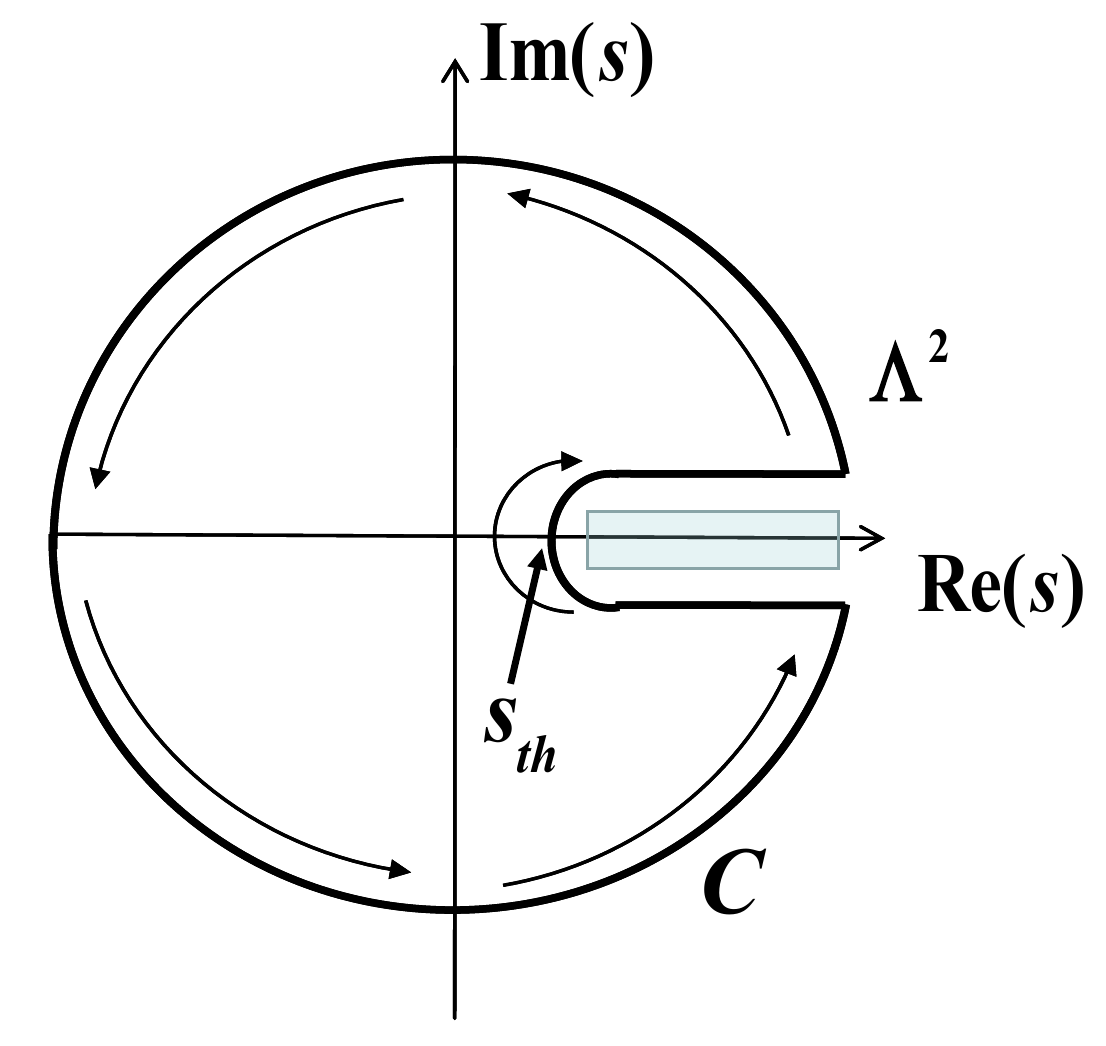} 
    \caption{Graphical representation of the Cauchy contour in the complex $s$ plane.}
    \label{fig:CauchyContour}
  \end{center}
\end{figure}

As a practical example, we consider a form factor $F(s)$, a function of a single Mandelstam variable $s$. 
(A similar discussion applies to scattering amplitudes, as we describe below.) 
In many cases, these form factors are real below some threshold, $s < \sth$, while above threshold, $s > \sth$, they have both real and imaginary parts, the latter due to the propagation of on-shell intermediate states.
Analyticity allows us to relate the real part of the form factor to its discontinuity or imaginary part. 
To fully exploit these properties one needs to analytically continue $s$ into the complex plane where the discontinuity is represented as a branch cut along the positive real axis, for $s > \sth$, as shown in Fig.~\ref{fig:CauchyContour}. 
The form factor is then a complex-valued function $F(s)$ of complex argument $s$, which has the following properties:
(1)~$F(s)$ is real along the real axis for  
$s < \sth$ (below threshold)
and (2)~$F(s)$ is analytic in the entire complex plane 
except along the branch cut.
The sign of the imaginary part of $F$ along the cut is fixed by the convention 
$
F(s+ i \epsilon) = \Re F(s) + i\,\Im F(s) 
$, 
with $\epsilon$ a positive infinitesimal quantity. 

To obtain a dispersion relation, we start from Cauchy's integral formula
\beq
F(s) = \frac{1}{2 \pi i} \oint_\mathcal{C} \diff s' \frac{F(s')}{s'-s} \,,
\eeq
and perform the integral on the contour shown in Fig.~\ref{fig:CauchyContour} to obtain
\begin{align}
F(s) 
&=  \frac{1}{2 \pi i} \left(\int_{\sth}^{\Lambda^2} \diff s' \frac{F(s'+i\epsilon)-F(s'-i\epsilon)}{s'-s} +  \int_{|s'|=\Lambda^2} \diff s' \frac{F(s')}{s'-s} 
\right) \nonumber \\
&=  \frac{1}{2 \pi i} \left(\int_{\sth}^{\Lambda^2} \diff s' \frac{\disc F(s')}{s'-s} +  \int_{|s'|=\Lambda^2} \diff s' \frac{F(s')}{s'-s} 
\right) \,.
\label{eq:Cauchy}
\end{align}
If Schwartz's reflection principle applies, i.e., $F(z^*)=F^*(z)$, we obtain
\begin{equation}
\disc F(s) =  F(s+i\epsilon)-F(s-i\epsilon) = F(s+i\epsilon)-F^*(s+i\epsilon) = 2 i \,\Im F(s+i\epsilon)\,,  
\end{equation}
and Eq.~\eqref{eq:Cauchy} becomes 
\begin{equation}
F(s) =  \frac{1}{\pi} \int_{\sth}^{\Lambda^2} \diff s' \frac{\Im F(s')}{s'-s-i \epsilon} + 
 \frac{1}{2 \pi i}\int_{|s'|=\Lambda^2} \diff s' \frac{F(s')}{s'-s}\,.
 \label{eq:CauchyIm}
\end{equation}
If the second integral vanishes in the limit $\Lambda\to\infty$,
we obtain an unsubtracted dispersion relation:
\begin{equation}
F(s) =  \frac{1}{\pi} \int_{\sth}^{\infty} \diff s' \frac{\Im F(s')}{s'-s-i \epsilon}\,.
\end{equation}
This relation given by analyticity is very powerful: it implies that the form factor $F(s)$ can be reconstructed anywhere in the complex plane 
provided we know its absorptive part along the branch cut, which is in turn given by unitarity. 

On the other hand, if $F(s)$ does not approach zero fast enough at infinity, 
we can perform subtractions at $s=s_0 < \sth$. If we perform one such subtraction, Eq.~\eqref{eq:CauchyIm} becomes
\begin{align}
F(s) -F(s_0) &=  \frac{1}{\pi} \int_{\sth}^{\Lambda^2} \diff s' \frac{\Im F(s')}{s'-s-i \epsilon} + 
 \frac{1}{2 \pi i}\int_{|s'|=\Lambda^2} \diff s' \frac{F(s')}{s'-s} - \left( \frac{1}{\pi} \int_{\sth}^{\Lambda^2} \diff s' \frac{\Im F(s')}{s'-s_0} + 
 \frac{1}{2 \pi i}\int_{|s'|=\Lambda^2} \diff s' \frac{F(s')}{s'-s_0} \right) \nonumber\\
\Rightarrow \quad F(s) &= F(s_0) + \frac{s-s_0}{\pi} \int_{\sth}^{\Lambda^2} \frac{\diff s'}{s'-s_0} \frac{\Im F(s')}{s'-s-i \epsilon} + \frac{s-s_0}{2 \pi i}\int_{|s'|=\Lambda^2} \diff s' \frac{F(s')}{(s'-s_0)(s'-s)}  \,.
\end{align}
The last term now contains one more power of $s'$ in the denominator, ensuring better convergence when $\Lambda \to\infty$. 
If this term vanishes, we obtain a once-subtracted dispersion relation:
\begin{equation}
F(s) = F(s_0) + \frac{s-s_0}{\pi} \int_{\sth}^{\infty} \frac{\diff s'}{s'-s_0} \frac{\Im F(s')}{s'-s-i \epsilon} \,.
\end{equation} 
An $n$-times-subtracted dispersion relation at $s=s_0$ will take the form
\begin{equation}
F(s)= P_{n}(s-s_0) + \frac{(s-s_0)^n}{\pi} \int_{\sth}^{\infty} \frac{\diff s'}{(s'-s_0)^n} \frac{\Im F(s')}{s'-s-i \epsilon}\,,
\end{equation}
with $P_{n}(s-s_0)$ a polynomial of power $n-1$ in ($s-s_0$). Note that we can perform subtractions in different points provided they are on the real axis 
to the left of the branch cut. 

\subsection{Unitarity}
One fundamental property of the $S$-matrix is its unitarity:
\begin{equation}
S^\dagger S = 1\,.
\end{equation}
The decomposition of the $S$-matrix into the identity and the nontrivial scattering matrix $T$, 
$S = 1 + i\,T$,
then immediately implies
\begin{equation}
-i (T-T^\dagger) = T^\dagger T\,. 
\end{equation}
Evaluating this relation between initial and final states, inserting a complete set of intermediate states
on the right-hand side, and assuming time reversal invariance, we arrive at the well-known optical theorem
\beq
 \Im T_{fi} = \frac{1}{2}\sum_n (2 \pi)^4 \delta^{(4)} \left(P_n - P_i \right) T^*_{nf} T_{ni}\,,
 \label{eq:OptTheo}
\eeq
where $\langle f | T | i \rangle \equiv (2 \pi)^4 \delta^{(4)}(P_f - P_i ) T_{fi}$.

Let us consider the scattering of two incoming and two outgoing particles and assume that we are in an energy region where only elastic final-state rescattering is allowed. 
This means that in Eq.~\eqref{eq:OptTheo}, the only intermediate state is $|n \rangle = |f \rangle$ and the completeness sum reduces to an integral over the intermediate momenta:
\begin{equation}
\Im T_{fi} = \frac{(2\pi)^4}{2S} \int \frac{d^3q_1 d^3q_2} {2 E_1 (2 \pi)^3 2 E_2  (2 \pi)^3 } \delta^{(4)} \left( p_i -q_1-q_2 \right) 
T^*_{ff} T_{fi}\,.
\end{equation}
The symmetry factor $S$ is equal to $2$ for indistinguishable particles and $1$ otherwise. $q_1 = (E_1,\vec{q}_1)$ and 
$q_2 = (E_2, \vec{q}_2)$ denote the on-shell four-momenta of the two intermediate particles, and 
the total initial and final four-momenta are $p_i = k + k' = p_f = p + p' = (\sqrt{s},\vec{0})$ in the center-of-mass frame. 
Defining $\theta = \measuredangle({\vec{k},\vec{p}})$, $\theta' = \measuredangle({\vec{k},\vec{q}_1})$, and $\theta'' = \measuredangle({\vec{q}_1,\vec{p}})$
and integrating over the delta function we find
\begin{equation}
\Im T_{fi} (s, \theta) = \frac{1}{8 (2\pi)^2 S} \frac{|\vec{q}_1|}{\sqrt{s}}
  \int T^*_{ff}(s, \theta') T_{fi}(s, \theta'') \diff\Omega\,,
  \label{eq:unitarity}
\end{equation}
with $\diff\Omega \equiv \sin \theta' \diff\theta' \diff\phi $ and 
$
|\vec{q}_1| = \sqrt{{s}/{4}- M_\pi^2}
$
for the example of an intermediate state of two pions. A unitarity relation very similar to Eq.~\eqref{eq:unitarity}
results if, instead of a $T$-matrix element $T_{fi}$, we consider a production amplitude or a form factor that produces
the final state $f$ in the elastic regime.

Let us now consider the pion vector form factor defined as 
\begin{equation}
\langle \pi^+ (p') \pi^-(p) | j_\mu(0) |0 \rangle = (p'-p)_\mu \, F_\pi^V (s)\,,
\end{equation}
where
\begin{equation}
j_\mu = \frac{2}{3}\bar u \gamma_\mu u-\frac{1}{3}\bar d\gamma_\mu d-\frac{1}{3}\bar s\gamma_\mu s \label{eq:EMquarkcurrent}
\end{equation}
denotes the electromagnetic vector current for the light quarks.
The unitarity relation~\eqref{eq:unitarity} turns into
\begin{equation}
 \Im F_\pi^V(s) = \sigma(s) \, \left(t^{I = 1}_{J=1} (s)\right)^* F_\pi^V(s) \times \theta \big(s-4 M_\pi^2\big)\,,
 \label{eq:UnitarityFF}
\end{equation}
where 
$
\sigma(s) = \sqrt{1-{4 M_\pi^2}/{s}}
$,
and $t^{I = 1}_{J=1}(s)$ is the $\pi \pi$ $P$-wave isospin $I=1$ scattering amplitude.
\begin{figure}
  \begin{center}
    \includegraphics[width=0.55\linewidth]{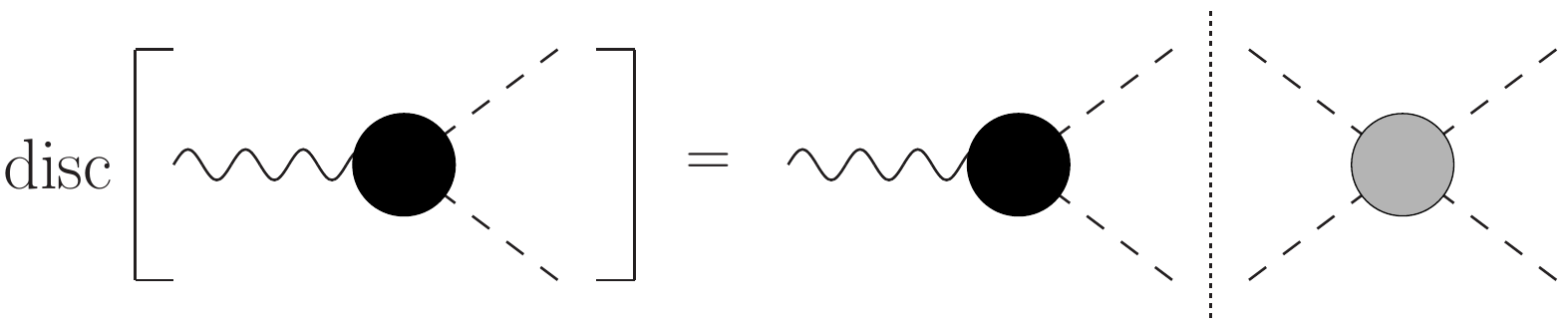} 
    \caption{Graphical representation of the discontinuity relation for pion form factors.
    The black disc denotes the form factor, while the gray disc denotes the pion--pion scattering $T$-matrix,
    projected onto the appropriate partial wave.}
    \label{fig:FFunit}
  \end{center}
\end{figure}
Equation~\eqref{eq:UnitarityFF} is displayed graphically in Fig.~\ref{fig:FFunit}.
Applying now the unitarity relation to $t^{1}_{1}(s)$, we find
\beq
 \Im t^{1}_{1} (s) 
  = \sigma(s) \, \left|t^{1}_{1}(s)\right|^2 \times \theta \big(s-4 M_\pi^2\big) \,.
\eeq
Writing $t^{1}_{1} (s) = \left| t^{1}_{1}(s) \right| e^{i \delta^{1}_{1}}(s)$ leads to
\beq
\left| t^{1}_{1}(s) \right| = \frac{\sin\delta^{1}_{1} (s)}{\sigma(s)}\,;
\eeq
inserting this into the form factor unitarity relation~\eqref{eq:UnitarityFF}, we arrive at
\beq
\Im F_\pi^V(s) 
= \sin\delta_1^1(s) e^{-i\delta_1^1(s)} F_\pi^V(s) \times \theta \big(s-4 M_\pi^2\big)\,. 
\label{eq:UnitarityFF2}
\eeq
If $\phi_V$ is the phase of the vector form factor, $F_\pi^V(s) = |F_\pi^V(s) | e^{i \phi_V(s)}$,
the unitarity equation implies $\phi_V(s) = \delta^{1}_{1}(s)$
in order for the imaginary part of $F_\pi^V(s)$ to be real:
unitarity forces the phase of the form factor to be equal to the $\pi \pi$ (elastic) scattering phase. 
This is the so-called Watson theorem~\cite{Watson:1954uc} that holds only in the elastic region. 

\subsection{Omn\`es formalism}\label{sec:Omnes}
The solution to the form factor unitarity relation Eq.~\eqref{eq:UnitarityFF} is readily obtained in terms of the so-called Omn\`es function $\Omega_1^1(s)$~\cite{Omnes:1958hv},
\begin{equation}
F_\pi^V(s) = R(s) \Omega_1^1(s) \,, \quad
\Omega_1^1(s)=\exp\Bigg\{\frac{s}{\pi}\int_{4M_\pi^2}^\infty \diff s'\frac{\delta_1^1(s')}{s'(s'-s-i\eps)} \Bigg\} \,,
\label{eq:Omnes}
\end{equation}
where $R(s)$ is a function free of (right-hand) cuts up to the first inelastic threshold. 
At low energies, $R(s)$ can be approximated by a 
polynomial whose coefficients need to be determined by other methods, e.g., by matching
to chiral perturbation theory near $s=0$.  The Omn\`es function is entirely given in terms of the appropriate pion--pion
phase shift, which is particularly useful as we today have excellent information on pion--pion scattering
at our disposal~\cite{Ananthanarayan:2000ht,Colangelo:2001df,GarciaMartin:2011cn,Caprini:2011ky}.
The Omn\`es function then represents the exact resummation of elastic two-body rescattering to all orders.
The pion vector form factor $F_\pi^V(s)$ as extracted from $\tau^-\to\pi^-\pi^0\nu_\tau$ decays~\cite{Fujikawa:2008ma}, for example, can be described 
very accurately up to $\sqrt{s}=1\GeV$ by a representation~\eqref{eq:Omnes} with a linear polynomial $R(s) = 1+\alpha_V s$. 
(At higher energies, the nonlinear
effects of higher, inelastic ($\rho'$, $\rho''$) resonances become important~\cite{Hanhart:2013vba}.)
For the pion vector form factor as measured in 
$e^+e^-\to\pi^+\pi^-$~\cite{Achasov:2006vp,Akhmetshin:2006bx,Aubert:2009ad,Ambrosino:2010bv,Babusci:2012rp,Ablikim:2015orh}, 
the isospin-violating mixing effect with the 
$\omega$-meson needs to be taken into account; see, e.g., the extensive discussion in Ref.~\cite{Hanhart:2016pcd} and references therein. 
More refined representations parameterizing inelastic effects beyond roughly $1\GeV$
have employed conformal polynomials instead, which also allows for better high-energy asymptotic behavior
of the form factor representation~\cite{Leutwyler:2002hm,Colangelo:2003yw,Colangelo:2018mtw}.
Equations~\eqref{eq:UnitarityFF}, \eqref{eq:Omnes} have also been generalized and employed frequently to describe coupled channels---e.g., $\pi\pi \leftrightarrow \bar{K}K$ scalar form 
factors~\cite{Donoghue:1990xh,Moussallam:1999aq,DescotesGenon:2000ct,Daub:2012mu,Celis:2013xja,Daub:2015xja,Winkler:2018qyg,Ropertz:2018stk}---by promoting the Omn\`es function to a matrix with a coupled-channel $T$-matrix as input. 
However, the coupled-channels description does not allow for a similarly compact closed form as in Eq.~\eqref{eq:Omnes}.

When describing more complicated amplitudes such as four-point functions (either in $2\to2$ scattering processes or in $1\to3$ decays),
a more complex unitarity relation is to be considered because of the presence of left-hand cuts.  These are a consequence of crossing
symmetry and unitarity in the crossed channel: e.g., the pion--pion scattering amplitude possesses not only a cut in the 
$s$-channel for $s\geq 4M_\pi^2$, but also for $t,u\geq 4M_\pi^2$.
As a consequence, after projection onto $s$-channel partial waves,
the crossed-channel unitarity cuts translate into another discontinuity along the negative real axis for $s\leq 0$.
If we separate right- and left-hand cuts into individual functions $f_J^I(s)$ and $\hat{f}_J^I(s)$ of ($s$-channel) isospin $I$ and 
angular momentum $J$, the unitarity condition Eq.~\eqref{eq:UnitarityFF} for $f_J^I(s)$ becomes
\begin{equation}
\Im f_J^I(s) = \sin \delta_J^I(s) e^{-i \delta_J^I(s)}  \left( f_J^I(s) + \hat{f}_J^I(s) \right ) \theta \big(s-4 M_\pi^2\big) \,,
\label{eq:InOm}
\end{equation}
where the \textit{inhomogeneity}
\begin{equation}
 \hat f_J^I(s) =\sum_{n,I',J'} \int_{-1}^1 \diff \cos\theta~\cos^{n} \theta \, c_{n}^{II'JJ'} f_{J'}^{I'}\big(t(s,\cos\theta)\big) 
\end{equation}
is a consequence of the singularities in the $t$- and $u$-channels, and 
arises from their projection onto the $s$-channel partial wave.
The two terms on the right-hand side of Eq.~\eqref{eq:InOm} 
are depicted in Fig.~\ref{fig:InOm}:
\begin{figure}
\centering
\includegraphics[width=0.6\linewidth]{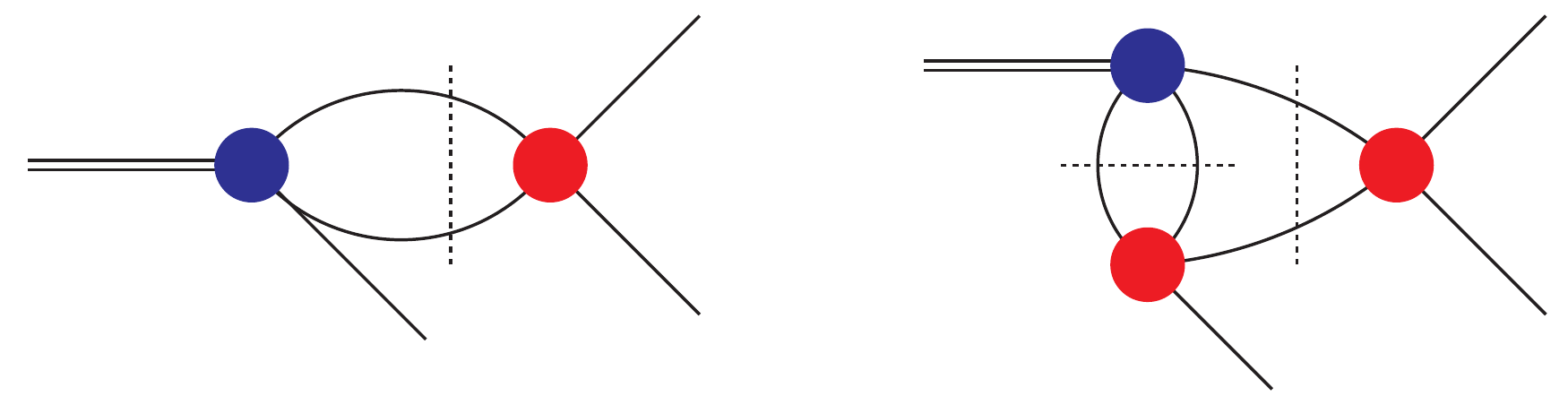}
\caption{Pictorial representation of the inhomogeneous unitarity relation of Eq.~\eqref{eq:InOm}:
the homogeneous term analogous to a form factor unitarity relation (left), plus the projection of a typical diagram
representing crossed-channel dynamics (right), resulting in the inhomogeneous Omn\`es problem.
The double line represents a heavy particle with its three-body decay partial wave denoted by the blue dot, 
the single lines depict the three outgoing decay products that rescatter elastically (red dots).}\label{fig:InOm}
\end{figure}
$t(s, \cos \theta)$ is the appropriate crossed-channel Mandelstam variable written as a function of center-of-mass energy squared $s$ and scattering angle $\theta$, and $c_{n}^{I I'JJ'}$ are process-dependent coefficients. 
Equation~\eqref{eq:InOm} represents the inhomogeneous Omn\`es problem.

Since both $f_J^I(s)$ and the Omn\`es function $\Omega_J^I(s)$ are analytic on the entire complex plane except on the real axis when $s> \sth$, $g_J^I(s)=f_J^I(s)/\Omega_J^I(s)$ has the same analytic properties. We can compute the imaginary part of $g_J^I(s)$ on the upper rim of the cut: 
\begin{align}
 \Im g_J^I(s) &= \frac{\Im f_J^I(s)~\Re \Omega_J^I(s)-\Re f_J^I(s)~\Im \Omega_J^I(s) }{\left|\Omega_J^I(s)\right|^2}= \frac{\Im f_J^I(s)~\Omega_J^I(s)-f_J^I(s)~\Im \Omega_J^I(s) }{\left|\Omega_J^I(s)\right|^2} \nonumber\\
 &=\frac{\left( f_J^I(s) + \hat{f}_J^I(s) \right ) e^{-i \delta_J^I(s)} \sin \delta_J^I(s)~\Omega_J^I(s)-f_J^I(s)~\Omega_J^I(s)~e^{-i \delta_J^I(s)}~\sin \delta_J^I(s)}{\left|\Omega_J^I(s)\right|^2}
= \frac{\hat{f}_J^I(s)~\sin \delta_J^I(s)}{\left|\Omega_J^I(s)\right|}\,.
\end{align}
This allows us to write an $n$-times subtracted dispersion relation for 
$g_J^I(s)$, which, solved for $f_J^I(s)$,
results in the solution of the inhomogeneous Omn\`es problem 
\begin{equation}
  f_J^I(s)= \Omega_J^I(s) \left(P_{n}(s-s_0) + \frac{(s-s_0)^n}{\pi} \int_{\sth}^{\infty}  \frac{\hat{f}_J^I(s)~\sin \delta_J^I(s)}{\left|\Omega_J^I(s)\right|(s'-s_0)^n (s'-s-i \epsilon)}\diff s'\right)\,.  
\end{equation}

As Fig.~\ref{fig:InOm} suggests, the inhomogeneous Omn\`es problem 
provides a possible dispersion-theoretical description of 
three-body decays in the form of Khuri--Treiman equations~\cite{Khuri:1960zz,Aitchison:1966lpz}, which we will describe in more
detail in Sect.~\ref{sec:eta3pi_theo} for the decay $\eta\to3\pi$.  
Alternatively, unitarity relations of the form Eq.~\eqref{eq:InOm} 
have frequently been used to model left-hand cuts in partial waves of four-point amplitudes
through the exchange of narrow resonances in the crossed channel.  We will see this at work 
in the description of $\ep\to\pi^+\pi^-\gamma$, see Sect.~\ref{sec:eta-pipigamma}, 
but other applications of this technique 
include $\gamma\gamma\to\pi\pi$~\cite{GarciaMartin:2010cw} or decays of heavy mesons~\cite{Kang:2013jaa}
and heavy quarkonia~\cite{Chen:2015jgl,Chen:2016mjn,Chen:2019mgp} involving pion pairs.

In principle dispersion relations provide us with a rigorous, model-independent method to describe intermediate-energy regions: beyond the chiral regime, yet mostly limited to the range where elastic unitarity applies. 
In practice the use of these techniques is limited by our knowledge of the experimental input. 
However, as we will see in the next sections dispersive techniques used in combination with $\chi$PT have been shown to be successful to describe various $\eta$ and $\eta'$ decays very accurately.

\subsection{Roy equations}
A central piece of input to many of the dispersive analyses discussed throughout this review are pion--pion phase shifts.  Pion--pion scattering constitutes the most important final-state interaction in many $\eta$ and $\eta'$ decays; see, e.g., the Omn\`es function in Eq.~\eqref{eq:Omnes}.  As a lot of progress in the description of light-meson decays are based on the precision with which we nowadays know the $\pi\pi$ phases, we here wish to indicate very briefly how these have been constrained so rigorously, by the means of so-called Roy equations~\cite{Roy:1971tc}.

Roy equations correspond to a coupled system of partial-wave dispersion relations, as such based on analyticity of the scattering amplitude, which make maximal use of crossing symmetry, unitarity, and isospin.  To construct this set of equations, one considers twice-subtracted dispersion relations at fixed Mandelstam $t$:
\beq
T(s,t) = c(t) + \frac{1}{\pi} \int_{4M_\pi^2}^\infty \diff s' \bigg\{
\frac{s^2}{s'^2(s'-s)} + \frac{u^2}{s'^2(s'-u)} \bigg\} \Im T(s',t) \,,
\label{eq:Roy:fixed-t}
\eeq
where the two terms inside the integral correspond to the right- and the left-hand cut, respectively,
and we have omitted isospin indices for simplicity.
The subtraction function $c(t)$ can be determined from crossing symmetry.  
Expanding the imaginary parts into and projecting the resulting amplitude onto partial waves
leads to~\cite{Roy:1971tc}
\beq
t_J^I(s) = k_J^I(s) + \sum_{I'=0}^2 \sum_{J'=0}^\infty  \int_{4M_\pi^2}^\infty \diff s'
K_{JJ'}^{II'}(s,s') \Im t_{J'}^{I'}(s') \,, \label{eq:Roy:pwsum}
\eeq
where isospin dependence is recovered in the notation.  The kernels $K_{JJ'}^{II'}(s,s')$
inside the dispersion integrals are known analytically; they contain in particular a singular 
Cauchy kernel (diagonal in both isospin and angular momentum).  The subtraction polynomial
$k_J^I(s)$ depends on the two $\pi\pi$ $S$-wave scattering lengths $a_0^{I=0,2}$ only.
As long as \textit{elastic} unitarity is sufficient, we can write the partial waves $t_J^I(s)$
in terms of phase shifts $\delta_J^I(s)$, 
\beq
t_J^I(s) = \frac{e^{i\delta_J^I(s)}\sin\delta_J^I(s)}{\sigma(s)} \,,
\eeq
and turn Eq.~\eqref{eq:Roy:pwsum} into a set of coupled integral equations for the phase shifts. 

The Roy equations thus derived are not applicable at arbitrarily high energies: the convergence
of the partial-wave expansion of the imaginary parts in Eq.~\eqref{eq:Roy:pwsum} limits their
validity to $s \leq s_{\text{max}} = (1.15 \GeV)^2$ if Mandelstam analyticity of the scattering 
amplitude is assumed (see, e.g., Ref.~\cite{Roy:1990hw}).  In practice, the phase shifts are therefore
only solved for at low energies below a certain matching point $s_{\text{m}}$, above which experimental
input is required.  Similarly, the solution is typically limited to the lowest partial waves
$J \leq J_{\text{max}}$, higher ones similarly serve as input.  Furthermore, the existence and uniqueness
of the solution depends on the values of the phase shifts at the matching point, summarized in 
a characteristic multiplicity index~\cite{Gasser:1999hz}.  However, the low-energy
$\pi\pi$ scattering phase shifts are hardly sensitive to the high-energy and high-angular-momentum input, 
and can be determined with remarkable accuracy~\cite{Ananthanarayan:2000ht}, and the matching of the 
subtraction constants, i.e., the $S$-wave scattering lengths, to $\chi$PT further strengthens the 
constraints~\cite{Colangelo:2001df}, leading to the celebrated predictions of the $\pi\pi$ scattering 
lengths~\cite{Colangelo:2000jc} 
\beq
a_0^0 = 0.220(5) \,, \qquad a_0^2 = - 0.0444(10) \,.
\eeq
Variants of the Roy equations with a different subtraction scheme have also been derived and used
as constraints in a fit to $\pi\pi$ scattering data~\cite{Kaminski:2006qe,GarciaMartin:2011cn}, 
resulting in similarly accurate determinations of phase shifts that have widely been used 
in dispersive analyses reviewed here.

Roy equations have been generalized to processes that are less symmetric under crossing than 
pion--pion scattering.  In such cases, the use of Roy--Steiner equations based on hyperbolic
(instead of fixed-$t$) dispersion relations has proven advantageous~\cite{Hite:1973pm} to make maximal
use of the interconnection between crossed reactions.  Such Roy--Steiner equations have been constructed
and solved in detail in particular for pion--kaon~\cite{Buettiker:2003pp,Pelaez:2018qny}, 
photon--pion~\cite{Hoferichter:2011wk}, and
pion--nucleon scattering~\cite{Ditsche:2012fv,Hoferichter:2015hva}.

\section{\boldmath Hadronic $\eta$ and $\eta'$ decays}
\label{sec:hadronic}

\subsection{$\eta \to 3 \pi$ and extraction of the light quark mass ratio}
\label{sec:eta-3pi}
\subsubsection{Motivation: light quark masses}
\label{sec:eta3pi-lightquarkmasses}

The masses of the light quarks, i.e., the up, down, and strange quarks, are 
not directly accessible
to experimental determination due to confinement, which prevents quarks from appearing
as free particles. 
Since these masses are much smaller than the typical hadronic mass scale $\sim 1\GeV$, 
their contribution to hadron masses is typically small. There is,
however, a prominent exception to this rule, formulated in the famous Gell-Mann--Oakes--Renner
relation~\cite{GellMann:1968rz}. It states that the masses of the lightest mesons are determined by the combined effects
of spontaneous and explicit chiral symmetry breaking, that is, by the chiral quark condensate and
the light quark masses. 
At leading order in the quark mass expansion and including $\pi^0$--$\eta$ mixing
and first-order electromagnetic corrections, one finds:
\begin{align}\begin{aligned}
	\mpiz^2 &= B_0 (m_u+m_d) + \frac{2 \epsilon}{\sqrt{3}} B_0 (m_u - m_d) + \ldots \eolc
	&\mpip^2 &= B_0 (m_u+m_d) + \Delta_\textrm{em}^\pi + \ldots \eolc \\
	\mKz^2 &= B_0 (m_d+m_s) + \ldots \eolc
	&\mKp^2 &= B_0 (m_u+m_s) + \Delta_\textrm{em}^K + \ldots \eolc \\
	\meta^2 &= \frac{B_0}{3} (m_u+m_d + 4 m_s) - \frac{2 \epsilon}{\sqrt{3}} B_0 (m_u - m_d) + \ldots \eolp
	\label{eq:GMOR}
\end{aligned}\end{align}
The parameter 
\beq \label{pi0-eta-mixing-eps}
\epsilon = \frac{\sqrt{3} (m_d-m_u)}{4(m_s-\hat m)} \approx 0.012 \, , 
\eeq
where $\hat m = (m_u+m_d)/2$, is the (leading-order) $\pi^0$--$\eta$ mixing angle.
Provided the electromagnetic corrections are known, one can determine the light quark mass ratios from the meson masses. 
This is the so-called Weinberg formula~\cite{Weinberg:1977hb}: 
\beq
	\frac{m_d}{m_u} \approx \frac{\mKz^2 - \mKp^2 + \mpip^2}{\mKp^2 - \mKz^2 - \mpip^2 + 2 \mpiz^2} \approx 1.79 \eolc \qquad
	\frac{m_s}{m_d} \approx \frac{\mKp^2 + \mKz^2 - \mpip^2}{\mKz^2 - \mKp^2 + \mpip^2} \approx 20.2 \eolc
	\label{eq:WeinbergRatios}
\eeq
derived using current algebra and Dashen's 
theorem~\cite{Dashen:1969eg} that relates the electromagnetic corrections 
to the pion and the kaon mass differences,
$\Delta_{\textrm{em}}^{\pi} = \Delta_{\textrm{em}}^{K} + \Order(e^2 m_q)$. 
However, the $\Order(e^2 m_q)$ terms in the electromagnetic mass contributions are known to be sizeable, 
and it has been shown that these quark mass ratios are subject to important higher-order 
corrections~\cite{Gasser:1984pr,Kaplan:1986ru,Leutwyler:1996qg}.

\begin{figure}
\centering
\includegraphics*[width=0.54\textwidth]{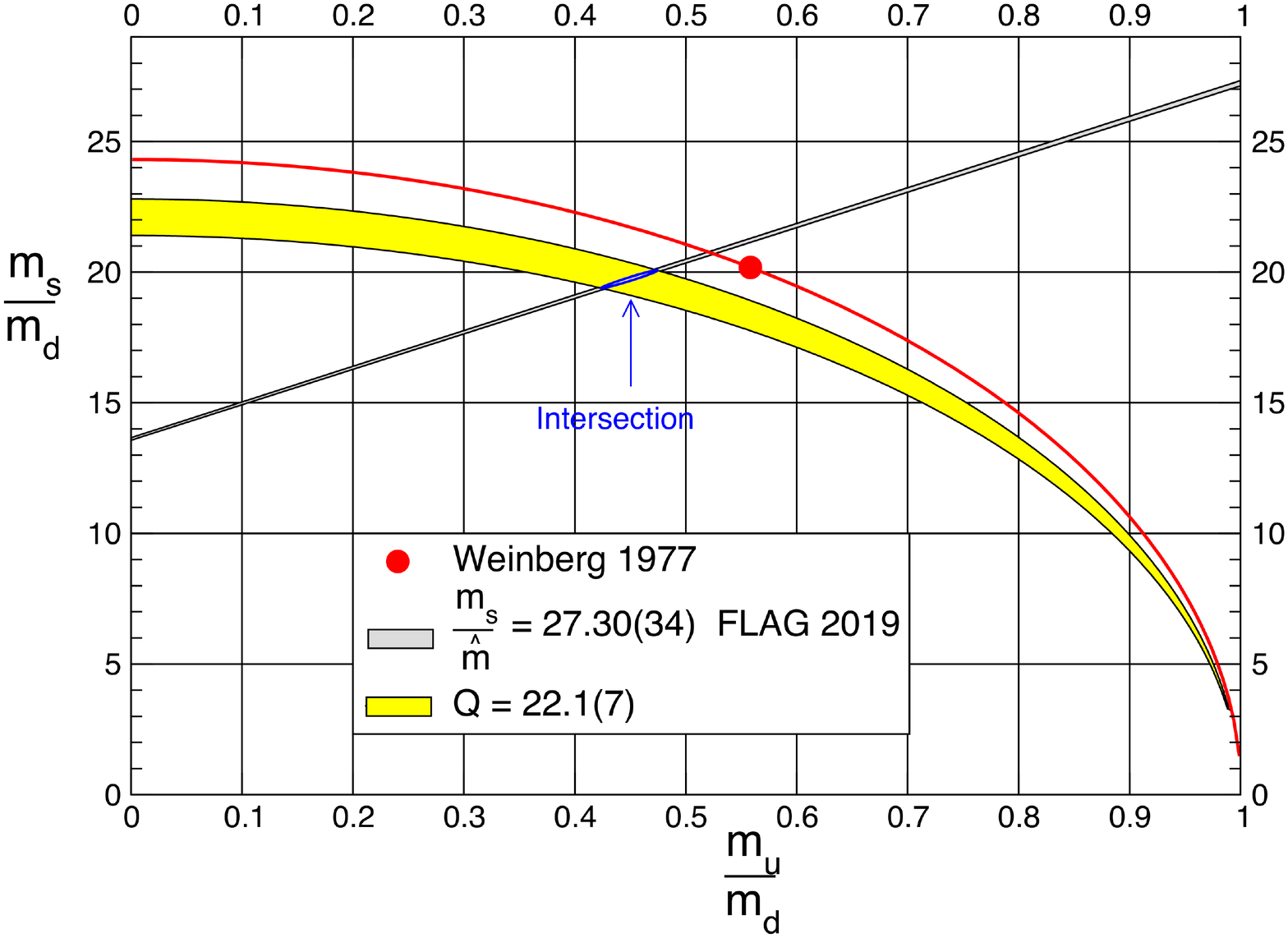}
\caption{\label{fig:Q} Constraints on the quark mass ratios from $\eta \to 3\pi$~\cite{Colangelo:2016jmc,Colangelo:2018jxw} (yellow band) and from lattice QCD~\cite{Aoki:2019cca} (gray band). The current algebra result is shown in red. Figure adapted from Ref.~\cite{Colangelo:2018jxw}.}
\end{figure}

Studying $\eta \to 3 \pi$ offers a unique way to extract $(m_u - m_d)$ from an experimental process in a reliable fashion. 
Indeed, this decay is forbidden by isospin symmetry: three pions cannot combine 
to a system with vanishing angular momentum, zero isospin, and even $C$-parity.
Isospin breaking contributions can arise in the Standard Model either from electromagnetic or strong interactions. $\eta \to 3 \pi$ is unique because, according to Sutherland's low-energy theorem~\cite{Sutherland:1966zz,Bell:1996mi}, the electromagnetic contributions vanish at leading order in $\chi$PT. 
Furthermore, even higher-order corrections of $\Order(e^2m_q)$ are constrained by a soft-pion theorem, which forbids potentially large terms of $\Order(e^2m_s)$ at the kinematical point of vanishing $\pi^0$ momentum.
Practical calculations confirm the terms of $\Order\big(e^2\hat{m}\big)$~\cite{Baur:1995gc} and $\Order\big(e^2(m_d-m_u)\big)$~\cite{Ditsche:2008cq} to be very small.
The decay is then driven almost exclusively by the following isospin-breaking operator 
in the QCD Lagrangian, 
\begin{align}
	\mathcal{L}_\textrm{IB} = - \frac{m_u - m_d}{2} \big(\bar u u - \bar d d\big) \eolp
	\label{eq:LIB}
\end{align}
As a consequence of being generated by this $\Delta I = 1$ operator, the decay amplitude must be proportional to \mbox{$(m_u - m_d)$} and can be used to extract this quantity. 
The decay width can be seen as a measure for the size of isospin breaking in QCD. 
More precisely, one extracts the quark mass double ratio $Q$, which does not receive corrections up to $\Order\big(m_q^3\big)$: 
\begin{align} 
	\A_{\eta \to 3 \pi} \propto B_0 (m_u - m_d) = -\frac{1}{Q^2} \frac{M_K^2(M_K^2 - M_\pi^2)}{M_\pi^2} + \O\left(m_q^3\right)
\quad \text{with} \quad
	Q^2 = \frac{m_s^2 - \hat{m}^2}{m_d^2 - m_u^2} \,.
\label{eq:QRprefactor}
\end{align}
Neglecting a tiny term proportional to $(\hat{m}/m_s)^2$, $Q$
gives an elliptic constraint in the plane spanned by $m_s/m_d$ and $m_u/m_d$~\cite{Leutwyler:1996qg}:
\begin{equation}
\left( \frac{m_u}{m_d} \right)^2 + \frac{1}{Q^2} \left( \frac{m_s}{m_d} \right)^2 = 1 \,,
\end{equation}
see Fig.~\ref{fig:Q}. 
Since the decay width is basically given by the phase space integral over the amplitude squared,
\begin{equation}
\Gamma(\eta \to 3 \pi) \propto \int \diff s \,\diff u|\A_{\eta \to 3 \pi}(s,t,u) |^2 \propto \frac{1}{Q^4}  \eolc
\end{equation}
an experimentally measured value of the decay width, $\Gamma(\eta \to 3 \pi)$, translates into a value of $Q$. As a reference value, one often compares to the number $Q_D$ extracted from Eq.~\eqref{eq:WeinbergRatios} 
relying on Dashen's theorem,
\beq
Q_D^2 = \frac{\Big(\mKz^2 + \mKp^2 - \mpip^2 + \mpiz^2\Big)\Big(\mKz^2 + \mKp^2 - \mpip^2 - \mpiz^2\Big)}
{4\mpiz^2\Big(\mKz^2 - \mKp^2 + \mpip^2 - \mpiz^2\Big)} = (24.3)^2 \,. \label{eq:QD}
\eeq
The aim of theoretical analyses is to compute the normalized amplitude $\M(s,t,u)$, defined as 
\begin{equation}
 \A_{\eta \to \pi^+\pi^-\pi^0}(s,t,u) \equiv \A_c(s,t,u)
 = -\frac{1}{Q^2} \frac{M_K^2 \big(M_K^2-M_\pi^2\big)}{3 \sqrt{3} M_\pi^2 F_\pi^2} \M(s,t,u) \,,
\label{eq:Q}
\end{equation}
with the best possible accuracy. Here, the Mandelstam variables are defined according to
$s=(p_{\pi^+}+p_{\pi^-})^2=(p_\eta-p_{\pi^0})^2$,
$t=(p_{\pi^-}+p_{\pi^0})^2=(p_\eta-p_{\pi^+})^2$, and
$u=(p_{\pi^+}+p_{\pi^0})^2=(p_\eta-p_{\pi^-})^2$. 
At first order in isospin breaking, i.e., $\Delta I = 1$, the amplitude in the neutral channel can be computed from the amplitude in the charged channel according to 
\begin{equation} \label{eq:isoRel}
 \A_{\eta \to 3\pi^0}(s,t,u) \equiv \A_n(s,t,u) = \A_c(s,t,u) + \A_c(t,u,s) + \A_c(u,s,t) \eolp
\end{equation}

\subsubsection{Experimental measurements}
To extract $Q$ via Eq.~\eqref{eq:Q}, we need to know the $\eta \to 3 \pi$ decay width as well as the shape of the amplitude experimentally. We have 
\begin{equation}
\Gamma\big(\eta \to \pi^+ \pi^- \pi^0\big) = \frac{1}{Q^4} \frac{M_K^4 (M_K^2-M_\pi^2)^2}{6912 \pi^3 M_\eta^3 M_\pi^4 F_\pi^4} \int_{s_{\rm min}}^{s_{\rm max}} \diff s 
\int_{u_-(s)}^{u_+(s)} \diff u \left|\M(s,t,u) \right|^2
\label{eq:Qexp}
\end{equation}
with the integration boundaries $s_{\min} =4 \mpip^2$, $s_{\max} = (\meta - \mpiz)^2$, and 
\begin{equation} 
u_{\pm} (s)=\frac{1}{2} \left ( 3 s_0 -s \pm \sigma \, \lambda^{1/2}\big(\meta^2,\mpiz^2,s\big) \right) 
\eolc
\label{eq:ubound}
\end{equation}
where $s_0 = \frac{1}{3}\left(M_\eta^2 + 2 M_{\pi^+}^2+M_{\pi^0}^2\right)$, $\sigma = \sqrt{1-4\mpip^2/s}$, and 
\beq \label{eq:lambda}
\lambda(a,b,c) = a^2+b^2+c^2-2(ab+ac+bc)
\eeq
is the standard K\"all\'en function.
To determine $\Gamma(\eta \to 3 \pi)$ experimentally, one needs to know the total $\eta$ decay width, which is determined by the measured decay width of $\eta$ into two photons and the corresponding branching ratio that is known at better than 1\% accuracy~\cite{Tanabashi:2018oca}. 
Thus, our knowledge of $\Gamma(\eta \to 2 \gamma)$ will have a direct impact on the experimental value of $\Gamma(\eta \to 3 \pi)$ and consequently on the light quark mass ratio determination.
The results of $\Gamma(\eta \to 2 \gamma)$ measured from $e^+e^-$ collisions and the ones via the Primakoff process have been in disagreement by a $\sim 3\sigma$ deviation over the past four decades. 
The systematic difference between these two types of experimental results on $\Gamma(\eta \to 2 \gamma)$ potentially contributes significant uncertainty in $Q$; see Sect.~\ref{sec:eta-2g} for a more detailed discussion.  A new measurement with improved precision is clearly of high significance. 

\begin{table}[t!]
\centering
\renewcommand{\arraystretch}{1.3}
\begin{tabular}{lllllll} \toprule
Experiment & Events  & $-a$  & $b$ & $d$ & $f$ & $-g$  \\ \midrule
Gormley(1970)~\cite{Gormley:1970qz} & 7.2k &$1.17(2)$ & $2.1(3)$ & $6(4)$ & --- & ---\\
Layter(1973)~\cite{Layter:1973ti} & 81k &$1.080(14)$ & $0.34(27)$ & $4.6(3.1)$ & --- & ---\\
CBarrel(1998)~\cite{Abele:1998yj} & 3.2k &$1.22(7)$ & $2.2(1.1)$ & $6$(fixed) & ---&---\\
KLOE(2008)~\cite{Ambrosino:2008ht} & 1.34M & $1.090\big({}^{~9}_{20}\big)$ & $0.124(12)$  & $0.057\big({}^{~9}_{17}\big)$ & $0.14(2)$ & --- \\ 
WASA(2014)~\cite{Adlarson:2014aks} & 174k & $1.144(18)$ & $0.219(51)$  & $0.086(23)$ & $0.115(37)$ & --- \\ 
BESIII(2015)~\cite{Ablikim:2015cmz} & 79.6k & $1.128(17) $ & $0.153(17)$
& $0.085(18)$  & $0.173(35)$ & --- \\ 
KLOE(2016)~\cite{Anastasi:2016cdz} & 4.7M & $1.095(4)$ & $0.145(6)$  & $0.081\big({}^{7}_{6}\big)$ & $0.141\big({}^{10}_{11}\big)$ & $0.044\big({}^{15}_{16}\big)$ \\ \midrule
Theory & & & & & &  \\ \midrule
NLO $\chi$PT~\cite{Gasser:1984pr} & --- &$1.33$ & $0.42$ & $0.08$ & \\
U-$\chi$PT~\cite{Borasoy:2005du} & --- & $1.054(25)$ & $0.185(15)$ & $0.079(26)$ & $0.064(12)$ & --- \\ 
NNLO $\chi$PT~\cite{Bijnens:2007pr} & --- & $1.271(75)$ & $0.394(102)$ & $0.055(57)$ & $0.025(160)$ & ---  \\
NREFT~\cite{Schneider:2010hs} & --- & $1.213(14)$ & $0.308(23)$ & $0.050(3)$& $0.083(19)$ & $0.039(2)$ \\
JPAC~\cite{Guo:2016wsi} & ---& $1.075(28)$ & $0.155(6)$ & $0.084(2)$ & $0.101(3)$  & $0.074(3)$  \\
CLLP~\cite{Colangelo:2018jxw} & ---& $1.081(2)$ & $0.144(4)$ & $0.081(3)$ & $0.118(4)$ & $0.069(4)$ \\
AM single-ch.~\cite{Albaladejo:2017hhj} &---& $1.156$ & $0.200$ & $0.095$ & $0.109$ & $0.088$ \\
AM coupled-ch.~\cite{Albaladejo:2017hhj} & ---& $1.142$ & $0.172$ & $0.097$ & $0.122$ & $0.089$ \\
\bottomrule
\end{tabular}
\renewcommand{\arraystretch}{1.0}
\caption{Dalitz distribution parameters determined from $\eta\rightarrow \pi^+\pi^-\pi^0$ experiments. In the lower half of the table, theoretical results are shown for comparison.}
\label{tab:dalitz}
\end{table}

\begin{figure}
\centering
\includegraphics[width=0.54\textwidth]{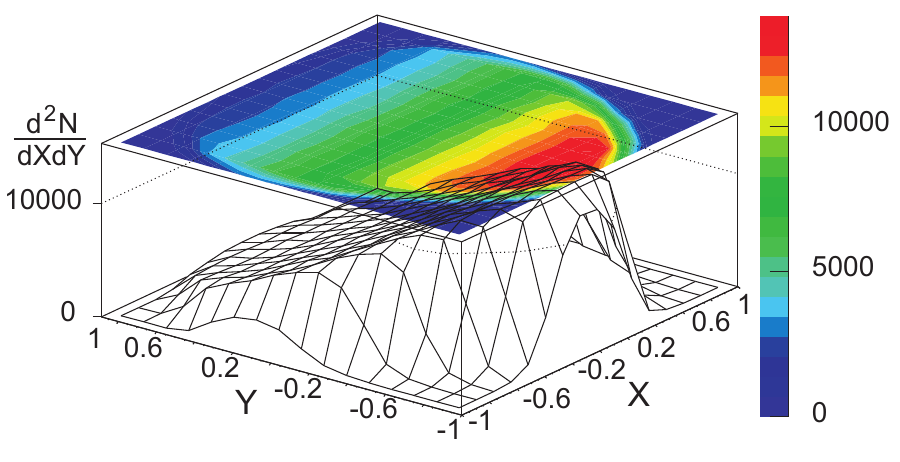} \hfill
\includegraphics[width=0.45\textwidth]{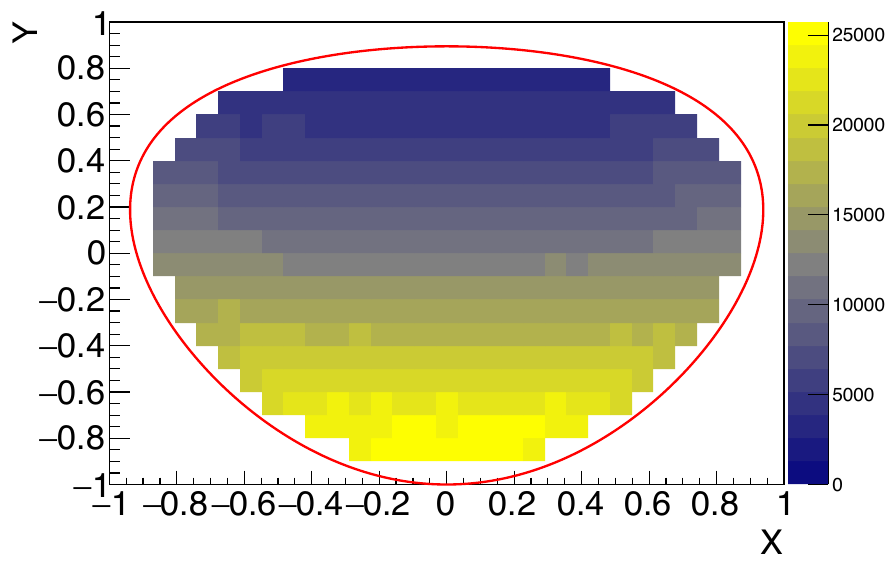}
\caption{\label{fig:KLOEDalitz} Dalitz plots for $\eta \to \pi^+ \pi^- \pi^0$ measured by KLOE in 2008~\cite{Ambrosino:2008ht} (left) and the latest KLOE result in 2016~\cite{Anastasi:2016cdz} (right). 
Reprinted from Refs.~\cite{Ambrosino:2008ht,Anastasi:2016cdz}.}
\end{figure}

Another important input is the shape of the amplitude $\M(s,t,u)$ in Eq.~\eqref{eq:Q}.
It can be assessed by measurements of the Dalitz plot distribution. 
The latter is the standard representation of the momentum dependence of a three-particle decay,
and is given in terms of the squared absolute value of the amplitude.
In experimental analyses, the latter is typically
expanded as a polynomial around the center of the Dalitz plot in terms
of symmetrized coordinates. For the charged decay channel one uses
\begin{align}
X &= \sqrt{3}\,\frac{T_+-T_-}{Q_c} = 
\frac{\sqrt{3}}{2 M_\eta Q_c} (u-t)\,,
\quad
Y = \frac{3 T_0}{Q_c}-1\,
= \frac{3}{2 M_\eta Q_c}\left\{\left(M_\eta-M_{\pi^0}\right)^2-s\right\}-1
\, ,\quad
Q_c = M_\eta - 2 M_{\pi^+} - M_{\pi^0}
\,, \label{eq:XYdef}
\end{align}
where $T_{+/-/0}$ and $Q_c$ are the kinetic energies of the $\pi^{+/-/0}$ and the sum of all three, respectively, in the $\eta$ rest frame. 
The Dalitz plot parameterization then reads
\begin{equation}
\label{eq:DalitzCharged}
	\Gamma_c(X,Y) = |\A_c(s,t,u)|^2 \propto 1 + aY + bY^2 + cX + dX^2 + eXY + fY^3 +  gX^2Y + hXY^2 + lX^3 + \ldots \eolc
\end{equation}
where the coefficients $a,\,b,\,\ldots$ are called Dalitz plot parameters. 
Charge conjugation symmetry
requires terms odd in $X$ to vanish, such that $c = e = h = l = 0$; we will comment on potential 
$C$-violating effects in $\eta\to\pi^+\pi^-\pi^0$ in Sect.~\ref{sec:BSMCV}. 
The values of the Dalitz plots parameters as experimentally measured compared to various theory predictions 
are presented in Table~\ref{tab:dalitz}.
The most recent KLOE(2016) result~\cite{Anastasi:2016cdz} has the smallest uncertainties in both statistics and systematics. It is largely consistent with the previous measurements by KLOE(2008)~\cite{Ambrosino:2008ht}, WASA(2014)~\cite{Adlarson:2014aks}, and BESIII(2015)~\cite{Ablikim:2015cmz} within error bars, with small tensions beyond one standard deviations for $b$ in comparison to WASA(2014) and $d$ compared to KLOE(2008) only. In addition, the modern experimental results for the $b$ parameter show rather large deviations from the $\chi$PT predictions. For the first time, KLOE(2016)~\cite{Anastasi:2016cdz} indicates the need for a nonzero $g$ parameter in the $\eta\rightarrow\pi^+\pi^-\pi^0$ Dalitz distribution. 
The Dalitz plots measured by KLOE in 2008~\cite{Ambrosino:2008ht} and 2016~\cite{Anastasi:2016cdz} are shown in Fig.~\ref{fig:KLOEDalitz}. 
One should note that the $\eta \to 3 \pi$ Dalitz plots 
do not exhibit prominent resonance bands, as the $f_0(500)$ is very broad and, in addition, the available phase space too small.

\begin{figure}[t!]
\centering
\includegraphics[width=0.35\textwidth]{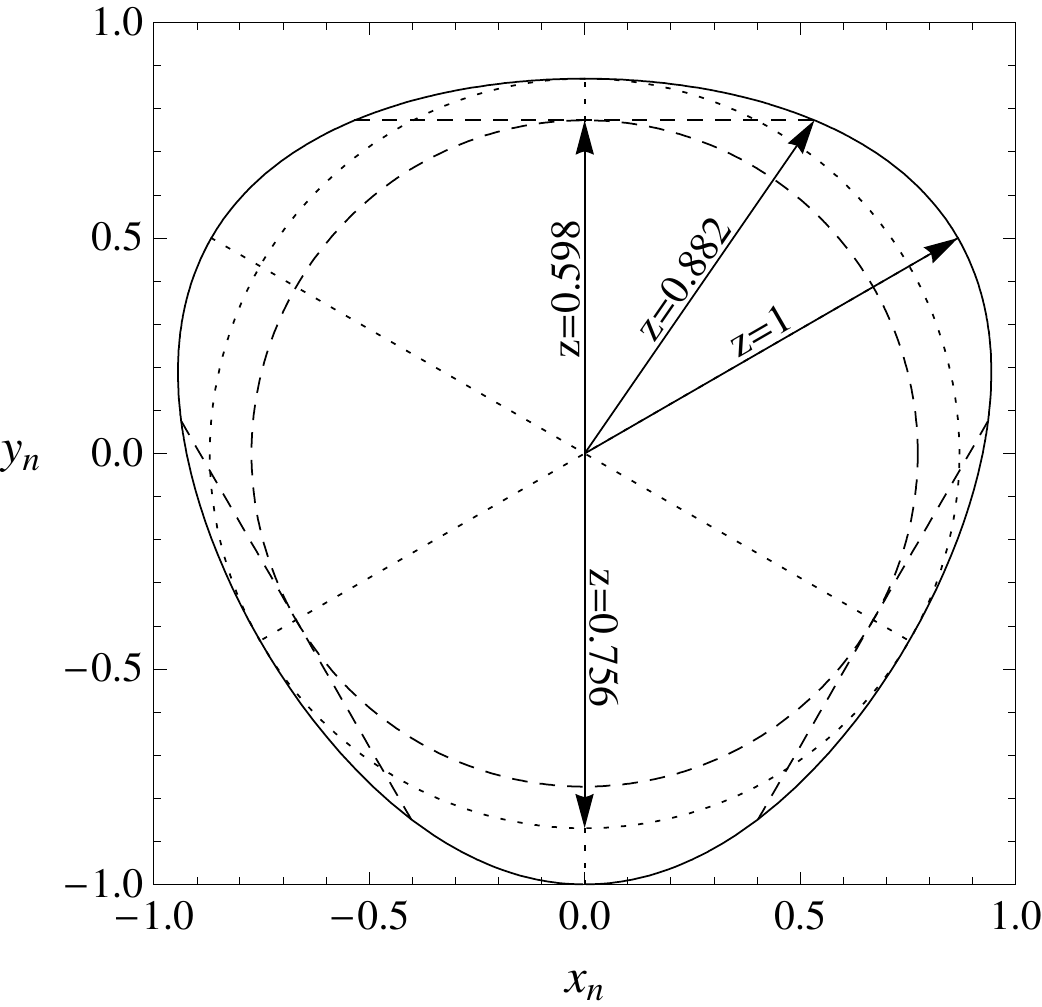}
\caption{\label{fig:Dalitz} Representation of the $\eta\to3\pi^0$ Dalitz plot. 
Symmetry axes and the biggest fully enclosed circle (at $Z=0.756$) are shown as dotted lines. 
The cusps at $s_i=4\mpc^2$ hit for values between $Z=0.598$ and $Z=0.882$ are shown in dashed style. 
Reprinted from Ref.~\cite{Schneider:2010hs}.}
\end{figure}

A polynomial expansion analogous to the one in Eq.~\eqref{eq:DalitzCharged} can also be performed for the neutral channel $\eta\to 3\pi^0$: 
\begin{align}
	\Gamma_n(X,Y) = |\A_n(s,t,u)|^2 \propto 1 + 2 \alpha Z + 2 \beta Y \left( 3X^2 - Y^2 \right) + 2 \gamma Z^2 +\ldots \,,
	\label{eq:alpha}
\end{align}
with $Z=X^2+Y^2$, and $X$ and $Y$ defined in analogy to Eq.~\eqref{eq:XYdef}, with $\mpip \to \mpiz$. 
The main difference compared to the charged channel is that the Dalitz plot distribution is nearly flat. The deviation of the normalized distribution from 
unity is by a few percent only. This renders measurements as well as theoretical predictions rather difficult, and for a long time theory failed to reproduce the experimental results. 
In addition, since the invariant mass of 2$\pi^0$ from the $\eta\to 3\pi^0$ decay extends to the region below the mass of two charged pions, a cusp structure~\cite{Bissegger:2007yq,Gullstrom:2008sy} is expected in the 2$\pi^0$ invariant mass distribution around the mass of two charged pions, corresponding to the $\pi^+ \pi^- \to \pi^0 \pi^0$ transition (cf.\ also Ref.~\cite{Kampf:2019bkf} for a reconstruction of the chiral two-loop amplitude including the pion mass difference); see the schematic representation in Fig.~\ref{fig:Dalitz}.   
In order to properly represent this cusp in the Dalitz plot description, an additional term $+2\delta \sum_{i=1}^{3} \rho(s_i)$ needs to be added to Eq.~\eqref{eq:alpha}, where $\rho(s)=\theta(4\mpip^2-s)\sqrt{1-s/4\mpip^2}$~\cite{Prakhov:2018tou}.

\begin{figure}[t!]
\centering
\includegraphics[width=0.7\textwidth]{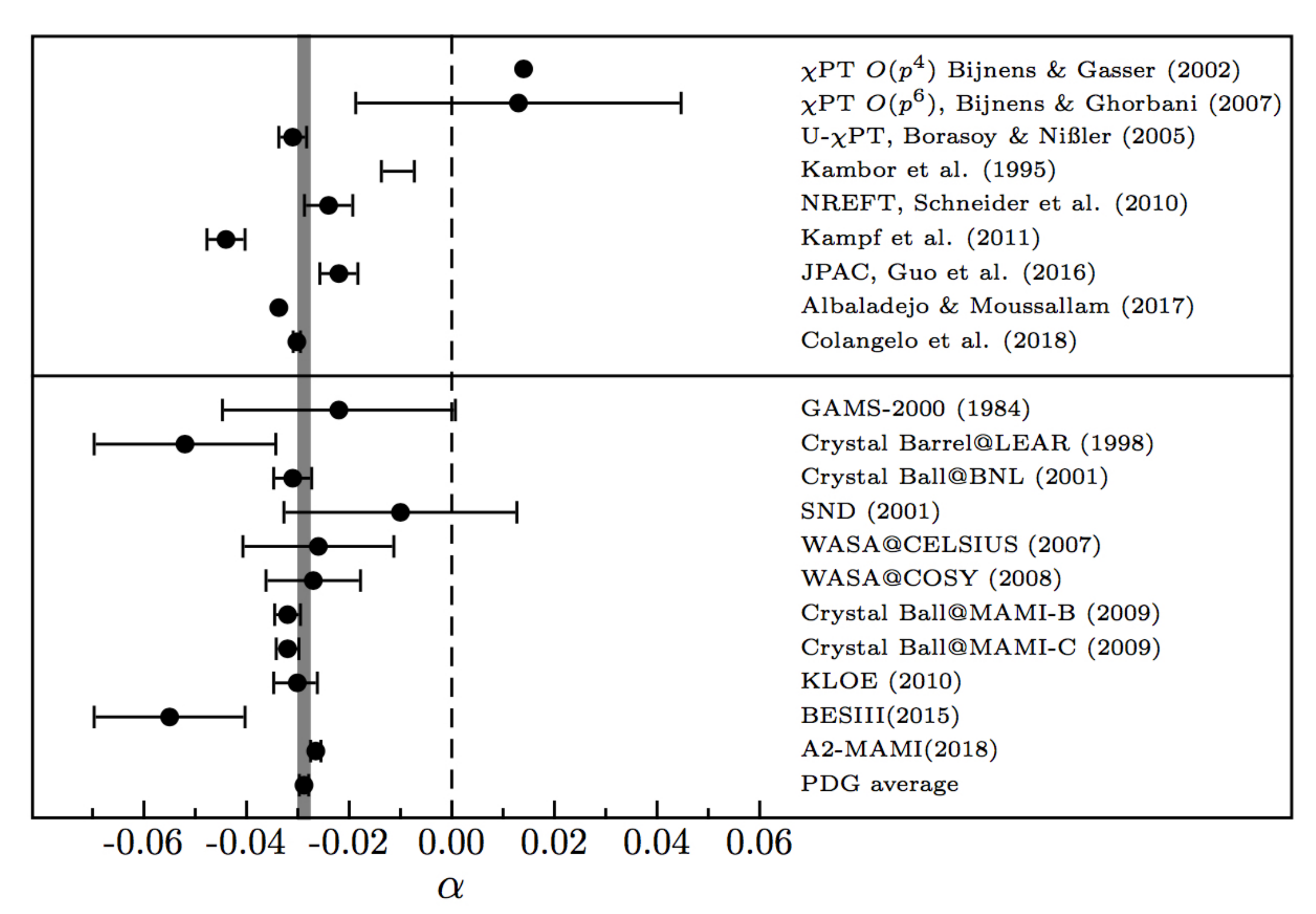}
\caption{\label{fig:alpha} Summary of determinations of the slope parameter $\alpha$. Figure adapted from Ref.~\cite{Colangelo:2018jxw}.}
\end{figure}

The final-state $\pi^0\pi^0$ rescattering in $\eta\to 3\pi^0$ is dominated by the $S$-wave due to the small energies of the pions. Except for the most recent A2 measurement~\cite{Prakhov:2018tou}, all experimental Dalitz plot investigations were based on the leading-order parameterization $\Gamma_n(Z) \propto 1 + 2 \alpha Z$. The slope parameter $\alpha$ was extracted by fitting the deviation of measured $Z$ distributions from the simulated phase space. Note that it was shown in Ref.~\cite{Colangelo:2018jxw} that the value for $\alpha$ employing the linear approximation only is very sensitive to the fit range.
A comparison of the experimental and theoretical results for $\alpha$ is shown in Fig.~\ref{fig:alpha}. 
Some of the recent experimental determinations of the Dalitz plot parameters along with theoretical calculations are listed in Table~\ref{tab:exp-3pi-n}.
As one can see, the experimental precision for the slope parameter $\alpha$ has been significantly improved in recent years, leading to the very precise  PDG average $\alpha = -0.0288(12)$~\cite{Tanabashi:2018oca}. 
The most accurate results all came from the A2 collaboration: 
A2-MAMI-B(2009)~\cite{Unverzagt:2008ny},  A2-MAMI-C(2009)~\cite{Prakhov:2008ff}, and  A2-MAMI(2018)~\cite{Prakhov:2018tou}. They are consistent with each other and agree with the measurements (with less statistics) by the KLOE~\cite{Ambrosino:2008ht} and WASA-at-COSY~\cite{Adlarson:2014aks} collaborations, but all find smaller values for $\alpha$ than BESIII~\cite{Ablikim:2015cmz}. 
All experimental values for the slope parameter $\alpha$ have a negative sign. The $\chi$PT prediction~\cite{Bijnens:2007pr}, however, fails to produce a negative slope, at least as its central value. The theoretical calculations 
using unitarized $\chi$PT~\cite{Borasoy:2005du} show better agreement with the experiments. As pointed out in Ref.~\cite{Kolesar:2016jwe}, the discrepancies in $\eta\rightarrow 3\pi$ Dalitz plot parameters between $\chi$PT and experiments suggest a slow or irregular convergence of the chiral expansion. The final-state $\pi\pi$ rescattering contributes significantly to the negative value of $\alpha$. It can be accounted for to all orders using dispersion theory and precisely known $\pi\pi$ phase shifts~\cite{Kampf:2011wr,Guo:2015zqa,Colangelo:2016jmc,Colangelo:2018jxw,Albaladejo:2017hhj}.

The most recent A2 measurement of the $\eta\to3\pi^0$ Dalitz plot~\cite{Prakhov:2018tou} is based on an analysis of data taken in both 2007 and 2009 with an improved cluster algorithm.  The obtained high statistics of $7\times 10^6$ $\eta\to 3\pi^0$ events enabled an exploration of terms beyond $\alpha$ for the first time. Several interesting features were observed in this study: it clearly demonstrated that the $\alpha$ term alone is not sufficient to describe the $\eta\to 3\pi^0$ decay, the extracted magnitude for $\beta$ differed from zero by $\sim 5.5\sigma$. A large correlation was found between $\alpha$ and $\gamma$ so that the parameter $\gamma$ could not be determined reliably from this analysis.
The best $\chi^2/\text{ndof}$ value was achieved by fitting to the experimental Dalitz distribution with three parameters ($\alpha$, $\beta$, and $\delta$).  The magnitude of the cusp term was found to be close to 1\% with $\delta=-0.018(7)(7)$, which is consistent with a prediction based on nonrelativistic effective field theory (NREFT)~\cite{Gullstrom:2008sy}, even though the experimental uncertainty for $\delta$ is larger than 50\%. In addition, the inclusion of higher-order terms led to a smaller value in $\alpha$ by about 20\% compared to the results obtained with the leading term alone. 
The precision on the slope parameter, $\alpha$, now reaches to a level of 5\%~\cite{Tanabashi:2018oca}.   
Compared to the charged decay channel, however, the neutral channel provides altogether weaker constraints on theory due to identical final-state particles. 

All existing experiments described above were performed with little or unboosted $\eta$ mesons, where the energy of the $\eta$ decay products was relatively small in the lab frame;
hence the detection efficiencies for these measurements were rather sensitive to the detector threshold. For the ongoing GlueX experiment in Hall~D  at Jefferson Lab, highly boosted $\eta$ mesons are produced by a $\sim 12\GeV$ tagged-photon beam on a liquid-hydrogen target.  The decay products are highly energetic with average energies above $1\GeV$. Consequently, the detection efficiencies will be less sensitive to the detector thresholds. In addition, the GlueX apparatus can measure both charged and neutral channels, and has 
nearly flat acceptance and detection efficiency over the phase space of $\eta \to 3\pi$, which will greatly help to control systematic 
uncertainties. The phase-I low-luminosity GlueX run was completed in the end of 2018. About 20\% of this data has been analyzed for the charged channel $\eta\to \pi^+\pi^-\pi^0$. Based on this analysis, 
the full phase-I data set will yield $\sim 2\times 10^6$ $\eta\to \pi^+\pi^-\pi^0$ events.
A similar yield is expected for the neutral channel. The phase-II GlueX run with high luminosity started data collection in fall 2019. An order of magnitude more statistics will be collected in the next 2--3~years. The measurements from GlueX will offer improved systematics and high statistics for $\eta\to 3\pi$.
Such cross checks on the systematics are important for understanding the uncertainty on the quark mass ratio $Q$. 

\begin{table}[t!]
\centering
\renewcommand{\arraystretch}{1.3}
\begin{tabular}{llccc} \toprule
Experiment &  Events  & $\alpha$ & $\beta$ & $\gamma$ \\ \midrule
WASA-at-COSY(2009)\cite{Adolph:2008vn} & 120k & $-0.027(8)(5)$ & --- & ---\\ 
A2-MAMI-B(2009)\cite{Unverzagt:2008ny} & 1.8M & $-0.032(2)(2)$ & --- & ---\\ 
A2-MAMI-C(2009)\cite{Prakhov:2008ff} & 3M & $-0.0322(12)(22)$ & --- & ---\\ 
KLOE(2010)\cite{Ambrosinod:2010mj} & 512k & $-0.0301(35)\big({}^{22}_{35}\big)$ & --- & ---\\ 
BESIII(2015)\cite{Ablikim:2015cmz} & 33k & $-0.055(14)(4)$ & --- & ---\\ 
A2-MAMI(2018)\cite{Prakhov:2018tou} & 7M & $-0.0265(10)(9)$ & $-0.0074(10)(9)$ & ---\\ 
\midrule
Theory &  &   \\ \midrule
$\chi$PT NLO~\cite{Gasser:1984pr,Bijnens:2002qy} & --- & $+0.014$ & --- & ---\\ 
$\chi$PT NNLO~\cite{Bijnens:2007pr} & --- & $+0.013(32)$ & --- & ---\\ 
U-$\chi$PT~\cite{Borasoy:2005du} & --- & $-0.031(3)$ & --- & ---\\
KWW~\cite{Kambor:1995yc}& --- & $-0.014$ $\ldots$ $-0.007$ & --- & ---\\
NREFT~\cite{Schneider:2010hs}  & --- & $-0.025(5)$ & $-0.0042(7)$ & $0.0013(4)$\\
KKNZ~\cite{Kampf:2011wr} & --- & $-0.044(4)$ & --- & --- \\
JPAC~\cite{Guo:2016wsi} & ---& $-0.025(4)$ & $0.000(2)$ & --- \\
AM~\cite{Albaladejo:2017hhj} & ---&  $- 0.0337(12)$ & $-0.0054(1)$ & --- \\
CLLP~\cite{Colangelo:2018jxw} &---& $- 0.0307(17)$ & $-0.0052(5)$ & $0.0019(3)$\\
\bottomrule
\end{tabular}
\renewcommand{\arraystretch}{1.0}
\caption{Dalitz distribution parameters determined from the $\eta\rightarrow 3\pi^0$ experiments and theoretical calculations. \label{tab:exp-3pi-n}}
\end{table}

An interesting quantity to consider is the branching ratio
\begin{equation} \label{eq:Beta3pi}
B= \frac{\Gamma(\eta \to 3\pi^0)}{\Gamma(\eta \to \pi^+ \pi^- \pi^0)} \,. 
\end{equation}
Indeed by taking the ratio $B$ the overall normalization appearing in the rates $\Gamma(\eta \to \pi^+ \pi^- \pi^0)$ and 
$\Gamma(\eta \to 3\pi^0)$ cancels out as well as most of the theoretical uncertainties, such that
the dominant error in the theoretical prediction of $B$
is due to (higher-order) isospin breaking. 
The experimental values given by the PDG~\cite{Tanabashi:2018oca}
are $B = 1.426(26)$ for the value coming from the PDG fit (including a scale factor of $1.2$) 
and $B = 1.48(5)$ for the average. 
New experimental information to clarify and sharpen this branching ratio further would be very welcome.

\subsubsection{Theoretical description of $\eta \to 3 \pi$}
\label{sec:eta3pi_theo}
The main difficulty in the theoretical prediction of $\eta \to 3 \pi$ is the evaluation of rescattering effects among the final-state pions. 
This can be done perturbatively within $\chi$PT~\cite{Cronin:1967jq, Gasser:1984pr, Bijnens:2007pr}, 
but not with sufficient accuracy. 
This is clearly seen in the prediction of the decay width $\Gamma(\eta \to \pi^+ \pi^- \pi^0)$:
Gasser and Leutwyler found an enhancement of the current algebra result~\cite{Osborn:1970nn}
$\Gamma(\eta \to \pi^+ \pi^- \pi^0)_{\text{LO}} = 66\eV$ to 
$\Gamma(\eta \to \pi^+ \pi^- \pi^0)_{\text{NLO}} = 160(50)\eV$ due to chiral one-loop corrections, 
both based on $Q=Q_D$ extracted from Dashen's theorem, see Eq.~\eqref{eq:QD}.
Including two-loop contributions, Bijnens and Ghorbani~\cite{Bijnens:2007pr} quote the following chiral series for the same quantity,
expressed as a function of the $\pi^0$--$\eta$ mixing angle:
\beq
\Gamma(\eta \to \pi^+ \pi^- \pi^0)_{\{\text{LO},\,\text{NLO},\,\text{NNLO}\}} \approx \sin^2\eps \times\{ 0.572, \, 1.59, \, 2.68 \} \MeV
\approx \big\{ 1, 1.7^2, 2.2^2 \big\} \times \Gamma(\eta \to \pi^+ \pi^- \pi^0)_{\text{LO}}\,,
\eeq
hence demonstrating a rather slow convergence behavior. 
On the other hand, the branching ratio $B$, see Eq.~\eqref{eq:Beta3pi}, is remarkably stable under higher-order corrections: the current algebra prediction $B_{\text{CA}}=1.5$~\cite{Osborn:1970nn} is modified to $B_{\text{LO}}=1.54$ when taking into account the modified phase space due to different pion masses; higher-order corrections reduce this number to
$B_{\text{NLO}}=1.46$ or $B_{\text{NNLO}}=1.47$~\cite{Bijnens:2007pr}, to be compared to 
$B=1.48(5)$ as the PDG average~\cite{Tanabashi:2018oca}. 

Moreover, it has been shown that the two-loop calculation~\cite{Bijnens:2007pr} may lead to a precise numerical prediction 
only once the low-energy constants appearing in the amplitude are determined reliably. 
In particular, the role played by the $\O(p^6)$ LECs is not negligible and they are largely unknown. 
The associated convergence problem can also be seen in the corresponding Dalitz plot parameters: 
we already noted above that $\chi$PT alone fails to predict the correct sign for $\alpha$, the (small) curvature of the neutral Dalitz plot; see Fig.~\ref{fig:alpha}.

A more accurate approach relies on dispersion relations to evaluate rescattering effects to all 
orders~\cite{Anisovich:1993kn,Anisovich:1996tx,Kambor:1995yc,Walker:1998zz}. 
This is not completely independent of $\chi$PT, because the dispersive representation requires input for the subtraction constants---this input can be provided by $\chi$PT. 
There has been a resurgence of interest recently in analyzing $\eta\to3\pi$ dispersively, for the
following reasons: 
\begin{enumerate}
\item Earlier dispersive studies relied on rather crude input for $\pi \pi$ scattering, 
which is the essential ingredient in the calculation. Today a much more accurate representation for this amplitude is available~\cite{GarciaMartin:2011cn,Caprini:2011ky}. 
\item New measurements of the process $\eta \to \gamma \gamma$ entering the lifetime and therefore the $\eta\to3\pi$ decay rate are planned such as the PrimEx experiment~\cite{eta-PAC35,Gan:2015nyc} at JLab, see the discussion in Sect.~\ref{sec:eta-2g}, and call for an improved precision in the theoretical calculations. 
\item Recent measurements of the $\eta\to\pi^+\pi^-\pi^0$ Dalitz plot by KLOE~\cite{Ambrosino:2008ht,Anastasi:2016cdz} 
and BESIII~\cite{Ablikim:2015cmz} achieve an impressive level of precision. New measurements are planned by BESIII and at JLab by GlueX~\cite{Gan:2015nyc, JEF-PAC42} and CLAS~\cite{CLAS2016}, with completely different systematics and even better accuracy. 
\end{enumerate}

As we have detailed in Sect.~\ref{sec:DispTheory}, 
the dispersive method relies on general properties from $S$-matrix theory: unitarity, analyticity, and crossing symmetry. It can be combined with various constraints: at low energy from chiral symmetry as well as low-energy theorems such as soft-pion theorems, and at high energy from the Froissart bound~\cite{Froissart:1961ux,Martin:1962rt} or Brodsky--Lepage asymptotics. 

The starting point for this framework is to construct a scattering amplitude for $\pi \eta \to \pi \pi$ and then analytically continue it to describe the decay $\eta \to 3\pi$.  This is the so-called Khuri--Treiman framework~\cite{Khuri:1960zz,Aitchison:1966lpz}. 
We decompose the amplitude $\pi \eta \to \pi \pi$ into single-variable functions with right-hand cuts only, which have definite quantum numbers of angular momentum, isospin, and parity. 
Neglecting discontinuities in $D$- and higher partial waves
results in the following expression~\cite{Stern:1993rg,Anisovich:1996tx}:
\begin{align} 
\label{eq:Mdecomp}
	\M(s, t, u) = M_0^0(s) + (s-u) M_1^1(t) + (s-t) M_1^1(u) + M^2_0(t) + M^2_0(u) - \frac{2}{3} M^2_0(s) \eolp
\end{align}
The functions $M^I_\ell(s)$ have isospin $I$ and angular momentum $\ell$. 
This partial-wave and isospin decomposition, commonly referred to as a \textit{reconstruction theorem}~\cite{Stern:1993rg,Knecht:1995tr,Ananthanarayan:2000cp,Zdrahal:2008bd}, relies on the observation that up to corrections of $\Order (p^{8})$ (or three loops) in the chiral expansion, partial waves of any meson--meson scattering process with angular momentum $\ell \geq 2$ contain no imaginary parts.\footnote{For an example of a reconstruction theorem retaining $D$-waves, see Refs.~\cite{Niecknig:2015ija,Niecknig:2017ylb}.}
As in the approximation~\eqref{eq:Mdecomp}, Bose symmetry implies the angular momentum corresponding to a given isospin in an unambiguous way, we will omit the subscript $\ell$ in the following and refer to $M^I_\ell$ by $M_I$.
The splitting of the full amplitude into these single-variable functions 
is not unique: there is some ambiguity in the distribution of the polynomial terms over the various $M_I$ due to 
$s+t+u$ being constant.
The decomposition~\eqref{eq:Mdecomp} is useful as it reduces the problem to a set of functions
of only {\it one} variable.  It is furthermore exact up to two-loop order in $\chi$PT; 
the functions $M_I(s)$ then contain the main two-body rescattering corrections.

Using analyticity and unitarity, we can construct dispersion relations for the single-variable functions $M_I(s)$ in $\pi \eta \to \pi \pi$ for an $\eta$ mass below the three-pion threshold.
Unitarity in the elastic region determines the discontinuity from the knowledge of $\pi \pi$ scattering. 
Here we encounter precisely the type of problem discussed in Sect.~\ref{sec:Omnes}:
the three-body decay contains crossed-channel singularities, which however can again be related to $\pi \pi$ scattering in an isospin $I=0$, $1$, or $2$ state. 
The discontinuities thus obey 
\begin{equation}
\disc M_I(s) = 2i \sin \delta_I(s) e^{-i\delta_I(s)} \left\{ M_I(s) + \hat M_I(s) \right\} \,, \quad \text{with} \quad \hat M_I(s) =\sum_{n,I'} \int_{-1}^1 \diff \cos\theta \, c_{n I I'} \cos^{n} \theta \,  M_{I'}\big(t(s,\cos\theta)\big) 
\end{equation}
a consequence of the singularities in the $t$- and $u$-channels. 
The explicit coefficients $c_{nII'}$ can be found in Refs.~\cite{Anisovich:1996tx,Walker:1998zz}. 
One then arrives at a coupled set of inhomogeneous Omn\`es solutions for the $M_I(s)$, see the discussion in Sect.~\ref{sec:Omnes}:
\begin{align} 
\label{eq:MIdisp}
	M_I(s) = \Omega_I(s) \left\{ P_I(s) + \frac{s^{n_I}}{\pi} \int _{4 \mpi^2}^{\infty} \frac{\diff s^\prime}{s^{\prime n_I}}
		\frac{\sin \delta_I(s^\prime) \hat{M}_I(s^\prime)}{|\Omega_I(s^\prime)| (s^\prime - s -i \epsilon)}
		\right\},
\end{align}
where the $P_I(s)$ refer to polynomials containing the subtraction constants. 
There are, however, two important technical subtleties specific to the decay kinematics of the four-point function discussed here: 
(1) after analytically continuing $M_\eta^2$ to its physical value, extra singularities appear;
(2) $\hat M(s)$ is needed for values of $s$ outside the physical domain, so singularities in the relation of $t$ with $s$ and 
$\cos\theta$ also need to be taken care of. 
The integration path has to be chosen to avoid these extra singularities, see the dedicated discussion in Ref.~\cite{Gasser:2018qtg}.
This set of equations can then be solved numerically, e.g., by iteration.
Its application in data fits simplifies considerably if one takes into account that the solutions of the dispersion relations are linear in the subtraction constants: this means that one can calculate a set of basis solutions that are independent of the numerical values of the latter. 

The second step then consists in determining the subtraction constants. 
The number of subtractions needed for the dispersion integral in Eq.~\eqref{eq:MIdisp}, or the value of $n_I$, is a rather delicate matter. From a purely mathematical point of view, subtracting the dispersion integral is simply a rearrangement of the equation. As the integral along the infinite circle has to vanish, we just need to subtract sufficiently often to make this happen. To subtract to an even higher degree does not change the equation anymore, because the change in the dispersion integral is absorbed in additional parameters in the subtraction polynomial, 
which need to be fixed.
From a physical point of view, however, there are further aspects to be taken into account: if only few subtractions are used, the high-energy behavior of the integrand is less suppressed. 
To study the convergence behavior of
the integrand we have to make assumptions as regards the
asymptotic behavior of the phase shifts. It is usually assumed that 
\begin{equation}
\delta_{0}^{0}(s) \to \pi \,,\quad \delta_{1}^{1}(s) \to \pi\,, \quad \text{and}\quad \delta^{2}_{0}(s) \to 0 \,, \quad \text{as} \quad s \to \infty \,.
\label{eq:asymdelta}
\end{equation}
An asymptotic behavior of $\delta(s) \to k\pi$ implies that the corresponding Omn\`es function behaves like $s^{-k}$ for high $s$. 
Therefore if the Froissart bound~\cite{Froissart:1961ux,Martin:1962rt} is assumed, implying $M_0(s), M_2(s) \to s$ and 
$M_1(s) \to {\rm const.}$, then four subtraction constants are required. 
Due to the relation between the Mandelstam variables, $s+t+u = M_\eta^2 + 3M_\pi^2$,
there exists a five-parameter polynomial transformation of the single-variable functions $M_I$
that leaves the amplitude $M(s,t,u)$ in Eq.~\eqref{eq:Mdecomp} invariant.
Therefore there is some freedom to assign the subtraction constants to the functions $M_I$. 
We can relax the Froissart bound and oversubtract the dispersive integrals~\eqref{eq:MIdisp} with the aim of being insensitive, in the physical region, to the high-energy inelastic behavior of the phase, which is unknown. The price to pay is that one has more subtraction constants to be determined. 
\begin{figure}
\centering
\includegraphics[width=0.5\textwidth]{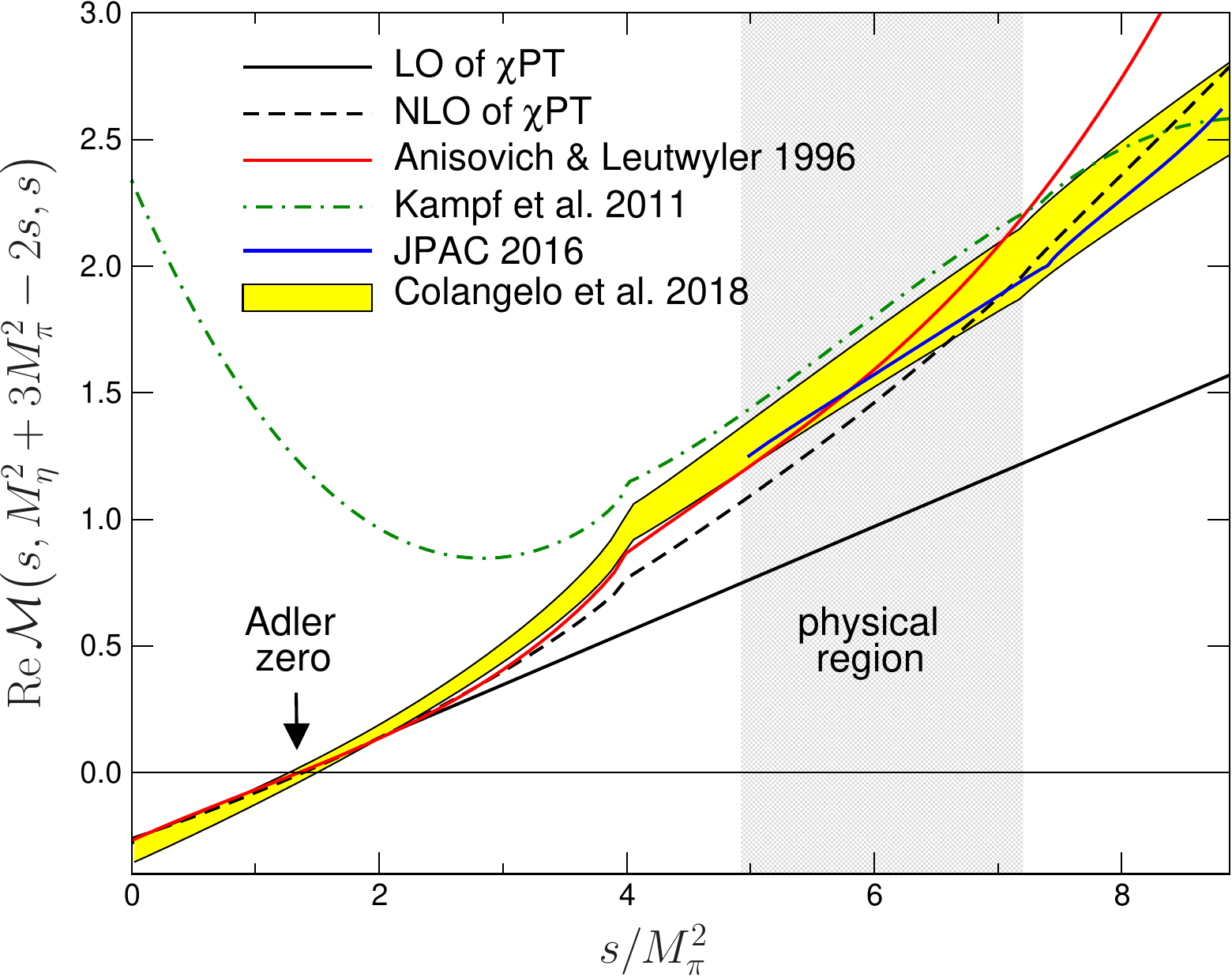}
\caption{\label{fig:Adler} 
Real part of the amplitude $\M(s,t,u)$ along the line $s=u$, with
the position of the Adler zero marked explicitly. 
The latter lies outside the physical region, which corresponds to the gray band. 
Full and dashed black lines correspond to the LO and NLO $\chi$PT amplitudes; in addition, various dispersive determinations are shown~\cite{Anisovich:1996tx,Kampf:2011wr,Guo:2016wsi,Colangelo:2018jxw}. One observes that the corrections are negligible around the Adler zero, as
the amplitude is protected by an $\SU(2) \times \SU(2)$ symmetry. Figure adapted from Ref.~\cite{Colangelo:2018jxw}.} 
\end{figure}
In some recent dispersive analyses~\cite{Kampf:2011wr,Colangelo:2016jmc,Colangelo:2018jxw}, six subtraction constants have been considered; for instance in Refs.~\cite{Colangelo:2016jmc,Colangelo:2018jxw}, the three dispersive integrals are written as 
\begin{align}
 M_0(s) &= \Omega_0(s)\Biggl\{\alpha_0+\beta_0 s +\gamma_0 s^2  + \delta_0 s^3+ \frac{s^4}{\pi}\int_{4\mpi^2}^{\infty}\frac{\diff s'}{{s'}^4} \frac{\sin\delta_0(s')\hat M_0(s')}{|\Omega_0(s')|(s'-s-i\eps)}\Biggr\}\,,\notag\\
 M_1(s) &= \Omega_1(t)\Biggl\{\beta_1 s + \gamma_1 s^2 + \frac{s^3}{\pi}\int_{4\mpi^2}^{\infty}
 \frac{\diff s'}{{s'}^3} \frac{\sin\delta_1(s')\hat M_1(s')}{|\Omega_1(s')|(s'-s-i\eps)}\Biggr\} \,, \notag\\
 M_2(s) &= \Omega_2(s)\Biggl\{ \frac{s^2}{\pi}\int_{4\mpi^2}^{\infty}\frac{\diff s'}{{s'}^2} \frac{\sin\delta_2 (s')\hat M_2(s')}{|\Omega_2(s')|(s'-s-i\eps)}\Biggr\}\,. \label{eq:DR6}
\end{align}

The subtraction constants are unknown and have to be determined using a combination of experimental information and theory input.  In particular since the overall normalization multiplies $1/Q^2$, the quantity that should be extracted from the analysis, it cannot be obtained from data alone and one has to match to $\chi$PT.  On the other hand, this matching has to be performed in such a way that the problematic convergence of the chiral expansion is not transferred directly to the dispersive representation.  This can be achieved as follows:
As shown in Fig.~\ref{fig:Adler}, the amplitude possesses so-called Adler zeros~\cite{Adler:1964um}. 
They are the result of a soft-pion theorem stating that in the $\SU(2)$ chiral limit, the decay amplitude has two zeros for $p_{\pi^+} \to 0$ ($s = u = 0$) 
and $p_{\pi^-} \to 0$ ($s = t = 0$). 
Moving away from the chiral limit (but keeping $\mpip = \mpiz \equiv \mpi$), 
the positions of the Adler zeros are shifted by a contribution of the order of $\mpi^2$ to 
$s = u = 4\mpi^2/3$, $t = \meta^2 + \mpi^2/3$ and $s = t = 4\mpi^2/3$, $u = \meta^2 + \mpi^2/3$
(at tree level).
As one expects from an $\SU(2)$ soft-pion theorem, the corrections to the position of the 
Adler zero are of order $\mpi^2$, since the symmetry forbids large $\O(m_s)$ contributions. 
At one-loop order, the real part has Adler zeros at $s = u = 1.35\,\mpi^2$ and $s = t = 1.35\,\mpi^2$, 
where also the imaginary part of the amplitude is small. 
As emphasized in~Refs.~\cite{Anisovich:1996tx,Walker:1998zz},
this is a particularly appropriate kinematical point where to match the amplitude to $\chi$PT.

The dispersive analyses are usually performed in the isospin limit except for the phase space integrals. However in the case of $\eta \to 3\pi$, at the experimental accuracy reached, the electromagnetic interaction cannot be ignored. In particular it was shown in Ref.~\cite{Ditsche:2008cq} that the electromagnetic self-energy of the charged pions has a nonnegligible effect on the amplitude in the isospin limit. These effects have been accounted for in the analyses of Refs.~\cite{Ditsche:2008cq,Schneider:2010hs}. In the latest dispersive analysis~\cite{Colangelo:2016jmc,Colangelo:2018jxw} it was shown that a substantial part of the electromagnetic corrections can be captured in constructing a kinematic map that takes the physical phase space of the decay $\eta \to \pi^+ \pi^- \pi^0$ into the isospin symmetric one where the dispersive analysis is performed. There electromagnetic effects have been accounted for using the one-loop $\chi$PT corrections~\cite{Ditsche:2008cq}. 

\begin{table}[t!]
	\renewcommand{\arraystretch}{1.3}
		\begin{center}\begin{tabular}{llr}	\toprule
									& \hspace{0.3em} $Q$						& \\ \midrule
		Gasser \& Leutwyler (1975)		& 30.2				& \cite{Gasser:1974wd} \\
		Weinberg (1977)				& 24.1				& \cite{Weinberg:1977hb} \\ 
		Gasser \& Leutwyler (1985)		& 23.2(1.8)			& \cite{Gasser:1984pr} \\
		Donoghue et al.	(1993)		& 21.8				& \cite{Donoghue:1993hj} \\ 
		Kambor et al.\ (1996)				& 22.4(9)				& \cite{Kambor:1995yc} \\ 
		Anisovich \& Leutwyler (1996)	  	& 22.7(8)				& \cite{Anisovich:1996tx} \\
		Walker (1998)					& 22.8(8)				& \cite{Walker:1998zz} \\ 
		Martemyanov \& Sopov (2005)		& 22.8(4)				& \cite{Martemyanov:2005bt} \\ 
		Bijnens \& Ghorbani	 (2007) 		& 22.9				& \cite{Bijnens:2007pr} \\   
		Kastner \& Neufeld (2008)			& 20.7(1.2)			& \cite{Kastner:2008ch} \\ 
		Kampf et al.\ (2011)				& 23.1(7)				& \cite{Kampf:2011wr} \\ 
		JPAC (2017)					& 21.6(1.1)			& \cite{Guo:2016wsi} \\ 
		Albaladejo \& Moussallam (2017) 	& 21.5(1.0) 			& \cite{Albaladejo:2017hhj} \\ 
		Colangelo et al.\ (2018)		& 22.1(7)				& \cite{Colangelo:2018jxw}\\
		FLAG ($N_f = 2$) (2019)		& 24.3(1.5)				& \cite{Aoki:2019cca} \\ 
		FLAG ($N_f = 2 + 1$) (2019)		& 23.3(5)				& \cite{Aoki:2019cca} \\ 
		FLAG ($N_f = 2 + 1 + 1$) (2019) 	& 24.0(8)			& \cite{Aoki:2019cca} \\ 
		\bottomrule
	\end{tabular} \end{center} 
	\caption{Theoretical results for the quark mass ratio $Q$ (statistical and systematic uncertainties added in quadrature). Note that the $\mathcal{O} (p^6)$
    value (Bijnens \& Ghorbani (2007)) does not agree with the number given in Ref.~\cite{Bijnens:2007pr}, which contains a misprint~\cite{Lanz:2018mhm}. Table adapted from Refs.~\cite{Lanz:2018mhm, Colangelo:2018jxw}.}  
	\label{tab:resultsQ}
\end{table}

\subsubsection{Results}
\label{sec:eta3pi_res}
\begin{figure}[t!]
\centering
\includegraphics[width=0.6\textwidth]{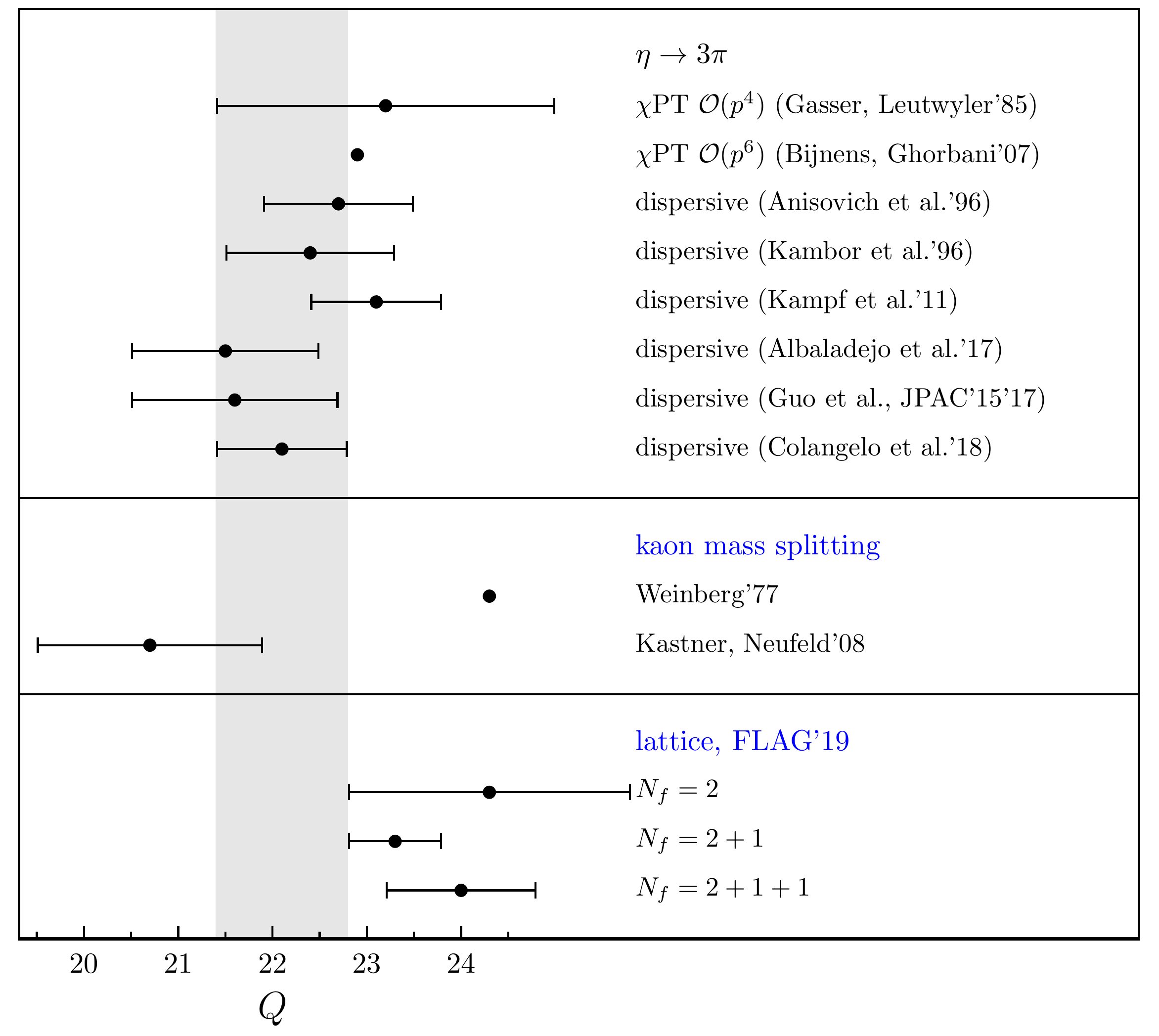}
\caption{\label{fig:results_etaQ}
Different determinations of the light quark mass double ratio, see text. Adapted from Ref.~\cite{Lanz:2018mhm}.}
\end{figure}

Several theoretical analyses of $\eta\to3\pi$ have been published over the last few years. 
Table~\ref{tab:resultsQ} and Fig.~\ref{fig:results_etaQ} summarize the results on the extraction of $Q$ from the different analyses; different theoretical determinations of the slope $\alpha$ in the neutral channel were already compared in Fig.~\ref{fig:alpha}.

In Ref.~\cite{Schneider:2010hs} a nonrelativistic effective field theory to two-loop accuracy has been used. This analysis is very well suited to study the dynamics of the final-state
interaction and allows one to include all isospin-violating and electromagnetic corrections in a consistent framework. 
Several dispersive analyses have been performed~\cite{Kampf:2011wr,Guo:2016wsi, Colangelo:2016jmc,Colangelo:2018jxw, Albaladejo:2017hhj}. 
They all rely on the ingredients described in Sect.~\ref{sec:eta3pi_theo} with some differences we comment on below. 
In Ref.~\cite{Kampf:2011wr} the dispersive representation was matched to $\chi$PT at the NNLO level for the first time. This matching was performed along the line $t=u$ in a region where the differences between NLO and NNLO amplitudes are small. 
The main drawback of this approach is that the fitted amplitude is quite far from fulfilling the Adler zero condition, see Fig.~\ref{fig:Adler}. 
The analysis of the JPAC group~\cite{Guo:2015zqa,Guo:2016wsi} uses a different technique to solve the dispersion relation, called the Pasquier inversion~\cite{Pasquier:1968zz,Aitchison:1978pw}.
Moreover, the left-hand cut is approximated using a Taylor series in the physical region and the number of subtraction constants is reduced from six to three.  
Their result is then matched to NLO $\chi$PT near the
Adler zero, see Fig.~\ref{fig:Adler}, to extract a value for $Q$. 
The analysis of Refs.~\cite{Colangelo:2016jmc,Colangelo:2018jxw} is a modern update of the approach of Anisovich and Leutwyler~\cite{Anisovich:1996tx}. There a matching to NLO and NNLO $\chi$PT has been performed. Moreover, electromagnetic and isospin-breaking corrections have been taken into account. Fits to experimental data by KLOE~\cite{Anastasi:2016cdz}, but also to the recent neutral-channel Dalitz plot by A2~\cite{Prakhov:2018tou} have been explored. 
\begin{figure}
\centering
\includegraphics[width=0.55\textwidth]{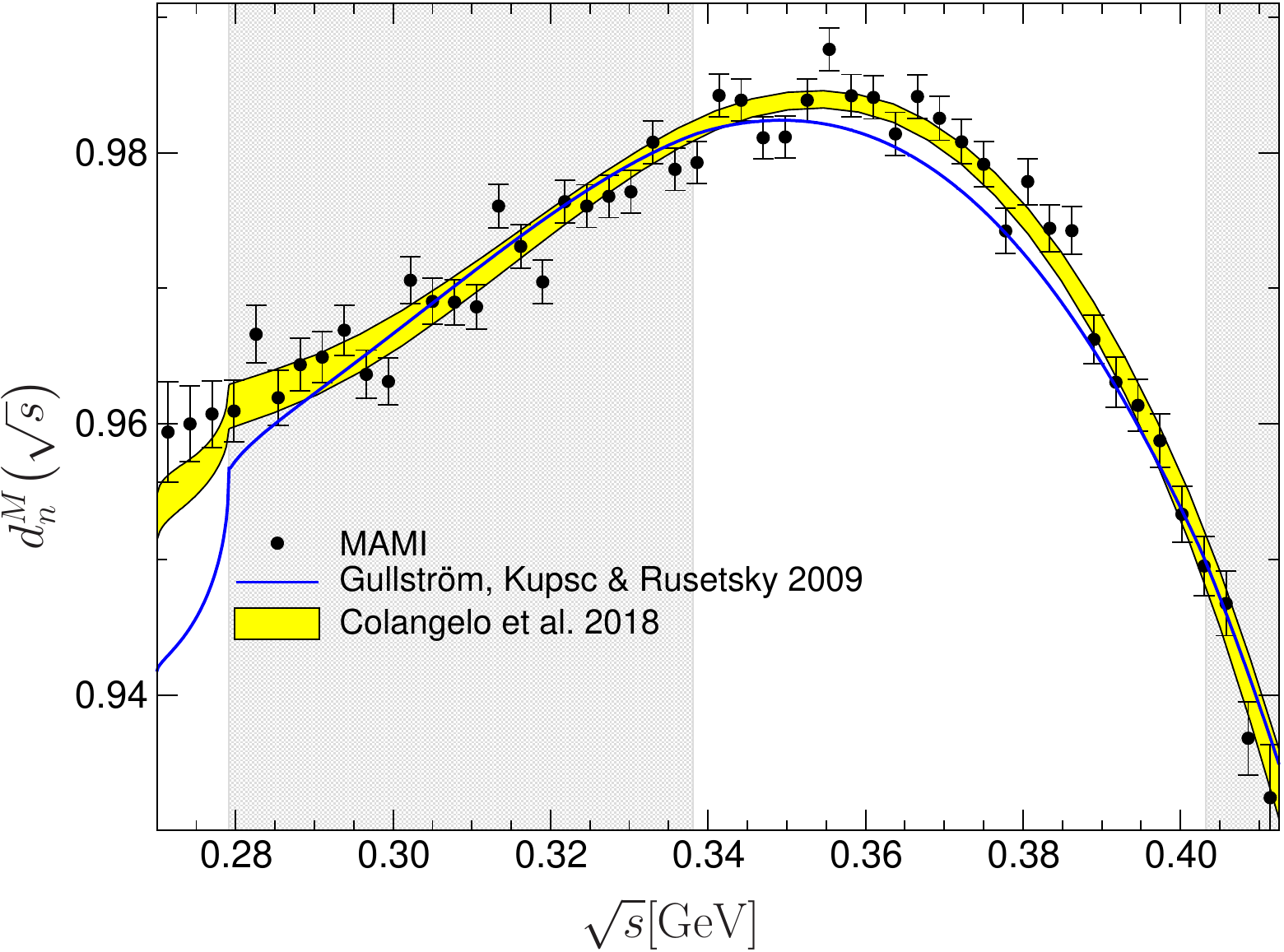}
\caption{\label{fig:results_eta3pi0} Neutral-channel distribution in the dipion invariant mass $\sqrt{s}$ in comparison to the A2 data~\cite{Prakhov:2018tou}, predicted from the analysis of Ref.~\cite{Colangelo:2018jxw}; see there for precise definitions. The shaded areas indicate the cusp-free regions. Figure adapted from Ref.~\cite{Colangelo:2018jxw}.}
\end{figure}
Figure~\ref{fig:results_eta3pi0} shows the prediction for the neutral channel and a comparison to the A2 data: the agreement is very good. 
Finally the analysis of Ref.~\cite{Albaladejo:2017hhj} studies the impact of inelasticities on the dispersive integrals, Eq.~\eqref{eq:DR6}. To this end the inelastic channels $\eta \pi$ and $K \bar K$ have been included. 
Overall, the extracted values of $Q$ agree very well between the different dispersive analyses, as can be seen in Fig.~\ref{fig:results_etaQ}.

The quark mass ratios can also be obtained from \textit{ab initio} calculations using lattice QCD. 
Until recently, lattice QCD simulations were obtaining only the average of the up and down quark masses at or near its physical value. In order to extract the quark mass ratio $Q$, one has to face the 
challenge of calculating the light-quark mass difference $m_u - m_d$. This quantity is more difficult to obtain since it is a small effect on hadron masses, expected to be of comparable size as the leading electromagnetic corrections that were usually not included in lattice simulations. 
Important progress has been achieved recently to include QED corrections. 
This is carried out following mainly two methods. In the first one, QED is added directly to the action and simulations are performed at a few values of the electric charge, as for example in Refs.~\cite{Horsley:2015vla,Fodor:2016bgu,Basak:2018yzz}. In the second method developed by RM123~\cite{deDivitiis:2013xla,Giusti:2017dmp}, electromagnetic effects are included at leading order in  
$\alpha_{\text{em}}$ through simple insertions of the fundamental electromagnetic interaction in quark lines of relevant Feynman graphs. 
Lattice results have reached remarkable precision; an extended discussion can be found in Ref.~\cite{Aoki:2019cca} and their averages for $Q$ are reported in Fig.~\ref{fig:results_etaQ} for $N_f = 2$, $N_f=2+1$, and $N_f = 2+1+1$ dynamical flavors. As emphasized in Ref.~\cite{Colangelo:2018jxw}, lattice estimates for $Q$ tend to be higher by $\sim 1.5 \sigma$, suggesting sizeable corrections $\Delta Q = \mathcal{O} \big( m^2_{\rm quark} \big)$ to the low-energy theorem relating $Q$ to the meson masses~\cite{Gasser:1984pr},
\begin{equation}\label{eq:DeltaQ}
 Q^2 (1+\Delta Q)= \frac{M_K^2}{M_\pi^2} \frac{(M_K^2-M_\pi^2)}{\big(M_{K^0}^2- M_{K^+}^2\big)_{\rm QCD}}  \,.
\end{equation}
This would be surprising in particular with respect to the chiral behavior of two other, closely related quark mass ratios, $S=m_s/\hat{m}$ and $R=(m_s-\hat{m})/(m_d-m_u)$, both of which can be expressed in terms of pion and kaon masses at leading order in analogy to Eq.~\eqref{eq:DeltaQ}.  The corresponding higher-order corrections $\Delta S$ and $\Delta R$ are small, of the order of $5\%$~\cite{Colangelo:2018jxw}, despite scaling linearly (not quadratically) with the quark masses.  Why $\Delta Q$ should be numerically enhanced with respect to these two while simultaneously being suppressed in the chiral expansion is an issue that remains to be investigated. 

\begin{figure}[t!]
\centering
\includegraphics[width=0.4\textwidth]{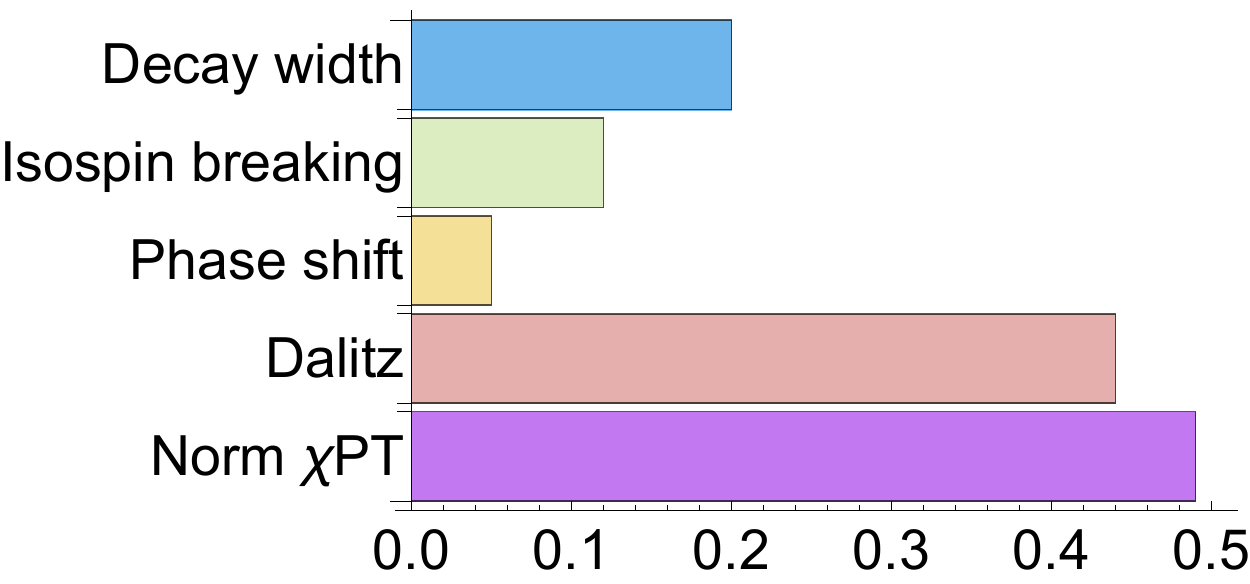}
\caption{\label{fig:results_eta-Qerror} Assessment of the different uncertainties on the quark mass double ratio $Q$, $\delta Q = \pm 0.72_{\rm total} = \pm 0.49_{{\rm Norm~}\chi{\rm PT}} \pm 0.44_{\rm Dalitz} \pm 0.05_{\rm Phase} \pm 0.12_{\rm IB} \pm 0.20_{\Gamma}$.
The final uncertainty due to the partial decay width, $\pm0.20_\Gamma$, 
is fully dominated by the one in the total width $\Gamma_\eta$, with the branching ratio $\BR(\eta\to\pi^+\pi^-\pi^0)=\Gamma(\eta\to\pi^+\pi^-\pi^0)/\Gamma_\eta$
giving a negligible contribution ($\pm 0.067_{\BR}$).}
\end{figure}

In Refs.~\cite{Colangelo:2016jmc,Colangelo:2018jxw} a careful analysis of all sources of uncertainties entering the extraction of $Q$ has been performed. Figure~\ref{fig:results_eta-Qerror} summarizes their finding. 
The authors find $Q= 22.04(72)$ from $\eta \to \pi^+ \pi^- \pi^0$. The main contribution to the uncertainty comes, in almost equal parts, from the $\chi$PT normalization ($\pm 0.49$) and the Gaussian error of the fit to the Dalitz distribution ($\pm 0.44$). The uncertainty coming from the phase shifts used as input is very small ($\pm 0.05$), while the one due to the treatment of isospin breaking for the charged channel is more important ($\pm 0.12$). This uncertainty is much smaller if we consider the neutral channel ($0.12 \to 0.04$). 
The uncertainty coming from the decay width is $\pm 0.20$. Note that for this evaluation the PDG average for $\Gamma(\eta \to 2 \gamma)$ has been used. It relies on the collider measurements of this partial width only.  If the Primakoff measurement were to be taken instead, this would amount to a shift by $\sim 2\sigma$ for the central value of $Q$ as we will see in Sect.~\ref{sec-12Pri}. A new and more precise Primakoff measurement will therefore be important to have. 
Moreover more precise experimental measurements of the $\eta \to 3\pi$ Dalitz plot distribution will help to reduce the uncertainty on $Q$. 

\subsection{$\eta'\to\eta\pi\pi$}
\label{sec:etap-etapipi}
\subsubsection{Motivation}
Due to the $U(1)_A$ anomaly, the $\eta'(958)$ is not a Goldstone boson in the chiral limit of QCD, and is significantly
heavier than the $\eta$ also in the real world.  The $\eta'$, in contrast to the 
$\eta$, decays 
strongly even in the isospin limit,
via the channels $\eta'\to\eta\pi\pi$ and $\eta'\to 4\pi$,
which renders the width of the $\eta'$ two orders of magnitude larger than the one of the $\eta$.
However, a combination of small phase space (for the $\eta\pi\pi$ decays) and suppression due to Bose symmetry, angular
momentum conservation, and high multiplicity of the final state (for the $4\pi$ decays) still leaves the $\eta'$ 
more long-lived than, e.g., the $\omega(782)$ or the $\phi(1020)$. 

The decay $\eta' \to \eta \pi \pi$ is interesting for several reasons. First, due to the quantum
numbers of the pseudoscalar mesons involved, the resonances featuring most prominently therein are the 
scalars: $G$-parity prevents vectors from contributing. Therefore, this
process is particularly suitable for an analysis of the properties of the scalar 
$f_0(500)$ (or $\sigma$) resonance,
even though the $a_0(980)$ is also present and, in fact, widely believed to be dominant. 
Second, the presence of both $\eta$ and $\eta'$ 
is ideal for studying the mixing properties of these two mesons.
Third, and more generally, this decay allows one to test $\chi$PT and its possible extensions such as
large-$N_c$ $\chi$PT and resonance chiral theory (R$\chi$T). 
In Sect.~\ref{sec:large-Nc}, we have briefly described how $\chi$PT needs to be generalized to a 
simultaneous expansion in small momenta, small quark masses, and $1/N_c$ to describe the
pseudoscalar nonet including the $\eta'$;
$\eta'\to\eta\pi\pi$ is a prime channel to test this framework. 
Finally, this decay is also very important to constrain $\pi \eta$ scattering. 
The $\eta'$ mass is sufficiently small so that the extraction of $\pi \eta$ scattering is not polluted by intermediate states other than $\pi \pi$ scattering in the crossed channel. 

Experimentally, this decay has been measured both in the charged channel $\eta' \to \eta \pi^+ \pi^-$ by VES~\cite{Dorofeev:2006fb} and \mbox{BESIII}~\cite{Ablikim:2010kp,Ablikim:2017irx}
and in the neutral channel $\eta' \to \eta \pi^0 \pi^0$ by GAMS-$4\pi$~\cite{Blik:2009zz},
A2~\cite{Adlarson:2017wlz}, 
and BESIII~\cite{Ablikim:2017irx}. At JLab the CLAS and GlueX 
collaborations will also be able to 
perform precise Dalitz plot measurements in the near future. 

In experimental analyses of the $\eta'\to\eta\pi\pi$ Dalitz plots, one 
defines symmetrized coordinates $X$ and $Y$ according to
\begin{align}\label{eq:xyetapdef}
 X=\frac{\sqrt{3}(t-u)}{2\metap \Qetap}\,,\qquad
 Y=(\meta+2\mpc)\frac{(\metap-\meta)^2-s}{2\mpc \metap \Qetap}-1\,,
\end{align}
where $Q_{\eta'} = M_{\eta'}-M_{\eta}-2M_{\pi}$. The squared decay amplitude is then expanded in terms of these variables,
\begin{align}\label{eq:dalitzparameterization}
 |\A_\mathrm{exp}(X,Y)|^2 =|\N_\mathrm{exp}|^2 \big\{1+aY+bY^2+cX+dX^2 + \ldots \big\}\,,
\end{align}
and the parameters $a$, $b$, $c$, $d$ are fitted to data.  Experimental determinations of these Dalitz plot parameters are summarized in Table~\ref{tab:etapexp}.  Note that Dalitz plot parameters odd in $X$ 
violate $C$-parity and are therefore required to be zero for strong (and electromagnetic) interactions.
\begin{table}
\centering
\renewcommand{\arraystretch}{1.3}
\begin{tabular}{c c c c c c c}
\toprule
		& VES~\cite{Dorofeev:2006fb}&BESIII~\cite{Ablikim:2010kp}& BESIII~\cite{Ablikim:2017irx} &
GAMS-$4\pi$~\cite{Blik:2009zz} & A2~\cite{Adlarson:2017wlz} & BESIII~\cite{Ablikim:2017irx}	\\
& $\eta'\to\eta\pi^+\pi^-$ & $\eta'\to\eta\pi^+\pi^-$ & $\eta'\to\eta\pi^+\pi^-$ & 
$\eta'\to\eta\pi^0\pi^0$ & $\eta'\to\eta\pi^0\pi^0$ & $\eta'\to\eta\pi^0\pi^0$ \\
\midrule
$a$		& $-0.127(16)(8)$&$-0.047(11)(3)$& $-0.056(4)(2)$
& $-0.067(16)(3)$  & $-0.074(8)(6)$ & $-0.087(9)(6)$			\\
$b$		& $-0.106(28)(14)$&$-0.069(19)(9)$ & $-0.049(6)(6)$
& $-0.064(29)(4)$ & $-0.063(14)(5)$ & $-0.073(4)(5)$			\\
$c$		& $+0.015(11)(14)$ &$+0.019(11)(3)$ & $\!\!+0.0027(24)(18)\!\!$
& --- & --- & --- 	\\
$d$		& $-0.082(17)(8)$&$-0.073(12)(3)$ & $-0.063(4)(3)$
& $-0.067(20)(3)$ & $-0.050(9)(5)$ & $-0.074(9)(4)$	\\
\midrule
\# events \hspace*{-2.5mm}	& $\simeq 8\,563$	&$43\,826$	& $351\,016$
& $15\,000$ & $124\,100$	& $56\,249$	 		\\
\bottomrule
\end{tabular}
\renewcommand{\arraystretch}{1.0}
\caption{Experimental results for Dalitz-plot parameters in $\eta'\to\eta\pi\pi$.
Errors refer to statistical and systematic uncertainties, in order. The number of 
$\eta'\to\eta\pi^+\pi^-$ events for the VES collaboration has been estimated from the total number of $\eta'$ events 
and the branching ratio $\BR(\eta'\to\eta\pi^+\pi^-)=42.6(7)\%$~\cite{Tanabashi:2018oca}.}
\label{tab:etapexp}
\end{table}

\subsubsection{Theoretical framework: chiral perturbation theory}
The transition amplitude and kinematic variables of the $\eta'\to\eta\pi\pi$ decay are defined in the usual manner, 
\begin{align}
\langle\pi^i(p_1)\pi^j(p_2)\eta(p_3)|T|\eta'(P_{\eta'})\rangle = (2\pi)^4\delta^{(4)}(P_{\eta'}-p_1-p_2-p_3) \delta^{ij}\A(s,t,u)\,,
\end{align}
where $i,\,j$ refer to the pion isospin indices.  In the strict isospin limit, the amplitudes for charged
and neutral final states are identical, and the partial width or branching ratio for $\eta\pi^+\pi^-$ is
twice that for $\eta\pi^0\pi^0$, cf.\ 
Table~\ref{tab:etaprime}.  
The corresponding Mandelstam variables are given by
\begin{equation}
s = (P_{\eta'} - p_3)^2\,,\quad t = (P_{\eta'} - p_1)^2\,,\quad u = (P_{\eta'} - p_2)^2\,, \quad
s+t+u = \metap^2 + \meta^2 + 2\mpc^2 \equiv 3\setap\,.
\end{equation}

Large-$N_c$ chiral perturbation theory allows the explicit inclusion
of the $\eta'$ meson in an effective-Lagrangian framework. 
Using $\L^{(0)}$, see Eq.~\eqref{L0}, the $\eta' \to \eta \pi \pi$ amplitude has been computed to be~\cite{Cronin:1967jq,Schwinger:1968zz,DiVecchia:1980vpx,HerreraSiklody:1999ss,Fajfer:1987ij,Schechter:1993tc,Singh:1975aq,Bijnens:2005sj}
\begin{equation}
\label{amplLO}
\A_{\eta^\prime\to\eta\pi\pi}^{\chi \rm PT}\Big |_{\rm LO}= \frac{M^2_\pi}{6F^2} \left[2\sqrt{2}\cos(2\theta_P)-\sin(2\theta_P)\right] \,,
\end{equation}
where $F=F_\pi$ (the pion decay constant) at this order.
The prediction at leading order is very far from the experimental measurement:
it results, e.g., in a partial width $\Gamma(\eta'\to\eta\pi^+\pi^-)_{\text{LO}} \approx 2\keV$,
less than 3\% of the experimental value $\Gamma(\eta'\to\eta\pi^+\pi^-)_{\text{exp}}= 83.5(4.1)\keV$.
This can be understood from the factor $M_\pi^2$ in Eq.~\eqref{amplLO}: the amplitude vanishes in the $\SU(2)$ chiral limit.
This suppression is lifted at higher orders; 
indeed the NLO prediction is of the form~\cite{Escribano:2010wt}
\begin{align}
\A_{\eta^\prime\to\eta\pi\pi}^{\chi\text{PT}}\Big|_{\rm NLO}(s,t,u)&=
\frac{c_{qq}}{F^2}\left[\frac{M_\pi^2}{2}+
\frac{2(3L_2+ L_3)}{F_\pi^2}\left(s^2+t^2+u^2-M^4_{\eta^\prime}-M^4_{\eta}-2 M^4_{\pi}\right)
\right.\nonumber\\
& \left.
-\frac{2 L_5}{F_\pi^2}\left(M^2_{\eta^\prime}+M^2_{\eta}+2 M^2_{\pi}\right)M^2_{\pi}
+\frac{24 L_8}{F_\pi^2}M^4_{\pi}+\frac{2}{3}\Lambda_2 M^2_{\pi}\right]
+\frac{c_{sq}}{F^2}\frac{\sqrt{2}}{3}\Lambda_2 M^2_{\pi} \,,
\label{amplLN}
\end{align}
where $c_{qq}$ and $c_{sq}$ are functions of the octet and singlet decay constants $F_{8/0}$, as well as the two mixing angles $\theta_{8/0}$ required in the 
$\eta$--$\eta'$ mixing scheme at NLO~\cite{Kaiser:2000gs}:
\begin{align}
c_{qq}&=
-\frac{F^2}{3 F_8^2 F_0^2 \cos^2(\theta_8-\theta_0)}
\left[2F_8^2\sin(2\theta_8) - F_0^2 \sin(2\theta_0) - 2\sqrt{2} F_8 F_0\cos(\theta_8+\theta_0)\right] \,,
\nonumber\\
c_{sq} &=
-\frac{F^2}{3 F_8^2 F_0^2 \cos^2(\theta_8-\theta_0)}
\left[\sqrt{2} F_8^2\sin(2\theta_8) + \sqrt{2} F_0^2\sin(2\theta_0) +
F_8 F_0\cos(\theta_8+\theta_0)\right] \,.
\label{eq.cqq-mixing}
\end{align}
Due to the smallness of the pion mass, the NLO amplitude Eq.~\eqref{amplLN} is dominated by the term $\propto(3L_2+ L_3)$.
The resulting prediction of the partial width in large-$N_c$ $\chi$PT thus depends strongly on the precise values for these LECs;
typically it is enhanced over the LO prediction by an order of magnitude~\cite{Escribano:2010wt}, but still too small compared 
to experiment.
As for $\eta \to 3 \pi$, see Sect.~\ref{sec:eta-3pi}, this problem can partly be traced back to the strong rescattering of the final-state particles. 
Therefore either a unitarization of the amplitude or a dispersive approach seems appropriate to describe this decay accurately. 
Alternatively, the enhancement of the experimental results compared to the $\chi$PT prediction may be thought of as
the low-energy tails of (scalar)
resonance contributions, and one may try to learn about their nature via precise decay data. 
To this end, $\eta' \to \eta \pi \pi$ can also be described using the R$\chi$T framework,
which includes resonances and their couplings to the Goldstone bosons explicitly~\cite{Escribano:2010wt}. 

In Ref.~\cite{Escribano:2010wt} the unitarization of the NLO amplitude has been performed using a $K$-matrix formalism, 
while in Ref.~\cite{Isken:2017dkw} a dispersive 
approach following the method described in Sect.~\ref{sec:eta3pi_theo}, including $\pi \pi$ and $\pi\eta$ rescattering, 
has been pursued. We briefly describe the latter in the following subsection. 
In Ref.~\cite{Gonzalez-Solis:2018xnw} an $N/D$ approach for the unitarization has been used. 
While this framework restricts the analysis to pairwise meson rescattering, 
in contrast to the Khuri--Treiman approach~\cite{Isken:2017dkw}, it 
allows the authors to include $\pi \pi$ $D$-wave rescattering that was suggested 
to be important for this process~\cite{Gonzalez-Solis:2018xnw}. 

\subsubsection{Theoretical framework: dispersive representation}\label{sec:etap-etapipi-dispersive}

The decomposition of the $\eta'\to\eta\pi\pi$ decay amplitude can be performed 
in analogy to Eq.~\eqref{eq:Mdecomp}, following the reconstruction theorem.  If we neglect the $\pi\eta$ $P$-wave,
as it has exotic quantum numbers and hence final-state-interaction effects are expected to be very small at low energies,
it simply reads
\begin{equation}
	\label{eq:decompositionSWaves}
	\A_{\eta'\to\eta\pi\pi}(s,t,u) = \M_0^0(s) + \M_0^1(t) + \M_0^1(u) \, ,
\end{equation}
where the single-variable functions $\M_\ell^I$ denote amplitudes of angular momentum $\ell$ and isospin $I$, 
hence $I=0$ refers to the $\pi\pi$ and $I=1$ to the $\pi\eta$ amplitude.
The minimally-subtracted Omn\`es-type solution for the functions $\M_0^I$ contains three free subtraction constants and 
is given by~\cite{Isken:2017dkw}
\begin{align}
 \M_0^0(s) &= \Omega_0^0(s)\Biggl\{\alpha+\beta \frac{s}{\metap^2}  + \frac{s^2}{\pi}\int_{4\mpi^2}^{\infty}\frac{\diff s'}{{s'}^2} \frac{\hat \M_0^0(s')\sin\delta_0^0(s')}{|\Omega_0^0(s')|(s'-s-i\eps)}\Biggr\}\,,\notag\\
 \M_0^1(t) &= \Omega_0^1(t)\Biggl\{\gamma \frac{t}{\metap^2} + \frac{t^2}{\pi}\int_{(M_\pi+M_\eta)^2}^{\infty}\frac{\diff t'}{{t'}^2} \frac{\hat \M_0^1(t')\sin\delta_0^1(t')}{|\Omega_0^1(t')|(t'-t-i\eps)}\Biggr\} \,. \label{eq:etap-etapipi_disp}
\end{align}
The precise form of the inhomogeneities $\hat \M_0^I$ is given in Ref.~\cite{Isken:2017dkw}.
The required $\pi\eta$ $S$-wave phase shift is not known with the same precision as the low-energy
$\pi\pi$ scattering phase shifts; here, it is taken from an analysis of the $\pi\eta$ scalar form 
factor~\cite{Albaladejo:2015aca},
built on a $\pi\eta$--$K\bar K$ coupled-channel $T$-matrix with chiral constraints as well as experimental information 
on the $a_0(980)$ and $a_0(1450)$ resonances.
An improvement of our theoretical understanding of $\pi\eta$ scattering, e.g., using data 
from heavy-meson decays~\cite{Albaladejo:2016mad}, would be highly desirable.
The three-parameter scheme of Eq.~\eqref{eq:etap-etapipi_disp} describes the VES~\cite{Dorofeev:2006fb}
and BESIII~\cite{Ablikim:2010kp} Dalitz plots very well, with $\chi^2/\text{ndof}$ of $0.92$ and $1.06$, respectively;
this is a nontrivial result, as three parameters reproduce four pieces of experimental information: the three Dalitz-plot
parameters $a$, $b$, and $d$, as well as the partial decay width.  This result can be expressed in terms 
of a constraint between the three leading Dalitz-plot parameters~\cite{Isken:2017dkw}.  
On the other hand, the uncertainties in the subtraction 
constants extracted from fits to experimental data are very large, 
and dominated by the error bands assigned to the effective ($\pi\pi$ and $\pi\eta$) phase input
above the $K\bar K$ threshold.  To suppress this source of uncertainty more efficiently, an alternative subtraction scheme
can be introduced~\cite{Isken:2017dkw}:
\begin{align}
 \M_0^0(s) &= \Omega_0^0(s)\Biggl\{\alpha_0+\beta_0 \frac{s}{\metap^2} +\gamma_0 \frac{s^2}{\metap^4}  + \frac{s^3}{\pi}\int_{4\mpi^2}^{\infty}\frac{\diff s'}{{s'}^3} \frac{\hat \M_0^0(s')\sin\delta_0^0(s')}{|\Omega_0^0(s')|(s'-s-i\eps)}\Biggr\}\,,\notag\\
 \M_0^1(t) &= \Omega_0^1(t)\Biggl\{\gamma_1 \frac{t^2}{\metap^4} + \frac{t^3}{\pi}\int_{(M_\pi+M_\eta)^2}^{\infty}\frac{\diff t'}{{t'}^3} \frac{\hat \M_0^1(t')\sin\delta_0^1(t')}{|\Omega_0^1(t')|(t'-t-i\eps)}\Biggr\} \,, \label{eq:etap-etapipi_disp2}
\end{align}
where, despite the higher degree of subtraction, only one additional constant needs to be introduced.
Fits with the amplitude representation~\eqref{eq:etap-etapipi_disp2} are therefore less predictive, but the 
resulting uncertainty is then dominated by the experimental data employed.

\begin{figure}
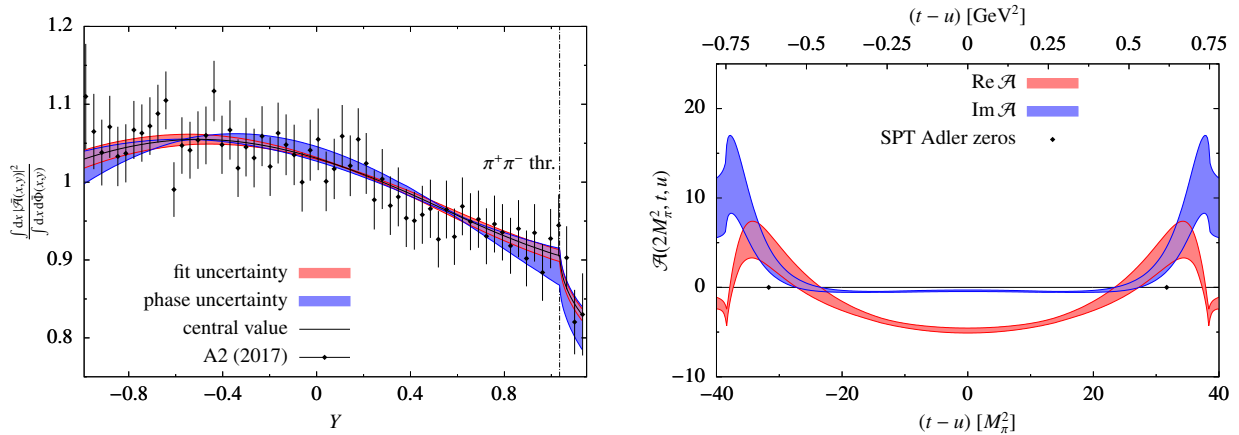

\centering\large
\scalebox{0.64}{\input{BES_Yproj_neutral_DR3}} \hfill
\scalebox{0.64}{\input{BES_adler_zero_DR3}}
\caption{\textit{Left:} Prediction of the (phase-space-normalized) \new{$Y$}-projection of the $\eta'\to\eta\pi^0\pi^0$ 
Dalitz plot distribution~\cite{Isken:2017dkw},
in comparison to the A2 data~\cite{Adlarson:2017wlz}.  
The three subtraction constants according to Eq.~\eqref{eq:etap-etapipi_disp} have been fitted to BESIII data
on the charged final state~\cite{Ablikim:2010kp}.
The vertical dash-dotted line marks the $\pi^+\pi^-$ threshold.
\textit{Right:} Real (red) and imaginary (blue) parts of the $\eta'\to\eta\pi\pi$ decay amplitude as a function of 
$(t-u)$ for fixed $s=2M_\pi^2$. The existence of two Adler zeros slightly shifted from the original soft-pion points
is clearly visible.  
Figures courtesy of T.~Isken.}\label{fig:etap-etapipi}
\end{figure}

As the dispersive amplitude representation does not lead to a polynomial Dalitz plot distribution, it allows one 
to predict higher-order (cubic, quartic) terms in the expansion~\eqref{eq:dalitzparameterization}.  A clear 
hierarchy was observed, with the two cubic Dalitz plot parameters suppressed by almost an order of magnitude
compared to $a$, $b$, and $d$, and the three quartic terms even more strongly; future high-statistics experiments
ought to test these predictions.  Furthermore, fits based on data for $\eta'\to\eta\pi^+\pi^-$~\cite{Dorofeev:2006fb,Ablikim:2010kp} allow for predictions for the neutral channel $\eta'\to\eta\pi^0\pi^0$~\cite{Adlarson:2017wlz,Ablikim:2017irx},
including the leading isospin-breaking effects such as the cusp in the $\pi^0\pi^0$ invariant mass distribution
at the $\pi^+\pi^-$ threshold~\cite{Kubis:2009sb}; see Fig.~\ref{fig:etap-etapipi} (left).
The cusp strength allows one to extract the $\pi\pi$ scattering length combination
$a_0^0-a_0^2$, much in analogy to what was done in $K^+\to\pi^+\pi^0\pi^0$ decays
(see Ref.~\cite{Gasser:2011ju} and references therein). 
Keeping the polynomial Dalitz plot parameters fixed, Ref.~\cite{Adlarson:2017wlz}
finds $a_0^0-a_0^2 = 0.255(90)$ and $a_0^0-a_0^2 = 0.262(58)$, based on different data subsets,
to be compared to the theoretical value $a_0^0-a_0^2 = 0.264(5)$~\cite{Colangelo:2001df}; hence 
the perturbation of the $\pi^0\pi^0$ spectrum due to the unitarity cusp is perfectly in agreement with 
theoretical expectations, though not yet competitive in accuracy with the best determinations of the 
scattering lengths.

Finally, the dispersive representation fitted to data predicts the existence of an Adler zero in the amplitude
at the unphysical point $s=2M_\pi^2$, $t=\metap^2$, $u=\meta^2$ (as well as the symmetric point $t\leftrightarrow u$).
The amplitude at these soft-pion points is protected by $\SU(2)\times \SU(2)$ symmetry, however, the survival of 
the corresponding amplitude suppression has been disputed due to the close proximity of the $a_0(980)$ 
resonance~\cite{Deshpande:1978iv}.  The dispersive analysis shows that the position of the Adler zeros is merely
slightly shifted~\cite{Isken:2017dkw}, see Fig.~\ref{fig:etap-etapipi} (right), where real and imaginary part
are plotted as a function of $(t-u)$ for fixed $s=2M_\pi^2$.  Given that the Dalitz plot is restricted
to values $|t-u|\lesssim 10 M_\pi^2$ (and $s\geq4M_\pi^2$), the stability of the extrapolation beyond the physical region
is quite remarkable and a great asset of the dispersive representation.

The attempt to match the subtraction constants determined from experiment to the next-to-leading-order large-$N_c$ $\chi$PT
or the R$\chi$T amplitude proved to be inconclusive so far, and may require further theoretical work.

\subsection{$\eta'\to3\pi$}
\label{sec:etap-3pi}

Due to the close similarity of $\eta$ and $\eta'$ in all properties except their masses, much of what has been said
about the decays $\eta\to3\pi$ applies equally to $\eta'\to3\pi$: they violate conservation of isospin; 
electromagnetic effects that could provide isospin breaking are strongly suppressed due to Sutherland's 
theorem~\cite{Sutherland:1966zz,Bell:1996mi}; as a consequence, they are almost exclusively caused by the 
light quark mass difference $m_u-m_d$.  Indeed, the partial decay widths of all four decays are of the 
same order of magnitude (of a few hundred eV).  However, as the width of the $\eta'$ is larger than the one of the $\eta$
by about a factor of 150, the branching ratios for $\eta'\to3\pi$ are much smaller, and hence precise 
experimental investigations of these decays are much more recent than in the case of the $\eta$
(the decay $\eta'\to\pi^+\pi^-\pi^0$ was only established by the CLEO collaboration in 2008~\cite{Naik:2008aa}).
Detailed investigations of the decay dynamics are exclusively due to the BESIII collaboration, 
who have measured the relative branching ratios most precisely~\cite{Ablikim:2015cob},
studied the $\eta'\to3\pi^0$ Dalitz plot for the first time~\cite{Ablikim:2015cmz},
and performed an amplitude analysis for both charged and neutral final states~\cite{Ablikim:2016frj}.
Interestingly enough, the branching ratio into $3\pi^0$ still shows some tension: while the
PDG average is $\BR(\eta'\to 3\pi^0) = 3.57(26)\times 10^{-3}$~\cite{Tanabashi:2018oca}, dominated by
Ref.~\cite{Ablikim:2016frj}, the PDG fit of various $\eta'$ branching ratios is quoted as
$\BR(\eta'\to 3\pi^0) = 2.54(18)\times 10^{-3}$; in particular measurements of 
$\BR(\eta'\to 3\pi^0)/\BR(\eta'\to \eta\pi^0\pi^0)$ by GAMS-$4\pi$~\cite{Blik:2008zz} seem to point towards smaller values.
An independent confirmation of the BESIII result would therefore be most desirable.
The branching ratio into the charged final state,
$\BR(\eta'\to \pi^+\pi^-\pi^0) = 3.61(18)\times 10^{-3}$~\cite{Tanabashi:2018oca}, has no such issues.

In Ref.~\cite{Ablikim:2015cmz}, a polynomial distribution analogous to Eq.~\eqref{eq:alpha} has been fitted to 
the $\eta'\to3\pi^0$ Dalitz plot, yielding a slope parameter $\alpha = -0.640(46)_{\text{stat}}(47)_{\text{syst}}$, 
consistent with (but significantly more precise than) a previous GAMS-$4\pi$ determination
$\alpha = -0.59(18)$~\cite{Blik:2008zz}.  The slope therefore has the same sign as in $\eta\to3\pi^0$, but is
larger in magnitude by more than one order.
Ref.~\cite{Ablikim:2016frj} performed a combined amplitude analysis of both final states, employing an isobar model
for the isospin $I=0$ $\pi\pi$ $S$-wave and the $I=1$ $P$-wave.  
This work established a nonvanishing $P$-wave contribution for the first time, expressed as a branching fraction $\BR(\eta'\to\rho^\pm\pi^\mp) = 7.4(2.3)\times10^{-4}$.
However, in comparison to Khuri--Treiman analyses of $\eta\to3\pi$ (see Sect.~\ref{sec:eta3pi_theo}), no phase universality for the final-state interactions is enforced, no third-pion rescattering is accounted for, and the (nonresonant) $I=2$ $S$-wave is omitted altogether.

Unfortunately, no sophisticated theoretical analysis of these latest experimental results exists to date.
The most advanced calculation has been performed in Ref.~\cite{Borasoy:2005du}
in the framework of unitarized $U(3)$ chiral effective field theory.
Final-state interactions are implemented solving a Bethe--Salpeter equation in the spectator approximation, 
i.e., taking into account pairwise rescattering only.  The rather large number of low-energy constants has been 
fitted simultaneously to pion--pion scattering phase shifts
as well as decay data on $\eta\to3\pi$, $\eta'\to\eta\pi\pi$, and $\eta'\to3\pi$, where no data on the last channel
except a partial decay rate for $\eta'\to3\pi^0$ (far less accurate than today) was available at the time.
The fits only produced different clusters of Dalitz plot parameters for $\eta'\to3\pi$ and no very clear picture.
A later update by the same authors including the $\eta'\to\eta\pi^+\pi^-$ Dalitz plot data by the VES
collaboration~\cite{Dorofeev:2006fb} led to a prediction of the $\eta'\to\pi^+\pi^-\pi^0$ partial width
in sharp contradiction with the experimental information we have nowadays~\cite{Borasoy:2006uv}.

Obviously, as the decay $\eta'\to3\pi$ is triggered by the light quark mass difference, the question is whether
independent information on the latter can directly be inferred from the decay dynamics, similar to what we have 
described in detail in Sect.~\ref{sec:eta-3pi} for $\eta\to3\pi$. 
There was a time-honored suggestion that the ratio of branching ratios 
$\BR(\eta'\to\pi^+\pi^-\pi^0)/\BR(\eta'\to\eta\pi^+\pi^-)$ would allow us to access the 
$\pi^0$--$\eta$ mixing angle $\eps = (\sqrt{3}/4)(m_u-m_d)/(m_s-\hat m)$ 
directly~\cite{Gross:1979ur}.  This relied on two crucial assumptions:
\begin{enumerate}
\item The decay amplitude for $\eta'\to\pi^+\pi^-\pi^0$ is given entirely by $\eta'\to\eta\pi^+\pi^-$, followed
by $\pi^0$--$\eta$ mixing: that is, \mbox{$\A_{\eta'\to\pi^+\pi^-\pi^0} = \eps\times \A_{\eta'\to\eta\pi^+\pi^-}$}.
\item Both amplitudes are ``essentially constant'' over the whole Dalitz plot.
\end{enumerate}
Both assumptions have been refuted in Ref.~\cite{Borasoy:2006uv}.  While the $\eta'\to\eta\pi^+\pi^-$ Dalitz plot is 
indeed relatively flat, experimental data now conclusively demonstrates that this is not the case for
$\eta'\to\pi^+\pi^-\pi^0$, which is not very surprising given that the phase space is large enough to allow for both 
the $f_0(500)$ and the $\rho(770)$ resonances to show up within the Dalitz plot.

An extraction of light quark mass ratios therefore likely has to proceed along similar lines as the far more involved 
procedure explained for $\eta\to3\pi$, combining Khuri--Treiman equations with a matching to chiral effective field theories.  
While an amplitude decomposition for $\eta'\to3\pi$ can be written down in strict analogy to Eq.~\eqref{eq:Mdecomp}, the dispersion relations for the different isospin amplitudes cannot be constructed based solely on elastic $\pi\pi$ rescattering in the final state.
Since $\eta\pi$ intermediate states are allowed to go on their mass shell within the physical
decay region, they can also play an important role and contribute via decays $\eta'\to\eta\pi\pi$, followed by isospin-breaking rescattering 
$\eta\pi\to\pi\pi$~\cite{Isken:2017dkw}.
Work along such lines is still in progress~\cite{Isken:2019tdj}.
In addition, the subsequent matching will be complicated by the fact that chiral effective theories involving the $\eta'$
are by far not as firmly established as $\chi$PT for $\eta\to3\pi$.  Further theoretical developments are therefore
required for a quark-mass-ratio extraction from $\eta'$ hadronic decays.

\section{\boldmath Anomalous $\eta$ and $\eta'$ decays and the physics of transition form factors}
\label{sec:radiative}
Precision measurements of radiative decays and transitions of $\pi^0$, $\eta$, and $\eta^\prime$ mesons, arising through the axial anomaly, not only probe low-energy QCD, but provide inputs necessary to characterize many other processes. 
The key quantity for these processes is the pseudoscalar transition form factor (TFF) $F_{P\gamma^*\gamma^*}(q_1^2,q_2^2)$ describing the coupling of $P=\pi^0,\,\eta,\,\eta'$ to two
(virtual) photons.
This is defined by the matrix element
\begin{equation}
i \int \diff^4x \, e^{iq_1x}  \left\langle 0 \left| T \big\{ j_\mu(x)j_\nu(0) \big\} \right| P (q_1+q_2)\right\rangle
= \epsilon_{\mu\nu\alpha\beta} q_1^\alpha q_2^\beta \, F_{P\gamma^*\gamma^*}(q_1^2,q_2^2) \,, \label{eq:defTFF}
\end{equation}
where $j_\mu$ is the electromagnetic current, $q_{1,2}$ represent the photon momenta, and our convention is $\eps^{0123}=+1$.
The simplest decays involving two real photons, $P \to \gamma\gamma$, are important not only from a foundational perspective---establishing the chiral anomaly of QCD---but remain an active target for both theory and experiment toward a precision description of light neutral mesons.
For the $\eta$ meson, $\Gamma(\eta \to \gamma\gamma)$ is used to extract the $\eta$--$\eta'$ mixing angles, as well as serving as a normalization for many other $\eta$ partial widths.
However, there remains a long-standing tension between different experimental methods for determining $\Gamma(\eta \to \gamma\gamma)$---Primakoff production at fixed target experiments versus $e^+e^-$ colliders---that requires further study.

For processes involving virtual photons, TFFs generalize the usual notion of elastic form factors as Fourier transforms of charge distributions, thereby probing hadronic substructure~\cite{Landsberg:1986fd}.\footnote{Due to charge conjugation symmetry, the elastic Coulomb form factors of neutral pseudoscalar mesons ($\gamma^*\to PP$) vanish.}
It has long been known, however, that $\chi$PT alone does not provide an accurate treatment of TFFs, due to the strong influence of vector mesons~\cite{Bijnens:1988kx, Bijnens:1989jb}, and must be supplemented with models~\cite{Ametller:1991jv}, such as R$\chi$T~\cite{Roig:2014uja,Guevara:2018rhj} and quark substructure models~\cite{Landsberg:1986fd,Ito:1991pn,Roberts:1994hh,Frank:1994gc,Anisovich:1996hh,Anisovich:1997dz,Musatov:1997pu,Maris:1999ta,Eichmann:2017wil}.
Below, we mainly focus on dispersive representations, along with VMD, to describe TFFs.
The usefulness of dispersion theory is that the inputs to TFFs are correlated with other physical cross sections and can be treated in a largely data-driven way~\cite{Czerwinski:2012ry,Colangelo:2014pva}.

\begin{figure}
\centering
\includegraphics*[width=0.5\linewidth]{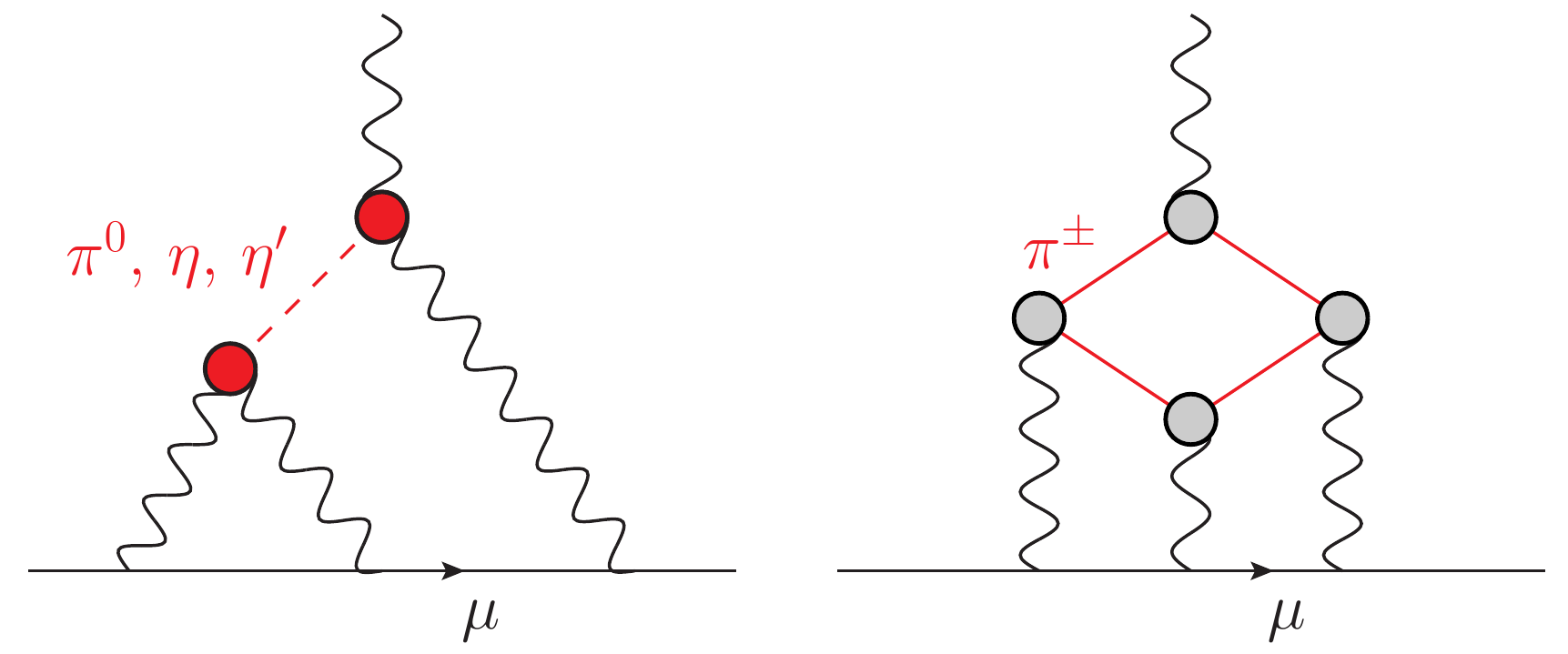}
\caption{Dominant contributions to hadronic light-by-light scattering: pseudoscalar pole terms (left)
and charged-pion loop graphs (right). The red blobs denote pseudoscalar TFFs, while the gray blobs
stand for the pion vector form factor.}\label{fig:hlbl}
\end{figure} 
In recent years, TFFs for $\pi^0$, $\eta$, and $\eta'$ mesons have come to the fore due to their prominence in Standard Model quantum corrections to $(g-2)_\mu$, the anomalous magnetic moment of the 
muon~\cite{Jegerlehner:2009ry,Aoyama:2020ynm}.
These mesons are among the lightest intermediate states to contribute to the so-called hadronic 
light-by-light-scattering in $(g-2)_\mu$, shown diagrammatically in Fig.~\ref{fig:hlbl} (left), where the TFFs enter via the red vertices.
The study of these contributions is part of a broader dispersive analysis of hadronic light-by-light scattering~\cite{Colangelo:2014dfa,Colangelo:2014pva,Colangelo:2015ama}, which includes in parallel pion-loop contributions, including rescattering 
effects~\cite{Colangelo:2017qdm,Colangelo:2017fiz,Hoferichter:2019nlq,Danilkin:2018qfn,Danilkin:2019opj}, shown in Fig.~\ref{fig:hlbl} (right).

\subsection{$\pi^0, \eta, \eta' \to \gamma\gamma$}
\label{sec:eta-2g}
\subsubsection{Theory: matrix elements and $\eta$--$\eta'$ mixing}
The decays of $\pi^0$, $\eta$, and $\eta^\prime$ mesons into two real photons, $P\to\gamma\gamma$, are governed by the chiral anomaly~\cite{Wess:1971yu,Witten:1983tw}, which fixes the overall TFF coupling $F_{P\gamma\gamma} \equiv F_{P\gamma^*\gamma^*}(0,0)$ at $q_1^2 = q_2^2 = 0$.
The decay widths are given by
\beq
\Gamma\left(P\to\gamma\gamma\right) = \frac{\pi\,\alpha_{\text{em}}^2 M_P^3}{4}\big|F_{P\gamma\gamma}\big|^2 \,.
\eeq 
The textbook example is $\pi^0$ decay,
for which a low-energy theorem predicts~\cite{Adler:1969gk,Bell:1969ts,Bardeen:1969md},
\beq
F_{\pi^0\gamma\gamma}=\frac{1}{4\pi^2 \Fpi} \quad \Rightarrow \quad
\Gamma\big(\pi^0\to\gamma\gamma\big)  
= \frac{\alpha_{\text{em}}^2 \mpiz^3}{64\pi^3F_\pi^2} \,; 
\label{eq:Gamma-pi0gg}
\eeq
see also Ref.~\cite{Bernstein:2011bx} for an extensive review.  Higher-order chiral corrections,
including isospin breaking 
(largely due to mixing of the $\pi^0$ with $\eta$ and $\eta'$)~\cite{Bijnens:1988kx,Goity:2002nn} 
and electromagnetic corrections~\cite{Ananthanarayan:2002kj,Kampf:2009tk}, have been worked out.
In fact, the numerically dominant corrections result from $\pi^0$--$\eta$--$\eta'$ mixing, which occur already at leading order in the $\SU(3)$ version of the theory~\cite{Moussallam:1994xp,Goity:2002nn}.
All other corrections are small, partly due to the absence of chiral logarithms in the NLO corrections~\cite{Donoghue:1986wv,Bijnens:1988kx,Kampf:2009tk}.
Altogether, these effects increase the leading-order prediction~\eqref{eq:Gamma-pi0gg} from
$\Gamma(\pi^0\to\gamma\gamma) = 7.75\eV$ to $\Gamma(\pi^0\to\gamma\gamma) = 8.09(11)\eV$,
which (as we discuss below) shows slight tension with the most recent determination from the PrimEx experiment, 
$\Gamma(\pi^0\to\gamma\gamma) = 7.80(12)\eV$~\cite{Larin:2020bhc}.
The improvement over the earlier PrimEx measurement, 
$\Gamma(\pi^0\to\gamma\gamma) = 7.82(22)\eV$~\cite{Larin:2010kq}, 
has already helped to make the prediction of the $\pi^0$ pole term to the hadronic light-by-light scattering 
contribution to the anomalous magnetic moment of the muon more 
precise~\cite{Hoferichter:2018dmo,Hoferichter:2018kwz}.

Two-photon decays of the $\eta$ and $\eta^\prime$ can be similarly predicted in the chiral and large-$N_c$ limits.
However, the situation becomes more complex once
$\SU(3)$ breaking due to nonvanishing and different quark masses is taken into account. 
The $\SU(3)$ breaking is primarily manifested by the $\eta$ mixing with the
$\eta^{\prime}$, which needs to be tightly controlled for
a rigorous theoretical description of the matrix elements $F_{\ep\gamma\gamma}$, following Sect.~\ref{sec:large-Nc}.
They are given in terms of the singlet and octet decay constants as well as mixing angles as
\beq
F_{\eta\gamma\gamma} = \frac{1}{4\sqrt{3}\cos(\theta_8-\theta_0)\pi^2} 
\bigg[ \frac{\cos\theta_0}{F_8} - \frac{2\sqrt{2}\sin\theta_8}{F_0} \bigg] \,, \qquad 
F_{\eta'\gamma\gamma} = \frac{1}{4\sqrt{3}\cos(\theta_8-\theta_0)\pi^2} 
\bigg[ \frac{\sin\theta_0}{F_8} + \frac{2\sqrt{2}\cos\theta_8}{F_0} \bigg] \,.  \label{eq:eta-gg-mixing}
\eeq
Here, a $1/N_c$-suppressed, Okubo--Zweig--Iizuka-(OZI-)rule-violating correction that amounts 
to a replacement $F_0 \to F_0/(1+\Lambda_3)$~\cite{Leutwyler:1997yr} has been omitted; 
it is theoretically required to cancel the scale-dependence in the singlet decay constant $F_0$,
but is assumed to be negligible in most phenomenological analyses of $\eta$--$\eta'$ 
mixing~\cite{Feldmann:1998vh,Feldmann:1998sh,Feldmann:1999uf,Escribano:2005qq}.
In the single-angle flavor-mixing scheme, Eq.~\eqref{eq:eta-gg-mixing} translates into
\beq
F_{\eta\gamma\gamma} = \frac{1}{12\pi^2} 
\bigg[ \frac{5\cos\phi}{F_q} - \frac{\sqrt{2}\sin\phi}{F_s} \bigg] \,, \qquad 
F_{\eta'\gamma\gamma} = \frac{1}{12\pi^2} 
\bigg[ \frac{5\sin\phi}{F_q} + \frac{\sqrt{2}\cos\phi}{F_s} \bigg] \,.  \label{eq:eta-gg-mixing2}
\eeq 
The partial widths $\Gamma(\ep\to\gamma\gamma)$ are a prime source for experimental information on the 
decay constants and mixing angles, and therefore of high theoretical interest.

\subsubsection{Experimental activities for $\pi^0, \eta, \eta' \to \gamma\gamma$ \label{sec-12Pri}}

\begin{figure}
\begin{center}
\includegraphics[width=0.5\linewidth]{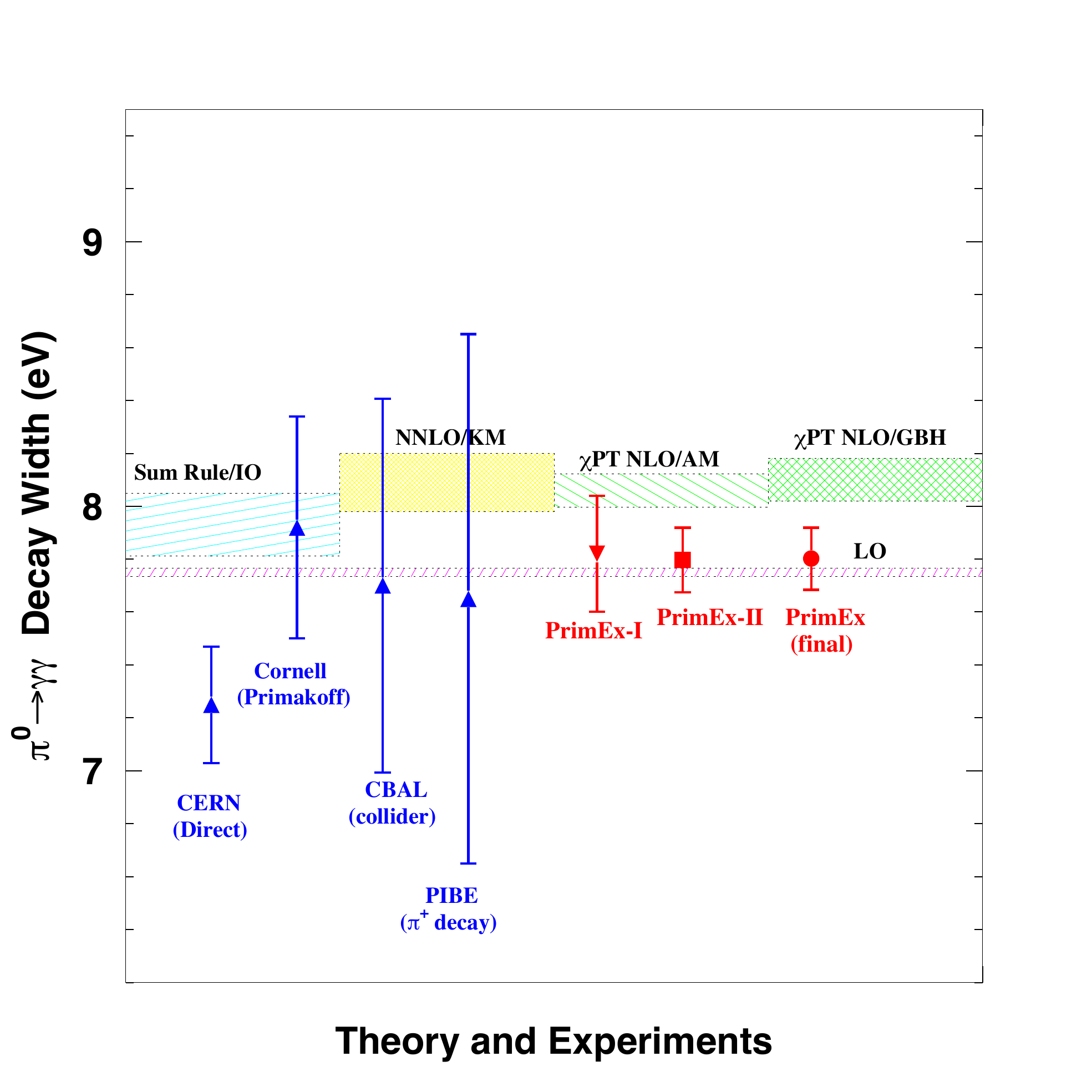}
\caption{Theoretical predictions and experimental results of the $\pi^0$ radiative  decay width. Theory: 
LO (chiral anomaly)~\cite{Adler:1969gk,Bell:1969ts,Bardeen:1969md} (pink~band); sum rule/IO~\cite{Ioffe:2007eg} (blue~band); 
NNLO/KM~\cite{Kampf:2009tk}  (yellow~band); $\chi$PT NLO/AM~\cite{Ananthanarayan:2002kj} (light-green~band); 
$\chi$PT  NLO/GBH~\cite{Goity:2002nn} (green~band). 
Experiments included in the current PDG ~\cite{Tanabashi:2018oca}: CERN (direct)~\cite{Atherton:1985av}; 
Cornell (Primakoff)~\cite{Browman:1974cu};
CBAL (collider)~\cite{Williams:1988sg}; 
 PIBE ($\pi^+$ decay)~\cite{Bychkov:2008ws}; \mbox{PrimEx-I}~\cite{Larin:2010kq}. 
New results: PrimEx-II and PrimEx-I and -II combined~\cite{Larin:2020bhc}.
}
\label{pi0wid_results}
\end{center}
\end{figure}

With the measurement of $\pi^0\to \gamma\gamma$ being an important precision test of low-energy QCD, its current theoretical and experimental status is presented in 
Fig.~\ref{pi0wid_results}. 
The chiral anomaly prediction in Eq.~\eqref{eq:Gamma-pi0gg} 
gives $\Gamma(\pi^0 \rightarrow \gamma \gamma) = 7.750(16)\eV$ (horizontal pink~band). 
Its width corresponds to its uncertainty, due to the experimental uncertainty from the pion decay constant, $F_{\pi}=92.277(95)\MeV$~\cite{Tanabashi:2018oca}, extracted from the charged-pion decay. 
Since the $\pi^0$ is the lightest hadron,  higher-order corrections to the anomaly prediction due to nonvanishing quark masses are small and can be calculated accurately.
Fig.~\ref{pi0wid_results} shows four theoretical calculations in colored bands.
The light-green, green, and yellow bands are calculated in the framework of $\chi$PT up to next-to-leading order (NLO)~\cite{Ananthanarayan:2002kj, Goity:2002nn} and next-to-next-to-leading order (NNLO)~\cite{Kampf:2009tk}, respectively; they are consistent within an uncertainty of $\sim$1\% with the average value $\sim$4\% higher than the leading-order anomaly prediction. 
Another result based on QCD sum rules~\cite{Ioffe:2007eg} (blue~band) is nearly 2\% below the three $\chi$PT results, a difference which may be due to the incomplete treatment of $\pi^0$ mixing with both $\eta$ and $\eta'$ in that work, as pointed out in Ref.~\cite{Bernstein:2011bx}.  
In any case, $\Gamma(\pi^0\to\gamma\gamma)$ thus offers a rare opportunity for a true precision test of QCD within the nonperturbative regime. 
Experimental measurements for $\Gamma(\pi^0\to\gamma\gamma)$ are shown by data points in Fig.~\ref{pi0wid_results}:
the first five points from the left are those listed by the Particle Data Group~\cite{Tanabashi:2018oca}, and the last two points are the recent PrimEx-II and final PrimEx (combined PrimEx-I and -II) results. 
The value for the first data point (CERN) is $7.25(18)_{\textrm{stat}}(14)_{\textrm{syst}}\eV$~\cite{Atherton:1985av}. This is the only result using the direct method by measuring the $\pi^0$ decay length. The $\pi^0$ meson has a short lifetime of $\sim 8.5\times 10^{-17}\,\text{s}$. Even though a $450\GeV$ proton beam was used to produce the boosted $\pi^0$, its decay length was still in the range of $5$--$250\,\mu\text{m}$ in the lab frame. Measuring such small distances represented a big challenge to this experiment. Another drawback of this result comes from the unknown $\pi^0$ momentum spectrum.  The second point (Cornell) is an earlier Primakoff measurement by the Cornell collaboration with the value of $7.92(42)\eV$~\cite{Browman:1974cu}. This experiment used an untagged photon beam to produce $\pi^0$ and a Pb-glass calorimeter with moderate experimental resolution to detect the $\pi^0$ decay photons. An earlier theoretical calculation for the form factors~\cite{Faeldt:1972db} was used in Ref.~\cite{Browman:1974cu} to extract the Primakoff amplitude, where the photon shadowing effect was not considered. The third point (CBAL) is  $7.70(72)\eV$~\cite{Williams:1988sg}, measured via the collision of  $e^+e^-\to e^+e^-\gamma^*\gamma^*\to e^+e^-\pi^0\to e^+e^-\gamma\gamma$. Only  two photons from the decay of the $\pi^0$ were measured, while both final-state lepton $e^+$ and $e^-$ were not detected due to small scattering angles.  The unknown virtualities of the two merging virtual photons and the knowledge of the luminosity limited the experimental accuracy. Detection of  out-going leptons near the beam line is important for any future improvement. 
The fourth result (PIBE) is $7.74(1.02)\eV$~\cite{Bychkov:2008ws}. It was indirectly extracted from the charged-pion radiative (weak) decay $\pi^+\to e^+\nu_e\gamma$ under the assumption of the conserved vector current to relate it to the $\pi^0$. Even if one takes into account isospin-violating effects, further improvement will still be limited by a small branching ratio of this decay channel ($\sim 7.4\times 10^{-7}$). The last three points are results for  PrimEx-I~\cite{Larin:2010kq}  [$7.82(14)_{\textrm{stat}}(17)_{\textrm{syst}}\eV$], PrimEx-II [$7.798(56)_{\textrm{stat}}(109)_{\textrm{syst}}\eV$], and the combination of PrimEx-I and -II [$7.802(52)_{\textrm{stat}}(105)_{\textrm{syst}}\eV$]~\cite{Larin:2020bhc}, measured via the Primakoff effect at JLab.

High-precision measurements of the $P\to \gamma\gamma$ decay width ($P=\pi^0, \,\eta, \,\eta'$) are part of the ongoing Primakoff program at JLab (discussed in Sect.~\ref{exp-pri}).  
The production of mesons in the Coulomb field of a nucleus via the Primakoff effect by real photons 
is essentially the inverse of the $P \to \gamma \gamma$ decay, and the
Primakoff cross section thus provides a measure of the 
two-photon decay width.
For unpolarized incident photons, the Primakoff cross section on a zero-spin nuclear target is given by~\cite{Bellettini:1970th}
\begin{equation}
\frac{\diff\sigma_P}{\diff\Omega}=\Gamma(P\to\gamma \gamma)\frac{8{\alpha_{\text{em}}}Z^2}{M_P^3}\frac{\beta^3{E^4}}{Q^4}\big|F_{\text{em}}\big(Q^2\big)\big|^2 \sin^{2}\theta \,,
\label{eq-primakoff}
\end{equation}
where $Z$ is the atomic
number of the target nucleus, $M_P$, $\beta$, and $\theta$ are the mass, velocity, and production 
angle of the meson, $E$ is the energy of the incoming photon, $Q^2=-q^2$ with $q$ the 
4-momentum transferred to the nucleus,
and $F_{\text{em}}(Q^2)$ is the nuclear electromagnetic form 
factor, corrected for the initial-state interaction of the incoming photon and the final-state interaction of the outgoing mesons in the nuclear medium. 

The Primakoff effect is not the only mechanism for meson photoproduction at high energies.
For a nuclear target, there is coherent background from strong production, 
an interference between the strong and Primakoff
 production amplitudes, and the incoherent nuclear process.
The full cross section is given by
\begin{equation}
\frac{\diff\sigma}{\diff\Omega} = \frac{\diff \sigma_P}{\diff\Omega} +
\frac{\diff\sigma_C}{\diff \Omega} + 
2  \sqrt{\frac{\diff\sigma_{P}}{\diff\Omega} \times 
\frac{\diff\sigma_{C}}{\diff\Omega}} \cos\phi
 +\frac{\diff \sigma_I}{\diff\Omega} \,,
\end{equation}
where ${\diff \sigma_P}/{\diff\Omega}$, ${\diff \sigma_C}/{\diff\Omega}$, and ${\diff \sigma_I}/{\diff\Omega}$ are the Primakoff, nuclear coherent, and incoherent cross sections, respectively.
The relative phase between the Primakoff and nuclear coherent amplitudes is given by $\phi$. The classical method of extracting the Primakoff amplitude is by fitting the measured total differential 
cross sections in the forward direction based on the different characteristic behaviors of the production mechanisms with respect to the production angles. 
\begin{figure}
\centering
\includegraphics[width=0.75\linewidth]{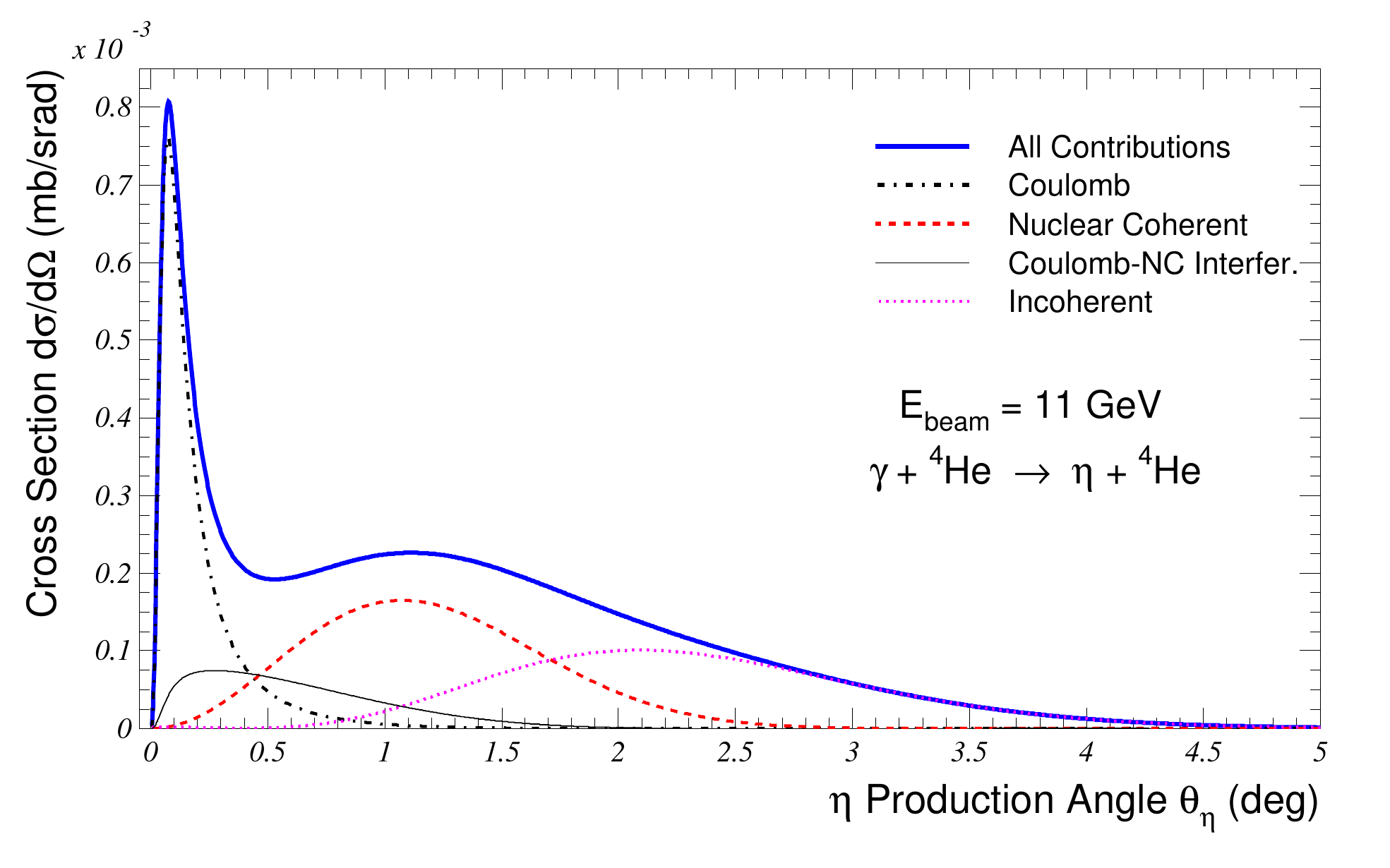}
\caption{Differential cross sections (electromagnetic and nuclear) for the 
$\gamma \, {}^4\text{He} \to \eta \, {}^4\text{He} $ reaction in small angles at $11\GeV$.
 The solid blue line is
the total differential cross section from all processes; the dot-dashed black curve is from the Primakoff process; the dashed red curve is 
from the nuclear coherent process; the black line is from the interference between the Primakoff and nuclear coherent; and the dotted pink curve is from the nuclear incoherent process.
}
\label{fig:cros}
\end{figure}
As shown in Fig.~\ref{fig:cros}, the Primakoff cross section has a sharp maximum at a very small angle, 
and falls off quickly at larger angles. The nuclear coherent cross section has a broad maximum outside the 
angular region of the Primakoff cross section and falls off at larger angles. Two types of background contributions are under the Primakoff peak---the extended tail of the 
nuclear coherent, and the interference of Primakoff and nuclear coherent production. Therefore, an extraction of the Primakoff amplitude from the measured cross section requires good experimental information on the nuclear amplitude outside of the Primakoff region. 
Control of the experimental systematics for such experiments requires: (1) measurement of the beam energy, (2) the luminosity, (3) good energy and angular resolutions for forward angles, (4) selection of compact targets, 
and (5) theoretical calculations of the angular distributions of electromagnetic and hadronic production off the target nucleus.
In addition, independent verifications of overall systematics on the measured cross section are important as well. 
The PrimEx experiments were optimized to 
make significant improvements in all those areas over the previous Primakoff experiment~\cite{Browman:1974cu}.

Two experiments (PrimEx-I and PrimEx-II) were performed for the $\pi^0$ in Hall~B using a tagged photon beam with an energy of $\sim 5\GeV$. A state-of-the-art, PbWO$_4$ and Pb-glass hybrid calorimeter was developed by the PrimEx collaboration to provide excellent energy and position resolutions with a large acceptance. 
 The tagging efficiency was measured by a total absorption counter at low beam intensity and was monitored by a pair production spectrometer at high beam intensity during the physics production. 
The overall systematic uncertainty on the measured cross section was verified by dedicated measurements of two QED processes: electron Compton scattering and $e^+e^-$ pair production using the same apparatus; the precision on the Compton cross section was achieved at the level of 1.5\%--2.0\% depending on the target~\cite{Ambrozewicz:2019xly}. The theoretical input used by PrimEx for fitting the measured total differential cross section to extract the Primakoff amplitude are greatly advanced compared to the one used by the previous Primakoff experiment~\cite{Browman:1974cu}:
a new theoretical calculation~\cite{Gevorkyan:2009ge} for the electromagnetic and strong form factors for the target nucleus has been employed, where the effects of final-state interactions, corrections for light nuclei, contributions from
 nuclear collective excitations, and photon shadowing in nuclei were incorporated; 
two independent approaches, one based on Glauber theory~\cite{Gevorkyan:2009mh} and another based on the cascade model~\cite{Rodrigues:2008zza}, were used for the incoherent production and the difference in the extracted $\Gamma(\pi^0\to\gamma\gamma)$ was less than 0.2\%;  
and the pion transition form factor~\cite{Hoferichter:2014vra} was included in fitting for the PrimEx-II analysis, which had negligible effect on the extracted $\Gamma(\pi^0\to\gamma\gamma)$. 
PrimEx-I was performed on two 5\% radiation length (R.L.) targets, $^{12}$C and $^{208}$Pb; the result was published in 2011~\cite{Larin:2010kq} with a total uncertainty of 2.8\% for $\Gamma(\pi^0\to\gamma\gamma)$. In order to reach the ultimate goal of testing QCD higher-order corrections, a second experiment (PrimEx-II) was carried out on an 8\% R.L.\ $^{12}$C target and a 10\% R.L.\ $^{28}$Si target with improved beam quality, better charged background rejection, and a factor of six more statistics. The PrimEx-II result further improved the precision by nearly a factor of two over PrimEx-I.  Combining PrimEx-I and -II, the final PrimEx result achieved an accuracy of 1.50\%~\cite{Larin:2020bhc}.  Its central value agrees with the chiral anomaly prediction and is about $2\sigma$ below the average of theoretical calculations with higher-order corrections, see Fig.~\ref{pi0wid_results}.  This is clearly a significant result calling for theoretical interpretation. On the other hand, independent experimental verifications, such as a new direct measurement at the high-energy facilities or a double-tagged $e^+e^-$ collision measurement, are highly desirable. 

\begin{figure}
\centering
\includegraphics[height=8cm]{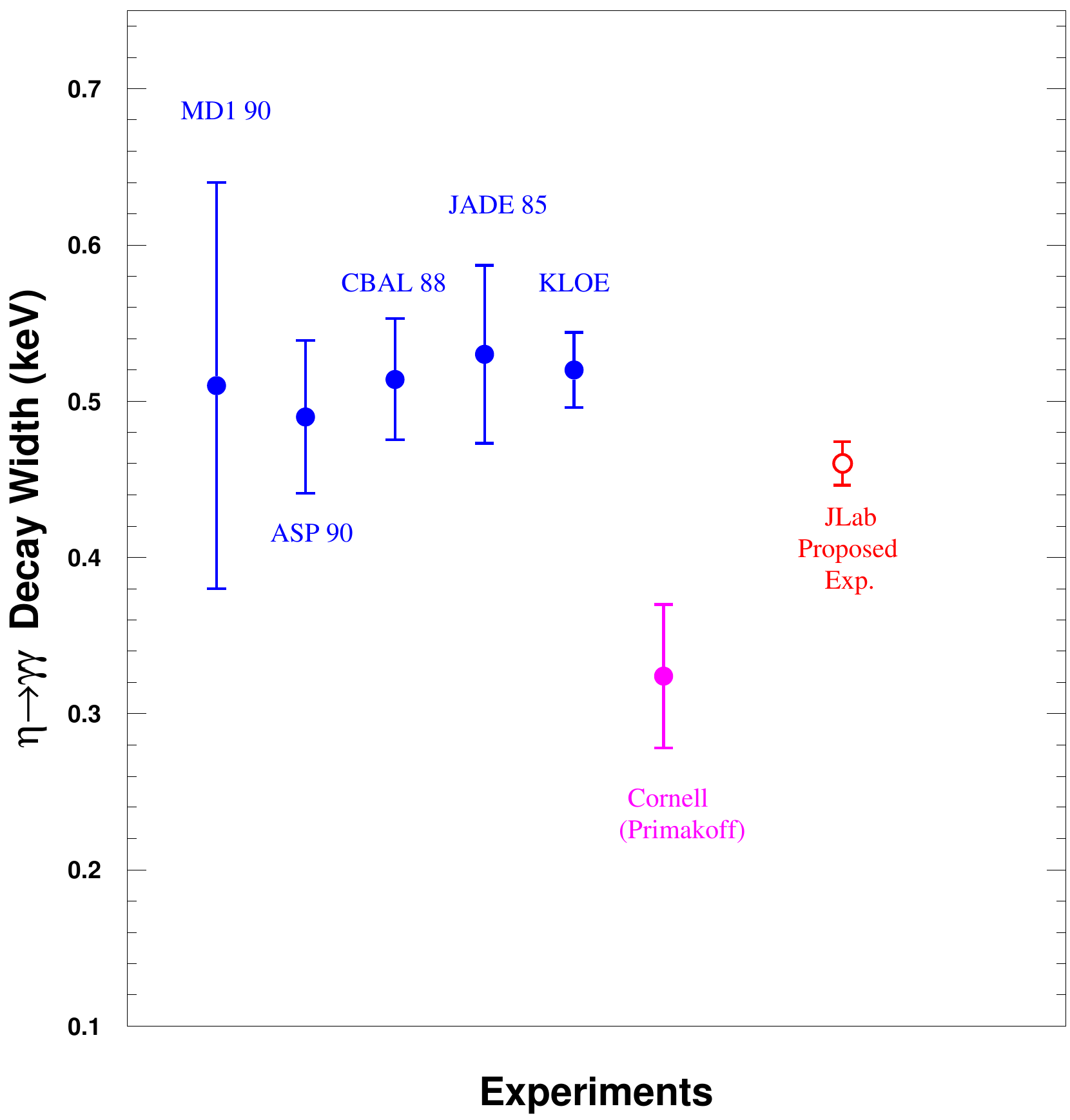} 
\caption{
Experimental status of
$\Gamma(\eta \to \gamma\gamma)$. The five points on the left are 
the results from collider experiments~\cite{Baru:1990pc,Roe:1989qy,Williams:1988sg,Bartel:1985zw,Babusci:2012ik}, point~6 represents the Cornell Primakoff measurement~\cite{Browman:1974sj}. 
Point~7 is the projected error for the PrimEx-eta measurement
 with a $\sim 3$\% total error, arbitrarily plotted to agree with the 
average value of previous measurements. Figure reprinted from Ref.~\cite{Gan:2015nyc}.
}\label{fig:etaLT+mixing} 
\end{figure}

Next, we turn to $\eta \to \gamma \gamma$. 
The experimental status of $\Gamma(\eta\to\gamma\gamma)$ is shown in Fig.~\ref{fig:etaLT+mixing}. 
Most existing results (blue points) come from $e^+e^-$ collisions via two-photon interactions ($e^+e^-\to\gamma^*\gamma^*e^+e^-\to\eta e^+e^-$)~\cite{Baru:1990pc,Roe:1989qy,Williams:1988sg,Bartel:1985zw,Babusci:2012ik}.
On the other hand, the most recent measurement based on Primakoff production was from 1974 by the Cornell collaboration~\cite{Browman:1974sj} (magenta point), for which there is a significant difference (at the $3\sigma$ level). 
The currently ongoing PrimEx-eta experiment (E12-10-011)~\cite{eta-PAC35,Gan:2013msa} 
aims to measure the Primakoff production cross section with projected precision of $\sim 3$\% (red point) to resolve this long-standing puzzle. 

The PrimEx-eta experiment is the first project among a series of measurements proposed in the JLab $12\GeV$ Primakoff program~\cite{Gan:2014pna,white-paper,CDR-12GeV}. It uses a tagged photon beam with an energy up to $11.7\GeV$ and the GlueX apparatus in Hall~D. The only addition to the standard GlueX apparatus is a calorimeter composed of a $12\times 12$ matrix of PbWO$_4$ crystal modules (CompCal) to measure the atomic electron Compton scattering
in parallel to the $\eta$ production to control the overall systematics. Compared to the Primakoff production of the $\pi^{0}$, the 
$\eta$ production has a smaller cross section and peaks at relatively 
larger production angles due to the larger mass of the $\eta$. Thus it is more challenging to separate the Primakoff process 
from hadronic backgrounds, as demonstrated in the earlier Primakoff experiment by the Cornell group~\cite{Browman:1974sj}. 
Two experimental techniques will be applied in the PrimEx-eta experiment to ameliorate this problem. One is to go to higher 
photon energies, which, in addition to increasing the Primakoff cross section
[$\sigma_P\sim Z^2\log(E)$],
will help better separating different processes by pushing the Primakoff peak to smaller angles [$\theta_P\sim {\meta^2}/(2 E^2)$]  as compared to the nuclear coherent production peaked at $\theta_{\rm NC} \sim 2/(E R)$~\cite{Engelbrecht:1964zz}, where $R$ is the nuclear radius ($R\sim A^{1/3}/M_\pi$).
As such, a higher-energy beam in the JLab $12\GeV$ era is vital for this measurement. 
The second is to use lighter targets, $^{1}$H and $^4$He, which are more compact
compared to heavier nuclei, thereby enhancing
coherency as well as offering less distortion to the physics signals due to the initial- and final-state interactions in the nuclear medium.  Since form factors for lighter nuclei fall slowly with increasing momentum transfer, 
 the nuclear coherent mechanism is peaked at larger angles for
lighter nuclei, which helps to separate it from Primakoff production. The PrimEx-eta experiment collected the first two data sets in spring 2019 and in fall 2021 on a liquid $^4$He target.  More data will be expected from the third run in  2022.  

\begin{figure}
\centering
\includegraphics*[width=12.cm, clip]{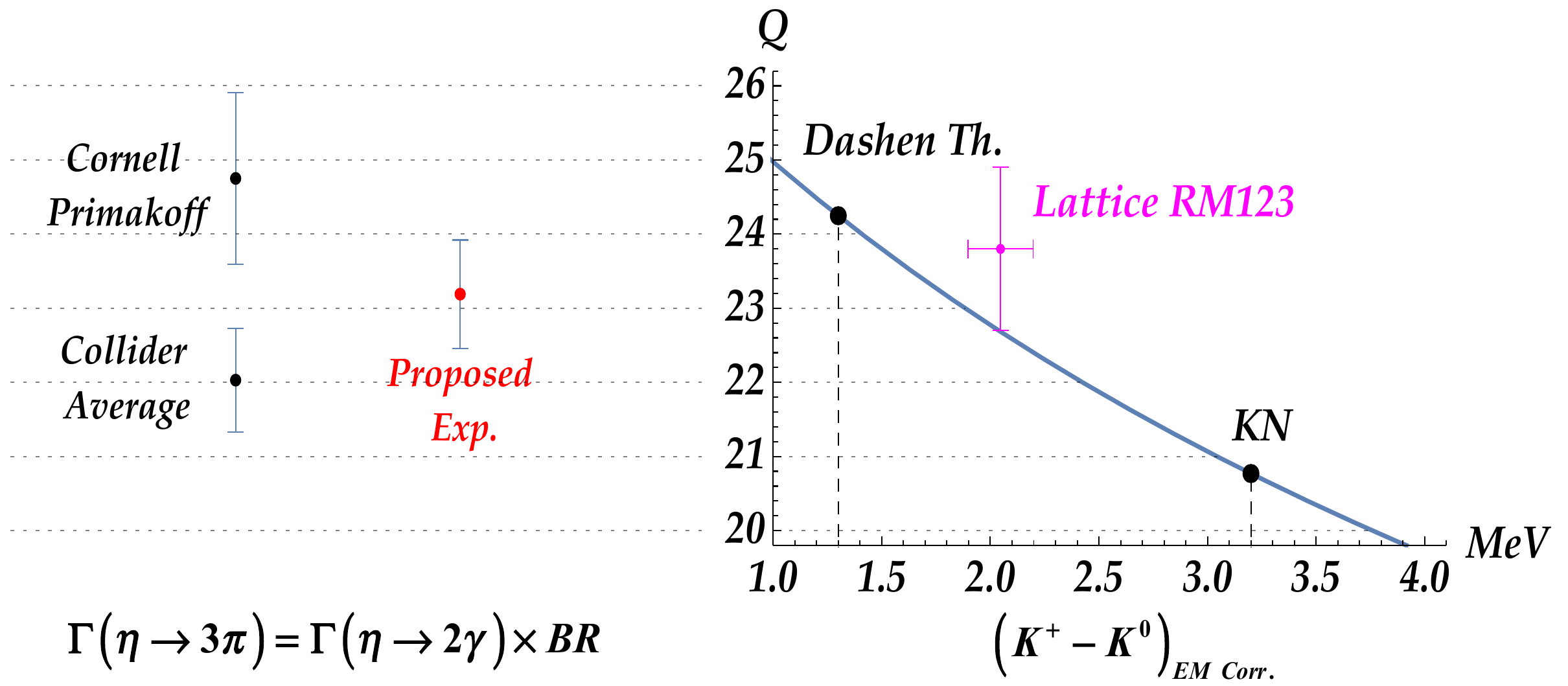}
\caption{Light quark mass ratio determined by two different methods.
The left-hand side indicates the values of $Q$ calculated from the $\eta\to 3\pi$ decay corresponding to the Primakoff~\cite{Browman:1974sj} and collider average~\cite{Tanabashi:2018oca} experimental results for $\Gamma(\eta\to \gamma\gamma)$ as input, as well as the PDG averages for $\BR(\eta\to\pi^+\pi^-\pi^0)$ and $\BR(\eta\to\gamma\gamma)$, see Table~\ref{tab:eta}. The right-hand side shows the results for $Q$ obtained from the kaon mass difference, see Eq.~\eqref{eq:DeltaQ}, with theoretical estimates for the electromagnetic corrections based on Dashen's theorem~\eqref{eq:QD}, Ref.~\cite{Kastner:2008ch} (KN), or the lattice~\cite{Giusti:2017dmp}. Figure adapted from \cite{Leutwyler:1996qg}. 
\label{fig:quark_mass_ratio}}
\end{figure}

The precision measurement of the $\eta$ radiative decay width will offer a sensitive probe into low-energy QCD. One example is the extraction of the $\eta$--$\eta'$ mixing angle.
In addition, an improvement in $\Gamma(\eta\to\gamma\gamma)$ 
will also have a broad impact on all other $\eta$ partial decay 
widths in the PDG listing, as they are determined by using the 
$\eta\to \gamma\gamma$ decay width and their corresponding experimental branching ratios. 
This holds true in particular for the $\eta\to 3\pi$ decay (discussed in Sect.~\ref{sec:eta-3pi}) used for an accurate determination of the quark mass double ratio $Q$~\cite{Leutwyler:1996qg, Bijnens:2002qy}. As shown in Fig.~\ref{fig:quark_mass_ratio}, a new Primakoff result from the PrimEx-eta experiment (the red point) will make an impact on $Q$ by resolving the systematic difference between the results determined by using collider and previous Primakoff measurements.

Lastly, we discuss $\eta^\prime \to \gamma\gamma$. All existing measurements of $\Gamma(\eta'\to\gamma\gamma)$ were carried out by using $e^+e^-$ collisions, with experimental uncertainty for each individual experiment in the range of 
7.3\%--27\%~\cite{Tanabashi:2018oca}. 
A planned new experiment with GlueX, an extension of PrimEx-eta, will perform the first Primakoff measurement with a projected uncertainty of 4\% for $\Gamma (\eta^{\prime}\rightarrow \gamma\gamma)$. 
This precision measurement, coupled with theory, will provide further input for global analyses of the $\eta$--$\eta^\prime$ system to determine their mixing angles and decay constants.
Moreover, it will further pin down the $\eta^\prime$ contribution to light-by-light scattering in $(g-2)_\mu$.

\subsection{$\pi^0, \eta, \eta'$ transition form factors}
\label{sec:pi0-eta-etapTFF}

The general two-photon couplings for the lightest flavor-neutral pseudoscalar mesons $P=\pi^0, \eta, \eta^\prime$ are described by $F_{P \gamma^* \gamma^*}(q_1^2, q_2^2)$, defined in Eq.~\eqref{eq:defTFF}.
Different experimental techniques can be used to access these TFFs in various kinematical regions, including both time-like and space-like momenta, which are related to one another by analytic continuation.
In the space-like case, it is customary to express the photon momenta in terms of the positive variables $Q_{1,2}^2 = -q_{1,2}^2 > 0$.

The general (doubly-virtual) TFFs are challenging both to predict theoretically and measure experimentally.
Consequently, most attention has focused on the singly-virtual TFF $F_{P\gamma^*\gamma}(q^2) \equiv F_{P\gamma^*\gamma^*}(q^2,0)$ involving one real and one virtual photon.
Although we present more sophisticated treatments below, the approximate behavior of this function can be understood simply within the context of VMD, which predicts a parameterization of the form 
\begin{equation}
F_{P\gamma^*\gamma}(q^2) \equiv F_{P\gamma^*\gamma^*}(q^2,0) = \frac{F_{P\gamma\gamma}}{1 - q^2/\Lambda_P^2} \,, \label{eq:FPgg-VMD}
\end{equation}
where the pole parameter is expected to be of the order of a vector-meson (typically $\rho$, $\omega$, $\phi$) mass, $\Lambda_P\approx m_V$.  At very low $q^2$, the transition form factor is expanded according to
\begin{equation}
F_{P\gamma^*\gamma}(q^2) = F_{P\gamma\gamma} \Big[ 1 + b_P q^2 + \Order(q^4) \Big] \,,
\label{eq:TFFslope}
\end{equation}
where the slope parameter $b_P$ 
\begin{equation}
b_P=\frac{1}{F_{P\gamma\gamma}}\frac{\diff F_{P\gamma^*\gamma}(q^2)}{\diff q^2}\Bigg|_{q^2=0}
\label{eq:def-bP}
\end{equation}
is a measure of the $P \to \gamma^* \gamma$ interaction radius. 
Obviously, the VMD model~\eqref{eq:FPgg-VMD} leads to the expectation $b_P \approx 1/m_V^2$.
Therefore, a measurement of the transition form factor at small $q^2$ plays a critical role in determining 
the electromagnetic interaction radius of a neutral pseudoscalar meson. 

The remainder of this section is organized as follows.
First, we summarize the experimental status for all TFF measurements.
This includes not only pseudoscalar decays $P \to \ell^+ \ell^- \gamma$, but also a variety of other processes to access a wide range of kinematic regions.
Second, we present a theoretical description of TFFs based on dispersion theory (including additional input from VMD and perturbative QCD).
Since these arguments are most advanced for the case of the $\pi^0$ TFF, we discuss this particular case in detail.
The parallel discussion for $\eta,\eta^\prime$ TFFs is given in subsequent sections below in the context of specific $\eta,\eta^\prime$ decay channels.

\subsubsection{Experiment}

Experimentally, measuring the doubly-virtual TFFs $F_{P\gamma^*\gamma^*}(q_1^2,q_2^2)$ as functions of both $q_1^2$ and $q_2^2$ is very challenging. 
In the time-like region ($q_{1,2}^2>0$), the TFF can be measured in principle through the double-Dalitz decay, $P \to e^+e^-e^+e^-$, where the photon virtualities are in the range of $4m_e^2< q_{1,2}^2<(M_P-2m_e)^2$. 
However, the difficulty lies in the fact that double-Dalitz decay has a small partial width. 
To date, only branching fractions or upper limits have been reported for the $\pi^0$~\cite{Tanabashi:2018oca, Abouzaid:2008cd} and $\eta$~\cite{KLOE2:2011aa,Berlowski:2007aa}, while double-Dalitz decay for the $\eta^\prime$ has not been observed so far (discussed further in Sect.~\ref{sec:doubleDalitz}).
Alternatively, vector meson decays $V\to P l^+l^-$ can be used to indirectly measure  $F_{P\gamma^*\gamma^*}(q_1^2,q_2^2)$ for the time-like region $q_1^2=m_V^2$ and $4m_l^2<q_2^2<(m_V-M_P)^2$, where $m_V$ is the mass of the vector meson. 
Several recent results are from NA60 on $\omega\to\pi^0 \mu^+\mu^-$~\cite{Arnaldi:2016pzu}, A2 on $\omega\to \pi^0 e^+e^-$~\cite{Adlarson:2016hpp}, KLOE-II on $\phi \to \pi^0 e^+e^-$~\cite{Anastasi:2016qga} and $\phi \to \eta e^+e^-$~\cite{Babusci:2014ldz}, as well as BESIII on $J/\psi \to P e^+e^-$ ($P=\pi^0,\,\eta,\,\eta^\prime$)~\cite{Ablikim:2014nro}.
For the $\eta^\prime$, one can also extract $F_{\eta^\prime\gamma^*\gamma^*}(q_1^2=m_{\omega}^2,q_2^2)$ from $\eta^\prime\to\omega e^+e^-$ for $4m_e^2<q_2^2<(M_{\eta^\prime}-m_\omega)^2$. The experimental result from this channel, however, is still lacking except a measurement of the branching ratio reported by BESIII~\cite{Ablikim:2015eos}. More discussion about $\eta^\prime\to\omega e^+e^-$ can be found in Sect.~\ref{sec:eta-omegag}.
Various collaborations have presented such (indirect) results on $F_{\pi^0\gamma^*\gamma^*}(q_1^2=m_\omega^2,q_2^2)$~\cite{Achasov:2000wy,Akhmetshin:2003ag,Achasov:2013btb,Achasov:2016zvn,TheBaBar:2017vzo} for $q_2^2>(m_\omega+M_\pi)^2$ through the production process $e^+e^- \to \omega \pi^0$. BESIII recently reported on $e^+e^-\to\phi\eta^\prime$ for $4.2\GeV^2<q_2^2<9.5\GeV^2$~\cite{Ablikim:2020coo} and on $e^+e^-\to J/\psi\eta^\prime$ for $17.5\GeV^2<q_2^2<21.2\GeV^2$~\cite{Ablikim:2019bwn}.

In the space-like region ($q_{1,2}^2=-Q_{1,2}^2 < 0$), the doubly-virtual TFF can be measured at the $e^+e^-$ collider facilities by the two-photon fusion reaction $e^+e^-\rightarrow \gamma^*\gamma^*e^+e^-\rightarrow P e^+e^-$, where the virtualities of the photons are measured by detecting the outgoing leptons.
In the double-tag mode where both outgoing leptons are detected, the cross section of the reaction offers a measurement of $F_{P\gamma^*\gamma^*}(Q_1^2, Q_2^2)$. Such cross sections, however, are very small.  The only doubly-virtual space-like TFF was recently reported for the $\eta^\prime$ by the BaBar collaboration~\cite{Gary:2019hzp,BaBar:2018zpn}. It covers values from $2\GeV^2$ to  $60\GeV^2$ for both $Q_1^2$ and $Q_2^2$. 

Another way to measure the space-like TFF is via the virtual Primakoff effect, $e^-A\rightarrow \gamma^*\gamma^*e^-A \to Pe^-A$, where $A$ represents a nuclear target. As the Primakoff cross section is peaked at extremely small $Q^2$, this process offers a sensitive probe for the space-like TFF in the small-$Q^2$ region. The vertex of two-photon fusion into a meson has two radiated photons, one by a high-energy electron and the other by an atomic nucleus (the second photon is nearly real).
This approach was initially investigated by Hadjimichael and Fallieros~\cite{Hadjimichael:1989ks}, who calculated the Primakoff cross section in the Born approximation. F\"aldt conducted a more comprehensive study for both Primakoff and hadronic productions with a Glauber model approach~\cite{Faldt:2010jh,Faldt:2012hz}. This work considered the shadowing effect where the initial photon is first converted into a $\rho$ meson, which in a subsequent collision with another nucleon creates the meson of interest. These theoretical studies have laid the foundations for precision measurements of TFFs via the Primakoff effect. Currently, a proposal for a measurement of $F_{\pi^0\gamma^*\gamma^*}(Q_1^2,Q_2^2)$ via the Primakoff effect is under development at JLab. The expected kinematical coverage will  be $0.001<Q_1^2<0.1\GeV^2$ and  $Q_2^2<0.01\GeV^2$. 

For the singly-virtual TFF, in contrast, significant experimental progress has been made in recent years.
Single-Dalitz decay $P\rightarrow l^+l^-\gamma$ accesses the time-like TFF $F_{P\gamma^*\gamma}(q^2)$ for small $q^2$ in the range $4m_l^2<q^2< M_P^2$.
Two recent results were reported for the $\pi^0$ by the NA62~\cite{TheNA62:2016fhr} and the A2~\cite{Adlarson:2016ykr} collaborations. 
As for $\eta$,
the NA60 result published in 2016~\cite{Arnaldi:2016pzu} was obtained in proton--nucleus ($p$--$A$) collisions using a $400\GeV$ proton beam at the CERN SPS. The $\eta$ TFF for a $q^2$ range of $0.21$--$0.47\GeV^2$ was measured using $\sim 1.8\times 10^5$ $\mu^+\mu^-$ pairs from the $\eta\rightarrow \mu^+\mu^-\gamma$ decay, which is a factor of 10 more statistics than their earlier publication~\cite{Arnaldi:2009aa}. Another new high-statistics result on the single-Dalitz decay $\eta\rightarrow e^+e^-\gamma$  was reported by the A2 collaboration at MAMI~\cite{Adlarson:2016hpp}. About $5.4\times 10^4$ $\eta\rightarrow e^+e^-\gamma$ events were reconstructed from a total of $5.87\times 10^7$ $\eta$ mesons produced via the photoproduction reaction $\gamma p\rightarrow\eta p$, using a $1508$--$1557\MeV$ tagged photon beam. 
The $q^2$ range of the $\eta$ TFF reported in Ref.~\cite{Adlarson:2016hpp} is $1.2\times 10^{-3}$--$0.23\GeV^2$; this is the only experimental result quoting systematic uncertainties for every individual $q^2$ bin.  The A2 experiment~\cite{Adlarson:2016hpp} reached smaller $q^2$ than NA60~\cite{Arnaldi:2016pzu} due to the smaller mass of the electron compared to the muon.
This new A2 result is more accurate than two older measurements~\cite{Aguar-Bartolome:2013vpw, Berghauser:2011zz}. 
The WASA-at-COSY collaboration also reported the $\eta$ TFF in a $q^2$ range of $0.02$--$0.38\GeV^2$~\cite{Pszczel:2019sdk} recently, an analysis of $\eta\rightarrow e^+e^-\gamma$ based on a data sample of $\sim 10^8$  $\eta$ mesons produced in proton--proton collisions. 
The first observation of the Dalitz decay $\eta^{\prime}\rightarrow e^+e^-\gamma$ was reported by the BESIII collaboration~\cite{Ablikim:2015wnx}, based on a data sample of $1.3\times 10^9$ $J/\psi$ events. The $\eta^{\prime}$ mesons were produced via $J/\psi\rightarrow \eta^{\prime}\gamma$. They observed $864(36)$ $\eta^{\prime}\rightarrow e^+e^-\gamma$ events  in the $q^2$ range of $2.5\times 10^{-3}$--$0.56\GeV^2$.  
This result has improved precision compared to the previous result published by the Lepton-G collaboration in 1979 from the $\eta^{\prime}\rightarrow\mu^+\mu^-\gamma$ decay~\cite{Dzhelyadin:1979za}.

For larger time-like momenta ($q^2>M_P^2$),  the singly-virtual TFF in the time-like region is accessible through lepton annihilation $e^+e^-\rightarrow\gamma^*\rightarrow P\gamma$ at the collider facilities.  There are two results on the $\pi^0$ presented by SND at the Novosibirsk VEPP-2000, covering the $q^2$ range of 0.60--$1.38\GeV^2$~\cite{Achasov:2016bfr} and 1.075--$2\GeV^2$~\cite{Achasov:2018ujw}, respectively. Future result from BESIII will cover $q^2$ range of 4--$21\GeV^2$~\cite{Lenz:2019vcx}.
The measurements of $e^+e^-\to \eta^{(\prime)}\gamma$ were carried out by 
\mbox{CMD-2}~\cite{Akhmetshin:1999zv,Akhmetshin:2001hm,Akhmetshin:2004gw} and
SND~\cite{Achasov:2000zd,Achasov:2003ed,Achasov:2007kw}. Results for $\eta$ and $\eta'$ were also reported by the BaBar collaboration for $q^2 = 112\GeV^2$~\cite{Aubert:2006cy}, as well as by CLEO for $\pi^0$, $\eta$, and $\eta'$ at $q^2 = 14.2\GeV^2$~\cite{Pedlar:2009aa}.  

Lastly, we discuss the singly-virtual TFF in the space-like region ($q^2=-Q^2<0$).
This can be measured through  $e^+e^-\rightarrow \gamma^*\gamma^*e^+e^-\rightarrow P e^+e^-$ at the $e^+e^-$ collider facilities.
Only one outgoing lepton is detected (single-tag mode). The other untagged photon is almost real (where the associated lepton is not detected).
All existing measurements were carried out by using this method. 
CLEO~\cite{Gronberg:1997fj} covered the $Q^2$ range from $1.5\GeV^2$ to $9$, $20$, and $30\GeV^2$ for $\pi^0$, $\eta$, and $\eta'$, respectively. Smaller $Q^2$ values were reported by the CELLO collaboration~\cite{Behrend:1990sr} from $0.62\GeV^2$ up to  $2.23\GeV^2$ for $\pi^0$ and $\eta$, and up to $7\GeV^2$ for the $\eta'$. 
An earlier result by the TPC/Two-Gamma collaboration~\cite{Aihara:1990nd} at SLAC was in the range of $0.1\GeV^2<Q^2<7\GeV^2$ but with poor statistics (only $38.0(9.4)$ reconstructed $\eta$ and $159.2(14.7)$ constructed $\eta^{\prime}$ events). 
A measurement for the $\eta'$ was also report by the L3 collaboration~\cite{Acciarri:1997yx} for $0.06\GeV^2<Q^2<4.14\GeV^2$ with poor $Q^2$ resolution. 

\begin{figure}
\begin{center}
\includegraphics[width=7.5cm]{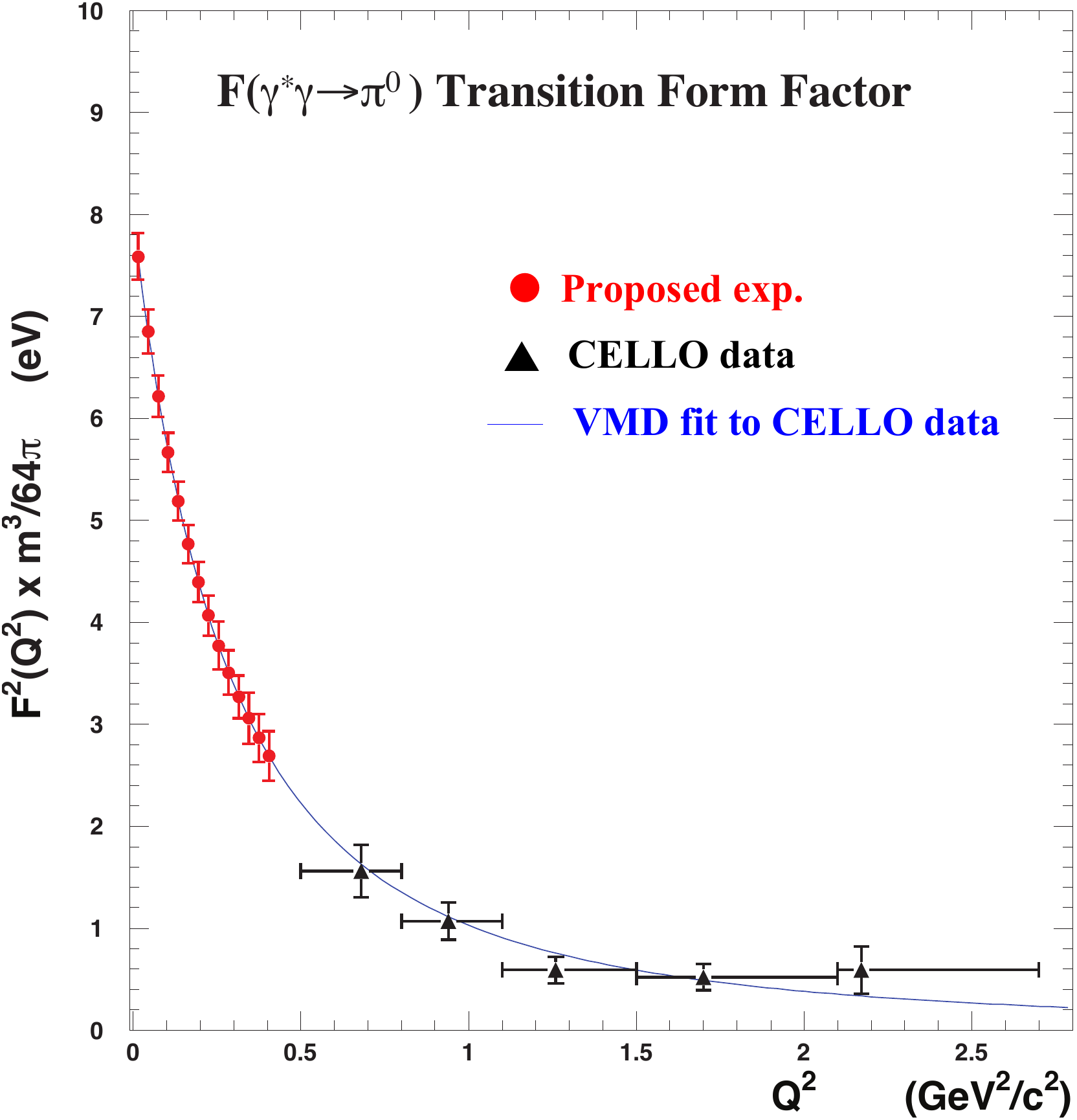} \hfill
\includegraphics[width=7.5cm]{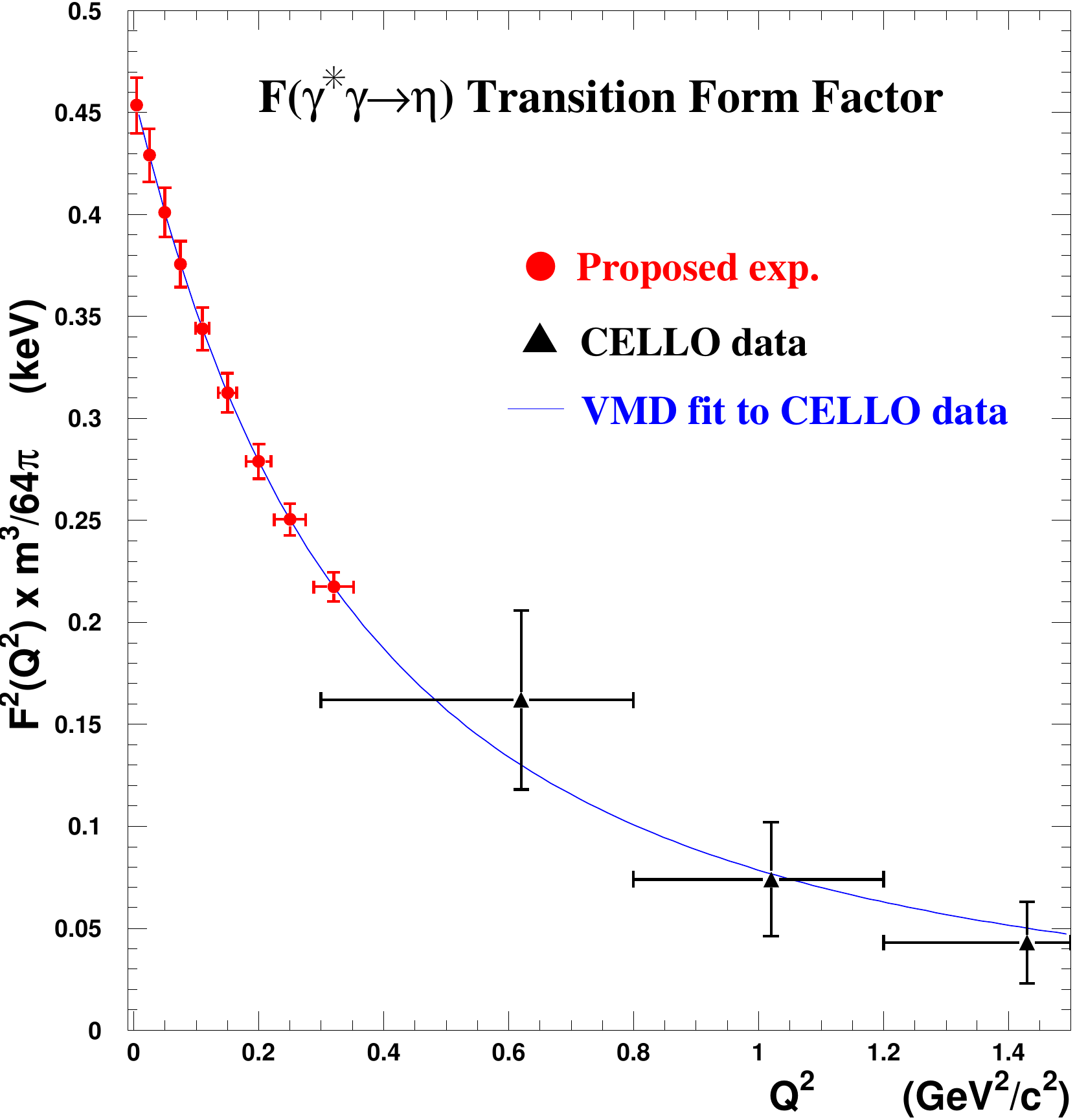}
\end{center}
\caption{The space-like $\pi^0$ and $\eta$ transition form factor measurements via the Primakoff effect at JLab. 
The proposed points are projected to the VMD predictions with expected total errors in comparison with CELLO data~\cite{Behrend:1990sr}. Figures adapted from Ref.~\cite{Gan:2015nyc}.
\label{fig:pi+eta_form}}
\end{figure}

Results for larger $Q^2$ were obtained by the $B$ factories.
The BaBar collaboration measured TFFs for $\pi^0$, $\eta$, and $\eta'$~\cite{Aubert:2009mc,BABAR:2011ad} in a $Q^2$ range of $4$--$35\GeV^2$, while the $\pi^0$-TFF was also reported by Belle in a similar $Q^2$ range of $4$--$40\GeV^2$~\cite{Uehara:2012ag}. 
It should be mentioned that the distribution of $Q^2F_{\pi^0\gamma^*\gamma}(-Q^2)$ measured by the BaBar collaboration is observed to be above the asymptotic limit~\cite{Aubert:2009mc} at large $Q^2$, which disagrees with the result reported by the Belle collaboration in a similar $Q^2$ range~\cite{Uehara:2012ag}. On the other hand, the $Q^2$ dependencies of $Q^2F_{P\gamma^*\gamma}(-Q^2)$ published by BaBar for $\eta$ and $\eta^{\prime}$~\cite{BABAR:2011ad} rise at best logarithmically with $Q^2$, and about three times weaker than what was observed for the $\pi^0$~\cite{Aubert:2009mc}. Future results from Belle-II will be important to clarify the large-$Q^2$ behavior of these form factors.

 The space-like TFFs at low $Q^2$ ($<0.5\GeV^2$) for $\pi^0$ and $\eta$ are totally lacking. This situation will be improved soon by two ongoing activities. One is 
 an analysis of data taken at the center-of-mass energy of $\sim 4\GeV$ by the BESIII collaboration~\cite{Redmer:2017fhg,Redmer:2018gah}.  The projected coverage for $Q^2$ is $0.3\GeV^2<Q^2<3.1\GeV^2$~\cite{Lenz:2019vcx} with statistical accuracy compatible with the published results of CELLO~\cite{Behrend:1990sr} and CLEO~\cite{Gronberg:1997fj}.  The second one is by the KLOE-II collaboration. The KLOE-II data collection was completed in March 2018 with a total luminosity of $5.5\,\text{fb}^{-1}$ at a center-of-mass energy near the mass of the $\phi$ meson. The space-like TFFs for $\pi^0$ and $\eta$ in the $Q^2$ range of $0.015\GeV^2<Q^2<0.1\GeV^2$~\cite{Giovannella:2018nib, PerezdelRio:2019ycb} will be extracted from this data set. 

As a part of the JLab Primakoff program~\cite{white-paper,CDR-12GeV}, the PrimEx collaboration will measure the space-like singly-virtual TFFs for $\pi^0$, $\eta$, and $\eta^{\prime}$ for a $Q^2$ range of $0.001$--$0.5\GeV^2$, via the Primakoff effect. 
The $Q^2$ range projected by the JLab Primakoff experiment for $\pi^0$ and $\eta$ is shown in Fig.~\ref{fig:pi+eta_form}. 
It will push the lower limit of $Q^2$ by more than one order of magnitude smaller than what will be expected from BESIII and KLOE-2 in the space-like region, in addition to providing independent measurements with different systematics. The meson transition form factors at small and intermediate $Q^2$ dominate the contributions to the hadronic light-by-light corrections to the anomalous magnetic moment of the muon~\cite{Knecht:2001qf,Nyffeler:2016gnb}; see also the following theory section. 
The JLab Primakoff experiments will fill the unexplored small $Q^2$ range 
and are complementary to other existing measurements.

\begin{figure}[t]
\includegraphics[width=0.45\linewidth]{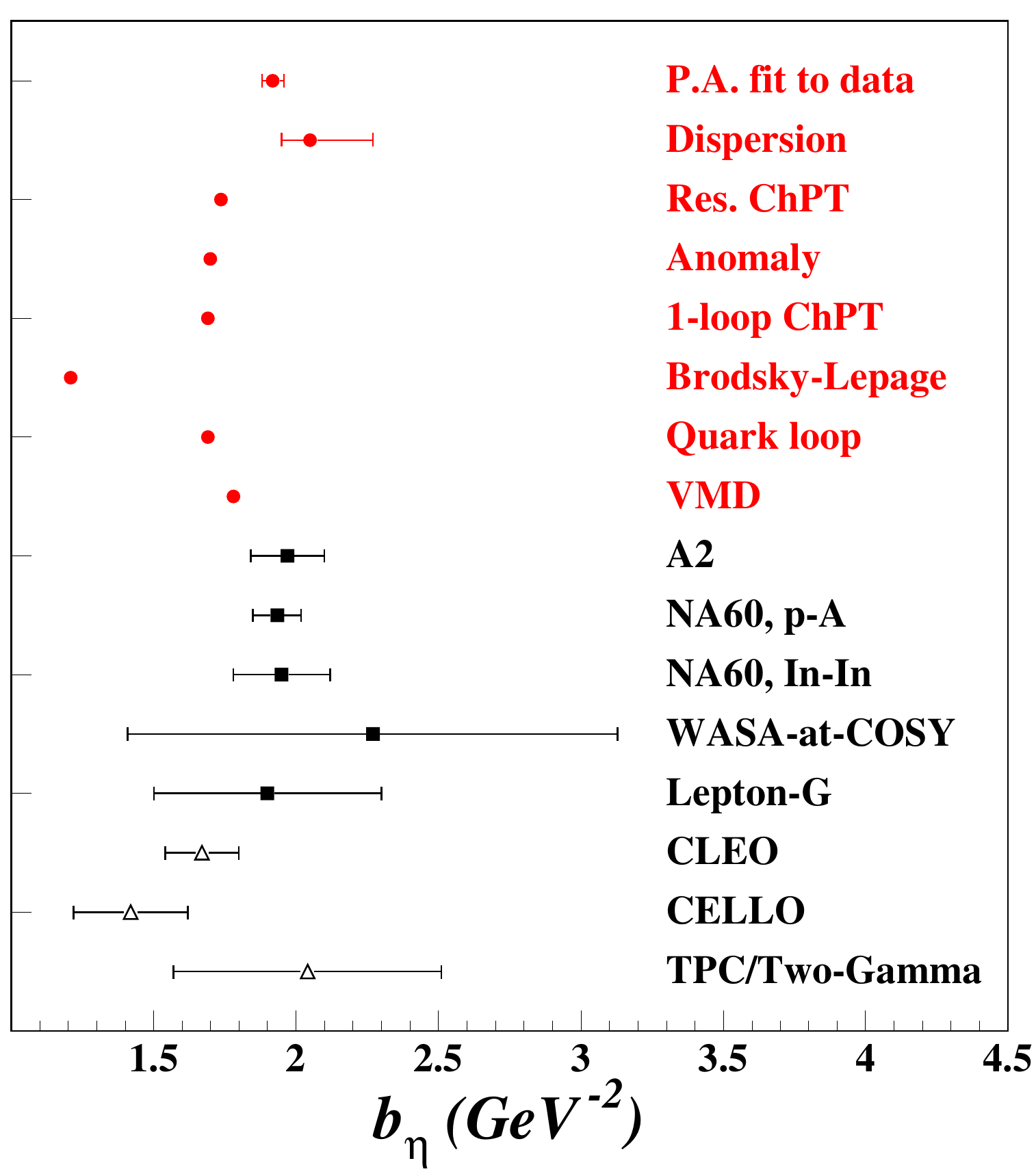} \hfill
\includegraphics[width=0.45\linewidth]{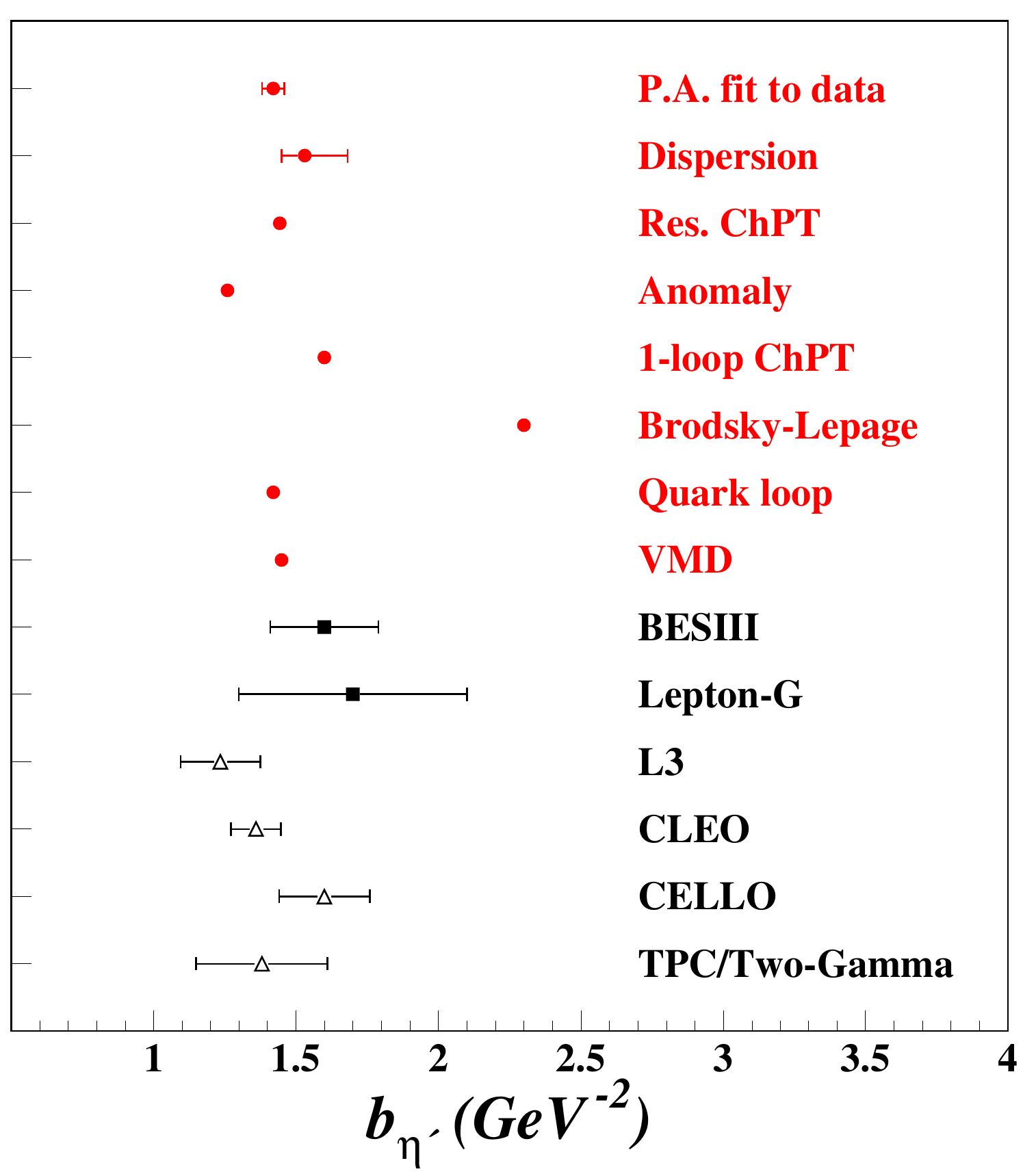}
\caption{Slope parameters of the $\eta$ (left) and $\eta'$ (right) TFFs 
extracted from different experiments and calculated by different theoretical models. 
For the $\eta$, the points refer to, from bottom to top, 
space-like measurements~\cite{Aihara:1990nd,Behrend:1990sr,Gronberg:1997fj} (three black triangles), 
time-like measurements~\cite{Dzhelyadin:1979za,Hodana:2012rc,Arnaldi:2009aa,Arnaldi:2016pzu,Adlarson:2016hpp} (five black squares),
and theoretical calculations~\cite{Bramon:1981sw,Ametller:1983ec,Pich:1983zk,Brodsky:1981rp,Ametller:1991jv,Czyz:2012nq,Klopot:2013laa,Hanhart:2013vba,Escribano:2015nra} (eight red dots). 
Similarly for the $\eta'$, we show, from bottom to top, 
the space-like~\cite{Aihara:1990nd,Behrend:1990sr,Gronberg:1997fj, Acciarri:1997yx} (four black triangles)
and time-like measurements~\cite{Dzhelyadin:1979za, Ablikim:2015wnx} (two black squares),
as well as theoretical calculations~\cite{Bramon:1981sw,Ametller:1983ec,Pich:1983zk,Brodsky:1981rp,Ametller:1991jv,Czyz:2012nq,Klopot:2013laa,Hanhart:2013vba,Escribano:2015yup}.}
\label{fig:eta-etap-slopes}
\end{figure}

The TFFs of $\pi^0$, $\eta$, and $\eta^{\prime}$ at low-momentum transfer are particularly important to extract the low-energy parameters of these mesons, such as the slope parameter $b_P$ in Eq.~\eqref{eq:def-bP}.
The current status of the slope parameters for $\eta$ and $\eta^{\prime}$ is shown in Fig.~\ref{fig:eta-etap-slopes}. All experimental determinations were obtained by fitting the TFF data with a normalized single-pole term as in Eq.~\eqref{eq:FPgg-VMD}. No radiative corrections were applied to any of those experimental results.  As the TFFs are analytic functions at $q^2=0$, the slope has to be unique, independently of whether this limit is approached from the space- or time-like regions.
As one can see in Fig.~\ref{fig:eta-etap-slopes} (left), however, the most precise $b_{\eta}$ extracted from time-like measurements by NA60~\cite{Arnaldi:2016pzu, Arnaldi:2009aa} and A2~\cite{Adlarson:2016hpp} are systematically larger than the results obtained from the space-like data by CLEO~\cite{Gronberg:1997fj} and CELLO~\cite{Behrend:1990sr}. A similar situation is also seen in Fig.~\ref{fig:eta-etap-slopes} (right) for $b_{\eta^{\prime}}$, where its value extracted by BESIII~\cite{Ablikim:2015wnx} in the time-like region is larger than the L3~\cite{Acciarri:1997yx} and CLEO~\cite{Gronberg:1997fj} results obtained from the space-like $\eta^{\prime}$ TFF.
In recent years, a data-driven approach that used Pad\'e approximants to fit $\eta^{(\prime)}$ TFF data from both space- and time-like regions globally~\cite{Escribano:2015nra, Escribano:2015yup} has demonstrated a significant improvement on the precision of the slope parameters, shown as the top red points in Fig.~\ref{fig:eta-etap-slopes}. Clearly, more precise measurements are desirable for understanding the electromagnetic radii of $\eta$ and $\eta^{\prime}$, particularly at small $Q^2$ for the $\eta$ TFF in the space-like region and for the $\eta^{\prime}$ TFF in both space- and time-like regions.

\subsubsection{Theory}

$\chi$PT alone is not very predictive for TFFs due to the strong
influence of vector mesons therein~\cite{Bijnens:1988kx, Bijnens:1989jb,Bickert:2020kbn}; hence already early on, the combination with 
models has been considered~\cite{Ametller:1991jv}, and data analyses are rather performed using
extensions such as R$\chi$T~\cite{Roig:2014uja,Guevara:2018rhj}.
For combined analyses of space-like data, completely model-independent methods using
rational (Pad\'e) approximants have been advocated~\cite{Escribano:2013kba}, although their 
extension to the time-like region~\cite{Escribano:2015nra,Escribano:2015yup} is at least doubtful
due to the absence of the physical cuts and resonance poles in the corresponding amplitudes. 
We will therefore concentrate on the prospects of high-precision analyses of the $\pi^0$, $\eta$, and $\eta'$ TFFs constructed using dispersion relations.
Presently, we mainly focus on the $\pi^0$ case, following~\cite{Hoferichter:2012pm,Hoferichter:2014vra,Hoferichter:2017ftn,Hoferichter:2018kwz,Hoferichter:2018dmo}, for which these methods are best established. 
The application to $\eta$ and $\eta^\prime$ TFFs follows in later sections.

To analyze dispersion relations for
$F_{P\gamma^*\gamma^*}(q_1^2,q_2^2)$, we first decompose them in terms of isospin. 
The $\pi^0$ always decays into one isovector and one isoscalar
photon ($vs$), combined in both possible ways, while the $\eta$ and $\eta'$ decays to photon pairs that are either isovector--isovector ($vv$) or isoscalar--isoscalar ($ss$), which must be summed:
\begin{equation}
F_{\pi^0\gamma^*\gamma^*}(q_1^2,q_2^2) = F_{vs}(q_1^2,q_2^2) + F_{vs}(q_2^2,q_1^2) \,, \qquad
F_{\ep\gamma^*\gamma^*}(q_1^2,q_2^2) = F_{vv^{(\prime)}}(q_1^2,q_2^2) + F_{ss^{(\prime)}}(q_1^2,q_2^2) \,. \label{eq:TFF-isodecomp}
\end{equation}
We need to identify the dominant hadronic intermediate states at low energies, where high precision
is of paramount importance.
\begin{figure}
\includegraphics*[width=0.32\linewidth]{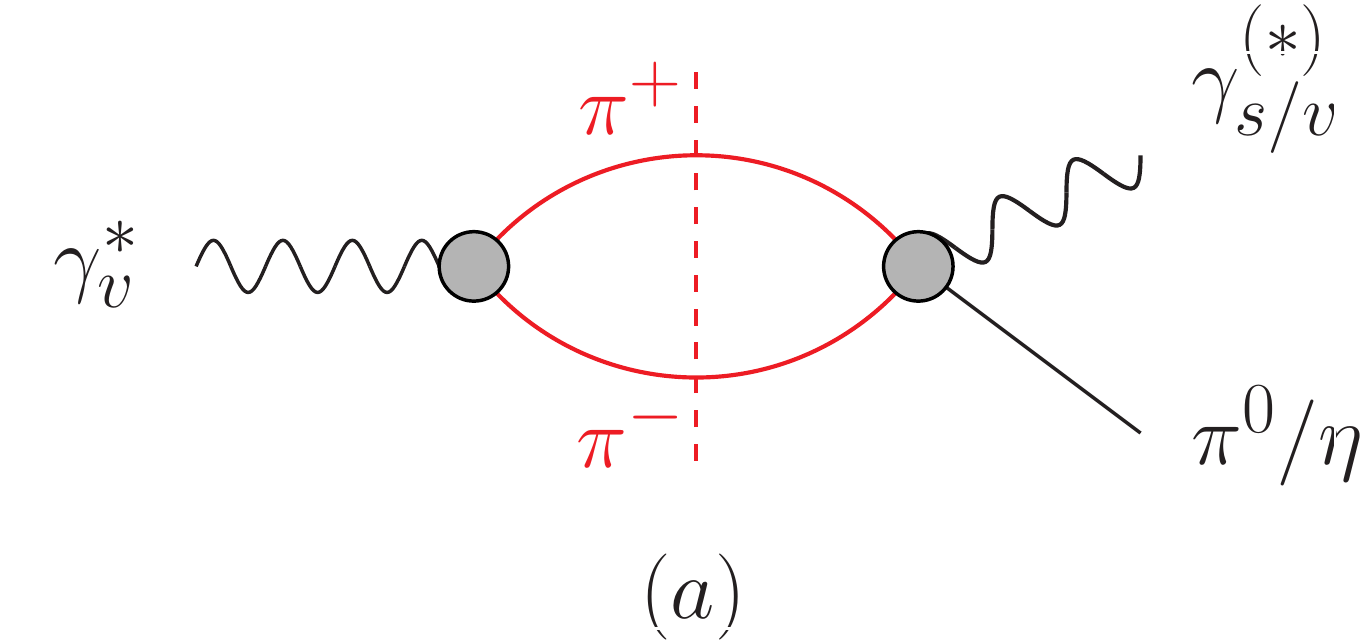} \hfill
\includegraphics*[width=0.32\linewidth]{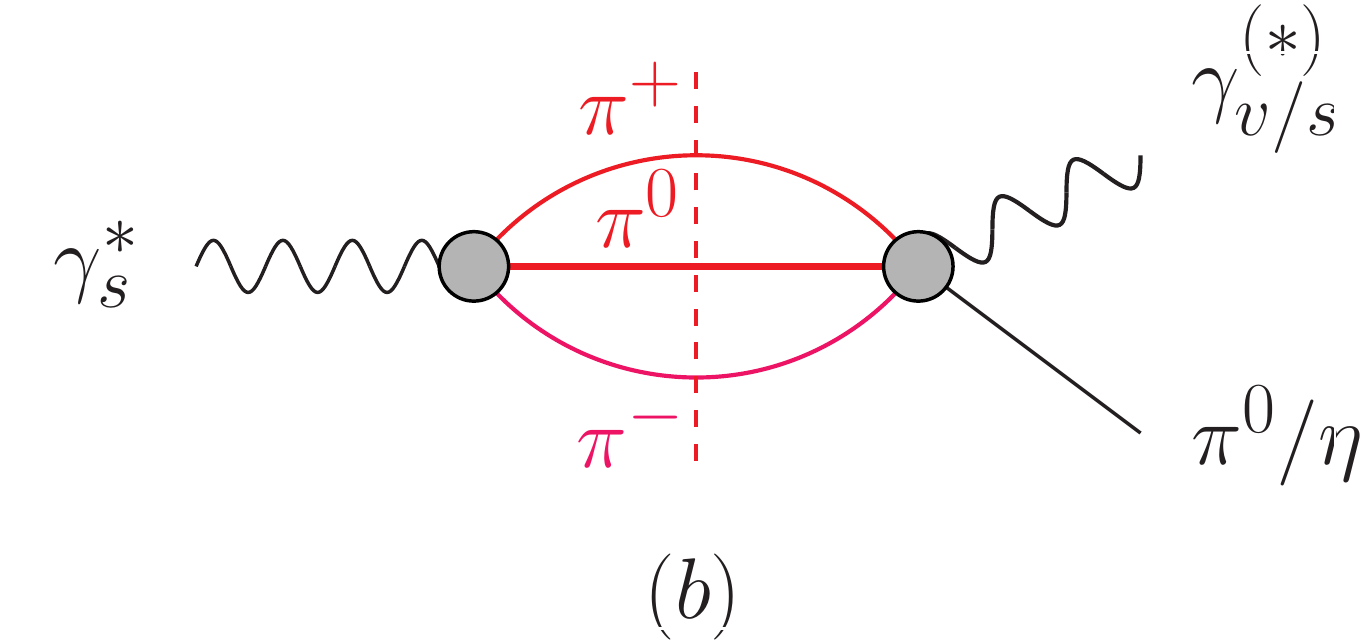} \hfill
\includegraphics*[width=0.32\linewidth]{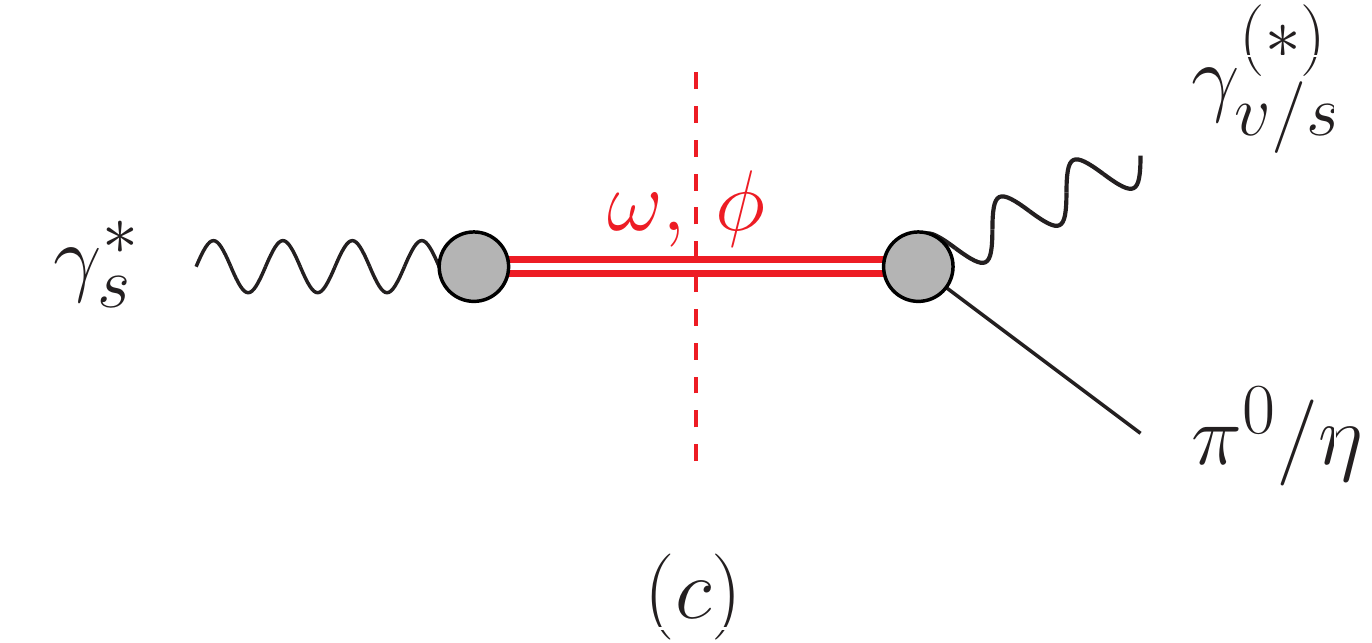}
\caption{Leading intermediate hadronic states in $\pi^0$ and $\eta$ transition form factors:
(a) due to two pions for an isovector photon, (b) due to three pions for an isoscalar photon,
(c) the three-pion intermediate state approximated by the lightest isoscalar vector resonances
$\omega$ and $\phi$.} \label{fig:2pi3pi}
\end{figure}
These are depicted in Fig.~\ref{fig:2pi3pi}.  For the isovector photons, the most important intermediate
state consists of a pair of charged pions, while for the isoscalar ones, three pions are the lightest option.
The isovector dispersion relation will then express the TFF in terms of the product
of the pion vector form factor $F_\pi^V(s)$ and 
the amplitude $\gamma^{(*)}\pi^0\to\pi^+\pi^-$ in the case of the $\pi^0$,
or the decay amplitudes $\ep\to\pi^+\pi^-\gamma^{(*)}$ for the $\ep$.
We will discuss the latter in detail in Sect.~\ref{sec:eta-pipigamma}.  
The most general description of the three-pion intermediate state for the isoscalar
photons, in contrast, would be a lot more complicated, and certain approximations at least for the transitions
$3\pi\to \gamma^{(*)}\pi^0/\ep$ are almost unavoidable (see, e.g., the corresponding discussion in Refs.~\cite{Hoferichter:2014vra,Hoferichter:2018dmo}).
Fortunately, the vector--isoscalar spectral function at low energies is strongly dominated by the narrow 
$\omega$ and $\phi$ resonances, see Fig.~\ref{fig:2pi3pi}(c), 
such that a VMD approximation is justified here to a large extent.
In the context of the $\pi^0$, the corresponding $\omega\to\pi^0\gamma^*$ and $\phi\to\pi^0\gamma^*$ transition form factors 
have hence been treated dispersively~\cite{Niecknig:2012sj,Schneider:2012ez,Danilkin:2014cra}
(see Ref.~\cite{Kubis:2014gka} for an extension even to the $J/\psi\to\pi^0\gamma^*$ transition), 
while a pure VMD description for 
$F_{ss}(q_1^2,q_2^2)$ was so far deemed sufficient for the $\ep$ TFFs.

An unsubtracted dispersion relation for the function $F_{vs}(q_1^2,q_2^2)$ specifying the 
$\pi^0$ TFF, see Eq.~\eqref{eq:TFF-isodecomp}, is given by~\cite{Hoferichter:2014vra}
\beq
F_{vs}^\text{disp}(q_1^2,q_2^2)=
\frac{1}{12\pi^2}\int^{\infty}_{4\mpi^2}\diff x\frac{q_{\pi}^3(x)\big(F_\pi^{V}(x)\big)^{*}f^\pi_1(x,q_2^2)}{x^{1/2}(x-q_1^2-i\eps)} \,.
\label{eq:pi0TFF-disp}
\eeq
Here, $q_\pi(s)=\sqrt{s/4-M_\pi^2}$, $F_\pi^{V}(s)$ is the pion vector form factor, as before, and $f^\pi_1(s,q^2)$ denotes 
the $P$-wave amplitude for $\gamma^*(q)\pi\to\pi\pi$:
\beq
\langle 0|j_\mu(0)|\pi^+(p_+)\pi^-(p_-)\pi^0(p_0)\rangle =-\epsilon_{\mu\nu\alpha\beta}\, p_+^\nu p_-^\alpha p_0^\beta \, \F_\pi(s,t,u;q^2) \,, \quad
f_1^\pi(s,q^2) = \frac{3}{4}\int_{-1}^1\diff z_s \,(1-z_s^2)\, \F_\pi(s,t,u;q^2) \,,
\eeq
where the Mandelstam variables are defined according to 
$s=(p_++p_-)^2$, $t=(p_-+p_0)^2$, and $u=(p_++p_0)^2$, and $z_s = \cos\theta_s$ is related to the $s$-channel center-of-mass
scattering angle $\theta_s$.  
The partial wave $f^\pi_1(s,q^2)$ 
is constructed based on solutions of Khuri--Treiman equations~\cite{Khuri:1960zz}. 
It contains a subtraction (normalization) function $a(q^2)$
that parameterizes the dependence on the three-pion invariant mass squared:
prominent ingredients are (dispersively improved) Breit--Wigner poles for $\omega$ and $\phi$, 
and the energy region beyond the $\phi$ is described by both
higher ($\omega'$) resonances and a conformal polynomial~\cite{Hoferichter:2018kwz}.
All free parameters are fitted to $e^+e^-\to3\pi$ cross section data up to 
$q^2 = (1.8\,\text{GeV})^2$~\cite{Achasov:2002ud,Achasov:2003ir,Aubert:2004kj} (cf.\ also Ref.~\cite{Hoferichter:2019gzf}).
At the real-photon point, $f^\pi_1(s,q^2=0)$ describes the photon--pion reaction $\gamma\pi\to\pi\pi$~\cite{Hoferichter:2012pm,Hoferichter:2017ftn}, while Dalitz plot distributions on $\omega\to3\pi$~\cite{Adlarson:2016wkw,Ablikim:2018yen} and $\phi\to3\pi$~\cite{Aloisio:2003ur,Akhmetshin:2006sc} can be used to constrain it on the narrow isoscalar vector resonances~\cite{Niecknig:2012sj}.  

A subtracted variant of Eq.~\eqref{eq:pi0TFF-disp} was studied in detail in Ref.~\cite{Hoferichter:2014vra} to predict
the singly-virtual $\pi^0$ TFF both for time-like and space-like momenta up to about $1\GeV^2$, and evaluate a precise
sum rule for the slope parameter $b_\pi$.  However, as an input to the evaluation of the $\pi^0$ pole contribution
to the anomalous magnetic moment of the muon, the TFF is integrated over all possible space-like momenta up to 
infinity, and hence a more consistent matching of the form factor representation onto its correct asymptotic form
is highly desirable.

At large $Q^2$, perturbative QCD (pQCD) predicts that the TFFs can be written as convolutions of a calculable hard scattering amplitude for $\gamma^*\gamma^*\rightarrow q\bar{q}$ with a nonperturbative meson distribution amplitude $\phi_P(x)$,
\begin{equation}
F_{P\gamma^*\gamma^*}(-Q_1^2,-Q_2^2) = \sum_a 2\langle\mathcal{Q}^2\lambda^a\rangle F_P^a \int_0^1 \diff x \frac{\phi_P^a(x)}{x Q_1^2 +(1-x)Q_2^2} + \Order\big(Q_i^{-4}\big) \,, \label{eq:distamp}
\end{equation}
where $\mathcal{Q}$ denotes the quark charge matrix for the three light flavors, $\mathcal{Q} = \text{diag}(2,-1,-1)/3$,
and $\lambda^a$ are the Gell-Mann matrices.  Asymptotically, the distribution amplitudes approach
$\phi_P^a(x) = 6x(1-x)$~\cite{Lepage:1980fj,Brodsky:1981rp}, which leads to 
\beq
F_{P\gamma^*\gamma^*}(-Q_1^2,-Q_2^2) = \sum_a 4\langle\mathcal{Q}^2\lambda^a\rangle F_P^a \frac{f(\omega)}{Q_1^2+Q_2^2} + \Order(Q_i^{-4})\,, \qquad
f(\omega) = \frac{3}{2\omega^2} \bigg(1-\frac{1-\omega^2}{2\omega}\log\frac{1+\omega}{1-\omega}\bigg) \,, \label{eq:fomega}
\eeq
with $\omega = (Q_1^2-Q_2^2)/(Q_1^2+Q_2^2)$.  Equation~\eqref{eq:fomega} predicts the asymptotic behavior for arbitrary
$Q_1^2+Q_2^2 \to \infty$.  In particular, $f(\omega=0)=1$ corresponds to the symmetric limit $Q_1^2=Q_2^2\to\infty$
given rigorously by the operator product expansion~\cite{Gorsky:1987,Manohar:1990hu}, while formal evaluation in 
the singly-virtual limit $f(\omega=\pm1)=3/2$ leads to the so-called Brodsky--Lepage limit.
In both limits, the anomalous dimension of the singlet axial current~\cite{Leutwyler:1997yr} 
leads to $\alpha_s$ corrections that are important when studying the asymptotic behavior of 
$\eta$ and $\eta'$ TFFs~\cite{Agaev:2014wna,Escribano:2015nra,Escribano:2015yup},
\begin{align}
\lim_{Q^2\to\infty}Q^2F_{P\gamma^*\gamma^*}(-Q^2,0) &= \sum_a 6\langle\mathcal{Q}^2\lambda^a\rangle F_P^a\left(1-\delta^{a0}\frac{2N_f}{\pi\beta_0}\alpha_s(\mu_0)\right) \,, \nonumber \\
\lim_{Q^2\to\infty}Q^2F_{P\gamma^*\gamma^*}(-Q^2,-Q^2) &= \sum_a 2\langle\mathcal{Q}^2\lambda^a\rangle F_P^a\left(1-\delta^{a0}\frac{2N_f}{\pi\beta_0}\alpha_s(\mu_0)\right) \,,
\label{eq:TFFOPE}
\end{align}
where $N_f$ refers to the number of effective active flavors at each scale, and $\beta_0 = 11N_c/3 -2N_f/3$.
It is obvious from the first line of Eq.~\eqref{eq:TFFOPE} that a careful study of the large-$Q^2$ behavior of the 
singly-virtual $\eta$ and $\eta'$ TFFs contains important information on $\eta$--$\eta'$ mixing (in addition to 
the real-photon decay widths discussed in the previous section).  Combined studies to extract decay constants and 
mixing parameters have been performed in Refs.~\cite{Escribano:2013kba,Escribano:2015nra,Escribano:2015yup}, 
using Pad\'e approximants to describe the data.  Here, we make use of the asymptotic distribution amplitude 
for the $\pi^0$.

For the continuation of the representation~\eqref{eq:pi0TFF-disp} into the (doubly-)space-like region 
as required, e.g., for the evaluation of the $(g-2)_\mu$ loop integration, 
it is advantageous to rewrite it in terms of a double-spectral representation~\cite{Hoferichter:2018dmo,Hoferichter:2018kwz},
\beq
F_{vs}^\text{disp}(-Q_1^2,-Q_2^2) =
\frac{1}{\pi^2} \int_{4M_\pi^2}^\infty \diff x \int_{s_\text{thr}}^\infty \diff y \frac{\rho(x,y)}{\big(x+Q_1^2\big)\big(y+Q_2^2\big)} \,, \qquad
\rho(x,y) =\frac{q_\pi^3(x)}{12\pi\sqrt{x}}\Im \Big[\big(F_\pi^{V}(x)\big)^*f^\pi_1(x,y)\Big] \,.
\label{eq:piTFF-doublespectral}
\eeq
In practice, this unsubtracted representation based on two- and three-pion intermediate states only is not sufficiently accurate yet: a resulting sum rule for the form factor normalization is only fulfilled at the 90\% level.  The original analysis of the singly-virtual TFF was therefore based on a subtracted dispersion relation~\cite{Hoferichter:2014vra}, whose high-energy asymptotics were, as a consequence, distorted.  Alternatively, in Refs.~\cite{Hoferichter:2018dmo,Hoferichter:2018kwz}, the form factor~\eqref{eq:pi0TFF-disp} [or, equivalently, the double-spectral function~\eqref{eq:piTFF-doublespectral}] was supplemented with two further components,
\beq
F_{\pi^0\gamma^*\gamma^*}=F_{\pi^0\gamma^*\gamma^*}^\text{disp}+F_{\pi^0\gamma^*\gamma^*}^\text{eff}+F_{\pi^0\gamma^*\gamma^*}^\text{asym} \,.
\label{eq:pi0TFF-complete}
\eeq
The low-energy dispersive part is evaluated up to maximal cutoffs.  The second term is
an effective pole that subsumes higher intermediate states as well as higher energies in the two- and three-pion continua, 
and is of the form
\begin{equation}
F_{\pi^0\gamma^*\gamma^*}^\text{eff}(-Q_1^2,-Q_2^2)=\frac{g_{\text{eff}}}{4\pi^2F_\pi}\frac{M_{\text{eff}}^4}{(M_{\text{eff}}^2+Q_1^2)(M_{\text{eff}}^2+Q_2^2)} \,.
\end{equation}
The effective coupling $g_{\text{eff}}$ is chosen to fulfill the sum rule for the form factor normalization, while
the effective mass $M_{\text{eff}}$ can be adapted to the Brodsky--Lepage limit in Eq.~\eqref{eq:TFFOPE}.  
The third, asymptotic, piece in Eq.~\eqref{eq:pi0TFF-complete} is given by
\beq
F_{\pi^0\gamma^*\gamma^*}^\text{asym}(-Q_1^2,-Q_2^2)
 = 2F_\pi\int_{s_m}^\infty \diff x \frac{Q_1^2Q_2^2}{(x+Q_1^2)^2(x+Q_2^2)^2} \,,
\eeq
which is derived from the double-spectral representation of the distribution amplitude~\eqref{eq:distamp}, but does not
contribute in singly-virtual kinematics.
In this way, the representation~\eqref{eq:pi0TFF-complete} fulfills all asymptotic constraints from pQCD (at leading order),
reproduces the high-$Q^2$ singly-virtual data, and is faithful to the analytic structure at low-energies with high accuracy.

\begin{figure}
\includegraphics[width=0.51\linewidth]{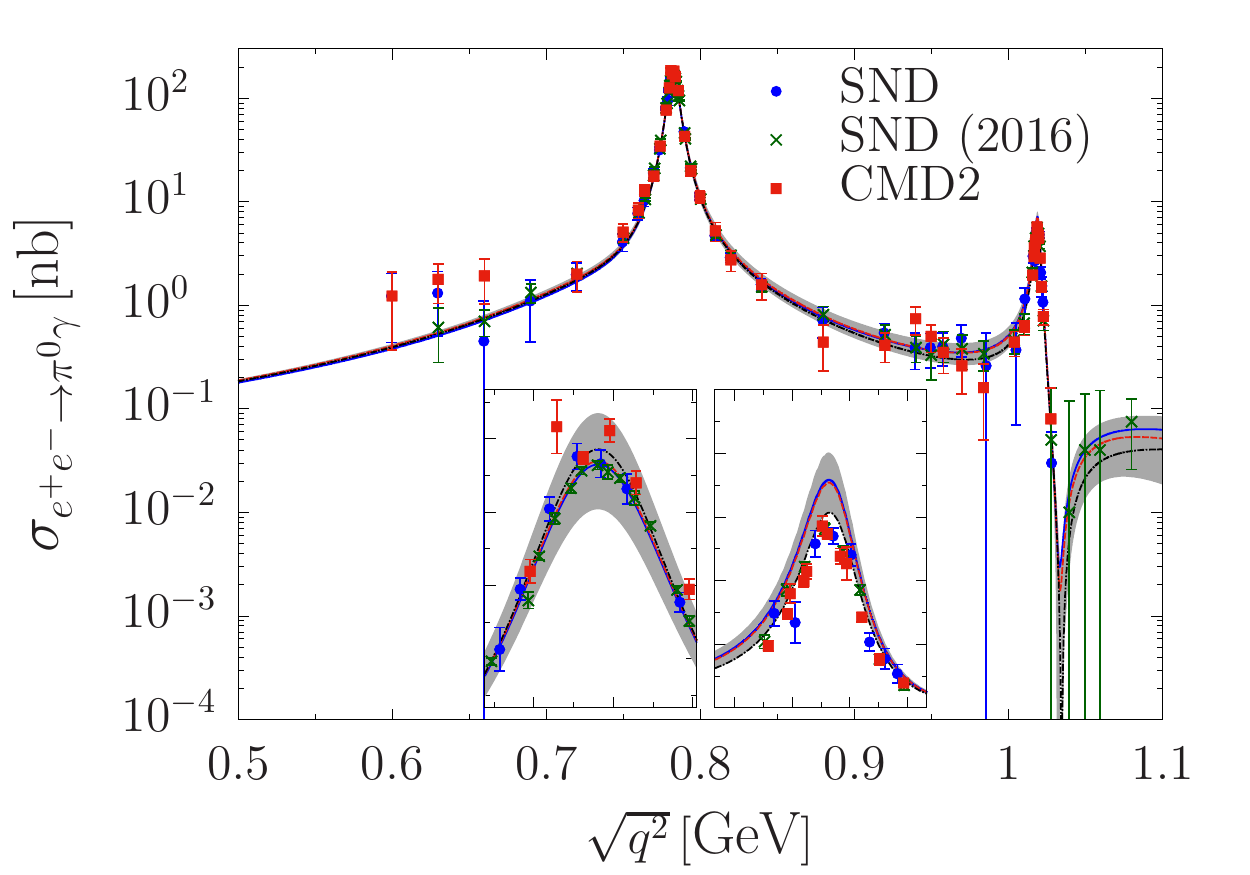} \hfill
\includegraphics[width=0.48\linewidth]{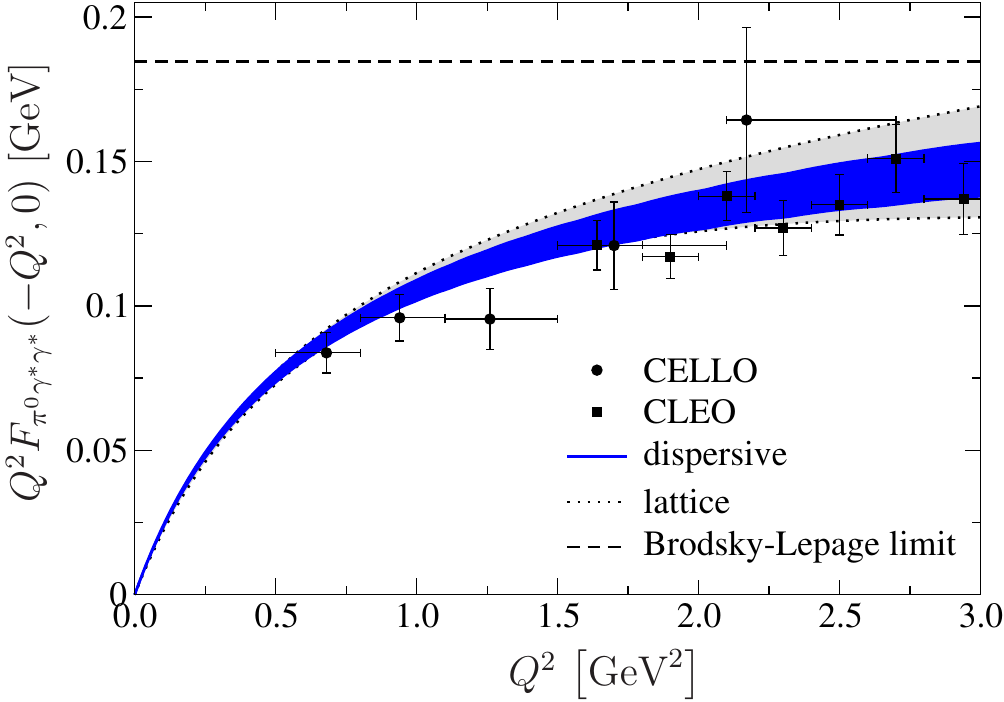} 
\caption{\textit{Left:} $e^+e^-\to\pi^0\gamma$ cross section of Refs.~\cite{Hoferichter:2014vra,Hoferichter:2018kwz} 
in comparison to data of SND~\cite{Achasov:2000zd,Achasov:2003ed}, CMD-2~\cite{Akhmetshin:2004gw}, and SND (2016)~\cite{Achasov:2016bfr}.  Figure courtesy of B.-L.~Hoid, adapted from Ref.~\cite{Hoferichter:2018kwz}.
\textit{Right:} Space-like TFF of Refs.~\cite{Hoferichter:2018dmo,Hoferichter:2018kwz} (blue), compared to 
lattice QCD~\cite{Gerardin:2019vio} (gray), as well as the CELLO~\cite{Behrend:1990sr} and CLEO~\cite{Gronberg:1997fj} data.}
\label{fig:pi0TFF}
\end{figure}
In Fig.~\ref{fig:pi0TFF}, we show both the cross section 
\beq
\sigma_{e^+e^- \to \pi^0 \gamma}(q^2) = \frac{2\pi^2\alpha_\text{em}^3}{3} \frac{(q^2-M_{\pi^0}^2)^3}{q^6} \, \big|F_{\pi^0\gamma^*\gamma^*}\big(q^2,0\big)\big|^2 
\eeq
that is based on the time-like $\pi^0$ TFF, 
and the space-like $\pi^0$ TFF at low energies, compared to data and
to a recent lattice calculation~\cite{Gerardin:2019vio}.  Agreement with the existing data is excellent 
throughout.  The theoretical prediction for the slope parameter $b_\pi$, Eq.~\eqref{eq:TFFslope},
is~\cite{Hoferichter:2018kwz,Hoferichter:2018dmo}
\beq
b_\pi = 31.5(9)\times 10^{-3} \mpiz^{-2} = 1.73(5) \GeV^{-2} \,,
\eeq
where the uncertainty is dominated by the variation of the dispersive input (different phase shifts,
cutoffs, etc.).
Also higher terms in the polynomial low-energy expansion can be extracted from dispersive sum rules
with high precision.  This result can be compared to the 
experimental average $b_\pi = 33.5(3.1) \times 10^{-3} \mpiz^{-2} =  1.84(17)\GeV^{-2}$~\cite{Tanabashi:2018oca}
as well as to an extraction from data based on
Pad\'e approximants, $b_\pi = 32.4(2.2) \times 10^{-3} \mpiz^{-2} = 1.78(12)\GeV^{-2}$~\cite{Masjuan:2012wy},
both with appreciably larger uncertainties. 
Detailed radiative corrections~\cite{Husek:2015sma} have to be taken into account for precise experimental determinations
of this slope~\cite{Adlarson:2016ykr,TheNA62:2016fhr}.
Due to the smallness of the phase space, there is even a close relation between the form factor
slope and the branching ratio of the $\pi^0$ Dalitz decay, which can be scrutinized with much higher precision
including NLO QED effects~\cite{Husek:2018qdx}.

The dispersive representation~\eqref{eq:pi0TFF-complete} has been used to calculate the $\pi^0$ pole piece in
the hadronic light-by-light scattering contribution to the anomalous magnetic moment of the muon, 
\begin{equation}
\label{eq:g-2result}
 a_\mu^{\pi^0}=63.0(0.9)_{F_{\pi\gamma\gamma}}(1.1)_\text{disp}\big({}^{2.2}_{1.4}\big)_\text{BL}(0.6)_\text{asym}\times 10^{-11}
 =63.0\big({}^{2.7}_{2.1}\big)\times 10^{-11} \,.
\end{equation}
Here, the individual uncertainties are due to the form factor normalization, dispersive input, experimental 
uncertainty in the singly-virtual data around the Brodsky--Lepage limit, and the onset of the asymptotic
contribution to the double-spectral function.  This result makes the potential impact of new experimental 
measurements on the $(g-2)_\mu$ determination as transparent as possible; in fact, Eq.~\eqref{eq:g-2result}
has already been updated with respect to the published result~\cite{Hoferichter:2018kwz,Hoferichter:2018dmo}
by taking the more precise combined PrimEx-I and -II result on the $\pi^0\to\gamma\gamma$ width into account~\cite{Larin:2020bhc}.
Equation~\eqref{eq:g-2result} is perfectly consistent with the determination based on rational approximants,
$a_\mu^{\pi^0}=63.6(2.7)\times 10^{-11}$~\cite{Masjuan:2017tvw},
the latest lattice determination,
$a_\mu^{\pi^0}=62.3(2.3)\times 10^{-11}$~\cite{Gerardin:2019vio},
or evaluations based on Dyson--Schwinger equations,
$a_\mu^{\pi^0}=62.6(1.3)\times10^{-11}$~\cite{Eichmann:2019tjk} and 
$a_\mu^{\pi^0}=61.4(2.1)\times10^{-11}$~\cite{Raya:2019dnh}
(where in the last two cases, systematic 
uncertainties may be somewhat underestimated).

\subsection{$\eta, \eta' \to\pi^+\pi^-\gamma$}
\label{sec:eta-pipigamma}
As we have seen in Sect.~\ref{sec:pi0-eta-etapTFF}, one of the central input quantities
for a dispersion-theoretical
description of the $\eta$ transition form factor as measured, e.g., in $\eta\to\ell^+\ell^-\gamma$
is the decay $\eta\to \pi^+\pi^-\gamma$.  
Due to its much higher rate, or the less severe suppression of the decay in powers of the fine structure constant,
a dispersive reconstruction based on such data can be significantly more precise than a direct measurement 
of the Dalitz decay.

$\eta\to \pi^+\pi^-\gamma$ is a further process driven by the chiral anomaly~\cite{Wess:1971yu,Witten:1983tw}.  
We write the decay amplitude for $\eta(q) \to \pi^+(p_1)\pi^-(p_2)\gamma(k)$, 
assuming \textit{CP}-invariance,\footnote{We discuss $CP$-violating amplitudes for $\eta \to \pi^+ \pi^-\gamma$ transitions in Sect.~\ref{sec:eta-2pi}.}
in terms of a scalar function $\F_\eta(s,t,u)$ according to
\beq
\langle \pi^+(p_1)\pi^-(p_2) | j_\mu(0) | \eta(q) \rangle = \eps_{\mu\nu\alpha\beta} p_1^\nu p_2^\alpha q^\beta \, \F_\eta(s,t,u), \label{eq:Amp-eta-pipigamma}
\eeq
with the Mandelstam variables given as $s=(p_1+p_2)^2$, $t=(q-p_1)^2$, and $u=(q-p_2)^2$. 
$\F_\eta(s,t,u)$ in the chiral limit fulfills the low-energy theorem 
\beq
\F_\eta(0,0,0)=F_{\eta\pi\pi\gamma} = \frac{1}{4\sqrt{3}\pi^2F_\pi^3} \bigg( \cos\theta_P\,\frac{F_\pi}{F_8} - \sqrt{2} \sin\theta_P\, \frac{F_\pi}{F_0} \bigg) \times(1+\delta) ,
\eeq
using a single-mixing-angle scheme, where $\delta$ denotes corrections proportional to the quark masses.
The higher-order corrections to the anomaly have been evaluated in chiral 
perturbation theory~\cite{Bijnens:1989ff}, as well as unitarized versions thereof~\cite{Borasoy:2004qj}.
Comparison with data~\cite{Babusci:2012ft}, employing the theoretical energy dependence of the amplitude
as described below~\cite{Kubis:2015sga}, suggests quark mass corrections of reasonable size, $\delta = -0.17(3)$
(based on $\theta_P = -19.5^\circ$ and Eq.~\eqref{eq:mixing-Feldmann} for the decay constants).

The pion--pion partial-wave expansion is of the form
\beq \label{eq:t-pwa}
\F_\eta(s,t,u) = \sum_{\text{odd}~l} P'_l(z) f^\eta_l(s) , \qquad
z = \cos\theta = \frac{t-u}{\sigma(\meta^2-s)} , \qquad \sigma = \sqrt{1-\frac{4\mpi^2}{s}},
\eeq
where $P'_l(z)$ are the derivatives of the standard Legendre polynomials.
As $F$ and higher partial waves are strongly suppressed at low energies, 
the decay is totally dominated by the $P$-wave, which is obtained by angular projection
according to
\beq
f^\eta_1(s) = \frac{3}{4}\int_{-1}^1\diff z \big(1-z^2\big)\F_\eta(s,t,u).
\eeq
The differential decay rate with respect to the pion--pion invariant mass squared is then given by
\begin{align} \label{eq:dGdt}
\frac{\diff\Gamma}{\diff s} &= \Gamma_0(s) \times
\frac{3}{4}\int_{-1}^1 \diff z \big(1-z^2\big)|\F_\eta(s,t,u)|^2 
=  \Gamma_0(s) \times \Big( \big|f^\eta_1(s)\big|^2 + \ldots \Big) , \qquad 
\Gamma_0(s) = \frac{e^2 s \sigma^3(\meta^2-s)^3}{12(8\pi\meta)^3} ,
\end{align}
where the ellipsis represents neglected higher partial waves.

The power of the universality of final-state interactions lies in the fact that an Omn\`es representation similar
to the one employed for the pion vector form factor, see Eq.~\eqref{eq:Omnes},
will apply everywhere where two pions are produced from a point source in a relative $P$-wave; the process-dependence can be reduced to the 
coefficients of the multiplicative polynomial.  
It was pointed out that such a representation can in particular
be used for the decays $\ep\to\pi^+\pi^-\gamma$~\cite{Stollenwerk:2011zz}: 
for the decay of the $\eta$, an ansatz with a linear polynomial,
\begin{equation}
f^\eta_1(s) = P^\eta(s) \Omega_1^1(s) = A_\eta \bar P^\eta(s) \Omega_1^1(s) = A_\eta(1+\alpha_\eta s) \Omega_1^1(s) \,, \label{eq:f1}
\end{equation}
was shown to be sufficient to describe the data in the physical decay region~\cite{Adlarson:2011xb,Babusci:2012ft}.
However, there is one important difference to the vector form factor: 
\begin{figure}
\includegraphics[width=0.48\linewidth]{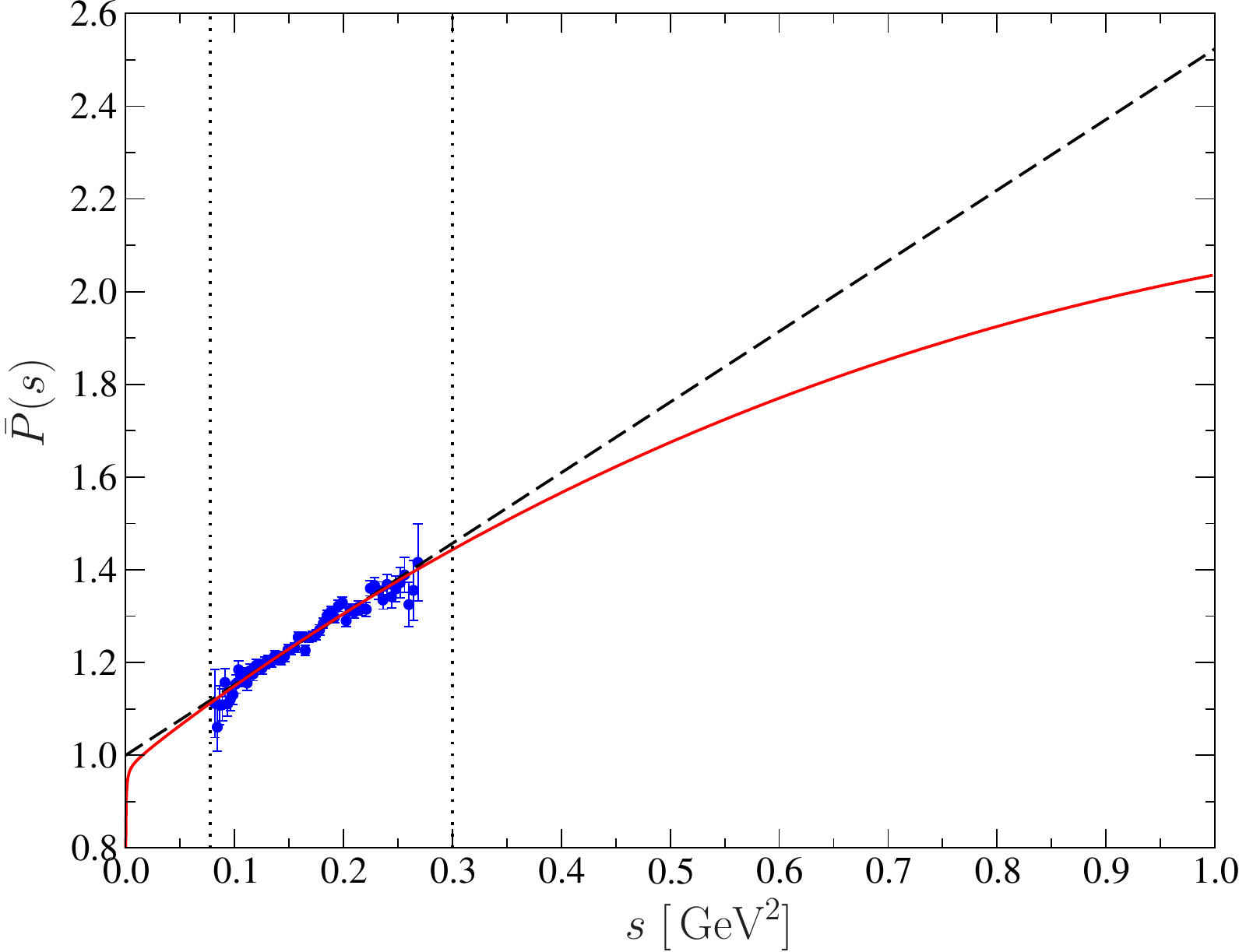} \hfill
\includegraphics[width=0.48\linewidth]{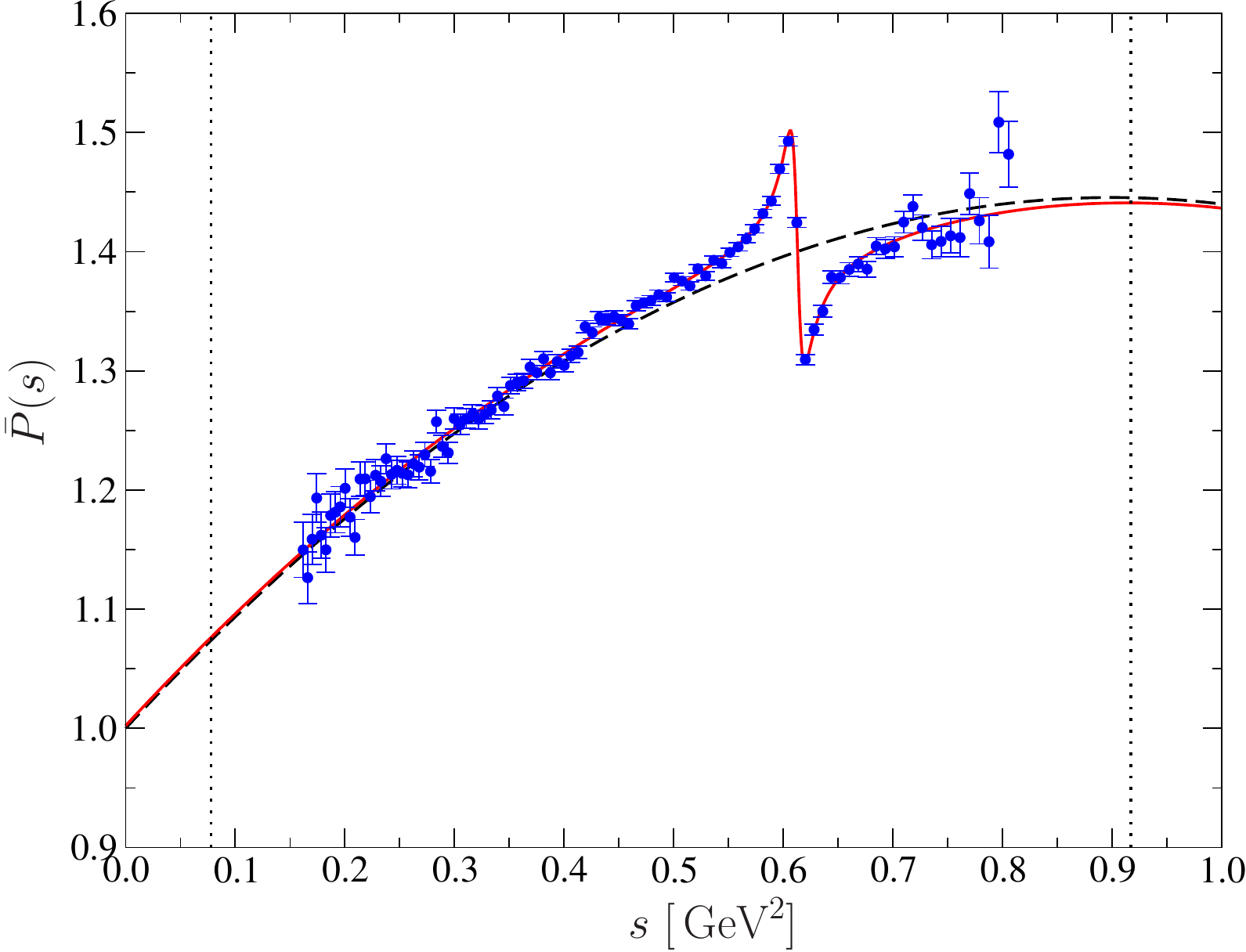} 
\caption{\textit{Left:} Representation of the decay distribution $\eta\to\pi^+\pi^-\gamma$ as measured
by KLOE~\cite{Babusci:2012ft}
to allow for comparison to the polynomial $\bar P^\eta(s)=1+\alpha_\eta s$ (dashed curve).
The full red curve includes the effects of $a_2$-exchange in addition.
The vertical dotted lines denote the boundaries of phase space at $s=4M_\pi^2$ and $s=M_\eta^2$.
\textit{Right:} The same for the decay distribution $\eta'\to\pi^+\pi^-\gamma$, showing pseudo-data 
generated according to preliminary BESIII results~\cite{Fang:2015vva}. 
The dashed curve represents the fit of a quadratic polynomial $\bar P^{\eta'}(s)=1+\alpha_{\eta'} s+\beta_{\eta'} s^2$, 
while the full red curve includes $\rho$--$\omega$ mixing. 
Figures adapted from Ref.~\cite{Kubis:2019tdz}.}\label{fig:KLOElinear}
\end{figure}
while the slope parameter $\alpha_V$ therein is relatively small, $\alpha_V \sim 0.1\GeV^{-2}$, here
$\alpha_\eta$ turns out to be large, 
$\alpha_\eta \sim 1.5\GeV^{-2}$~\cite{Babusci:2012ft,Kubis:2015sga}; see Fig.~\ref{fig:KLOElinear}.
While therefore $\alpha_V$ can consistently be described as the low-energy tail of higher resonances 
in the region $1\GeV \lesssim \sqrt{s} \lesssim 2\GeV$, such an interpretation seems implausible
for the $\eta\to\pi^+\pi^-\gamma$ decay.  Furthermore, it is known that a linear polynomial $P^\eta(s)$ 
can only be a low-energy approximation: with the pion--pion $P$-wave Omn\`es function dropping like $1/s$
for large energies, the polynomial multiplying it should rather approach a constant asymptotically.
We will discuss in Sect.~\ref{sec:DRetaTFF} that the $\eta\to\pi^+\pi^-\gamma$ $P$-wave $f^\eta_1(s)$ features
prominently in a dispersive analysis of the $\eta$ transition form factor.
Given that the corresponding dispersion integral formally extends to infinity, 
deviations from the linear rise in $P^\eta(s)$ may become important.  
It is therefore crucial to search for opportunities to investigate $f^\eta_1(s)$ outside the relatively
narrow $\eta$ decay region.

One remarkable property of the amplitude~\eqref{eq:f1} is the fact that it displays a zero 
for negative values of $s$ at 
$s = -1/\alpha_\eta \approx -0.66\GeV^2$, which therefore occurs at rather low energies.  This kinematic configuration
can be tested in the crossed process $\gamma\pi^-\to\pi^-\eta$, which can be investigated experimentally
in a Primakoff reaction, i.e., the scattering of highly energetic charged pions in the strong Coulomb field
of a heavy nucleus, 
e.g., at COMPASS; see Ref.~\cite{Kaiser:2008ss} for an overview of the COMPASS 
Primakoff program.  
The zero would then occur in the angular distribution
at fixed energies in the crossed center-of-mass system.  In order to judge how reliable such a prediction
is, we need to consider the possible effects of nontrivial $\pi\eta$ dynamics---something that has been entirely
neglected in the discussion of the decay $\eta\to\pi^+\pi^-\gamma$ so far.  
A $\pi\eta$ $S$-wave is forbidden in this anomalous process.  The $P$-wave is of exotic quantum numbers,
$J^{PC}=1^{-+}$, i.e., there are no resonances possible in a quark model of mesons.
The first hybrid candidate, the $\pi_1(1600)$~\cite{Adolph:2014rpp,Schott:2012wqa}, is a rather broad state 
($m_{\pi_1} = 1564(24)(86)\MeV$, $\Gamma_{\pi_1} = 492(54)(102)\MeV$~\cite{Rodas:2018owy})
that does not couple particularly strongly to $\pi\eta$,
hence the $P$-wave is expected to be very weak at low energies.
The first resonance to occur in $\gamma\pi^-\to\pi^-\eta$ is therefore the $D$-wave $a_2(1320)$.
At the same time, it provides the dominant left-hand-cut contribution to the decay $\eta\to\pi^+\pi^-\gamma$ that has not
been considered so far: 
\begin{figure}[t!]
\centering
\includegraphics[width=0.54\linewidth]{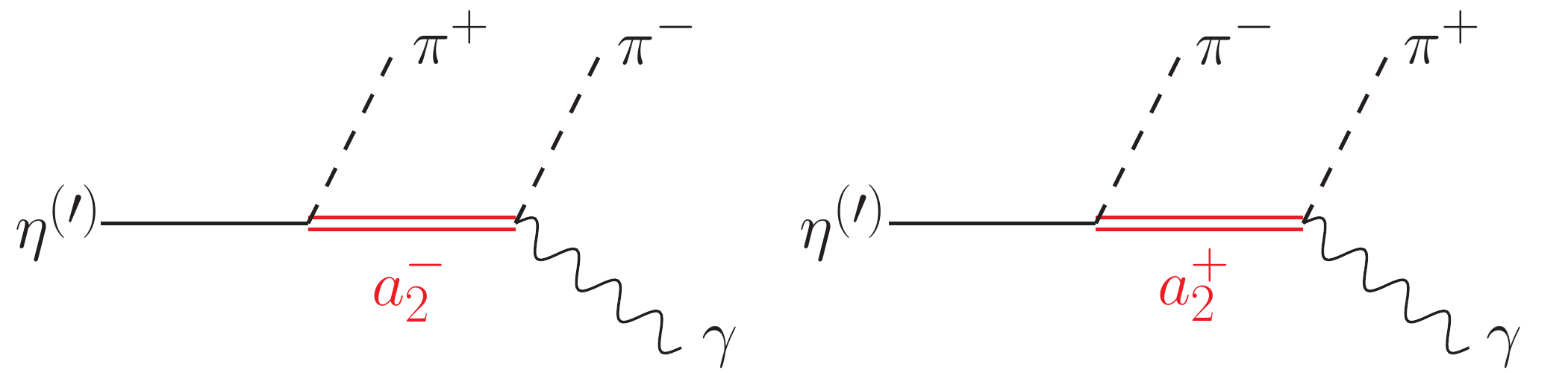}
\includegraphics[width=0.27\linewidth]{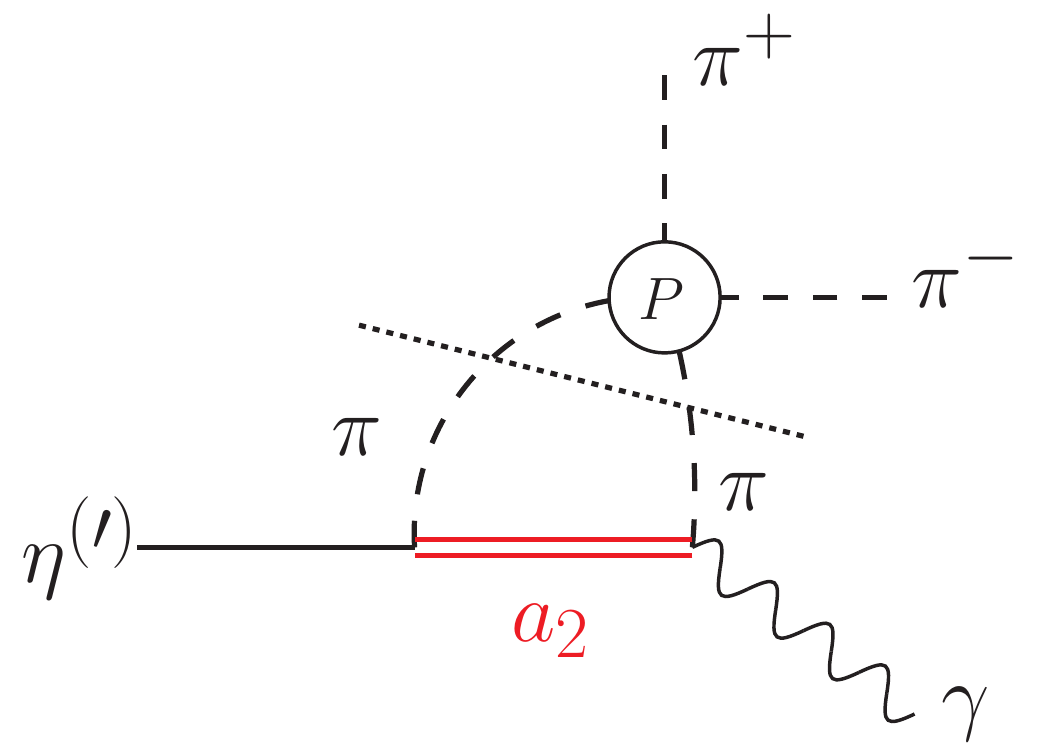}
\caption{$a_2$ contributions to $\ep\to\pi^+\pi^-\gamma$, both at tree level and including pion--pion $P$-wave rescattering.
The combination of both is required to preserve the final-state theorem.}\label{fig:a2}
\end{figure}
a dispersion relation including the effects of $a_2$-exchange, see Fig.~\ref{fig:a2}, is more complicated
than the form factor relation~\eqref{eq:UnitarityFF}, and requires a more complicated solution~\cite{Kubis:2015sga}.  
The full dispersive amplitude that respects the phase relation in the $s$-channel $P$-wave is of the form
\begin{align} \label{eq:full}
f^\eta_1(s) = \F(s)+\hat\G(s), \quad
\F(s) = \Omega_1^1(s)\Bigg\{ A_\eta(1+\alpha_\eta s) + \frac{s^2}{\pi}\int_{4\mpi^2}^\infty
\frac{\diff x}{x^2} \frac{\sin\delta_1^1(x) \hat\G(x)}{|\Omega_1^1(x)|(x-s-i\eps)} \Bigg\} .
\end{align}
$\hat\G(s)$ is the projection of the $a_2$ tree-level $t$- and $u$-channel exchange amplitudes
onto the $s$-channel $P$-wave.
Ref.~\cite{Kubis:2015sga} also discusses the predictions for the Primakoff reaction $\gamma\pi^-\to\pi^-\eta$: 
the zero of the simple amplitude representation~\eqref{eq:f1} partially survives in the form of a strong amplitude suppression
in backward direction in the energy region between threshold and the $a_2$ resonance, that can be explained as a 
$P$-$D$-wave interference effect. 
Here we only discuss the compatibility of this more refined model with the decay data: if the very precise
data on $\eta\to\pi^+\pi^-\gamma$ by the KLOE collaboration~\cite{Babusci:2012ft} was already accurately described
by the simpler amplitude of the form $P^\eta(s)\times\Omega_1^1(s)$, is this accuracy preserved when including $a_2$ effects?
The answer is shown in Fig.~\ref{fig:KLOElinear}: the left-hand cut induces a curvature in $P^\eta(s)$ that has only
very small effects in the decay region, resulting in a data fit of equal quality; however, it also demonstrates that the effect up to $\sqrt{s} \approx 1\GeV$ is quite 
nonnegligible, and can have a significant impact on the $\eta$ transition form factor via the
corresponding dispersion integral, see Sect.~\ref{sec:DRetaTFF}.

All the steps discussed above can be carried over 
to an analysis of the $\eta'\to\pi^+\pi^-\gamma$ decay in a straightforward manner.
The decay amplitude is written in the same form as Eq.~\eqref{eq:Amp-eta-pipigamma}, with the normalization
now given according to~\cite{Stollenwerk:2011zz}
\beq
\F_{\eta'}(0,0,0)=F_{\eta'\pi\pi\gamma} = \frac{1}{4\sqrt{3}\pi^2F_\pi^3} \bigg( \sin\theta_P\,\frac{F_\pi}{F_8} + \sqrt{2} \cos\theta_P\, \frac{F_\pi}{F_0} \bigg) \times(1+\delta') .
\eeq
Here, the quark-mass correction $\delta'$ in the amplitude normalization
as extracted phenomenologically from data~\cite{Hanhart:2016pcd}
seem to be even smaller, $\delta' = 0.03(5)$.  
While old data on $\eta'\to\pi^+\pi^-\gamma$ by the Crystal Barrel Collaboration~\cite{Abele:1997yi}
was not precise enough for an advanced analysis, a new measurement
by BESIII~\cite{Ablikim:2017fll} is in certain ways even more conclusive than the $\eta$ decay data, 
as the larger phase space
of the $\eta'$ decay allows one to see deviations from the assumption of a linear polynomial much more clearly.
It has been demonstrated in Ref.~\cite{Kubis:2015sga} that, despite the occurrence of a left-hand cut, 
in fact the ratio $f^{\eta'}_1(s)/\Omega_1^1(s)$ in the model including the $a_2$ can be approximated very accurately
by a \textit{quadratic} polynomial within the physical decay region, i.e., for $4M_\pi^2 \leq s \leq M_{\eta'}^2$.
The new BESIII data~\cite{Ablikim:2017fll} demonstrates the need of such a quadratic term to very high 
significance; moreover, they are so precise that even the isospin-breaking $\rho$--$\omega$-mixing effect is 
clearly discernible (see Ref.~\cite{Hanhart:2016pcd} for details on how to extract the mixing strength
and subsequently the partial width $\Gamma(\omega\to\pi^+\pi^-)$ from this decay, and Ref.~\cite{Dai:2017tew} for an analysis in R$\chi$T).
The full representation of the 
$\eta'\to\pi^+\pi^-\gamma$ $P$-wave amplitude is therefore of the form~\cite{Hanhart:2016pcd}
\begin{equation}
f^{\eta'}_1(s) = \bigg[A_{\eta'}\big(1+\alpha_{\eta'} s+\beta_{\eta'} s^2\big) + \frac{\kappa_2}{m_\omega^2-s-im_\omega \Gamma_{\omega}} \bigg] 
\times \Omega_1^1(s) \,.
\label{eq:etaprime}
\end{equation}
The fit to pseudo-data generated according to preliminary BESIII results~\cite{Fang:2015vva} is also shown in Fig.~\ref{fig:KLOElinear}.
The leading left-hand-cut contribution provided by $a_2$-exchange gives an estimate 
of the parameter $\beta_{\eta'} = -1.0(1)\GeV^{-4}$~\cite{Kubis:2015sga}, which yields the correct sign and order of magnitude, but 
is somewhat larger than what the new data suggest~\cite{Ablikim:2017fll,Hanhart:2016pcd}.
A preliminary analysis of data on the same decay taken by the CLAS collaboration~\cite{MbiandaNjencheu:2017vwy}
comes to qualitatively similar conclusions as far as the necessity of a curvature parameter $\beta_{\eta'}$ is concerned.
It will be interesting to see to what extent both high-statistics experiments, BESIII and CLAS, will ultimately agree
quantitatively on the precise decay distribution.

\subsection{Dispersive analysis of the singly-virtual $\eta$ and $\eta'$ transition form factors}\label{sec:DRetaTFF}
We now turn to the singly-virtual $\eta$ and $\eta'$ transition form factors and their dispersive description~\cite{Hanhart:2013vba}.
With the preparation of the previous section to provide a very precise representation of the $\ep\to\pi^+\pi^-\gamma$ decay amplitudes~\cite{Hanhart:2013vba,Kubis:2015sga,Hanhart:2016pcd} with few input parameters fixed from data~\cite{Babusci:2012ft,Ablikim:2017fll}, the main input to the isovector part of the dispersion relation is fixed.
Employing a once-subtracted dispersion relation 
one finds~\cite{Hanhart:2013vba}
\begin{equation} \label{eq:etaTFF-DR}
F_{\ep\gamma^*\gamma^*}(q^2,0) = F_{\ep\gamma\gamma} + 
\frac{q^2}{12\pi^2} \int_{4M_\pi^2}^{\infty} \diff x \frac{q_\pi^3(x)\big(F_\pi^{V}(x)\big)^{*}f^{\ep}_1(x)}{x^{3/2}(x-q^2-i\eps)} 
+ \Delta F^{I=0}_{\ep\gamma^*\gamma^*}(q^2,0) \,,
\end{equation}
where $F_{\ep\gamma\gamma}$ refers to the $\ep\to2\gamma$ anomalies~\eqref{eq:eta-gg-mixing}, and
the last term denotes the isoscalar contributions; see Ref.~\cite{Hanhart:2013vba} for details. 
The latter can be modeled well using the vector-meson-dominance assumption due to the narrowness
of $\omega$ and $\phi$, with the moduli of all necessary coupling constants extracted from 
the experimentally determined partial decay widths for 
$\omega\to\eta\gamma$, $\eta'\to\omega\gamma$, and $\phi\to\ep\gamma$, 
as well as $\omega,\,\phi\to e^+e^-$; see Sect.~\ref{sec:eta-omegag}.
In Ref.~\cite{Hanhart:2013vba}, $F^{I=0}_{\ep\gamma^*\gamma^*}(q^2,0)$ was thus expressed as
\beq
\Delta F^{I=0}_{\ep\gamma^*\gamma^*}(q^2,0) = F_{\ep\gamma\gamma}   \sum_{V=\omega,\phi} \frac{w_{\ep V \gamma}\,q^2}{m_V^2-q^2-im_V\Gamma_V}  \,,
\eeq
where the (real) weight factors are given as
\beq
w_{PV\gamma}^2 = \frac{9m_V^2M_P^3\,\Gamma(V\to e^+e^-) \,\Gamma(V\to P\gamma)}
{2\alpha_{\rm em}(m_V^2-M_P^2)^3\,\Gamma(P\to\gamma\gamma)} \,, \qquad
w_{\eta'\omega\gamma}^2 = \frac{3M_{\eta'}^6 \,\Gamma(\omega\to e^+e^-) \,\Gamma(\eta'\to \omega\gamma)}
{2\alpha_{\rm em}m_\omega(M_{\eta'}^2-m_\omega^2)^3\,\Gamma(\eta'\to\gamma\gamma)} \,,
\eeq
with the first relation valid for all cases with $m_V>M_P$, and we have neglected the electron mass throughout.  
The signs are obtained from comparison to VMD, which yields a negative $w_{\eta\phi\gamma}$ and positive weight 
factors otherwise.  This results in 
\beq
w_{\eta\omega\gamma} = 0.097(5) \,, \quad
w_{\eta\phi\gamma} = -0.188(3) \,, \quad
w_{\eta'\omega\gamma} = 0.074(2) \,, \quad
w_{\eta'\phi\gamma} = 0.154(4) \,. \quad
\eeq
We find that there is a strong cancellation in the $\omega$ and $\phi$ contributions to the $\eta$ TFF:
the isoscalar contribution therein is overall small.  This is different for the $\eta'$, where both add up, 
and the isoscalar part, e.g., of the form factor slope is more significant.

\begin{figure}
\centering
\includegraphics[width=0.5\linewidth]{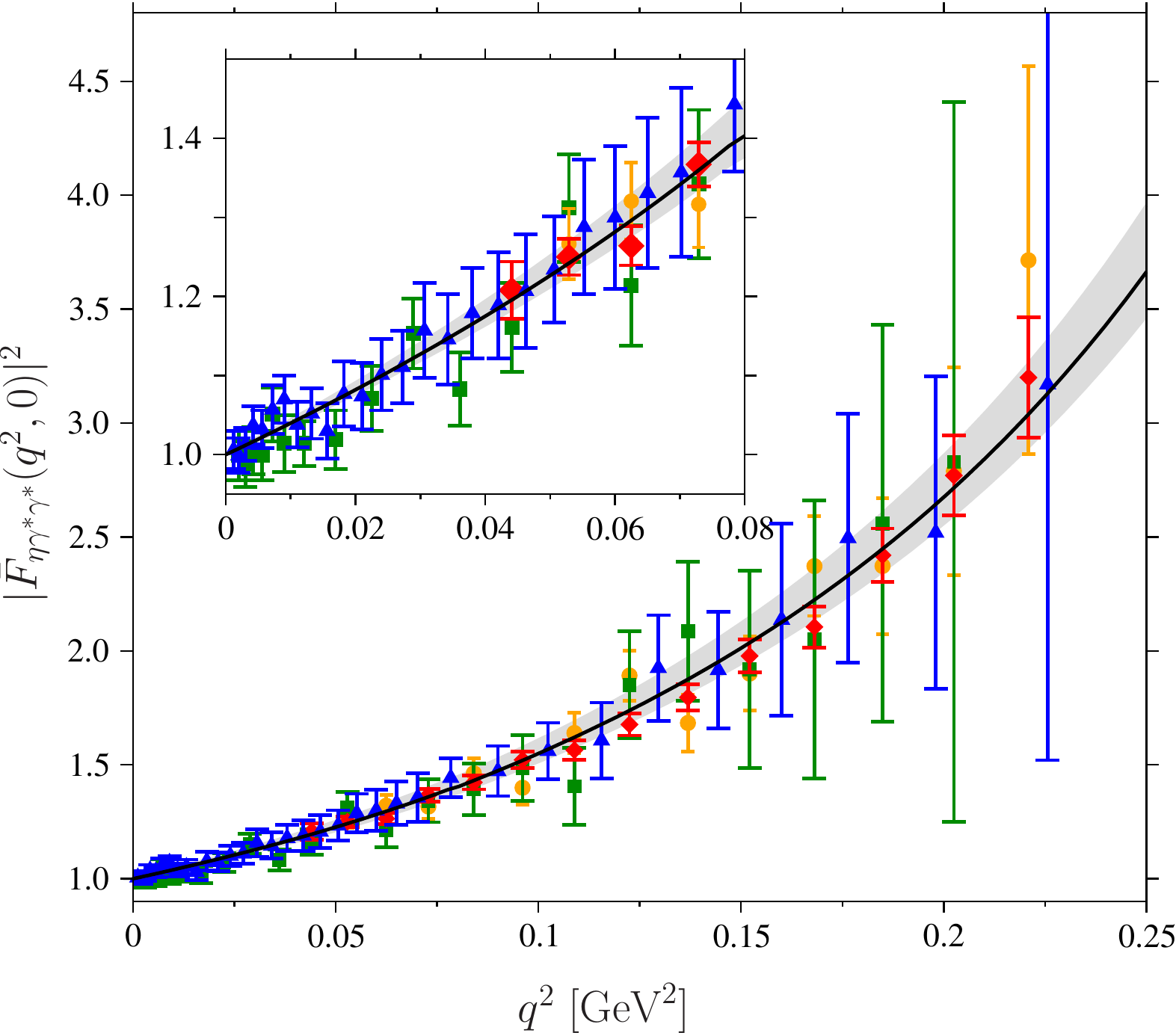}
\caption{Comparison of the dispersive prediction for the squared normalized form factor
$|\bar F_{\eta\gamma^*\gamma^*}(q^2,0)|^2 = |F_{\eta\gamma^*\gamma^*}(q^2,0)/F_{\eta\gamma\gamma}|^2$ (gray band)
to $\eta\to e^+e^-\gamma$ data from the A2 (green squares~\cite{Aguar-Bartolome:2013vpw}, blue triangles~\cite{Adlarson:2016hpp}) and $\eta\to \mu^+\mu^-\gamma$ data from the NA60 (orange circles~\cite{Arnaldi:2009aa}, red diamonds~\cite{Arnaldi:2016pzu}) collaborations.  The inset magnifies the low-$q^2$ region.}
\label{fig:etaTFF}
\end{figure}

The resulting prediction for the normalized time-like $\eta$ transition form factor $\bar F_{\eta\gamma^*\gamma^*}(q^2,0) = F_{\eta\gamma^*\gamma^*}(q^2,0)/F_{\eta\gamma\gamma}$ in Fig.~\ref{fig:etaTFF}
includes the propagated uncertainties both from the experimental input employed
and due to the high-energy continuation of the dispersion integral; see Ref.~\cite{Hanhart:2013vba} 
for a detailed discussion. It is compared
to the experimental data on the decays $\eta\to \ell^+\ell^-\gamma$, which have been obtained both for the 
electron--positron~\cite{Aguar-Bartolome:2013vpw,Adlarson:2016hpp} and the dimuon~\cite{Arnaldi:2009aa,Arnaldi:2016pzu} final states.  
We observe one of the main strengths of the dispersive approach:
by fixing the input to the singly-radiative decay $\eta\to\pi^+\pi^-\gamma$ (with the decay 
rate scaling according to $\propto \alpha_{\text{em}}$), 
we gain a huge statistical advantage over the direct measurements of 
$\eta\to\ell^+\ell^-\gamma$ (rate $\propto \alpha_{\text{em}}^3$).\footnote{We mention in passing that 
one further ingredient for future refinements of the theoretical description of these decays 
consists in radiative corrections, which have most recently and comprehensively been 
studied in Ref.~\cite{Husek:2017vmo}.}
A direct measurement of the transition form factor
of comparable precision to the theoretical calculation based on dispersion relations
will be enormously difficult.

\begin{figure}[t!]
\centering
\includegraphics[width=0.65\linewidth]{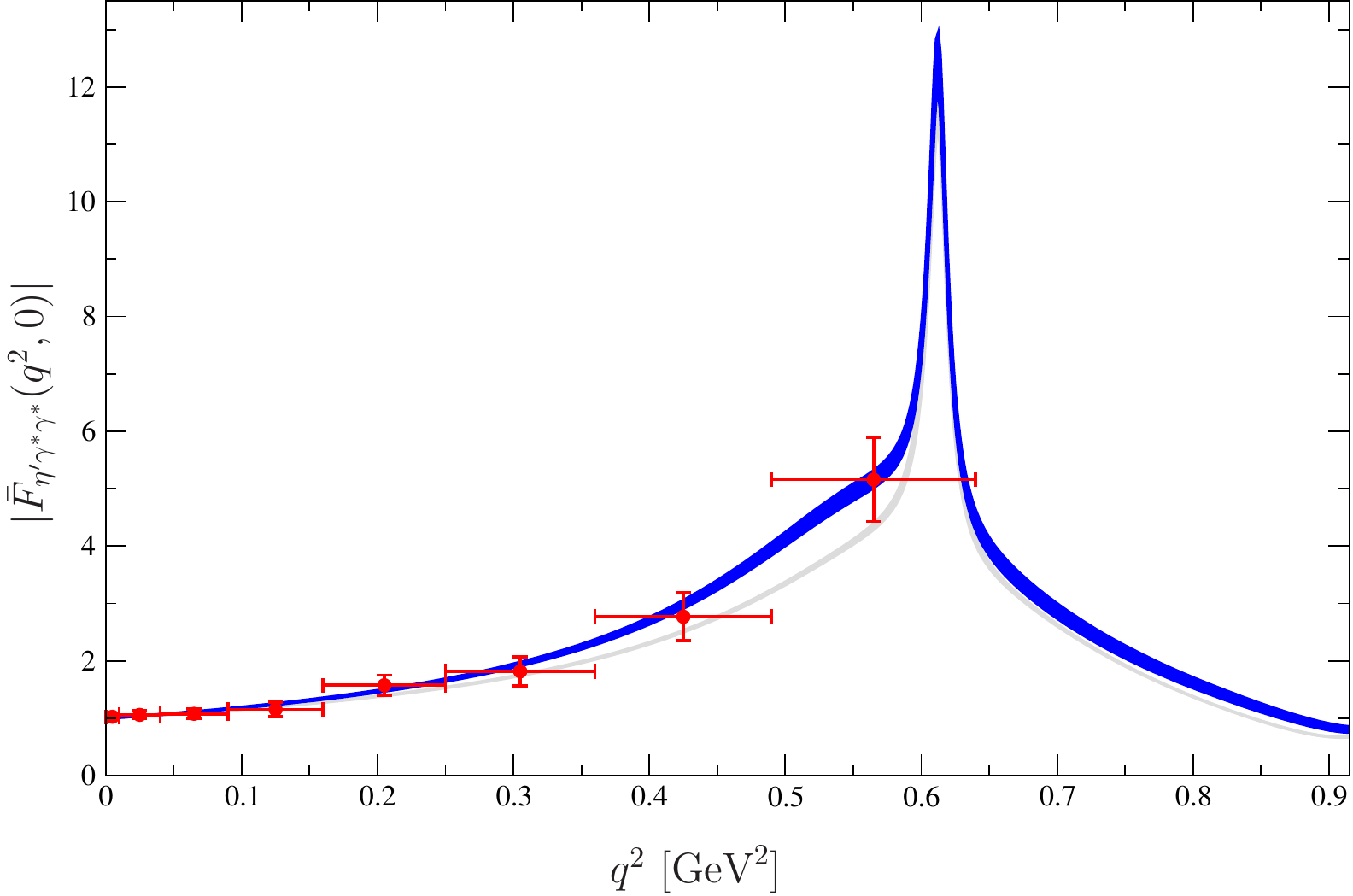}
\caption{Dispersion-theoretical prediction for the normalized, singly-virtual $\eta'$ transition form factor $|\bar F_{\eta'\gamma^*\gamma^*}(q^2,0)| = |F_{\eta'\gamma^*\gamma^*}(q^2,0)/F_{\eta'\gamma\gamma}|$ (blue band),
compared to a vector-meson-dominance model (gray), as well as the data points taken by the BESIII 
collaboration~\cite{Ablikim:2015wnx}. Figure courtesy of S.~Holz.}
\label{fig:etaprimeTFF}
\end{figure}

The combined prediction for the normalized singly-virtual time-like $\eta'$ transition form factor $\bar F_{\eta'\gamma^*\gamma^*}(q^2,0) = F_{\eta'\gamma^*\gamma^*}(q^2,0)/F_{\eta'\gamma\gamma}$ is shown in Fig.~\ref{fig:etaprimeTFF}
as the blue band~\cite{Holz:2016}.
Both the broad resonance shape due to the $\rho$ enhancement and the narrow $\omega$ peak
are clearly visible.  
The error band comprises the propagated uncertainties due to different pion form factor data sets, different
pion--pion scattering phase shifts, and different assumptions on the high-energy continuation of the input
to the dispersion integral.
However, despite these uncertainties 
we still note significant deviations of the dispersive prediction from a pure VMD model
(with a simple finite-width $\rho^0$ pole as the isovector contribution), 
which is due to the much more sophisticated treatment of the two-pion cut
contribution.  The available data on $\eta'\to e^+e^-\gamma$
from BESIII~\cite{Ablikim:2015wnx} is not yet precise enough to differentiate between
the two theoretical curves; new data also from other laboratories is eagerly awaited.

The slope parameters $b_{\ep}$ for the $\ep$ TFFs can be written in terms of a compact sum rule
for the isovector part, plus an analytic expression for the isoscalar contribution:
\beq
b_{\ep} = \frac{1}{12\pi^2 F_{\ep\gamma\gamma}} \int_{4M_\pi^2}^\infty \frac{\diff x}{x^{5/2}} q_\pi^3(x) \big(F_\pi^{V}(x)\big)^{*}f^{\ep}_1(x) + \sum_{V=\omega,\phi}\frac{w_{\ep V \gamma}}{m_V^2} \,.
\eeq
In practice, the upper limit of the sum rule has been replaced by a cutoff in the range 
$1\ldots2\GeV^2$~\cite{Hanhart:2013vba,Kubis:2015sga}, or the integrand guided to zero in a smooth way~\cite{Holz:2016}.  The resulting slopes as predicted by the dispersive analysis are
\beq
b_\eta = 1.95(9)\GeV^{-2} \,, \qquad b_{\eta'} = 1.45(5)\GeV^{-2} \,.  \label{eq:Nbeta}
\eeq
Here, the isoscalar contribution in the $\eta$ slope $b_\eta^{(I=0)} = -0.022(9)\GeV^{-2}$ 
is safely covered by the uncertainty in the isovector sum rule,
while $\omega$ and $\phi$ contribute roughly 20\% to $b_{\eta'}$, $b_{\eta'}^{(I=0)} = 0.269(5)\GeV^{-2}$.  
The values in Eq.~\eqref{eq:Nbeta}
agree nicely, e.g., with the extractions from TFF data using Pad\'e approximants,
$b_\eta = 1.92(4)\GeV^{-2}$~\cite{Escribano:2015nra} and
$b_{\eta'} = 1.43(4)\GeV^{-2}$~\cite{Escribano:2015yup}.  

In this section, we have largely confined our discussion to experimental tests of the time-like form factors that can be measured
in the decays $\ep\to \ell^+\ell^-\gamma$; we wish to emphasize that as the form factors by construction have
the correct analytic structure, they can be analytically continued into the space-like region without 
difficulties, i.e., Eq.~\eqref{eq:etaTFF-DR} is easily evaluated for $q^2<0$.
We expect this description to be reliable in a similar range of $|q^2|$ as the time-like momenta that
enter the description of the spectral function, $-q^2 \lesssim 1\GeV^2$.
However, the correct matching to the asymptotic behavior as for the $\pi^0$ transition form 
factor~\cite{Hoferichter:2014vra,Hoferichter:2018kwz,Hoferichter:2018dmo} has not been performed in practice yet.

\subsection{$\eta'\to4\pi$}
\label{sec:etap-4pi}
The decays of the $\eta'$ into four pions are interesting in that they form the probably most easily accessible set of processes
involving five (relatively) light pseudoscalar mesons, and are therefore of odd intrinsic parity.  
As we mentioned above, they constitute a second group of decays of the $\eta'$ that exists in isospin-symmetric QCD; 
however, the branching ratios measured recently are very small~\cite{Ablikim:2014eoc,Tanabashi:2018oca}:
\beq
\BR(\eta'\to2(\pi^+\pi^-)) = 8.4(9)(3)\times 10^{-5} \,, \qquad
\BR(\eta'\to\pi^+\pi^-2\pi^0) = 1.8(4)(1)\times 10^{-4} \,. \label{eq:BRexp-eta4pi}
\eeq
This can be understood (and was correctly predicted theoretically~\cite{Guo:2011ir}) as follows:
the anomalous nature of the decay forbids any two pseudoscalars to be in a relative $S$-wave, therefore 
$\eta'\to2(\pi^+\pi^-)$ and $\eta'\to\pi^+\pi^-2\pi^0$ are $P$-wave dominated; as Bose symmetry furthermore
forbids a pair of neutral pions to be in an odd relative partial wave, $\eta'\to4\pi^0$ will even be $D$-wave dominated
and therefore yet much more strongly suppressed.

The decay amplitudes for
\beq
\eta' \to \pi^+(p_1)\pi^-(p_2)\pi^+(p_3)\pi^-(p_4) \,, \qquad
\eta' \to \pi^+(p_1)\pi^0(p_2)\,\pi^-(p_3)\pi^0(p_4) 
\eeq
can be written in terms of six Mandelstam variables $s_{ij} = (p_i+p_j)^2$,
$i,\,j = 1,\ldots,4$, subject to the constraint 
$s_{12}+s_{13} +s_{14}+s_{23}+s_{24}+s_{34} = \metap^2 + 8M_\pi^2$.
Ref.~\cite{Guo:2011ir} employs the simple, single-angle $\eta$--$\eta'$ mixing scheme according to Eq.~\eqref{LOmixing},
with $\theta_P = \arcsin({-1}/{3})\approx -19.5^\circ$.  
The flavor structure of the Wess--Zumino--Witten anomaly~\cite{Wess:1971yu,Witten:1983tw}
does not allow for contributions to these decays at chiral order $p^4$, such that the leading amplitude in the chiral expansion
occurs at $\Order(p^6)$. These are of the form~\cite{Guo:2011ir}
\beq
\Amp(\eta_{0/8}\to\pi^+\pi^-\pi^+\pi^-) = - \Amp(\eta_{0/8}\to\pi^+\pi^0\pi^-\pi^0) 
= \frac{N_c\eps_{\mu\nu\alpha\beta}}{3\sqrt{3}F_\pi^5} p_1^\mu p_2^\nu p_3^\alpha p_4^\beta
\big[ \F_{0/8}(s_{12})+\F_{0/8}(s_{34})
     -\F_{0/8}(s_{14})-\F_{0/8}(s_{23}) \big] \,, 
\eeq
where both $\F_0(s_{ij})$ and $\F_8(s_{ij})$ have counterterm contributions from the anomalous chiral Lagrangian of $\Order(p^6)$~\cite{Bijnens:2001bb}
that are linear in the corresponding Mandelstam variables, and $\F_8(s_{ij})$ in addition has kaon loop contributions. 
The counterterms were subsequently estimated using vector-meson dominance (VMD), relying on a hidden-local-symmetry 
Lagrangian~\cite{Fujiwara:1984mp,Meissner:1987ge,Bando:1987br,Harada:2003jx}. 

It turned out that the kaon-loop contributions (at a reasonable scale) 
are negligible compared to the vector-meson amplitudes, while the latter lead to very significant corrections when retaining the full 
$\rho$-propagators in the (virtual) decay chains $\eta'\to 2\rho^0 \to 2(\pi^+\pi^-)$, $\eta'\to \rho^+\rho^- \to \pi^+\pi^0\pi^-\pi^0$.
This therefore goes beyond the approximation at chiral $\Order(p^6)$: the available phase space that extends up to
$\sqrt{s_{ij}^{\rm max}} = \metap-2\mpi$ contains too much of the tails of the $\rho$-resonances.
With the simplest coupling assignments (and in the standard mixing scenario), the VMD amplitudes are then of the form
\beq
\Amp_{V}(\eta' \to\pi^+\pi^-\pi^+\pi^-) 
= - \Amp_{V}(\eta'\to\pi^+\pi^0\pi^-\pi^0) 
= \frac{N_c\eps_{\mu\nu\alpha\beta}}{8\sqrt{3}\pi^2F_\pi^5}p_1^\mu p_2^\nu p_3^\alpha p_4^\beta
\Bigg[ \frac{m_\rho^4}{D_\rho(s_{12})D_\rho(s_{34})}
             - \frac{m_\rho^4}{D_\rho(s_{14})D_\rho(s_{23})} \Bigg] \,, 
\label{eq:Arho1}
\eeq
where $D_\rho(s) = m_\rho^2 -s - i \,m_\rho\Gamma_\rho(s)$ denotes the inverse $\rho$-propagator with 
an energy-dependent width $\Gamma_\rho(s)$.  Integrating these amplitudes squared over phase space and assigning a rather generic
model uncertainty of $\pm30\%$ leads to the predictions for the branching ratios~\cite{Guo:2011ir}
\beq
\BR(\eta'\to2(\pi^+\pi^-)) = 1.0(3)\times 10^{-4} \,, \qquad
\BR(\eta'\to\pi^+\pi^-2\pi^0) = 2.4(7)\times 10^{-4} \,
\eeq
(where deviations from the isospin relation $\BR(\eta'\to\pi^+\pi^-2\pi^0) /\BR(\eta'\to2(\pi^+\pi^-)) =2 $ are solely due to 
phase-space corrections according to different pion masses), which agrees with the measurements
Eq.~\eqref{eq:BRexp-eta4pi} within uncertainties.  A much earlier quark-model prediction~\cite{Parashar:1979js} 
had been ruled out already by previous experimental upper limits.

The BESIII collaboration has also determined dipion invariant mass distributions for $\eta'\to2(\pi^+\pi^-)$ and compared them
with Monte-Carlo simulations using the theoretical amplitude above as well as an event distribution generated according to phase space alone.
They conclude that the theoretical decay amplitude ``could provide a more reasonable description of data than the phase space 
events''~\cite{Ablikim:2014eoc}, however the preference is not very strong yet due to limited statistics.  
In particular due to the (potential) role of this process for an analysis of the doubly-virtual $\eta'$ transition form factor,
see Sect.~\ref{sec:doubleDalitz}, a more refined experimental study of the energy distribution of this decay would be highly welcome, ideally even one that is sensitive to corrections beyond the simple VMD picture described in this section.

As we have mentioned above, the fully neutral final state $\eta'\to 4\pi^0$ is yet more strongly suppressed due to the requirement
of all neutral-pion pairs in the final state to be at least in relative $D$-waves.  In Ref.~\cite{Guo:2011ir}, a model was constructed that
contains crossed-channel-$\rho$-exchange plus rescattering via the tail of the $f_2(1270)$.  In this way, the branching ratio was estimated
to be of the order of
\beq
\BR(\eta'\to4\pi^0) \sim 4 \times 10^{-8} \,, \label{eq:BRetap-4pi0}
\eeq
hence smaller by 3--4 orders of magnitude than the other four-pion final states, however still much larger than any branching ratio due to a \textit{CP}-violating
$S$-wave amplitude induced by the QCD $\theta$-term.  
The current experimental upper limit from the BESIII collaboration, 
$\BR(\eta'\to4\pi^0) < 4.94\times 10^{-5}$~\cite{Donskov:2014zja} 
(improving on a previous GAMS-$4\pi$ limit~\cite{Ablikim:2019msz} by approximately a factor of six),
is still three orders of magnitude above the estimate~\eqref{eq:BRetap-4pi0}.
We remark in passing that the decay $\eta \to 4\pi^0$ is the only four-pion decay
of the $\eta$ meson that is kinematically allowed.  It is not, strictly speaking, \textit{CP}-forbidden
(as are the decays $\eta^{(\prime)}\to 2\pi$), but only $S$-wave \textit{CP}-forbidden.  However, the tiny phase space
($\meta-4\mpii = 7.9\MeV$) combined with the $D$-wave characteristics described above make the \textit{CP}-conserving decay mechanism
predict a rate~\cite{Guo:2011ir}
\beq
\BR(\eta\to4\pi^0) \sim 3 \times 10^{-30} \,,
\eeq
such that any experimental signal for this decay, should one ever be found, can safely be interpreted as a sign of \textit{CP} violation 
all the same.

\subsection{$\eta, \eta' \to\pi^+\pi^-\ell^+\ell^-$ and $e^+e^-\to\eta\pi^+\pi^-$}\label{sec:eta-pipill}

Given the enormous experimental accuracy achieved for the radiative decays $\eta^{(\prime)}\to\pi^+\pi^-\gamma$, 
as well as the central role it has for an analysis of the singly-virtual $\eta^{(\prime)}$ transition form factors,
the closely related dilepton decays $\eta^{(\prime)}\to\pi^+\pi^-\ell^+\ell^-$ come into focus as a way to access
the doubly-virtual transition form factors, other than via double-Dalitz decays
(see Sect.~\ref{sec:doubleDalitz}) with their strong electromagnetic rate suppression.
These decays have been investigated in variants of vector-meson-dominance 
models~\cite{Picciotto:1993aa,Faessler:1999de}, as well as in a chiral unitary approach~\cite{Borasoy:2007dw}.
Experimentally, 
the $\eta\to\pi^+\pi^-e^+e^-$ branching ratio has been determined by the 
WASA@CELSIUS collaboration, 
$\BR(\eta\to\pi^+\pi^-e^+e^-) = 4.3(1.3)(0.4)\times10^{-4}$~\cite{Bargholtz:2006gz}, 
and subsequently improved upon by KLOE, 
$\BR(\eta\to\pi^+\pi^-e^+e^-) = 2.68(9)(7)\times10^{-4}$~\cite{Ambrosino:2008cp}.
The most accurate determination of the $\eta'\to\pi^+\pi^-e^+e^-$ decay rate has been provided
by the BESIII collaboration,
$\BR(\eta'\to\pi^+\pi^-e^+e^-) = 2.11(12)(15)\times10^{-3}$, 
which has also provided an upper limit on the corresponding dimuon rate,
$\BR(\eta'\to\pi^+\pi^-\mu^+\mu^-) < 2.9\times10^{-5}$~\cite{Ablikim:2013wfg}. The first observation of  $\eta^\prime\rightarrow\pi^+\pi^-\mu^+\mu^-$ was reported by BESIII recently, with the measured branching fraction of $\BR(\eta'\to\pi^+\pi^-\mu^+\mu^-) = 1.97(33)(18)\times 10^{-5} $~\cite{Ablikim:2020svz}. 
KLOE~\cite{Ambrosino:2008cp} has also measured the \textit{CP}-odd $\pi^+\pi^-$--$e^+e^-$ decay planes
angular asymmetry (with a result compatible with zero), which, given the strong suppression of 
\textit{CP}-violation within the Standard Model for this decay~\cite{Jarlskog:1995ww},
presents a test of new, unconventional \textit{CP}-odd mechanisms~\cite{Gao:2002gq}; see Sect.~\ref{sec:eta-2pi}.

In straightforward generalization of Sect.~\ref{sec:eta-pipigamma}, the $\eta^{(\prime)}(q) \to \pi^+(p_1)\pi^-(p_2) e^+(k_1)e^-(k_2)$
decay amplitude can be written as
\beq
\eps_{\mu\nu\alpha\beta} \big[\bar u(k_2)\gamma^\mu v(k_1)\big] \frac{e^2}{k^2} p_1^\nu p_2^\alpha q^\beta \, \F_\eta(s,t,u,k^2) \quad \longrightarrow \quad
\eps_{\mu\nu\alpha\beta} \big[\bar u(k_2)\gamma^\mu v(k_1)\big] \frac{e^2}{k^2} p_1^\nu p_2^\alpha q^\beta  
\Big(f^{\eta}_1(s,k^2) + \ldots \Big)
\,, \label{eq:eta-pipiee}
\eeq
with $s$, $t$, $u$ defined as in Sect.~\ref{sec:eta-pipigamma}, $k=k_1+k_2$, and the relation
$s+t+u=\meta^2+2\mpi^2+k^2$.  In the second step of Eq.~\eqref{eq:eta-pipiee}, we have inserted the ($\pi\pi$)
partial-wave expansion and retained the dominant $P$-wave only.  Both $\F_\eta(s,t,u,k^2)$ and its $P$-wave projection
$f^\eta_1(s,k^2)$ are defined such that they coincide with the corresponding quantities for the real-photon decay
for $k^2=0$.  

Obviously, for any fixed $k^2$, we should be able to parameterize $f^\eta_1(s,k^2)$ model-independently in a form 
analogous to Eqs.~\eqref{eq:f1} or \eqref{eq:etaprime}, hence as a polynomial multiplying a $\pi\pi$ $P$-wave
Omn\`es function; the polynomial coefficients will then depend on $k^2$.  A highly interesting question then is to 
investigate to what extent the dependencies on $k^2$ and $s$ factorize: 
\begin{equation}
f^{\eta^{(\prime)}}_1\big(s,k^2\big)  \sim  f^{\eta^{(\prime)}}_1(s) \times \bar F_{\eta^{(\prime)}\gamma\gamma^*} \big(k^2\big) \,.
\end{equation}
This question of factorization immediately translates into an equivalent statement concerning the doubly-virtual
transition form factors $\eta^{(\prime)}\to \gamma^*\gamma^*$, see Sect.~\ref{sec:doubleDalitz}: 
one assumption frequently employed in the description of doubly-virtual transition form factors 
at low-to-moderate energies is that the dependence on the two virtualities is simplified into the product of two functions
of a single variable each.  This approximation fails at asymptotically high energies, see Eq.~\eqref{eq:fomega}, 
however, to what extent there are relevant corrections already at low energies is largely unknown experimentally.
Future precision experiments on $\eta^{(\prime)}\to\pi^+\pi^-e^+e^-$ ought to investigate this question.

First studies along these lines have been performed for the \textit{crossed} reaction
$e^+e^-\to\eta\pi^+\pi^-$~\cite{Xiao:2015uva}.  The same argument concerning the investigation of transition form factors, 
using ``more hadronic'' reactions, holds here, as in $\eta$ decay kinematics:
the doubly-virtual (time-like) $\eta$ transition form factor would be directly
accessible experimentally in $e^+e^-\to\eta e^+e^-$, 
however, this process is strongly suppressed in the fine structure constant.
It is therefore a much more promising possibility to investigate $e^+e^-\to\eta\pi^+\pi^-$
instead, thus beating the suppression by $\alpha_{\text{em}}^2$, and subsequently using a dispersion relation
to reconstruct the dependence on the second (outgoing) photon's virtuality by means of the dominant
$\pi^+\pi^-$ intermediate state.
The question to investigate is therefore:
for fixed $\eta\pi^+\pi^-$ invariant mass squared $k^2$, 
factorization predicts the same dependence on the dipion invariant mass squared $s$ as observed
in the real-photon case, i.e., in the decay $\eta\to\pi^+\pi^-\gamma$.  
Ref.~\cite{Xiao:2015uva} has allowed for a similar functional form for $f^\eta_1(s,k^2)$ as in Sect.~\ref{sec:eta-pipigamma}, 
i.e., an Omn\`es function multiplied with a linear or a quadratic polynomial.
The subtraction constants have then be fitted to data and subsequently compared to the real-photon decay amplitude.
As $k^2$ is necessarily large---the reaction threshold is around $\sqrt{k^2} = 0.83\GeV$, and the cross section
is dominated by the $\rho'$ or $\rho(1450)$ resonance---the $k^2$ dependence $\bar F_{\eta\gamma\gamma^*}(k^2)$
was parameterized by a sum of two Breit--Wigner functions [for the $\rho(770)$ and the $\rho(1450)$].

\begin{figure}
\centering
\includegraphics*[width=\linewidth]{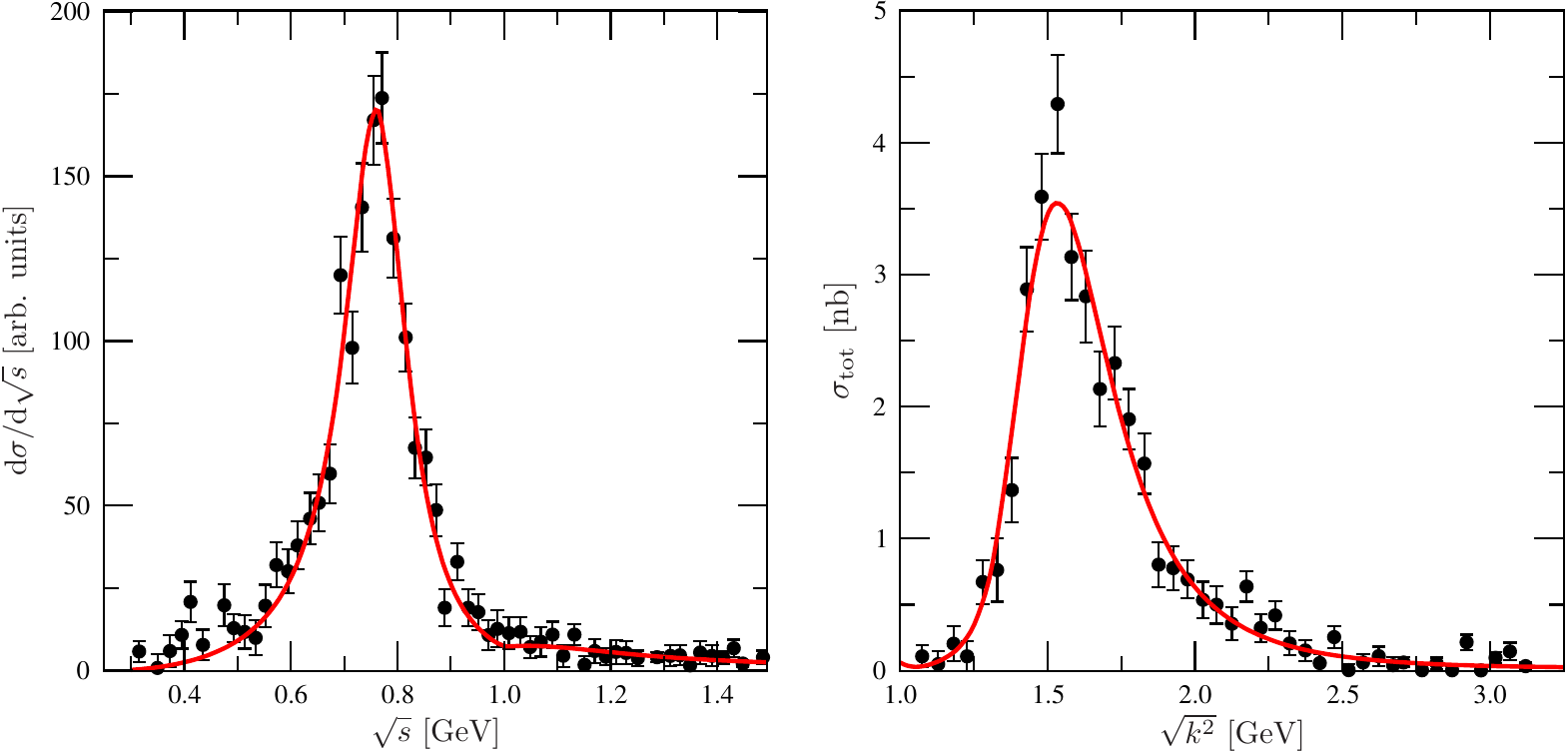}
\caption{Result of the best fit to both the $\pi\pi$ spectrum (left)
   and the total cross section (right) for the reaction $e^+e^-\to\eta\pi^+\pi^-$. 
   The data is taken from Ref.~\cite{Aubert:2007ef}.  Figure adapted from Ref.~\cite{Xiao:2015uva} (first arXiv version).}
\label{fig:ee-etapipi}
\end{figure}

Unfortunately, the BaBar data~\cite{Aubert:2007ef} that was analyzed for this purpose is not truly 
doubly differential in $s$ and $k^2$: only the total cross section $\sigma_{\rm tot}(k^2)$ 
as well as the $\pi\pi$ spectrum $\diff\Gamma/\diff\sqrt{s}$, integrated over 
$1\GeV \leq \sqrt{k^2} \leq 4.5\GeV$, are available.
The best fit to the data is shown in Fig.~\ref{fig:ee-etapipi}.
It was found that fitting a linear polynomial in $f^\eta_1(s)$ leads to polynomial 
parameters that are incompatible with the ones found in $\eta \to \pi^+\pi^-\gamma$, hence seemingly pointing towards
a large violation of the factorization assumption.  
In contrast, employing a quadratic polynomial with a quadratic term emulating the curvature effects
induced by crossed-channel $a_2$-exchange indicates that such a violation 
is much attenuated in this case~\cite{Xiao:2015uva}. 
However, the natural factorization-breaking corrections induced by the $a_2$ left-hand cut have not been 
included yet; investigations along these lines are still in progress, 
also taking into account more recent data on 
$e^+e^-\to\eta\pi^+\pi^-$~\cite{Achasov:2017kqm,TheBABAR:2018vvb,Gribanov:2019qgw}.

\subsection{Doubly-virtual transition form factors in $\pi^0, \eta, \eta' \to \ell^+\ell^-\ell^+\ell^-$ and related processes}\label{sec:doubleDalitz}

The light-by-light loop integral that determines the pseudoscalar pole contributions to the anomalous magnetic
moment of the muon also requires the doubly-virtual transition form factors, see Fig.~\ref{fig:hlbl}.
Obvious processes to access these directly in light-meson decays
are the double-Dalitz decays $P \to\ell^+\ell^-\ell^+\ell^-$.  
Experimental determinations of the branching fractions exist only for the electronic final states 
of $\pi^0$ and $\eta$: the PDG average 
$\BR\big(\pi^0\to 2(e^+e^-)\big)=3.34(16)\times10^{-5}$~\cite{Tanabashi:2018oca}
is dominated by a measurement of the KTeV collaboration~\cite{Abouzaid:2008cd}, 
while $\BR\big(\eta\to 2(e^+e^-)\big)= 2.4(2)_{\text{stat}}(1)_{\text{syst}}\times10^{-5}$
has been determined by KLOE~\cite{KLOE2:2011aa}. 
WASA-at-CELSIUS has determined upper limits on the muonic final states of the $\eta$, 
$\BR\big(\eta\to e^+e^-\mu^+\mu^-\big) < 1.6\times10^{-4}$ and
$\BR\big(\eta\to 2(\mu^+\mu^-)\big) < 3.6\times10^{-4}$~\cite{Berlowski:2007aa}.
No data exists to date for double-Dalitz decays of the $\eta'$.

\begin{figure}[t!]
\centering
\includegraphics[width=0.7\linewidth]{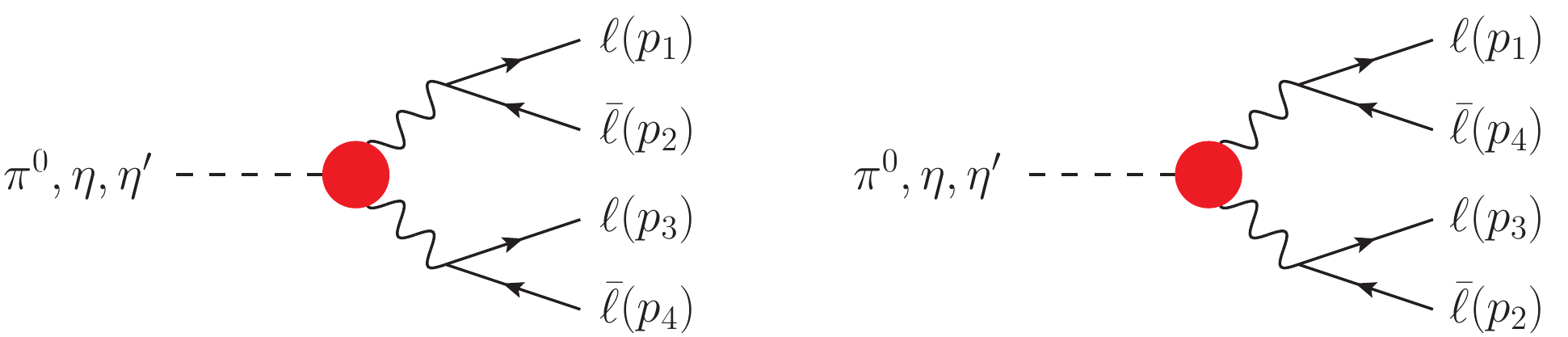}
\caption{Direct and exchange contributions to $P \to \ell^+\ell^-\ell^+\ell^-$.}
\label{fig:P-4l}
\end{figure}
The matrix element to describe the double-Dalitz decays is more complicated for decays into two
identical dilepton pairs, as in those cases, two different amplitudes contribute, a ``direct''
and an ``exchange'' contribution, see Fig.~\ref{fig:P-4l}, including an interference term in the
observable decay rates and distributions.  
Following the notation of Ref.~\cite{Kampf:2018wau}, the matrix elements 
(at leading order in the fine structure constant) are given by
\begin{align}
\M_D &= - e^4 \frac{F_{P\gamma^*\gamma^*}(s_{12},s_{34})}{s_{12}s_{34}}
\epsilon_{\mu\nu\alpha\beta}(p_1+p_2)^\mu(p_3+p_4)^\alpha \big[\bar{u}(p_1)\gamma^\nu v(p_2)\big]
 \big[\bar{u}(p_3)\gamma^\beta v(p_4)\big] \,,\notag\\
\M_E &= + e^4 \frac{F_{P\gamma^*\gamma^*}(s_{14},s_{23})}{s_{14}s_{23}}
\epsilon_{\mu\nu\alpha\beta}(p_1+p_4)^\mu(p_2+p_3)^\alpha \big[\bar{u}(p_1)\gamma^\nu v(p_4)\big]
 \big[\bar{u}(p_3)\gamma^\beta v(p_2)\big]  \label{eq:M-P-4l}
\end{align}
for the direct and exchange terms, respectively; see also Refs.~\cite{Petri:2010ea,Terschlusen:2013iqa,Escribano:2015vjz}.  The momentum assignments are as in Fig.~\ref{fig:P-4l}, and we have used
the dilepton squared invariant masses $s_{ij}=(p_i+p_j)^2$.  Note in particular that the relative
sign between $\M_D$ and $\M_E$ is a consequence of Fermi statistics.  No exchange terms exist
for the decays $\eta^{(\prime)}\to e^+e^-\mu^+\mu^-$.  

We refer to the literature for explicit expressions on the spin-averaged squared matrix elements
as well as the required four-body phase space integrals~\cite{Petri:2010ea,Escribano:2015vjz,Kampf:2018wau}.
Qualitatively, the following picture arises.  Due to the photon propagators, in particular the electron--positron distributions are strongly peaked at small invariant masses and, hence, only very mildly modified by form factor effects; this is true in particular for $\pi^0\to2(e^+e^-)$, for which the decay rate is enhanced at the sub-percent level due to deviations from a constant $\pi^0\to\gamma^*\gamma^*$ transition~\cite{Kampf:2018wau}.  The transition form factors change the rates of the $\eta$ and $\eta'$ decays with muons in the final states much more
significantly, but those are overall suppressed due to significantly smaller phase spaces.  Throughout, 
interference effects between direct and exchange terms are small, at the percent level.
Radiative corrections, that is next-to-leading-order effects in the fine structure constant, 
have been studied in great detail in Ref.~\cite{Kampf:2018wau} (completing and partially correcting earlier studies~\cite{Barker:2002ib}),
which might be relevant for the extraction of form factor effects, given the latter's smallness.
\begin{table}
\renewcommand{\arraystretch}{1.3}
\centering
\begin{tabular}{ccc}
\toprule
$\pi^0\to 2(e^+e^-)$ & $\eta \to 2(e^+e^-)$ & $\eta' \to 2(e^+e^-)$ \\ \midrule
$3.47\times 10^{-5}$~\cite{Terschlusen:2013iqa} & $2.71(2)\times10^{-5}$~\cite{Escribano:2015vjz} & $2.10(45)\times10^{-6}$~\cite{Escribano:2015vjz} \\
$3.36689(5)\times10^{-5}$~\cite{Escribano:2015vjz} & $2.701(14)\times 10^{-5}$~\cite{Kampf:2018wau} \\
$3.40(1)\times 10^{-5}$~\cite{Weil:2017knt}\\
$3.3919(13)\times 10^{-5}$~\cite{Kampf:2018wau}\\
 \midrule
& $\eta \to e^+e^-\mu^+\mu^-$ & $\eta' \to e^+e^-\mu^+\mu^-$ \\ \midrule
& $2.39(7)\times10^{-6}$~\cite{Escribano:2015vjz} & $6.39(91)\times10^{-7}$~\cite{Escribano:2015vjz}\\
& $2.335(12)\times10^{-6}$~\cite{Kampf:2018wau} \\ \midrule
& $\eta \to 2(\mu^+\mu^-)$ & $\eta' \to 2(\mu^+\mu^-)$ \\ \midrule
& $3.98(15)\times10^{-9}$~\cite{Escribano:2015vjz} & $1.69(36)\times10^{-8}$~\cite{Escribano:2015vjz} \\
& $3.878(20)\times10^{-9}$~\cite{Kampf:2018wau} \\
\bottomrule
\end{tabular}
\renewcommand{\arraystretch}{1.0}
\caption{Selected theoretical results for the branching ratios $\BR(P\to\ell^+\ell^-\ell^+\ell^-)$.}
\label{tab:P-4l:th}
\end{table}
--- Various theoretical predictions for the branching ratios are summarized in Table~\ref{tab:P-4l:th}.
They all agree with the available experimental data, which is not yet precise enough to discriminate between different 
calculations or form factor models.  

Obviously, we also wish to generalize the dispersive analysis of the $\eta$ and $\eta'$ transition form factors to the doubly-virtual case.
A complete dispersive reconstruction of the doubly-virtual $\eta$ and $\eta'$ transition form factors
is somewhat more complicated than for the $\pi^0$ precisely due to its different isospin structure,
see Eq.~\eqref{eq:TFF-isodecomp}.  The former are given in terms of \textit{two} independent
functions, not just a single one as for the $\pi^0$.  As a consequence, even with the isoscalar 
part fixed from $\omega$ and $\phi$ dominance, cf.\ Sect.~\ref{sec:eta-omegag}, the doubly-virtual behavior can be inferred much 
less rigorously than for the $\pi^0$ TFF, where the dominant part of the 
low-energy double-spectral function can be fixed from singly-virtual data alone.  

We have emphasized repeatedly how it is advantageous to use hadronic amplitudes, which can be measured
much more precisely due to the statistical advantage of lack of suppression in $\alpha_{\text{em}}$, 
as input to dispersion relations instead of 
measuring radiative amplitudes directly; therefore, we have based predictions for 
$\ep\to \ell^+\ell^-\gamma$ on precision data for $\ep\to\pi^+\pi^-\gamma$.  
In order to calculate the doubly-virtual transition form factor (or, more precisely, the isovector--isovector
contribution therein), we may therefore consider going one step further and base the
construction of a double spectral function on a description of the purely hadronic decay
$\eta'\to\pi^+\pi^-\pi^+\pi^-$, see Sect.~\ref{sec:etap-4pi}. 
Even based on the model of Ref.~\cite{Guo:2011ir}, one can attempt to assess
the size of nonfactorizing contributions to the doubly-virtual $\eta'$ transition form factor, 
\begin{figure}
\centering
\includegraphics*[width=0.7\linewidth]{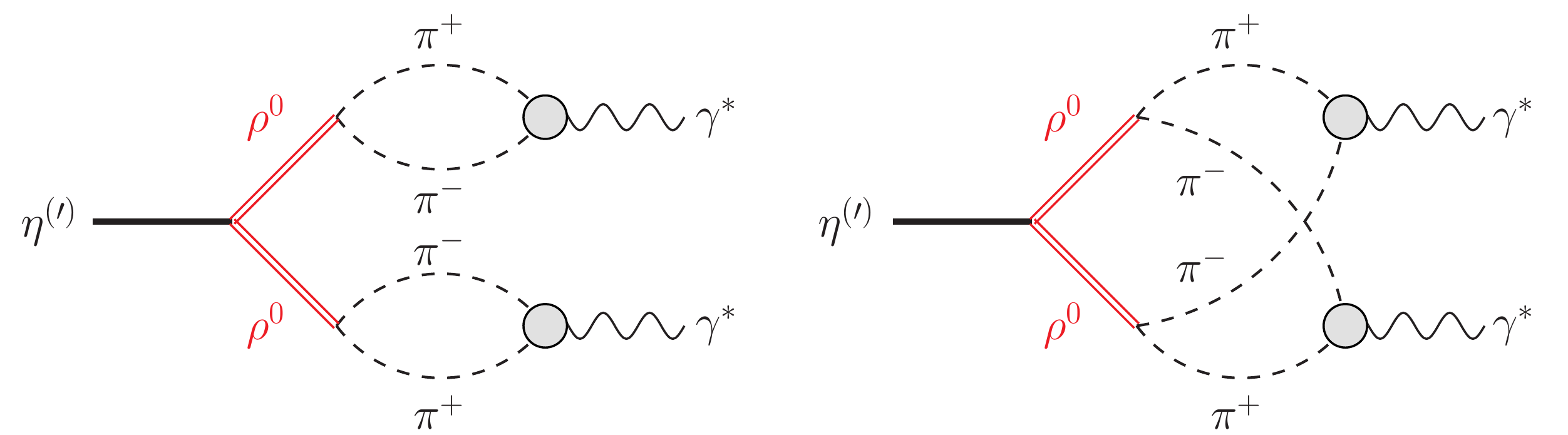}
\caption{Factorizing (left) and nonfactorizing (right) contributions to the doubly-virtual
$\ep$ transition form factors, based on a vector-meson-dominance model for the amplitude
$\ep\to \pi^+\pi^-\pi^+\pi^-$.}\label{fig:eta4pi}
\end{figure}
see Fig.~\ref{fig:eta4pi}, by combining $\pi^+\pi^-$ pairs into virtual photons in the sense of 
a dispersion relation:
it naturally yields a (dominant) factorizing term that closely resembles a VMD model, 
see the left diagram in Fig.~\ref{fig:eta4pi},
and a nonfactorizing one, where the charged pions stemming from different intermediate 
$\rho^0$ resonances are recombined [Fig.~\ref{fig:eta4pi} (right)].
First steps towards such an investigation indicate that these nonfactorizing terms
are small~\cite{Dato:2016}.

\subsection{$\eta'\to\omega\gamma$, $\eta'\to\omega e^+e^-$}\label{sec:eta-omegag}

The \textit{isoscalar} spectral functions of the $\eta$ and $\eta'$ singly-virtual transition form factors, 
see Sect.~\ref{sec:DRetaTFF}, are dominated by three-pion intermediate states, which however can 
be approximated to high precision by the narrow $\omega(782)$ and $\phi(1020)$ resonances---in contrast to the 
broad isovector $\rho(770)$, here a vector-meson-dominance picture is justified to high precision.
The decay $\eta'\to\omega\gamma$ therefore has an analogous role to $\eta'\to\pi^+\pi^-\gamma$ discussed
in Sect.~\ref{sec:eta-pipigamma}, however, it is described in a much simpler way in terms of a single coupling 
constant $g_{\eta'\omega\gamma}$ that is determined by the transition matrix element of the electromagnetic current~\cite{Hanhart:2016pcd}
\beq
  \langle \omega (p)| j_\mu(0) | \eta' (q)\rangle = g_{\eta'\omega\gamma} \, \epsilon_{\mu\nu\alpha\beta} \,q^\nu p^\alpha \epsilon^{\beta}(p)  \,,
  \label{eq:etap-omega}
\eeq
which results in the partial width
\beq
\Gamma(\eta'\to\omega\gamma)=\frac{\alpha_{\text{em}}\,g_{\eta'\omega\gamma}^2}{8}\left( \frac{M_{\eta'}^2-m_\omega^2}{M_{\eta'}} \right)^3 \,.
\label{eq:G:etap-omega}
\eeq
The PDG value for the branching ratio $\BR(\eta'\to\omega\gamma) = 2.62(13)\%$~\cite{Tanabashi:2018oca}
is dominated by a BESIII measurement~\cite{Ablikim:2015eos}, updated most recently 
to $\BR(\eta'\to\omega\gamma) = 2.489(76)\%$~\cite{Ablikim:2019wop}.  

All other coupling constants of the lightest isoscalar vectors to pseudoscalars have to be determined from vector-meson
decays $\phi\to\eta^{(\prime)}\gamma$, $\omega\to\eta\gamma$, where
\beq
\Gamma(V\to P\gamma)=\frac{\alpha_{\text{em}}\,g_{PV\gamma}^2}{24}\left( \frac{m_V^2-M_P^2}{m_V} \right)^3 \,,
\eeq
and the coupling constants are defined in analogy to Eq.~\eqref{eq:etap-omega}.  A single-angle $\eta$--$\eta'$
mixing scheme 
(assuming ideal mixing for the vector mesons)
leads to the approximate relations~\cite{Landsberg:1986fd,Hanhart:2013vba}
\beq
g_{\eta\omega\gamma} : g_{\eta\phi\gamma} : g_{\eta'\omega\gamma} : g_{\eta'\phi\gamma} 
~\approx~ \bigg(\frac{\cos\theta_P}{\sqrt{2}}-\sin\theta_P\bigg) : 
2\bigg(\cos\theta_P+\frac{\sin\theta_P}{\sqrt{2}}\bigg) : 
\bigg(\cos\theta_P + \frac{\sin\theta_P}{\sqrt{2}}\bigg) : 
2\bigg(-\frac{\cos\theta_P}{\sqrt{2}}+ \sin\theta_P\bigg)  \,, \label{eq:gPVgamma-VMD}
\eeq
which, when compared to the measured partial widths, suggests a significantly 
smaller mixing angle $\theta_P$ than the standard $\theta_P = \arcsin({-1}/{3})\approx -19.5^\circ$.
A more refined scenario is discussed, e.g., in Ref.~\cite{Escribano:2015nra}. 

The corresponding dilepton decays, in particular $\eta'\to\omega e^+e^-$ (as again the only accessible decay of a 
pseudoscalar), in principle offer spectral information relevant to the doubly-virtual transition form factors.
However, only a determination of the branching fraction is available (from BESIII) so far,
$\BR(\eta'\to\omega e^+e^-) = 1.97(34)_{\text{stat}}(17)_{\text{syst}}\times 10^{-4}$~\cite{Ablikim:2015eos}, 
and no form factor information.  In a VMD description, the latter would be given entirely by an $\omega$ pole term
(due to OZI suppression of a $\phi$ contribution), however, some theoretical models predict significant
deviations from such a picture~\cite{Terschlusen:2012xw}.
Large corrections to a simple VMD form are seen in some vector-meson conversion decays
such as $\omega\to\pi^0\ell^+\ell^-$~\cite{Arnaldi:2009aa,Arnaldi:2016pzu,Adlarson:2016hpp}
or $\phi\to\pi^0e^+e^-$~\cite{Anastasi:2016qga},
partly to an extent that they are hard to understand 
theoretically~\cite{Terschluesen:2010ik,Schneider:2012ez,Danilkin:2014cra,Ananthanarayan:2014pta,Caprini:2015wja};
but not in others such as $\phi\to\eta e^+e^-$~\cite{Babusci:2014ldz}.

There is an interesting relation of the amplitudes discussed above 
to direct-emission effects in $\eta\to\pi^+\pi^-\pi^0\gamma$~\cite{DAmbrosio:1999iog}.
Obviously, there is no $\omega$ resonance accessible in the $\pi^+\pi^-\pi^0$ invariant mass of the $\eta$
decay due to the latter's small mass; however, the very low-energy tail of the $\omega$ should contribute
to certain chiral low-energy constants describing this radiative process in the spirit of resonance saturation.
As bremsstrahlung off one of the final-state charged pions is proportional to the $\eta\to3\pi$ decay amplitude
and hence isospin-suppressed, see Sect.~\ref{sec:eta-3pi}, while direct emission via an $\omega$ intermediate state
is not, one might suspect that this is one of the rather rare cases where structure-dependent contributions in 
a radiative decay are dominant over bremsstrahlung and hence might lead to interesting new information.
However, Ref.~\cite{DAmbrosio:1999iog} has demonstrated that this is not the case: isospin-conserving direct emission
is strongly suppressed kinematically, such that generalized bremsstrahlung is dominant for almost all photon energies.
The central result is quoted as $\BR(\eta\to\pi^+\pi^-\pi^0\gamma; E_\gamma \geq 10\MeV)/\BR(\eta\to\pi^+\pi^-\pi^0) 
= 3.14(5)\times 10^{-3}$, where the results for different cuts on the minimal photon energy are quoted in numerical
tables in steps of $10\MeV$.  Direct-emission effects are well below 1\% for all realistic photon energy cuts.  
The experimental limit quoted by the PDG, $\BR(\eta\to\pi^+\pi^-\pi^0\gamma) < 5\times 10^{-4}$~\cite{Thaler:1973th}, 
refers to direct emission only, which was extracted with a cut on the $\pi^0\gamma$ invariant mass (not the 
photon energy in the $\eta$ rest frame).  

\subsection{$\pi^0, \eta, \eta' \to\ell^+\ell^-$}\label{sec:P-ll}

\begin{figure}
\centering
\includegraphics[width=0.35\linewidth]{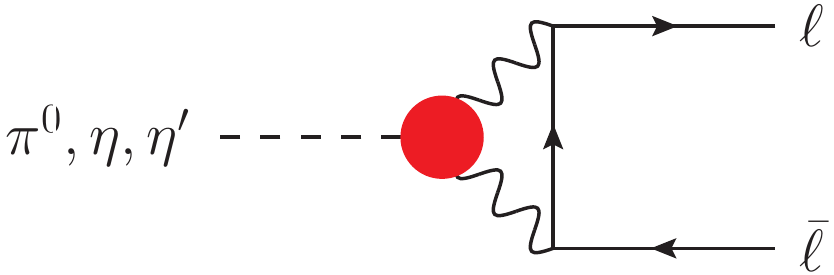}
\caption{Dominant decay mechanism for $P \to \ell^+\ell^-$, $P=\pi^0,\eta,\eta'$.  The red blob denotes the 
(doubly-virtual) transition form factor $P \to \gamma^*\gamma^*$.}
\label{fig:P-ll}
\end{figure}

The dilepton decays of the light pseudoscalars $\pi^0\to e^+e^-$ and $\eta,\eta'\to\ell^+\ell^-$, $\ell=e,\mu$
are strongly suppressed in the Standard Model and hence 
offer interesting opportunities to search for physics beyond it:
the dominant decay mechanism proceeds via two-photon intermediate states, see Fig.~\ref{fig:P-ll}, 
and is hence suppressed to loop level.  In addition, the decays are helicity-suppressed, i.e., branching fractions
into the electron--positron final state (the only one kinematically allowed for the $\pi^0$) are particularly small.
In this dominant decay mechanism, the dilepton decay channels are closely linked to the various radiative decays
discussed in the previous sections, as the (doubly-virtual) transition form factors also determine the dilepton
branching fractions, albeit inside loop integrals.  Hence, the transition form factors are integrated over (in principle)
all energy scales.

Conventionally, the decay rate for $P(q)\to\ell^+(p)\ell^-(q-p)$, $P=\pi^0,\eta,\eta'$, is expressed \textit{relative} to the diphoton
rate according to
\beq
\frac{\Gamma(P\to\ell^+\ell^-)}{\Gamma(P\to\gamma\gamma)} =
\frac{\BR(P\to\ell^+\ell^-)}{\BR(P\to\gamma\gamma)} = 
2 \bigg( \frac{\alpha_{\text{em}}}{\pi} \bigg)^2 \bigg(\frac{m_\ell}{M_P} \bigg)^2 \beta_\ell \big| \A(M_P^2)\big|^2 \,,
\label{eq:BR:P-ll}
\eeq
where $M_P$ and $m_\ell$ denote the masses of the decaying pseudoscalar meson and the final-state lepton, respectively, 
$\beta_\ell = \sqrt{1-4m_\ell^2/M_P^2}$, and the \textit{reduced amplitude} from the loop integration can be written as
\beq
\A(q^2) = \frac{2}{i\, \pi^2} \int \diff^4k \frac{q^2k^2-(qk)^2}{q^2k^2(q-k)^2\big(m_\ell^2-(p-k)^2\big)} 
\bar F_{P\gamma^*\gamma^*}\big(k^2,(q-k)^2\big) \,.
\label{eq:A:P-ll}
\eeq
Here, $\bar F_{P\gamma^*\gamma^*}\big(q_1^2,q_2^2\big) = F_{P\gamma^*\gamma^*}\big(q_1^2,q_2^2\big)/F_{P\gamma\gamma}$
denotes the normalized transition form factor.
Equation~\eqref{eq:A:P-ll} demonstrates that the loop integral is logarithmically divergent for a constant
transition form factor: it requires a form factor that vanishes for large arguments to render the 
expressions~\eqref{eq:BR:P-ll}, \eqref{eq:A:P-ll} finite.  Note that the reduced amplitude is well-defined in terms
of the physical, observable transition form factor $\bar F_{P\gamma^*\gamma^*}\big(q_1^2,q_2^2\big)$ only for 
on-shell kinematics $q^2 = M_P^2$.

\begin{table}
\renewcommand{\arraystretch}{1.3}
\centering
\begin{tabular}{ccccc}
\toprule
$\pi^0\to e^+e^-$ & $\eta \to e^+e^-$ & $\eta \to \mu^+\mu^-$ & $\eta' \to e^+e^-$ & $\eta' \to \mu^+\mu^-$  \\
\midrule
$7.48(38)\times 10^{-8}$~\cite{Abouzaid:2006kk} & 
$\leq 7\times 10^{-7}$~\cite{Achasov:2018idb}& 
$5.8(8)\times 10^{-6}$~\cite{Tanabashi:2018oca} & 
$\leq 5.6\times10^{-9}$~\cite{Akhmetshin:2014hxv,Achasov:2015mek} & 
--- \\
\bottomrule
\end{tabular}
\renewcommand{\arraystretch}{1.0}
\caption{Experimental results for the branching ratios $\BR(P\to\ell^+\ell^-)$.
The upper limit on $\BR(\eta\to e^+e^-)$ is deduced from the inverse reaction $e^+e^-\to\eta$
and supersedes the best direct limit on the decay~\cite{Agakishiev:2013fwl}.
$\BR(\eta\to\mu^+\mu^-)$ is an average of Refs.~\cite{Dzhelyadin:1980kj,Abegg:1994wx}.}
\label{tab:P-ll}
\end{table}
The experimental status is summarized in Table~\ref{tab:P-ll}.  Branching ratios have been established 
for two channels only: the determinations of $\BR_{\text{exp}}(\pi^0\to e^+e^-)$ 
are dominated by the KTeV measurement~\cite{Abouzaid:2006kk}, 
while the PDG average for $\BR_{\text{exp}}(\eta\to\mu^+\mu^-)$ is dominated by the most recent results of SATURNE~II~\cite{Abegg:1994wx}.
Upper limits have been established for $\BR_{\text{exp}}(\eta,\eta'\to e^+e^-)$, by the HADES collaboration~\cite{Agakishiev:2013fwl} 
in the case of the $\eta$, while the combined SND and CMD-3 limit on $\BR_{\text{exp}}(\eta'\to e^+ e^-)$ is derived from the inverse
reaction, i.e., a search for production $e^+e^-\to\eta'$~\cite{Akhmetshin:2014hxv,Achasov:2015mek}.
No experimental information has been gathered on $\eta'\to \mu^+ \mu^-$ to date.

The KTeV result $\BR_{\text{exp}}(\pi^0\to e^+e^-) = 7.48(38)\times 10^{-8}$~\cite{Abouzaid:2006kk} caused quite a stir:
as we will discuss below, the Standard Model predictions all yield values around $\BR_{\text{th}}(\pi^0\to e^+e^-) \simeq 6.2\times 10^{-8}$ with rather good precision, such that the experiment seemed to deviate from theory by more than $3\sigma$.
However, radiative corrections play an important role in the interpretation of the data,
and a renewed evaluation of these by the Prague group~\cite{Vasko:2011pi,Husek:2014tna} suggests that the tension is in fact
far smaller.  What KTeV has measured is in fact the quantity
\beq
\BR_{\text{exp}}(\pi^0\to e^+e^-(\gamma), x>0.95) = 6.44(25)_{\text{stat}}(22)_{\text{syst}}\times 10^{-8} \,,
\eeq
which is (necessarily) partially inclusive with respect to final-state photon radiation, with the experimental cut on the 
additional photon applied to the variable $x = m_{e^+e^-}^2/M_{\pi^0}^2$.  This is related to the theoretical (and theoretically
calculable) branching fraction, purified of radiative corrections, by
\beq
\BR_{\text{exp}}(\pi^0\to e^+e^-(\gamma), x>0.95) = \BR_{\text{LO}}(\pi^0\to e^+e^-)\times (1+\delta) \,,
\label{eq:P-ll-radcorr}
\eeq
where the subscript LO refers to the leading order in the fine structure constant, and 
$\delta = -6.0(2)\%$~\cite{Vasko:2011pi,Husek:2014tna,Hoferichter:2021lct} includes virtual QED corrections, real-photon radiation
beyond the soft-photon approximation, as well as background corrections due to Dalitz decays.  
Equation~\eqref{eq:P-ll-radcorr} then implies
\beq
\BR_{\text{LO}}(\pi^0\to e^+e^-) = 6.85(27)_{\text{stat}}(23)_{\text{syst}}\times 10^{-8} \,,
\eeq
which suggests a mild tension with theoretical predictions at best.  The official branching ratio quoted
in Table~\ref{tab:P-ll} was obtained by applying an old estimate of radiative corrections
that amounts to $\delta = -13.8\%$~\cite{Bergstrom:1982wk} (compare also Ref.~\cite{Dorokhov:2008qn}), implying
a very large $\BR_{\text{LO}}(\pi^0\to e^+e^-)$.  The difference was traced back to the assumption of a pointlike
$\pi^0 \to e^+e^-$ vertex in Ref.~\cite{Bergstrom:1982wk}, which was shown to be unreliable~\cite{Vasko:2011pi}.

\begin{figure}
\centering
\includegraphics[width=\linewidth]{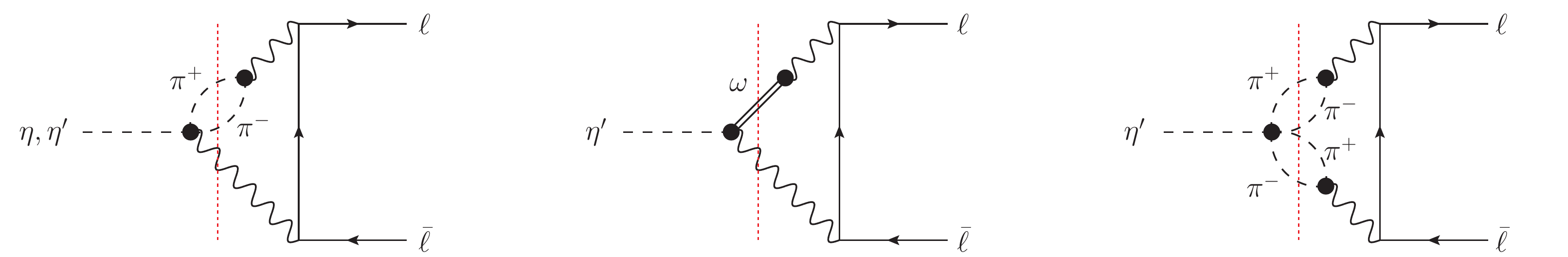}
\caption{Imaginary parts beyond the two-photon cut contributing to $\ep\to\ell^+\ell^-$.  The vertical red dotted lines denote the corresponding (partly) hadronic cuts.}
\label{fig:P-ll_cuts}
\end{figure}

According to Eq.~\eqref{eq:A:P-ll}, theoretical calculations of the dilepton decay rates are intimately linked to descriptions of the doubly-virtual transition form factors, although the loop integration for a given form factor is far from trivial.  Often, a \textit{unitarity bound} is quoted as a lower limit for the decay rate~\cite{Berman:1960zz}, based on the observation that the imaginary part due to the two-photon cut of the diagram displayed in Fig.~\ref{fig:P-ll} is determined by the form factor normalization only and hence trivially model-independent:
\beq
\Im \A(M_P^2) = \frac{\pi}{2\beta_\ell} \log\frac{1-\beta_\ell}{1+\beta_\ell} \,. 
\label{eq:P-ll:ImA}
\eeq
However, the two-photon cut represents the only relevant cut contribution only for the $\pi^0$ decay; for $\eta$ and $\eta'$, several (partly) hadronic cuts associated with cuts in the transition form factor also contribute~\cite{Masjuan:2015cjl} as should be obvious from several of the previous sections, such as $\eta,\eta'\to\pi^+\pi^-\gamma$, $\eta'\to\omega\gamma$, and $\eta'\to2(\pi^+\pi^-)$; see Fig.~\ref{fig:P-ll_cuts}.
In particular the corrections for the $\eta'$ are also numerically quite significant~\cite{Masjuan:2015cjl}.  The unitarity bound for $\pi^0\to e^+e^-$ results in $\BR(\pi^0\to e^+e^-) > 4.69\times 10^{-8}$.  Comparison with the experimental results demonstrates that the imaginary part generates a large part of the partial width, although a careful assessment of the real part is clearly required for a prediction of interesting accuracy.

$\chi$PT at leading order corresponds to constant pseudoscalar transition form factors (and, at any order, results in form factors that grow with momenta asymptotically), hence the loop integral needs to be regulated by a counterterm: the effective theory is not fully predictive here, but can only relate different dilepton decays to each other~\cite{Savage:1992ac,GomezDumm:1998gw}.  Vector-meson-dominance form factors~\cite{Ametller:1993we}, whether implemented as a plain model or motivated based on large-$N_c$ arguments~\cite{Knecht:1999gb,Husek:2015wta}, lead to a sufficient damping to render the loop integrals finite.  The particularly simple analytic form of such form factors even allows one to evaluate the loops analytically without approximations~\cite{Silagadze:2006rt}.  Based on Eq.~\eqref{eq:P-ll:ImA}, also the use of a dispersion relation for $\A(q^2)$ in $q^2$, i.e., de facto in the pion mass squared, has been suggested~\cite{Bergstrom:1983ay,Dorokhov:2007bd}.  However, this is not model-independent quite in the same way as the suggested use of dispersion relations elsewhere in this review: for $q^2 \neq M_P^2$, the transition form factor is not an observable.  Furthermore, the imaginary part is typically used in the approximation~\eqref{eq:P-ll:ImA} only, which is not a valid approximation for large pseudoscalar masses.  Nevertheless, these dispersion relations are mostly evaluated in series expansions in $(m_\ell/M_P)^2$, $(m_\ell/m_V)^2$, and $(M_P/m_V)^2$, where $m_V$ is the typical form factor scale (such as a vector-meson mass), and mass corrections beyond the leading order have been evaluated using Mellin--Barnes techniques~\cite{Dorokhov:2008cd,Dorokhov:2009xs}.

Among recent evaluations of $\BR(\pi^0\to e^+e^-)$ using rational form factors, we single out Ref.~\cite{Masjuan:2015lca} for using a systematic (and, in principle, systematically improvable) rational approximation scheme, so-called Canterbury approximants (two-variable generalizations of the more commonly-known Pad\'e approximants), that is adapted to all known asymptotic and data constraints on the $\pi^0$ transition form factor.  
\begin{table}[t!]
\renewcommand{\arraystretch}{1.3}
\centering
\begin{tabular}{ccc}
\toprule
$\pi^0\to e^+e^-$ & $\eta \to e^+e^-$ & $\eta' \to e^+e^-$  \\
\midrule
$6.23(5)\times 10^{-8}$~\cite{Masjuan:2015lca} & 
$5.31\big({}^{+0.14}_{-0.04}\big)\times 10^{-9}$~\cite{Masjuan:2015cjl}& 
$1.81(18)\times10^{-10}$~\cite{Masjuan:2015cjl} \\
$6.22(3)\times 10^{-8}$~\cite{Weil:2017knt} \\
$6.25(3)\times 10^{-8}$~\cite{Hoferichter:2021lct} \\
\midrule
& $\eta \to \mu^+\mu^-$ & $\eta' \to \mu^+\mu^-$ \\ \midrule
& $4.71\big({}^{+0.05}_{-0.21}\big)\times 10^{-6}$~\cite{Masjuan:2015cjl} 
& $1.36\big({}^{+0.29}_{-0.26}\big)\times 10^{-7}$~\cite{Masjuan:2015cjl} \\
\bottomrule
\end{tabular}
\renewcommand{\arraystretch}{1.0}
\caption{Selected theoretical results for the branching ratios $\BR(P\to\ell^+\ell^-)$. For completeness, we include the result from Ref.~\cite{Hoferichter:2021lct}, which appeared post completion of this review.}
\label{tab:P-ll:th}
\end{table}
In Table~\ref{tab:P-ll:th}, this is contrasted with a recent determination~\cite{Weil:2017knt} 
based on a form factor calculated
using Dyson--Schwinger equations~\cite{Eichmann:2017wil}, which is particularly noteworthy for a new evaluation 
of the loop in Eq.~\eqref{eq:A:P-ll}: while a naive Wick rotation seems impossible, the authors show how
to deform the integration path in such a way as to avoid all singularities and still arrive at a well-defined
Euclidean integral.  Remarkably, the central values for $\BR(\pi^0\to e^+e^-$) agree perfectly~\cite{Masjuan:2015lca,Weil:2017knt}.  While it is not entirely clear to what extent the uncertainty estimate of Ref.~\cite{Weil:2017knt} reliably reflects the truncation uncertainties in the Dyson--Schwinger approach, the one quoted in Ref.~\cite{Masjuan:2015lca} likely is a very good measure of the precision that can be achieved in a data-driven procedure, based on our combined experimental and theoretical understanding of the $\pi^0$ transition form factor.

Table~\ref{tab:P-ll:th} also includes the theoretical results for the $\eta$ and $\eta'$ dilepton decay rates, based on form factors evaluated using Canterbury approximants~\cite{Masjuan:2015cjl}.  These are largely compatible with many of the earlier determinations of some or all of these branching ratios~\cite{Savage:1992ac,GomezDumm:1998gw,Silagadze:2006rt,Ametller:1993we,Knecht:1999gb,Dorokhov:2007bd,Dorokhov:2009xs}, but probably contain the most reasonable assessment of the form factor uncertainties.  Furthermore, Ref.~\cite{Masjuan:2015cjl} contains an interesting suggestion for the evaluation of the loop integral~\eqref{eq:A:P-ll} based on form factors that obey a (double) dispersion relation: assuming a factorizing dependence on the two virtualities, the authors write the (normalized) form factor as
\beq
\bar F_{P\gamma^*\gamma^*}\big(q_1^2,q_2^2\big) = \frac{1}{\pi^2} \int_{s_{\text{th}}}^\infty \diff x \frac{\Im \bar F_{P\gamma^*\gamma^*}(x,0)}{x-q_1^2-i\eps} \int_{s_{\text{th}}}^\infty \diff y \frac{\Im \bar F_{P\gamma^*\gamma^*}(0,y)}{y-q_2^2-i\eps} \,,
\eeq
which allows them to reformulate Eq.~\eqref{eq:A:P-ll} as
\begin{align}
\A(q^2) &= \frac{1}{\pi^2} \int_{s_{\text{th}}}^\infty \diff x \int_{s_{\text{th}}}^\infty \diff y \,
\Im \bar F_{P\gamma^*\gamma^*}(x,0) \, \Im \bar F_{P\gamma^*\gamma^*}(0,y) \, K(x,y) \,, \notag\\
K(x,y) &=
\frac{2}{i\, \pi^2} \int \diff^4k \frac{q^2k^2-(qk)^2}{q^2k^2(q-k)^2\big(m_\ell^2-(p-k)^2\big)\big(x-k^2\big)\big(y-(q-k)^2\big)} \,.
\label{eq:A-disp:P-ll}
\end{align}
$K(x,y)$ can be expressed in terms of standard one-loop functions~\cite{tHooft:1978jhc}; it represents $\A(q^2)$ for a VMD form factor with squared vector-meson masses $x$ and $y$.  This loop amplitude is then weighted by the spectral functions for the two form factor virtualities.  Such a representation should prove very useful for future evaluations of $P\to\ell^+\ell^-$ decay rates based on dispersively constructed transition form factors that are, however, only available in numerical form.

\subsection{$\eta, \eta'$ decays to true muonium}\label{sec:muonium}

``True muonium'' ($\TM$) commonly refers to the electromagnetic, i.e., quasi-atomic,
bound states of a $\mu^+\mu^-$ pair~\cite{Bilenky:1969zd,Hughes:1971}.  While it remains undiscovered experimentally
due to low production rates to date, it has been advertised as another promising avenue to
further scrutinize the various potential anomalies involving muons~\cite{TuckerSmith:2010ra}, such
as the anomalous magnetic moment~\cite{Bennett:2006fi,Abi:2021gix,Albahri:2021ixb} or muonic hydrogen
spectroscopy~\cite{Pohl:2010zza,Antognini:1900ns}. In particular decay properties of $\TM$ may be affected in various ways by BSM physics contributions~\cite{CidVidal:2019qub}.

Studies of $\TM$ largely concentrate on the spin triplet $n{}^3S_1$ states, as the latters'
quantum numbers
allow them to be produced via a single virtual photon.
Their mass is given by $m_\TM=2m_\mu[1-\alpha_{\rm em}^2/(8n)]$ when accounting for the binding
energy at leading order.  The Standard Model decays are dominated by those into $e^+e^-$
(with a branching ratio of $\BR(\TM\to e^+e^-)\approx 98\%$), followed by those into $3\gamma$
($\BR(\TM\to 3\gamma)\approx 1.7\%$)~\cite{CidVidal:2019qub}.  
In a certain energy range not too far from its mass, $\TM$ has properties similar to those
of dark photons (see Sect.~\ref{sec:BSM-vector}), in that it couples to photons via kinetic
mixing.
Promising production channels hence include radiative meson decays with large branching fractions,
which makes $\eta\to\gamma\,\TM$ [$\BR(\eta\to\gamma\gamma)=39.41(20)\%$]
and $\eta'\to\pi^+\pi^-\TM$ [$\BR(\eta'\to\pi^+\pi^-\gamma)=28.9(5)\%$] particularly
fitting~\cite{Ji:2018dwx}.  Due to their simpler kinematic structure, the former has been
investigated in detail.

At leading order in $\alpha_{\rm em}$, the branching ratios for $\TM$ production in radiative $\ep$
decays are given by~\cite{Ji:2018dwx}
\beq
\frac{\BR(\ep\to \gamma\,\TM)}{\BR(\ep\to \gamma\gamma)} =
\frac{\alpha_{\rm em}^4\,\zeta(3)}{2}\Bigg(1-\frac{m_\TM^2}{M_{\ep}^2}\Bigg)^3
\Big|\bar F_{\ep\gamma^*\gamma^*}\Big(m_\TM^2,0\Big)\Big|^2 \,, \label{eq:BR-TM}
\eeq
where the summation over all principal quantum numbers is performed.
Eq.~\eqref{eq:BR-TM} depends on the singly-virtual $\ep$ transition form factors,
evaluated at $q^2=m^2_\TM$, and hence presents yet another application of the physics of TFFs discussed extensively throughout this chapter.
Compared to the corresponding photon final states, branching fractions into $\TM$ are generically suppressed by $\alpha_{\rm em}^4 \approx 2.8\times10^{-9}$ and are of order $\BR(\eta\to \gamma\,\TM) \approx 5 \times 10^{-10}$ and $\BR(\eta^\prime \to \gamma\,\TM) \approx 4 \times 10^{-11}$~\cite{Ji:2018dwx}.
Higher-order radiative corrections to Eq.~\eqref{eq:BR-TM} have also been calculated~\cite{Ji:2018dwx}, although they are unlikely to be of relevance in the comparison to experimental findings anytime soon.

Given these branching fractions, dedicated $\eta$ factories will require the statistics of the suggested REDTOP experiment to discover $\TM$.  
On the other hand, inclusive searches at LHCb, where $\eta\to\gamma\,\TM$ provides the dominant production mechanism, have been predicted to be able to observe $\TM$ with sufficient statistical significance~\cite{CidVidal:2019qub} in the upcoming Run-3 data~\cite{Bediaga:2018lhg}.

\section{\boldmath Probing scalar dynamics in $\epgg$}
\label{sec:eta-pi0gg}
\subsection{Theory}
\begin{figure}
\centering
\includegraphics[width=0.75\linewidth]{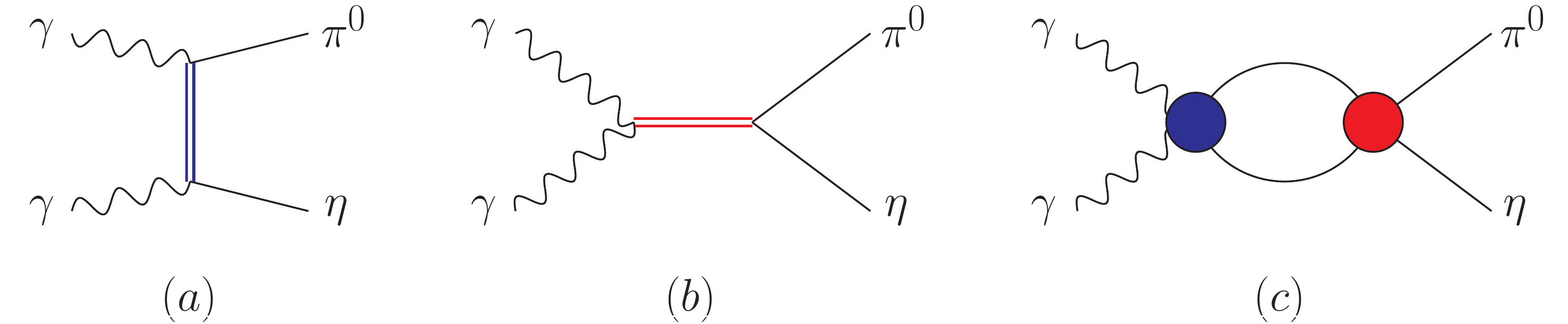}
\caption{Schematic form of mechanisms contributing to $\epgg$: 
(a) $t$-channel resonance exchange (e.g., of vectors); (b)
$s$-channel resonance exchange (predominantly $a_0(980)$ and $a_2(1320)$);
(c) a generic loop diagram/rescattering mechanism.
See text for details.\label{fig:diagrams}}
\end{figure}

The decay $\eta\to\pi^0 \gamma\gamma$ is one of the very rare meson decays in the Standard Model that proceeds via a polarizabi\-lity-type mechanism. 
It is interesting from the perspective of low-energy effective theories
as the usually very successful power counting of chiral perturbation theory seems to be bypassed:
leading contributions either vanish or are strongly suppressed, such that the main contributions 
to the decay amplitude are pushed to unusually high orders in the chiral expansion.
Our understanding of these high orders in terms of resonance degrees of freedom is based on the
interplay between vector and scalar resonances~\cite{Ametller:1991dp}.\footnote{See Ref.~\cite{Achasov:2001qm} for a historical overview of both theoretical and experimental studies of this channel, including references to early pre-$\chi$PT calculations.}
This interplay was first investigated comprehensively in Refs.~\cite{Oset:2002sh,Oset:2008hp},
where the crossed process $\ggpe$ was studied in parallel: the amplitudes of the two reactions are linked
by crossing symmetry and analytic continuation, the decay being restricted to diphoton invariant
masses in the range $0 \leq M_{\gamma\gamma}  \leq M_\eta-M_{\pi^0}$, while $\pi^0\eta$ production
in photon--photon fusion can be accessed above threshold, $ M_\eta+ M_{\pi^0} \leq M_{\gamma\gamma}$.
We will illustrate the various mechanisms for both processes by appealing to Fig.~\ref{fig:diagrams}.

Tree-level contributions to $\epgg$ vanish both at $\Order(p^2)$ (due to both $\eta$ and $\pi^0$ being uncharged)
and at $\Order(p^4)$ in the chiral expansion. 
The leading nonvanishing contribution that does exist at $\Order(p^4)$ are
meson loop graphs, with either $\pi^+\pi^-$ or $K^+K^-$ intermediate states, shown in Fig.~\ref{fig:diagrams}(c).
However, rescattering $\pi^+\pi^-\to\pi^0\eta$ requires isospin breaking and is strongly suppressed
(see Sect.~\ref{sec:eta-3pi}; an appreciable, enhanced signal in this crossed reaction of the decay
$\eta\to3\pi$ only occurs in the energy region of so-called $a_0$--$f_0$ mixing~\cite{Hanhart:2007bd}),
while kaon loops turn out to be small for the decay due to combinatorial factors and 
the large kaon mass in the denominator~\cite{Ametller:1991dp}.  
The smallness of pure loop contributions is maintained beyond one-loop order
as illustrated by the partial two-loop calculation in Ref.~\cite{Jetter:1995js}.
The bulk of the $\epgg$ decay width at $\Order(p^6)$ is reproduced by
two counterterms in the chiral Lagrangian (which corresponds to two-loop level).  

The size of counterterm contributions in chiral perturbation theory can be phenomenologically understood
by resonance exchanges~\cite{Ecker:1988te}.  The time-honored concept of vector-meson dominance is therein
resurrected by the observation that vector mesons tend to contribute most to those counterterms (where
allowed by quantum numbers).  This turns out to be true also for $\epgg$: the largest contribution 
to the decay width stems from $t$-channel $\rho$- and $\omega$-exchange (as in Fig.~\ref{fig:diagrams}(a)).
As has been pointed out in Ref.~\cite{Ametller:1991dp}, the details on how to implement those vector meson exchanges
matter significantly: the full vector meson propagators lead to a $\epgg$ width almost a factor of 2 larger than
what is found by employing resonance saturation in the strict sense, i.e., replacing the propagators by 
point interactions; the difference is of $\Order(p^8)$ in the chiral counting.
On the other hand, adjusting the coupling constants to the individual decays $\rho/\omega \to \pi^0/\eta \gamma$, as
opposed to using $\SU(3)$ symmetric couplings, reduces
the width by almost a factor of 2; updated measurements of those radiative vector decays
have allowed to somewhat reduce the error in the theory prediction, with the central value cut once more
by almost 20\%~\cite{Oset:2008hp}. 

However, the pure VMD prediction can be significantly modified by the $\pi\eta$ $S$-wave.  
Through coupled-channel effects with $K\bar K$, this includes the $a_0(980)$ resonance near the 
$K\bar K$ threshold. (The $a_0(980)$ is sometimes said to be ``dynamically generated'' by coupled-channel
meson--meson rescattering in the isospin $I=1$ $S$-wave.)  
In this way, no scalar resonance needs to be put into the calculation ``by hand''
(as in Fig.~\ref{fig:diagrams}(b)): it occurs naturally,
with the relative sign/phase of the corresponding amplitude fixed by the low-energy couplings of two photons to the
two channels $\pi^0\eta$ and $K\bar K$.  
In this way, scalar resonances and rescattering or loop effects are, as usual, intimately related.
This mechanism is also included in
Fig.~\ref{fig:diagrams}(c): the red vertex then denotes the two-channel rescattering matrix, which in 
Refs.~\cite{Oset:2002sh,Oset:2008hp} was calculated
in unitarized chiral perturbation theory (solving a Bethe--Salpeter equation with a momentum cutoff).
Production (as in the blue vertex in Fig.~\ref{fig:diagrams}(c)) was assumed to proceed via $\rho$ and $\omega$ 
for the $\pi^0\eta$ meson pair, and via $K^+$ pole terms, vector $K^*(892)$, and even axial-vector 
$K_1(1270)$ resonances for $K\bar K$~\cite{Oset:2002sh,Oset:2008hp}.
Nonstrange axial vectors ($b_1$ and $h_1$ resonances) coupling to $\pi^0$ and $\eta$ were not retained,
but included in the uncertainty estimate.  
$t$-channel loops with $\pi\pi$ and $K\bar K$ intermediate states, involving two anomalous vertices,
 were only included perturbatively, i.e., without $s$-channel rescattering.
The $a_0(980)$ resonance signal thus generated could be tested favorably against $\ggpe$ data~\cite{Antreasyan:1985wx,Oest:1990ki}, where the $D$-wave $a_2(1320)$ was added in a phenomenological way~\cite{Oller:1997yg}.  Note that Refs.~\cite{Oset:2002sh,Oset:2008hp} neglected the $a_2(1320)$ in the decay amplitude for $\epgg$.

A similar approach to $\ggpe$ and $\epgg$ (as well as other photon--photon fusion reactions) has been studied more
recently in Ref.~\cite{Danilkin:2012ua}, with a potentially more sophisticated unitarization procedure to generate 
the $a_0(980)$; however, only the tree-level amplitudes with vector-meson exchange (using a different Lagrangian scheme)
have been retained for the decay amplitude, such that the interplay with scalars is not made as transparent.

A more recent study of both the decay $\eta\to \pi^0\gamma\gamma$ and simultaneously the production reaction $\gamma\gamma\to\pi^0\eta$  has been performed in Ref.~\cite{Danilkin:2017lyn}.  The transition matrix element is decomposed into helicity amplitudes $H_{\lambda_1\lambda_2}$ according to
\beq
\langle \pi(p) \gamma(q_1,\lambda_1) \gamma(q_2,\lambda_2) | T | \eta(P) \rangle 
= (2\pi)^4 \delta^{(4)}(P-p-q_1-q_2) H_{\lambda_1\lambda_2}(s,t)  \,,
\eeq
for photon helicities $\lambda_{1,2} = \pm1$.  The helicity amplitudes are functions of two independent
Mandelstam variables out of $s=(q_1+q_2)^2$, $t=(p+q_1)^2$, and $u=(p+q_2)^2$; they can be partial-wave
expanded according to
\beq
H_{\lambda_1\lambda_2}(s,t) = \sum_{\text{even}~J}(2J+1) h_{\lambda_1\lambda_2}^{(J)}(s)d_{\lambda_1-\lambda_2,0}^J(\theta) \,,
\eeq
where $d_{\lambda_1-\lambda_2,0}^J(\theta)$ are the Wigner small $d$-functions, and $\theta$ denotes the reaction angle in the diphoton rest frame.  We note that the relevant Wigner $d$-functions are related to the standard Legendre polynomials by
\beq
d_{00}^J(\theta) = P_J(\cos\theta) \,, \qquad
d_{20}^J(\theta) = \frac{2P'_{J-1}(\cos\theta)-J(J-1)P_J(\cos\theta)}{\sqrt{(J-1)J(J+1)(J+2)}} \,,
\eeq
hence the partial-wave expansion of $H_{++}(s,t)$ includes all even partial waves, while the one
for $H_{+-}(s,t)$ begins with the $D$-waves only.

\begin{sloppypar}
In Ref.~\cite{Danilkin:2017lyn}, the $S$-wave is calculated in a coupled-channel ($\pi\eta$ and $K\bar{K}$) inhomogeneous Omn\`es formalism.  
The corresponding $\pi\eta \leftrightarrow K\bar{K}$ 
$T$-matrix was constructed using an $N/D$ approach, for which the left-hand cuts
were parameterized by a conformal expansion that is matched to a combination of tree-level $\chi$PT and 
vector-meson exchanges; this method was previously shown to describe pion--pion and pion--kaon scattering
sufficiently accurately~\cite{Danilkin:2011fz}.    
Left-hand cuts for $\gamma\gamma\to\pi\eta/K\bar{K}$ are approximated by kaon Born terms as well as vector-meson exchanges of $\rho$ and 
$\omega$ for $\pi\eta$ (the $\phi$ is omitted due to OZI suppression), 
and the $K^*(892)$ for the kaon channel: all couplings can be fixed from data and symmetry considerations.  
Two subtraction constants (one per channel) are fixed to the tree-level amplitude at $\pi\eta$ threshold.
An additional $D$-wave in the $\pi\eta$ production region
is modeled as a Breit--Wigner contribution due to the $a_2(1320)$ resonance.
\end{sloppypar}

Beyond the vector exchanges that are dominant in the decay region, 
Ref.~\cite{Danilkin:2017lyn}
includes the isospin-breaking $\pi^+\pi^-$ intermediate state perturbatively at the $\Order(p^4)$ level
in $\chi$PT; while individually very small, the effect of the threshold cusp in the diphoton spectrum might 
in principle leave observable traces.

\begin{figure}
\centering
\includegraphics[width=0.45\linewidth]{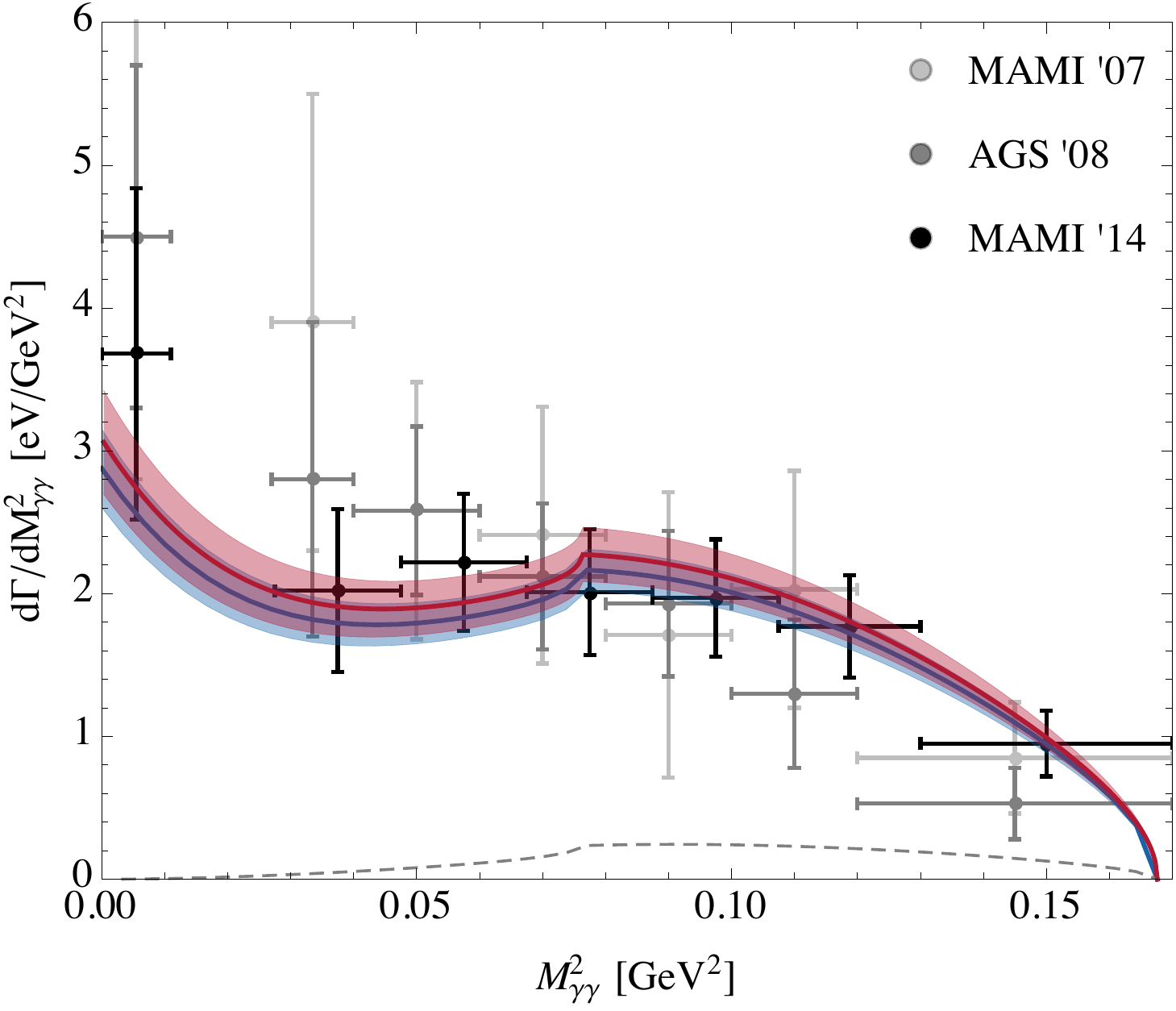} \hfill
\includegraphics[width=0.47\linewidth]{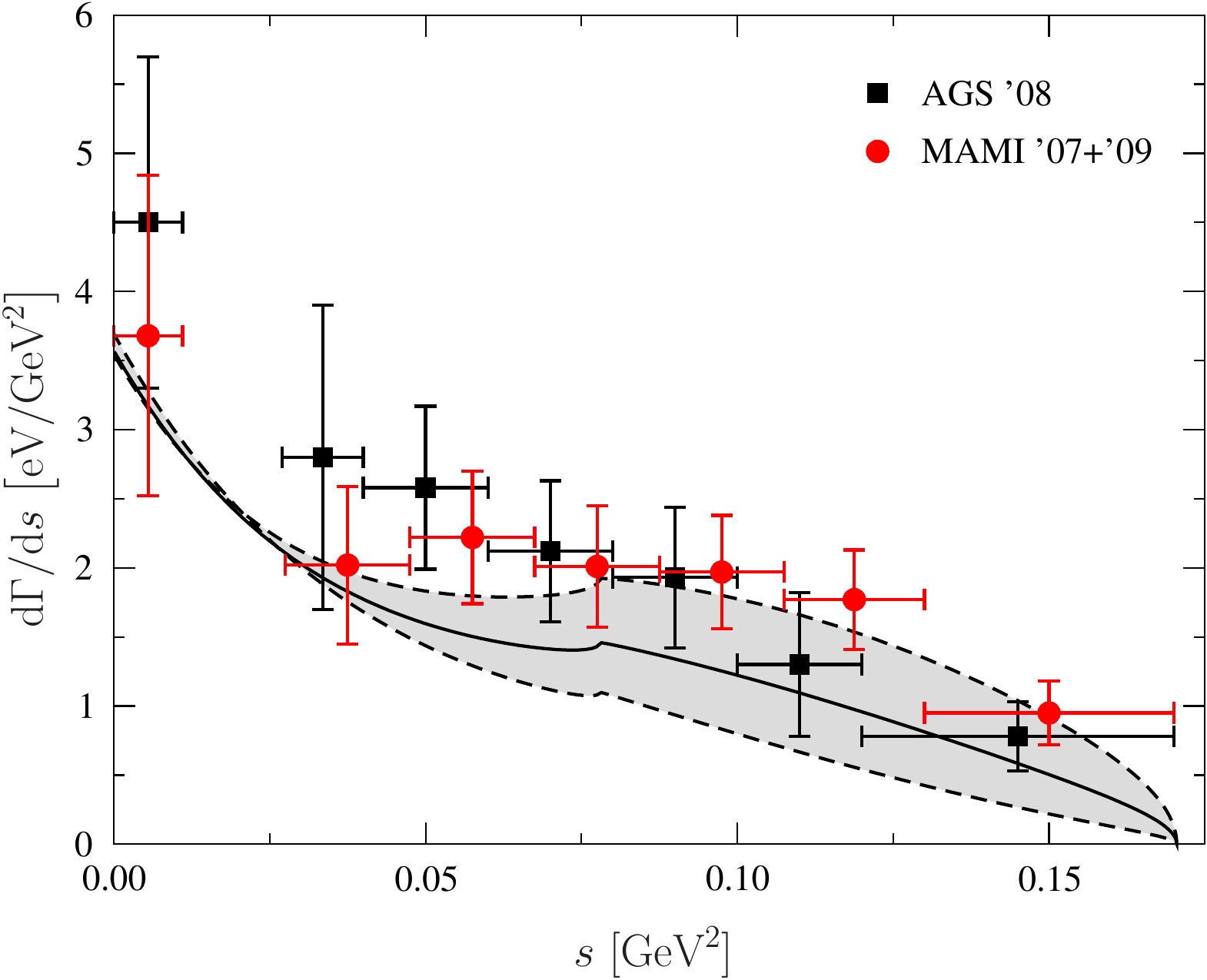}
\caption{Two-photon invariant mass distributions in $\epgg$.
Data is taken from Refs.~\cite{Prakhov:1900zz,Prakhov:2008zz,Nefkens:2014zlt}.
\textit{Left:} figure adapted from Ref.~\cite{Danilkin:2017lyn}.
The blue band is generated with vector-meson couplings individually adapted to data, 
while the red band relies on $\SU(3)$ symmetry; the error bands reflect the uncertainties
in either.  The dashed curve represents
the chiral prediction at $\Order(p^4)$, including the isospin-breaking two-pion intermediate state
inducing a threshold cusp at $M_{\gamma\gamma}^2 = 4M_\pi^2$.
\textit{Right:} figure adapted from Ref.~\cite{Lu:2020qeo}.
The band shows the influence of the exact position of the Adler zero $s_A$ on the decay spectrum:
while the central curve corresponds to $s_A=\meta^2$, its width represents the variation from 
$s_A=\meta^2+3\mpi^2$ (upper boundary) to $s_A=\meta^2-3\mpi^2$ (lower boundary).
We thank the authors of Refs.~\cite{Danilkin:2017lyn,Lu:2020qeo} for providing us with these figures.}
\label{fig:2g-spectrum}
\end{figure}
From the differential decay rate
\beq
\frac{\diff^2\Gamma}{\diff s\,\diff t} = \frac{1}{(2\pi)^3}\frac{1}{16M_\eta^3} \left(\left|H_{++}\right|^2+\left|H_{+-}\right|^2\right) \,,
\eeq
both diphoton invariant mass spectrum and partial width can be calculated.
The resulting spectrum for $\diff\Gamma/\diff M_{\gamma\gamma}^2 = \diff\Gamma/\diff s$ is shown in Fig.~\ref{fig:2g-spectrum} (left), and agrees with the existing data very well~\cite{Prakhov:1900zz,Prakhov:2008zz,Nefkens:2014zlt}. 
Unfortunately, $\pi\eta$ $S$-wave rescattering 
(claimed to lead to a significant enhancement at higher diphoton invariant masses~\cite{Oset:2008hp})
has not been retained explicitly in the decay region, but seems not to be required to reproduce the
available measurements. 
More accurate data is therefore certainly highly desirable to test theoretical calculations
and in particular scalar contributions more thoroughly and distinguish
them from the VMD mechanism alone.
The partial decay width is predicted to be~\cite{Danilkin:2017lyn}
\begin{equation}
\Gamma(\epgg) = 0.291(22)\eV\,, \label{eq:Gamma-etapi0gg-Danilkin}
\end{equation}
The uncertainty is given by the propagated errors on the vector meson coupling constants.
Equation~\eqref{eq:Gamma-etapi0gg-Danilkin} 
corresponds to $\BR(\epgg)=2.22(19)\times10^{-4}$, which agrees with the current PDG
average within uncertainties.

The latest combined study of $\epgg$ and $\gamma\gamma\to\pi^0\eta$, employing dispersion-theoretical methods
(with coupled channels)
matched to $\chi$PT, has been performed in Ref.~\cite{Lu:2020qeo}.  
While somewhat similar in spirit to Ref.~\cite{Danilkin:2017lyn}, several theoretical aspects have been 
further clarified and improved upon, in particular constraints due to soft-photon and soft-pion zeros.  
As a result of a soft-photon constraint, the helicity amplitudes $h_{++}^{(J)}(s)$ 
(Born-term subtracted in the case of the $\gamma\gamma\to K\bar{K}$ amplitudes entering the coupled-channel description)
have to vanish for $s=0$, explaining the dominance of $h_{+-}^{(2)}(s)$ in the diphoton invariant-mass spectrum
for small $s$.  While tree-level vector-meson exchange fulfills this constraint automatically, it needs to be built
into a dispersive representation of rescattering corrections explicitly.
Furthermore, a soft-pion theorem states that the amplitude for $\epgg$ vanishes in the $\SU(2)$ chiral limit
at the kinematical point $s=\meta^2$, $t=u=0$, implying in particular an Adler zero in the $S$-wave
$h_{++}^{(0)}(s_A) = 0$ at a point close to the $\SU(2)$ limit, $s_A = \meta^2 + \Order(\mpi^2)$.  

In addition, Ref.~\cite{Lu:2020qeo} includes the isospin-breaking $\pi\pi$ intermediate state fully dispersively,
going beyond the chiral one-loop approximation~\cite{Ametller:1991dp,Danilkin:2017lyn}.  This contribution is
constructed in turn from dispersive representations of $\gamma\gamma\to\pi\pi$~\cite{GarciaMartin:2010cw} and
$\pi\pi\to\pi\eta$~\cite{Albaladejo:2017hhj}, which increases the strength of the cusp at $s=4M_\pi^2$ 
in the diphoton spectrum significantly.  
This isospin-violating contribution to $\epgg$ is constrained by its own, separate, soft-pion theorem~\cite{Lu:2020qeo}.

A fit of the remaining free subtraction constant to Belle data on $\gamma\gamma\to\pi^0\eta$~\cite{Uehara:2009cf}
reveals the need to readjust the $\pi\eta \leftrightarrow (K\bar K)_{I=1}$ coupled-channel $T$-matrix 
representation of Ref.~\cite{Albaladejo:2015aca}, which subsequently allows for a good description 
of $\ggpe$ (as well as the coupled $K\bar{K}$ channels) up to about $1.4\GeV$ (the lowest-lying tensor resonances
are modeled as simple Breit--Wigner resonances).  The same amplitudes are then used to describe the decay $\epgg$,
and we show the resulting spectrum in Fig.~\ref{fig:2g-spectrum} (right).
An important conclusion is that the exact position of the Adler zero $s_A$ has a strong impact on the decay 
spectrum: a representation using $s_A \approx \meta^2+3\mpi^2$ seems to describe the differential data~\cite{Prakhov:1900zz,Prakhov:2008zz,Nefkens:2014zlt} more accurately.  Correspondingly, the predicted partial width~\cite{Lu:2020qeo}
\begin{equation}
\Gamma(\epgg) = 0.237^{+0.060}_{-0.043}\eV\,, \label{eq:Gamma-etapi0gg-Moussallam}
\end{equation}
based on a range $\meta^2-3\mpi^2 \leq s_A \leq \meta^2+3\mpi^2$, lies on the lower side of the experimental values
for lower values of $s_A$.  This description thus somewhat changes the view on the potential impact of future
precision measurements of $\epgg$: all curves in Fig.~\ref{fig:2g-spectrum} (right) take $S$-wave rescattering
(or scalar dynamics) into account, however a required suppression of the scalar partial wave for unphysical $s=s_A$,
between decay and production regions, limits the possible enhancement.

\subsection{Experimental perspectives}

Experimental measurements of the doubly radiative decay $\epgg$ have a history spanning more
than five decades~\cite{Achasov:2001qm}.  
This channel is sufficiently suppressed [$\BR(\epgg)=2.56(22)\times 10^{-4}$]
that, while it has been possible for all recent experiments to observe a nonzero signal, 
measurements accurate enough to challenge theory have proven elusive. 
About two dozen experiments have been performed to measure 
this decay width since 1966. The first significant result was published by the GAMS-2000 
collaboration~\cite{Alde:1984wj,Landsberg:1986fd} in 1984 yielding $\Gamma (\eta\to\pi^0\gamma\gamma)=0.84(18)\eV$, more than twice the $\chi$PT prediction~\cite{Oset:2008hp}. 
By contrast, more recent results from the Crystal Ball~\cite{Prakhov:2008zz}, 
A2~\cite{Nefkens:2014zlt}, and KLOE~\cite{DiMicco:2005stk} collaborations are significantly lower. 
The Crystal Ball result, $\Gamma (\eta\to\pi^0\gamma\gamma)=0.285(68)\eV$~\cite{Prakhov:2008zz}, is consistent with the 
prediction of Ref.~\cite{Oset:2008hp} as well as with Eq.~\eqref{eq:Gamma-etapi0gg-Danilkin}.
However, the preliminary result from KLOE~\cite{DiMicco:2005stk} is lower than the Crystal Ball result by a factor of 3.  The most recent result came from the A2 collaboration at MAMI~\cite{Nefkens:2014zlt}. 
The decay width, $\Gamma(\eta\rightarrow \pi^0\gamma\gamma)=0.33(3)\eV$, was determined from $1.2\times 10^3$ $\eta\rightarrow \pi^0\gamma\gamma$ decay events, which
was a significant improvement over previous measurements. 
The collaboration also determined the one-dimensional projection of the Dalitz distribution, $d\Gamma/dM^2_{\gamma\gamma}$,
which is even more important than the decay width for constraining the underlying dynamics.
As stated in Ref.~\cite{Nefkens:2014zlt}, the 21\% per bin experimental 
uncertainties on the new MAMI $M_{\gamma\gamma}$ distribution are still  too large to 
rule out any of the theoretical calculations. 

All existing experimental results~\cite{Alde:1984wj,Landsberg:1986fd,Prakhov:2008zz,DiMicco:2005stk,Nefkens:2014zlt} were limited by large backgrounds from $\eta\rightarrow 3\pi^0$ leaking into the 4$\gamma$ final state data sample and a nonresonant $2\pi^0$ continuum production (see, e.g., Fig.~2 of Ref.~\cite{Nefkens:2014zlt}). 
In the first case, for a $6\gamma$ process from the $\eta\rightarrow 3\pi^0$ decay
to be a background to a $4\gamma$ process, two photons must effectively go uncounted while  the reconstructed invariant mass remains close to the $\eta$ mass. There are two contributing mechanisms: (1) soft photons falling out of the geometrical acceptance or
 below the threshold of the detector, or
(2) two photons merging into what appears to be a single shower in the calorimeter.  These two mechanisms can be greatly suppressed by increasing the 
energy of the $\eta$ mesons in the lab frame while  maintaining sufficient granularity in the calorimeter. To reduce the second major background from  nonresonant $2\pi^0$ production, one can tag the $\eta$ by detecting recoil particles in the production reaction and use a state-of-art calorimeter with high energy and position resolutions to improve the signal selection cuts. A new experiment with a significantly improved reduction in backgrounds would provide 
greatly reduced statistical and systematic uncertainties leading to a definitive 
result for the $\eta\rightarrow\pi^0 \gamma\gamma$ decay width. More importantly, the two-photon invariant mass 
spectrum, $d\Gamma/dM_{\gamma\gamma}$, will provide key guidance for understanding the underlying 
dynamics.

\begin{figure}
\includegraphics[height=6cm]{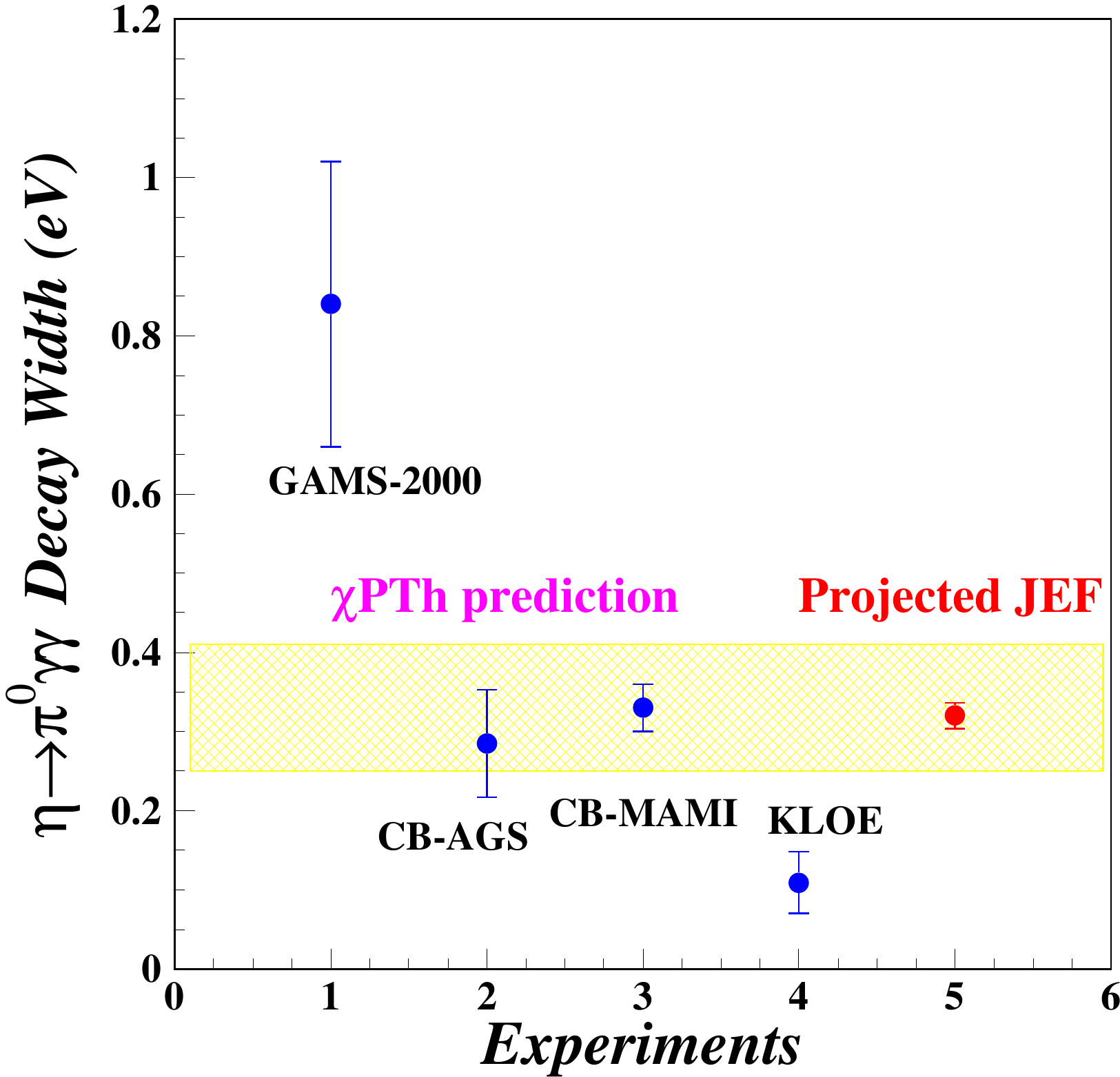} \hfill
\includegraphics[height=6cm]{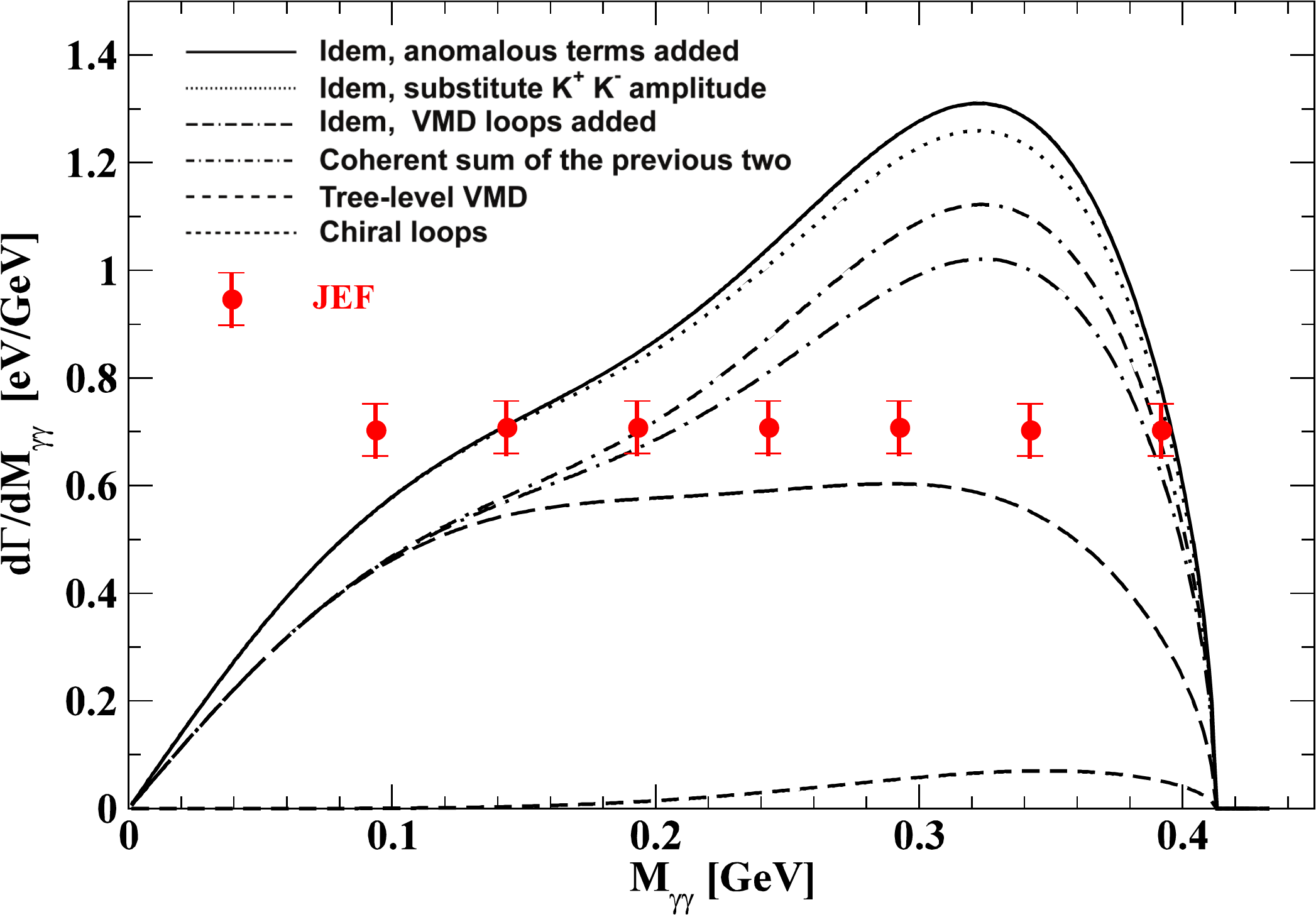}
\caption{\textit{Left:} Experimental results on the decay width of 
$\eta\to\pi^0\gamma\gamma$~\cite{Alde:1984wj,Prakhov:2008zz,Nefkens:2014zlt,DiMicco:2005stk}. 
The yellow band is $\Gamma = 0.33(8)\eV$ from the unitarized-$\chi$PT calculation of Refs.~\cite{Oset:2002sh,Oset:2008hp}.
The projected JEF measurement with a total error of 5\% (in red) for 100~days of beam time is arbitrarily 
plotted at the CB-MAMI value~\cite{Nefkens:2014zlt}. 
\textit{Right:}
Predicted two-photon invariant mass distributions from $\eta \to \pi^0 \gamma\gamma$ due to different mechanisms~\cite{Oset:2002sh,Oset:2008hp} 
and the projected JEF measurement with 100~days of beam time. Figures reprinted from Ref.~\cite{Gan:2015nyc}.
\label{fig:pi0gamma2}}
\end{figure}

The approved JEF experiment at JLab will measure the 
$\eta\to\pi^0 \gamma\gamma$ branching ratio and the 
Dalitz distribution as shown in Fig.~\ref{fig:pi0gamma2}, providing a determination of two $\Order(p^6)$ low-energy constants.
The decay of $\eta \to \pi^0\gamma\gamma$ has the striking feature that the shape of the two-photon invariant mass spectrum, $\diff\Gamma/\diff M_{\gamma\gamma}$, is rather sensitive to the role of scalar
dynamics.
 As clearly seen in Fig.~\ref{fig:pi0gamma2} (right) based on a comparison to the calculation in Ref.~\cite{Oset:2008hp}, the predicted full distribution including the scalar contributions (solid line) has a significant enhancement at high diphoton invariant masses as compared to the flatter pure VMD prediction (long-dashed). Therefore a precision measurement of this distribution offers a clean window towards a better understanding of the interplay of meson resonances and their impact on $\Order(p^6)$ 
$\chi$PT~\cite{Bijnens:2002qy}.
The projected JEF precision, shown in Fig.~\ref{fig:pi0gamma2}, would be sufficient to determine 
the scalar--VMD interference contribution and distinguish it from the VMD mechanism alone for the first time. It would provide a sensitive probe to test the ability of models such as meson resonance saturation to calculate many other unknown $\Order(p^6)$ low-energy constants.

\subsection{$C$-conserving $\eta\to\pi^0 \ell^+\ell^-$ decays}\label{sec:etapi0ll}

\begin{figure}
\centering
\includegraphics[width=0.6\linewidth]{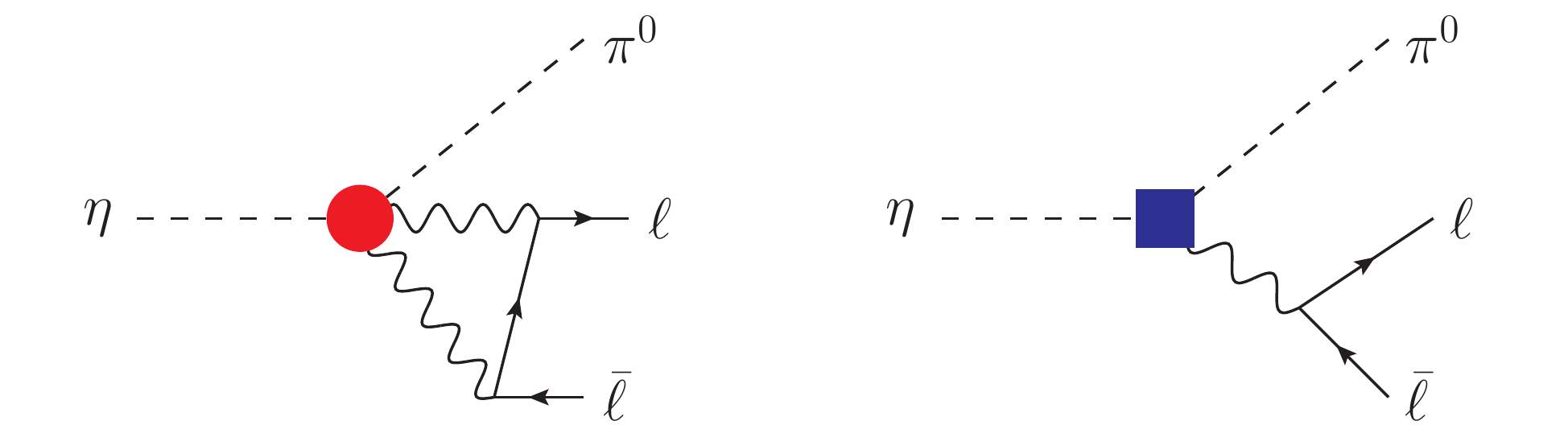}
\caption{Mechanisms for the decay $\eta\to\pi^0\ell^+\ell^-$. \textit{Left:} $C$-conserving via
a two-photon intermediate state; the red blob denotes the $\eta\to\pi^0\gamma^*\gamma^*$ amplitude.
\textit{Right:} via a $C$-violating form factor denoted by the blue square.
\label{fig:etapi0ll}}
\end{figure}

The decay $\eta \to \pi^0\gamma^* \to\pi^0\ell^+\ell^-$ via a single-photon transition is forbidden by $C$ and $CP$ conservation~\cite{Bernstein:1965hj}.  
However, the decay can proceed in the Standard Model via a two-photon intermediate state without $C$ or $CP$ violation (in analogy to the dilepton decays described in Sect.~\ref{sec:P-ll});
both mechanisms are depicted diagrammatically in Fig.~\ref{fig:etapi0ll}.
Therefore, it is important to calculate this two-photon rate since it is a background for BSM searches, whether due to single-photon exchange or new scalar mediators (see Sects.~\ref{sec:BSMfundsym} and \ref{sec:BSM-lightparticles}).
In the remainder of this section, we consider $\eta \to \pi^0\ell^+\ell^-$ exclusively as a two-photon process in the Standard Model.

Theoretical estimates for $\Gamma(\eta \to \pi^0 \ell^+ \ell^-)$ have been made in parallel with $\Gamma(\eta \to \pi^0 \gamma\gamma)$, using a model to describe the $\eta \to \pi^0 \gamma\gamma$ transition common to both.
Early calculations using VMD found $\BR(\eta\to\pi^0e^+e^-)/\BR(\eta\to\pi^0\gamma\gamma) \approx 10^{-5}$~\cite{Cheng:1967zza}, while an effective operator approach found a much smaller value $2 \times 10^{-8}$~\cite{LlewellynSmith1967}. 
For the latter result, the $\eta\pi^0\gamma\gamma$ vertex was assumed to be $S$-wave, which requires a lepton helicity flip for $\eta \to \pi^0 e^+ e^-$ (thereby suppressed by $m_e$) and likely represents only a subdominant contribution to this rate~\cite{Cheng:1967zza}.
The most recent results were obtained by Ng and Peters~\cite{Ng:1992yg,Ng:1993sc}, quoted as a unitarity bound.
This argument is based on the fact that $\eta \to \pi^0 \gamma\gamma$ is strictly related to the imaginary part of $\eta \to \pi^0 \ell^+ \ell^-$, where the photons are on-shell, while the real part of $\eta \to \pi^0 \ell^+ \ell^-$ is not calculable without an improved extension of $\eta\to\pi^0\gamma\gamma$ to virtual photons~\cite{Ng:1992yg}. 
Including only the imaginary part yields a lower limit on the rate.
Based on a VMD model including $a_0$ scalar exchange, Ref.~\cite{Ng:1992yg} found
\begin{align} \label{eq:BRetapi0ee}
\BR(\eta\to\pi^0e^+e^-)/\BR(\eta\to\pi^0\gamma\gamma) &\gtrsim 1.2\times 10^{-5} \,, \notag\\
\BR(\eta\to\pi^0\mu^+\mu^-)/\BR(\eta\to\pi^0\gamma\gamma) &\gtrsim 0.8\times 10^{-5} \,.
\end{align}
However, these bounds are likely nearly saturated since the real parts were reasonably estimated to be no more than 30\% of the imaginary parts.\footnote{We recall that the two-photon contributions to $\pi^0, \eta, \eta' \to\ell^+\ell^-$ are also dominated by their imaginary parts; see Sect.~\ref{sec:P-ll}.}
Consistent results were also found for a box-diagram calculation for the $\eta\pi^0\gamma\gamma$ vertex involving constituent quarks of mass around $300\MeV$~\cite{Ng:1993sc}.\footnote{While the box-diagram approach of Ref.~\cite{Ng:1993sc} may not seem particularly trustworthy, one must recall that experimental measurements of $\Gamma(\eta\to\pi^0\gamma\gamma)$ were a factor of 2 larger than VMD predictions at the time, calling into question the validity of VMD for $\eta \to \pi^0 \ell^+ \ell^-$.}

These decades-old theoretical predictions have been revisited most recently, based on an up-to-date analysis of the vector-meson couplings involved~\cite{Escribano:2020rfs}. 
The authors arrive at the predictions
\begin{align} \label{eq:BRetapi0ee-Escribano}
\BR(\eta\to\pi^0e^+e^-) &= 2.1(5)\times 10^{-9} \,, \notag\\
\BR(\eta\to\pi^0\mu^+\mu^-) &= 1.2(3)\times 10^{-9} \,.
\end{align}
These results are consistent with saturating the inequalities in Eq.~\eqref{eq:BRetapi0ee} provided one calculates the $\BR(\eta \to \pi^0 \gamma\gamma$) in a similar framework~\cite{Escribano:2018cwg}, which however does not agree well with the measured value (see next section).
Ref.~\cite{Escribano:2020rfs} also computes the related processes
\begin{align}
    \BR(\eta'\to\pi^0e^+e^- ) &= 4.5(8)\times10^{-9}\,, &
    \BR(\eta'\to\pi^0\mu^+\mu^- ) &= 1.8(4)\times10^{-9}\,, \nonumber\\
    \BR(\eta'\to\eta e^+e^- ) &= 3.9(9)\times10^{-10}\,, &
    \BR(\eta'\to\eta\mu^+\mu^- ) &= 1.6(2)\times10^{-10}\,. \label{eq:BRetaprime-Pll}
\end{align}

On the experimental side, the recent upper limit obtained by the WASA-at-COSY collaboration, 
$\BR(\eta\to\pi^0e^+e^-) < 7.5\times 10^{-6}$~\cite{Adlarson:2018imw},
is still several orders of magnitude away.
Nevertheless, it may be worthwhile to further refine this analysis in light of experimental measurements for the differential decay rate for
$\eta \to \pi^0 \gamma \gamma$, which could be used to reduce the model-dependence of the $\eta \to \pi^0 \ell^+ \ell^-$ calculation.
We quote the remaining experimental upper limits below in Sect.~\ref{sec:BSMCV}, Table~\ref{tab:CVdecays}, all of which are still several orders of magnitude above the predictions in Eq.~\eqref{eq:BRetaprime-Pll}.

\subsection{$\eta'\to\pi^0\gamma\gamma$, $\eta'\to\eta\gamma\gamma$}\label{sec:etap-Pgg}

There exist two decays of the $\eta'$ meson analogous to $\eta\to\pi^0\gamma\gamma$ discussed above:
$\eta'\to\pi^0\gamma\gamma$ and $\eta'\to\eta\gamma\gamma$.  
Theoretically, beyond model calculations based on VMD~\cite{Balytskyi:2018pzb,Balytskyi:2018uxb}, 
one comprehensive study has been reported~\cite{Escribano:2018cwg} (see also Refs.~\cite{Jora:2010zz,Escribano:2012dk}
for preliminary results of the same investigation).
Similar to what was found for $\eta\to\pi^0\gamma\gamma$, vector-meson exchange (by $\rho^0$, $\omega$, and $\phi$) was 
identified as the dominant mechanism; the main difference to the $\eta$ decay is that in particular the $\omega$ can go
(almost) on-shell (to the extent its width can be neglected), i.e., $\eta'\to\pi^0\gamma\gamma$ 
for instance can proceed as two real sequential two-body decays,
according to $\eta'\to\omega\gamma$ (with branching ratio $2.489(76)\%$~\cite{Ablikim:2019wop}) with subsequent
$\omega\to\pi^0\gamma$ (branching ratio $8.40(22)\%$).
The theoretical calculation of Refs.~\cite{Escribano:2018cwg} correspond to a
branching fraction that is inclusive with respect to $\omega$ and $\rho$ exchange.

In Ref.~\cite{Escribano:2018cwg}, 
the VMD amplitudes are implemented based on $U(3)$ symmetry and allow for (single-angle) $\eta$--$\eta'$ as well as 
$\omega$--$\phi$ mixing.  Different input values for the relevant coupling constants and mixing angles are tested.  
Very similar to the finding for $\eta\to\pi^0\gamma\gamma$, 
chiral one-loop contributions (formally leading in the chiral expansion, here amended by large-$N_c$ counting 
to include the $\eta'$) are found to be tiny in both decays.
A resummation of $s$-channel $S$-wave rescattering effects (where the Mandelstam variable $s$ refers to the
diphoton invariant mass squared) is included via
scalar-resonance contributions: by the $a_0(980)$ in $\eta'\to\pi^0\gamma\gamma$ 
as well as $f_0(500)$ and $f_0(980)$ in $\eta'\to\eta\gamma\gamma$. These are estimated using a linear $\sigma$-model,
which is implemented in a way consistent with chiral symmetry as manifest in the leading-order chiral amplitudes
for meson--meson (re)scattering.  In particular, this procedure does not require the introduction of 
independent couplings of the scalar resonances to two photons.

Identifying the $\omega$ exchange as the by far dominant mechanism
in $\eta'\to\pi^0\gamma\gamma$, 
the authors arrive at predictions slightly enhanced compared to what 
$\BR(\eta'\to\omega\gamma) \times \BR(\omega \to \pi^0\gamma) = 2.09(8)\times 10^{-3}$ would suggest: 
depending on whether the VMD couplings are fixed based on an elaborate mixing model or purely phenomenologically, they find
$\BR(\eta'\to\pi^0\gamma\gamma) = 3.57(25)\times 10^{-3}$ or $\BR(\eta'\to\pi^0\gamma\gamma) = 2.91(25)\times 10^{-3}$, respectively.  The diphoton spectrum measured by the BESIII collaboration~\cite{Ablikim:2016tuo} (see below) can be reproduced
in a satisfactory manner: the distribution as given by vector-meson exchange
is consistent with data, and scalar contributions due to the $a_0(980)$ are found to be entirely negligible in this channel.

The scalar isoscalar resonances are found to yield more noticeable corrections to the VMD picture for 
$\eta'\to\eta\gamma\gamma$.  In particular, the diphoton invariant mass spectrum is modified significantly above 
the two-pion threshold.
The theoretical predictions for the branching fractions of the latter process are found between
$\BR(\eta'\to\eta\gamma\gamma)_{\rm incl.} = 1.07(7)\times 10^{-4}$
and $\BR(\eta'\to\eta\gamma\gamma)_{\rm incl.} = 1.17(8)\times 10^{-4}$, 
depending on the input parameters~\cite{Escribano:2018cwg}.

On the experimental side, two earlier investigations determined upper limits on $\eta'\to\pi^0\gamma\gamma$,
\textit{excluding} the decay chain via $\omega\gamma$:
GAMS~2000 found $\BR(\eta'\to\pi^0\gamma\gamma) < 8 \times 10^{-4}$~\cite{Alde:1987jt}, 
slightly improved upon by GAMS-$4\pi$, $\BR(\eta'\to\pi^0\gamma\gamma) < 6 \times 10^{-4}$~\cite{Donskov:2015epm}.
These limits hence cannot be directly compared to the theoretical prediction.
The first determination of the branching ratio was achieved by BESIII, 
$\BR(\eta'\to\pi^0\gamma\gamma)_{\rm incl.} = 3.20(7)(23)\times 10^{-3}$~\cite{Ablikim:2016tuo},
which is consistent with Ref.~\cite{Escribano:2018cwg} and similarly slightly enhanced compared
to the dominant $\omega$-exchange mechanism whose strength was determined independently,
$\BR(\eta'\to\omega\gamma) \times \BR(\omega \to \pi^0\gamma) = 2.37(14)(18) \times 10^{-3}$,
compatible with the PDG averages~\cite{Tanabashi:2018oca}.
Ref.~\cite{Ablikim:2016tuo} also extracted
``nonresonant'' contributions to this decay, proceeding neither via intermediate $\omega$ nor $\rho$ exchanges,
which yield about one fifth of the total only,
$\BR(\eta'\to\pi^0\gamma\gamma)_{\rm NR} = 6.16(64)(67)\times 10^{-4}$.

For $\eta'\to\eta\gamma\gamma$, an upper limit excluding the $\omega$-exchange,
$\BR(\eta'\to\eta\gamma\gamma) < 8 \times 10^{-4}$~\cite{Donskov:2015epm}, has most recently 
been superseded by the first determination of the inclusive branching ratio by the 
BESIII collaboration~\cite{Ablikim:2019wsb},
$\BR(\eta'\to\eta\gamma\gamma) = 8.25(3.41)_{\text{stat}}(0.72)_{\text{syst}}\times10^{-5}$, which 
was alternatively translated into an upper limit at 90\% C.L., 
$\BR(\eta'\to\eta\gamma\gamma) < 1.33\times 10^{-4}$.  
This is consistent with the theoretical calculation~\cite{Escribano:2018cwg}.
We note that the extension of the VMD model in Eq.~\eqref{eq:gPVgamma-VMD} to the $\rho^0$ yields
\beq
\frac{g_{\eta'\rho\gamma}}{g_{\eta'\omega\gamma}} = \frac{g_{\eta\rho\gamma}}{g_{\eta\omega\gamma}} \approx 3 \,,
\eeq
hence among the two vector exchanges that can go ``on-shell'' in $\eta'\to\eta\gamma\gamma$, we
expect the $\rho^0$ to dominate: an enhancement in the couplings of almost two orders of magnitude
overwhelms the ratio of the widths, $\Gamma_\rho/\Gamma_\omega \approx 17.5$.  If we then naively identify
$\BR(\eta'\to\rho^0\gamma) \equiv \BR(\eta'\to\pi^+\pi^-\gamma)$, a crude estimate 
(analogous to $\omega$ dominance in $\eta'\to\pi^0\gamma\gamma)$ leads to
$\BR(\eta'\to\eta\gamma\gamma) \approx \BR(\eta'\to\rho^0\gamma) \times \BR(\rho^0 \to \eta\gamma)
= 8.7(6)\times 10^{-5}$~\cite{Tanabashi:2018oca}, 
perfectly compatible with the experimental result~\cite{Ablikim:2019wsb}.

Clearly, refined experimental measurements will be necessary to further elucidate the role
of scalar resonances in these decays, interfering with the dominant vector exchanges.

\section{\boldmath Weak decays of $\eta$ and $\eta'$}\label{sec:weak}

Despite the relative smallness of the total decay width of the $\eta$ (and the $\eta'$) and the suppression
of most of its decay channels by various (approximate) symmetries, weak-interaction decays are still
extremely rare.  None have been observed experimentally so far, and theoretical calculations  
predict branching ratios consistently below $10^{-10}$, usually many orders of magnitude below current
experimental upper limits.  This on the other hand may offer opportunities to find traces of BSM physics,
should any of the decays discussed in this section actually be observed.

For kinematical reasons, nonleptonic $\eta$ decays have to be flavor-conserving, as they
cannot involve kaons in the final state.  The simplest such decays are hence those into two pions,
which necessarily violate $CP$ and will be discussed separately below in Sect.~\ref{sec:eta-2pi}.
There exists, however, a $CP$-conserving nonleptonic $\eta'$ decay, $\eta'\to K^\pm\pi^\mp$.
In Ref.~\cite{Bergstrom:1987py}, the branching ratio was estimated by a simple rescaling of the
$K_S\to2\pi$ partial width with the ratio of decay masses, leading to 
$\BR(\eta'\to K^\pm\pi^\mp) \approx 2\times10^{-10}$.  A more refined analysis based on a chiral Lagrangian
in analogy to a tree-level analysis of $K\to\pi\pi$, on the other hand, arrived at 
$\BR(\eta'\to K^\pm\pi^\mp) \approx 2\times10^{-11}$~\cite{Shabalin:2002aa}, which still makes this by 
far the largest branching fraction of any weak decay discussed in this section.
The best experimental upper limit was obtained by the BESIII
collaboration and is given by $\BR(\eta'\to K^\pm\pi^\mp) < 4\times 10^{-5}$~\cite{Ablikim:2016bjc}.

The purely leptonic weak decays of $\eta$ and $\eta'$ via the exchange of a $Z$-boson lead to contributions
to the decays $\eta^{(\prime)}\to\ell^+\ell^-$, whose dominant decay mechanism via two photons
has been discussed in Sect.~\ref{sec:P-ll} (decays into neutrino pairs are helicity-forbidden for
massless neutrinos).  
As we have emphasized there, the strong suppression of the Standard Model mechanism
for the dilepton decays invites the search for potential effects beyond it.  At tree level, these decays
could be modified by the exchange of an axial vector $\A$ (of mass $m_\A$) 
or a pseudoscalar $\P$ (of mass $m_\P$), whose couplings to the 
Standard Model fermions are given in terms of an effective Lagrangian~\cite{Masjuan:2015lca,Masjuan:2015cjl}
\beq
\L = \frac{g}{4m_W} \sum_f \Big\{ m_\A c_f^\A \big(\bar f \gamma_\mu\gamma_5 f\big) \A^\mu
+ 2 m_f c_f^\P \big(\bar f i\gamma_5 f\big) \P \Big\} \,, \label{eq:effL-Pll-BSM}
\eeq
with the standard electroweak parameters $g$ and $m_W$ and dimensionless couplings $c_f^{\A,\P}$
to the lightest quarks and leptons, $f = \{u,d,s,e,\mu\}$.  
This formalism comprises in particular the effects of the Standard Model $Z$-boson, 
which are obtained from the axial-vector coupling in Eq.~\eqref{eq:effL-Pll-BSM} with $c_u^\A = - c_{d,s,e,\mu}^\A = 1$. 
In Ref.~\cite{Masjuan:2015cjl}, 
a very compact representation of the hadronic matrix elements of Eq.~\eqref{eq:effL-Pll-BSM} 
in terms of meson decay constants only is derived, which leads to a modification of the reduced amplitude
$\A(q^2)$ in Eq.~\eqref{eq:BR:P-ll} according to
\begin{align}
\A(q^2) &\to \A(q^2) + \frac{\sqrt{2}G_F}{4\alpha_\text{em}^2 F_{P\gamma\gamma}} \big(\lambda_P^\A+\lambda_P^\P\big) \,, \notag\\
\lambda_P^\A &= c_\ell^\A \bigg\{ F_P^3 \Big(c_u^\A-c_d^\A\Big)
+ F_P^q \Big(c_u^\A+c_d^\A\Big) + \sqrt{2} F_P^s c_s^\A \bigg\}  \,, \notag\\
\lambda_P^\P &= \frac{c_\ell^\P}{1-m_\P^2/M_P^2} \bigg\{ F_P^3 \Big(c_u^\P-c_d^\P\Big)
- F_P^q c_s^\P + \sqrt{2} F_P^s c_s^\P \bigg\} \,, \label{eq:Aq2-BSM}
\end{align}
where $F_P^3$ (that we have not employed so far) refers to the decay constant of the isovector current
in straightforward generalization of Eq.~\eqref{eq:def-decay-const}, and the isoscalar decay constants
are most easily expressed in terms of the flavor mixing scheme.  $G_F$ refers to the Fermi constant. 
The modifications of the resulting branching ratios
for all $P\to\ell^+\ell^-$ channels are summarized as~\cite{Masjuan:2015cjl}
\begin{align}
&\BR\big(\pi^0\to e^+e^-\big) \Bigg\{ 1+0.001 \Bigg[ c_\ell^\A\Big(c_u^\A-c_d^\A\Big)
+ c_\ell^\P \frac{c_u^\P-c_d^\P}{1-m_\P^2/\mpiz^2} \Bigg] \Bigg\} \,, \notag\\
&\BR\Big(\eta\to \genfrac{}{}{0pt}{1}{e^+e^-}{\mu^+\mu^-}\Big) 
\Bigg\{ 1+ \genfrac(){0pt}{1}{+0.001}{-0.002} 
\Bigg[ 0.84 c_\ell^\A\Big(c_u^\A+c_d^\A\Big) - 1.27 c_\ell^\A c_s^\A
- \frac{2.11 c_\ell^\P c_s^\P}{1-m_\P^2/\meta^2} \Bigg] \Bigg\} \,, \notag\\
&\BR\Big(\eta'\to \genfrac{}{}{0pt}{1}{e^+e^-}{\mu^+\mu^-}\Big) 
\Bigg\{ 1+ \genfrac(){0pt}{1}{+0.001}{+0.003} 
\Bigg[ 0.72 c_\ell^\A\Big(c_u^\A+c_d^\A\Big) + 1.61 c_\ell^\A c_s^\A
+ \frac{0.89 c_\ell^\P c_s^\P}{1-m_\P^2/\metap^2} \Bigg] \Bigg\} \,. 
\end{align}
In particular, the weak interactions modify the rates of $P\to\ell^+\ell^-$ at the permille level
(via $Z$-exchange).

In the remainder of this section, 
we describe the most relevant semileptonic $\eta$ and $\eta'$ decays.
We note that these are throughout not the most favorable processes to investigate the hadronic
matrix elements in question: as long as lepton flavor universality is assumed, they can be studied
over wider kinematic ranges, and with larger branching fractions, in $\tau$ lepton decays,
related to the semileptonic $\eta^{(\prime)}$ decays (with electrons or muons in the final state)
by crossing symmetry.

Semileptonic $\eta^{(\prime)}$ decays into nonstrange final states are induced by the Cabibbo-favored
Fermi interaction
\beq
\L_F = -\frac{G_F V_{ud}}{\sqrt{2}} \Big\{ \big[ \bar u\gamma_\mu (1-\gamma_5) d\big]\times
\big[\bar \ell \gamma^\mu (1-\gamma_5) \nu_\ell\big] + \text{h.c.} \Big\} \,. \label{eq:Fermi}
\eeq
For reasons of parity, only the vector current can contribute to the simplest semileptonic decay
$\eta\to \pi^\pm \ell^\mp \nu_\ell$.  The corresponding matrix element can be decomposed in terms 
of two form factors,
\beq
\langle \pi^+(p_\pi) | \bar u \gamma_\mu d | \eta(p_\eta) \rangle
= \sqrt{2} \Bigg\{ f_+(t) \bigg[ (p_\eta+p_\pi)_\mu - \frac{\meta^2-\mpc^2}{t} (p_\eta-p_\pi)_\mu \bigg]
+ f_0(t) \frac{\meta^2-\mpc^2}{t} (p_\eta-p_\pi)_\mu \Bigg\} \,, \label{eq:eta-l3}
\eeq
where $t=(p_\eta-p_\pi)^2$, and $f_+(t)$ and $f_0(t)$ refer to the vector and scalar form factors, 
respectively.  The differential decay rate for the $\eta$ decay is then given as 
(see, e.g., Ref.~\cite{Descotes-Genon:2014tla})
\beq
\frac{\diff \Gamma(\eta\to\pi^\pm \ell^\mp \nu_\ell)}{\diff t} = 
\frac{G_F^2 V_{ud}^2}{192\pi^3}\frac{\lambda^{1/2}\big(t,\meta^2,\mpc^2\big)}{\meta^3}\bigg(1-\frac{m_\ell^2}{t}\bigg)^2
\Bigg\{ \big|f_+(t)\big|^2 \lambda\big(t,\meta^2,\mpc^2\big)\bigg(2+\frac{m_\ell^2}{t}\bigg) 
+ 3 \big|f_0(t)\big|^2 \big(\meta^2-\mpc^2\big)^2\frac{m_\ell^2}{t}\Bigg\} \,.
\eeq
In the case of the electron/positron final state, the decay is almost entirely determined by the 
vector form factor.  The corresponding decay distribution for the inverse $\tau$ decay can also be
found in the literature~\cite{Descotes-Genon:2014tla}.

Equation~\eqref{eq:eta-l3} is a manifestation of a ``second-class current''~\cite{Weinberg:1958ut,Singer:1965vw}:
it vanishes in the limit of $G$-parity conservation.  The normalization of both vector and scalar
form factors is proportional to the light quark mass difference 
$m_u-m_d$~\cite{Tisserant:1982fc,Pich:1987qq,Bednyakov:1992af}; at leading order in the chiral expansion,
\beq
f_+(0) = f_0(0) = \eps + \Order(m_q) \,,
\eeq
where $\eps$ is the $\pi^0$--$\eta$ mixing angle.  
An observation of either 
$\eta\to \pi^\pm \ell^\mp \nu_\ell$ or $\tau^\pm \to \eta \pi^\pm \nu_\ell$ is therefore a clean 
manifestation of a second-class amplitude and a measure of an isospin-breaking quark mass ratio.
The vector and scalar form factors have 
been calculated to one loop in chiral perturbation theory~\cite{Neufeld:1994eg,Scora:1995sj},
as well as in the context of various resonance 
models~\cite{Nussinov:2008gx,Paver:2010mz,Volkov:2012be,Escribano:2016ntp}.
The chiral one-loop calculation of Ref.~\cite{Neufeld:1994eg} arrives at
\beq
\BR(\eta\to\pi^\pm e^\mp \nu_e) = (0.94 \ldots 1.16)\times 10^{-13} \,, \qquad
\BR(\eta\to\pi^\pm \mu^\mp \nu_\mu) = (0.68 \ldots 0.82)\times 10^{-13} \,, \label{eq:BR-etal3-ChPT}
\eeq
where both numbers have been multiplied by 2 to account for both charge combinations in the final state.

Of particular interest in the spirit of this review is the dispersion-theoretical analysis of 
the form factors in Ref.~\cite{Descotes-Genon:2014tla}.  
The discontinuity of the vector form factor is dominated by $\pi\pi$ intermediate states:
\beq
\disc f_+(t) = -2i \, \theta(t-4\mpc^2) \frac{t-4\mpc^2}{16\pi\lambda^{1/2}(t,\meta^2,\mpc^2)}
F_\pi^V(t) \big[\M^{J=1}_{\pi^+\pi^0\to\pi^+\eta}(t)\big]^* \,,
\eeq
where $F_\pi^V$ is the pion vector form factor, and $\M^{J=1}_{\pi^+\pi^0\to\pi^+\eta}$ denotes the $P$-wave
projection of the (crossed) $\eta\to\pi^+\pi^-\pi^0$ decay amplitude in Eq.~\eqref{eq:Mdecomp}.  
The latter is constructed via the Khuri--Treiman formalism, whose analytic continuation beyond the 
decay region is reliable at least throughout the kinematic range of the $\rho(770)$ resonance.
The unitarity relation for the $\pi\eta$ scalar form factor is dominated by the elastic part, 
and modeled in terms of a phase-dispersive Omn\`es representation
for the $S$-wave $\pi\eta$ intermediate state only (a theoretical construction subsequently
improved upon by the $\pi\eta \leftrightarrow K\bar{K}$ isospin $I=1$ coupled-channel scalar
form factor description in Ref.~\cite{Albaladejo:2015aca}; cf.\ also Ref.~\cite{Escribano:2016ntp}), 
which covers the effects of the $a_0(980)$ resonance.  
Both form factors are matched to the chiral one-loop representation near $t=0$.  

The results for the branching ratios amount to~\cite{Descotes-Genon:2014tla}
\beq
\BR(\eta\to\pi^\pm e^\mp \nu_e) \approx 1.40 \times 10^{-13} \,, \qquad
\BR(\eta\to\pi^\pm \mu^\mp \nu_\mu) \approx 1.02 \times 10^{-13} \,,  \label{eq:BR-etal3-disp}
\eeq
which is only very moderately enhanced over the $\chi$PT result of Eq.~\eqref{eq:BR-etal3-ChPT}.
The available experimental upper limit was obtained by the BESIII collaboration, 
$\BR(\eta\to \pi^\pm \ell^\mp \nu_\ell) < 1.7\times 10^{-4}$~\cite{Ablikim:2012vn}, hence
nine orders of magnitude above the theoretical estimates. 

Given the smallness of the prediction~\eqref{eq:BR-etal3-disp}, 
first-class scalar-type interactions (hence not suppressed by $G$-parity) or other, BSM, second-class vector-type currents
might enhance $\BR(\eta\to \pi^\pm \ell^\mp \nu_\ell)$ significantly.  
In Ref.~\cite{Herczeg:1991jp}, experimental information on nucleon and pion $\beta$-decays was used to limit such interactions, 
suggesting the semileptonic $\eta$ decay might still be enhanced up to the level of $10^{-9}\ldots 10^{-8}$ thereby; 
these limits seem not to have been updated since.
A general analysis of BSM effects in the framework of the Standard Model Effective Field Theory, investigating in particular
new scalar or tensor interactions, concentrates on the $\tau$ decays~\cite{Garces:2017jpz} in view of upcoming opportunities
at Belle-II~\cite{Kou:2018nap}.

The dispersive predictions for the corresponding $\tau^-\to\eta\pi^-\nu_\tau$ decay are much more uncertain.
This is, in the case of the vector form factor, 
largely due to uncertainties in the $\eta\to3\pi$ amplitude, whose
subtraction constants were obtained purely from matching to $\chi$PT~\cite{Descotes-Genon:2014tla}
(and not from a sophisticated fit to the high-precision KLOE(2016) data~\cite{Anastasi:2016cdz} 
not available yet at the time).  
For the scalar part, lack of knowledge of the precise $\pi\eta$ $S$-wave scattering phase 
limits the precision of the prediction.
In Ref.~\cite{Escribano:2016ntp}, coupled-channel descriptions of the $\pi\eta$ scalar form factor
have been constructed, taking the three channels $\pi\eta$, $\pi\eta'$, and $K\bar{K}$ into account
explicitly, and building on a $T$-matrix that has been constructed based on $U(3)$ one-loop $\chi$PT
plus explicit resonance exchanges; in this case, the range of potential scalar contributions
to the branching ratio is even wider.
Overall, the two parts of the branching fraction are predicted to be in the ranges~\cite{Descotes-Genon:2014tla,Escribano:2016ntp}
\begin{align}
\BR(\tau^-\to\eta\pi^-\nu_\tau)_{V} &= (1\ldots 4)\times 10^{-6} \,, &
\BR(\tau^-\to\eta\pi^-\nu_\tau)_{S} &= (1\ldots 6)\times 10^{-6} \,, \notag\\
\BR(\tau^-\to\eta\pi^-\nu_\tau)_{V} &= 2.6(2)\times 10^{-6} \,, &
\BR(\tau^-\to\eta\pi^-\nu_\tau)_{S} &= (0.7\ldots 19.7)\times 10^{-6} \,.
\end{align}
The second, scalar, part of Ref.~\cite{Descotes-Genon:2014tla} is smaller than most model calculations based on simple
$a_0(980)$ dominance~\cite{Tisserant:1982fc,Pich:1987qq,Neufeld:1994eg,Nussinov:2008gx}. 
In contrast to the $\eta$ decay results of Eq.~\eqref{eq:BR-etal3-disp}, 
these predictions are starting to be tested seriously by the most recent experimental upper limit
from BaBar, $\BR(\tau^-\to \eta\pi^-\nu_\tau) < 9.9 \times 10^{-6}$~\cite{delAmoSanchez:2010pc},
which already excludes part of the scalar form factor parameter space of Ref.~\cite{Escribano:2016ntp}.

The formalism to describe the corresponding $\eta'$ decays $\eta'\to\pi^\pm \ell^\mp \nu_\ell$,
$\tau^-\to\eta'\pi^-\nu_\tau$ is obviously the same; due to the added difficulty in describing 
the form factors of the $\eta'$, there exist somewhat less theoretical studies in this 
case~\cite{Nussinov:2009sn,Paver:2011md,Volkov:2012be,Escribano:2016ntp}.  
The pseudoscalar decays are again dominated by the vector form factor, as the scalar one is 
suppressed by $m_\ell^2$; for the $\eta'$ decay, the dominant feature of the vector form factor,
the $\rho(770)$ resonance, lies within the physical decay region.  Nevertheless,
Ref.~\cite{Escribano:2016ntp} predicts a branching ratio yet four orders of magnitude
smaller than for the $\eta$,
\beq
\BR(\eta'\to\pi^\pm e^\mp \nu_e) \approx \BR(\eta' \to\pi^\pm \mu^\mp \nu_\mu) \approx 1.7\times10^{-17} \,;
\eeq
as the authors predict the $\eta$ branching ratios a factor of 2 smaller than 
Ref.~\cite{Descotes-Genon:2014tla}, an uncertainty of that order might be appropriate to assume.
The experimental upper limit is 
$\BR(\eta' \to \pi^\pm e^\mp \nu_e) < 2.1\times 10^{-4}$~\cite{Ablikim:2012vn}.
On the other hand, the corresponding $\tau$ decay is predicted to be entirely dominated by the \textit{scalar}
contribution, as here the $\rho(770)$ lies well below the $\pi\eta'$ threshold.  
The theoretical uncertainty for the scalar $\pi\eta'$ form factor does not even fix the order of magnitude
of $\BR(\tau^-\to\eta'\pi^-\nu_\tau)$ convincingly, with predictions in the range
$5\times 10^{-11} \ldots 3\times 10^{-6}$~\cite{Escribano:2016ntp}.  The experimental upper limit
obtained by BaBar is $\BR(\tau^-\to \eta'\pi^-\nu_\tau) < 4.0 \times 10^{-5}$~\cite{Lees:2012ks}.

Semileptonic decays with charged kaons, replacing the pions in the preceding discussion,
are driven by the Cabibbo-suppressed analog of the Fermi interaction Eq.~\eqref{eq:Fermi}, 
with the $d$-quark replaced by an $s$-quark, and $V_{ud} \to V_{us}$.  The corresponding form factors, 
however, are not isospin suppressed, which results in a significantly larger $\tau$ decay rate
into the $\eta K^-$ final state:
$\BR(\tau^-\to \eta K^-\nu_\tau) = 1.55(8)\times 10^{-4}$~\cite{Tanabashi:2018oca},
where the average is dominated by Belle~\cite{Inami:2008ar} and BaBar~\cite{delAmoSanchez:2010pc}
measurements. 
For the corresponding decay with the $\eta'$, there is so far only an upper limit, $\BR(\tau^-\to \eta' K^-\nu_\tau ) <  2.4\times 10^{-6}$~\cite{Lees:2012ks}. 
However, the true value may not be far below; the prediction based on a coupled-channel
treatment for the $\pi K$, $\eta K$, and $\eta' K$ scalar form factors~\cite{Jamin:2000wn}, which describe well $\tau$-decay data for the first two channels~\cite{Escribano:2014joa},
results in $\BR(\tau^-\to \eta' K^-\nu_\tau ) = 1.0\big({}^{+0.4}_{-0.3}\big)\times 10^{-6}$~\cite{Escribano:2013bca}
(compare also Ref.~\cite{Volkov:2016gsi}).
However, none of these advanced theoretical studies seem to have converted their form factors into predictions
for the corresponding semileptonic $\eta^{(\prime)}$ decays.  The decay
$\eta\to K^\pm e^\mp \nu_e$ is suppressed by very small phase space, such that an estimate based on 
the form factor normalization only is probably warranted; this seems far less clear for 
$\eta'\to K^\pm e^\mp \nu_e$, where the dilepton invariant mass however is still constrained to well below any
resonances.  The resulting predictions are
$\BR(\eta\to K^\pm e^\mp \nu_e) \approx 3.5\times 10^{-15}$ and
$\BR(\eta' \to K^\pm e^\mp \nu_e) \approx 1.2\times 10^{-16}$~\cite{Shabalin:2002aa}.

The formalism to describe the next more complicated semileptonic $\eta^{(\prime)}$ decays,
$\eta^{(\prime)} \to \pi^\pm \pi^0 \ell^\mp \nu_\ell$, is analogous to the well-known $K_{\ell4}$
decays (see, e.g., Ref.~\cite{Colangelo:2015kha} for a recent dispersion-theoretical study)
and will not be repeated here in full.
The form factors of the axial-vector current once more break $G$-parity and are hence 
suppressed by $(m_u-m_d)$; only the (single) form factor of the vector current does not require isospin breaking.
It is determined, at leading order, by the chiral anomaly, and in fact related to the matrix element for $\eta\to\pi^+\pi^-\gamma^*$
by an isospin rotation.  This connection was used early to predict $\BR(\tau^-\to \eta\pi^-\pi^0\nu_\tau)$
based on data for $e^+e^-\to\eta\pi^+\pi^-$~\cite{Gilman:1987my,Eidelman:1990pb} (see also Ref.~\cite{Dumm:2012vb} 
for a $\chi$PT-plus-resonances calculation of both processes); 
the current experimental average 
$\BR(\tau^-\to \eta\pi^-\pi^0\nu_\tau) = 1.39(7)\times 10^{-3}$ is dominated by the Belle result~\cite{Inami:2008ar}.
Estimates of the corresponding $\eta^{(\prime)}$ decay probabilities seem not to have been performed based on this analogy, 
but rather on rougher relations to the $K_{\ell 4}$ rates, resulting in
$\BR(\eta \to \pi^\pm \pi^0 e^\mp \nu_e) \leq 1.5\times 10^{-16}$,
$\BR(\eta' \to \pi^\pm \pi^0 \ell^\mp \nu_\ell) \leq 4\times 10^{-15}$~\cite{Bramon:1994bd}.

\section{Tests of discrete spacetime symmetries and lepton flavor violation}
\label{sec:BSMfundsym}

Long before the theory of QCD was developed, Purcell and Ramsey first raised the possibility of $P$ violation (and also implicitly $T$ violation) in the strong interaction, manifesting as an electric dipole moment (EDM) for the neutron~\cite{Purcell:1950zz}.
Though their measurement did not find evidence for this effect~\cite{Smith:1957ht}, their work opened a new perspective to regard $C$, $P$, and $T$ invariance as experimental questions, rather than assumptions.
Kobzarev and Okun recognized the absence of $\eta \to 2 \pi$ decays as further evidence for $P$ and $CP$ invariance in strong interactions~\cite{Kobzarev}.

It was the weak force, not the strong, for which symmetry tests proved truly revelatory, beginning with the discovery of $P$ violation in $\beta$-decay by Wu et al.~\cite{Wu:1957my}.
Later, $CP$ violation was discovered in $K_L \to 2\pi$ decays~\cite{Christenson:1964fg} and the nature of this effect stirred many theoretical ideas.
One hypothesis proposed that $CP$ violation came from a second-order effect involving both the weak force and new $C$-violating interactions that preserved both $P$ and strangeness~\cite{Prentki:1965tt,Lee:1965zza,Lee:1965hi,Bernstein:1965hj}.
To test these latter interactions, the $\eta$ meson came to the fore as an eigenstate of $C$, whose decays could test $C$ violation in the absence of the weak force.
There is a long list of $C$-violating decays to study, including those that are strictly forbidden, such as $\eta \to 3\gamma$, or forbidden at via single photon exchange, such as $\eta \to \pi^0 e^+ e^-$~\cite{Bernstein:1965hj}.
Additionally, $C$ violation may be measured as a charge asymmetry in $\eta \to \pi^+ \pi^- \pi^0$~\cite{Lee:1965zza}.
Kobayashi and Maskawa's work showing how $CP$ violation is realized within the weak interaction itself, a consequence of a third generation of quarks, came almost a decade later as the Standard Model rose to prominence~\cite{Kobayashi:1973fv}.

In the Standard Model, strong and electromagnetic interactions preserve each of the discrete spacetime symmetries $C$, $P$, and $T$ separately.
However, weak interactions violate all three symmetries and only the combined operation $CPT$ remains an invariant of the Lagrangian, as expected on general grounds.
$C$ and $P$ violation arise since left- and right-handed chiral fermions have different $\SU(2)_L \times U(1)_Y$ gauge quantum numbers, while $T$ violation (or equivalently $CP$ violation) enters through the Kobayashi--Maskawa (KM) phase~\cite{Kobayashi:1973fv} in the quark mixing matrix.\footnote{For the present discussion, we take neutrinos to be massless in the Standard Model.
While the neutrino mixing matrix may contain one or more additional $CP$-violating phases, any contributions to $\eta$ or $\eta^\prime$ physics are utterly irrelevant since they are suppressed by the tiny neutrino masses.}
Since the KM phase is inextricably tied to mixing between all three generations, the strongest tests of $CP$ violation in the Standard Model come from the flavor sector ($K$, $D$, and $B$ mesons)~\cite{Isidori:2010kg}.

There is a strong motivation, however, that the Standard Model is not the end of the story as far as symmetries are concerned.
Explaining the cosmological origin of matter via baryogenesis requires both $C$ and $CP$ violation~\cite{Sakharov:1967dj}. 
The KM phase is insufficient for this purpose, suggesting \textit{CP} violation from BSM physics.
Searches for rare $\eta$ (and $\eta^\prime$) decays remain an active target for experimental tests of $C$, $P$, and $CP$ invariance in flavor-conserving systems.
Any $\eta,\eta^\prime$ channels violating these symmetries are expected to be highly suppressed within the Standard Model, since the weak interaction is flavor-violating, and any detection would be evidence for BSM physics.

\begin{table}
\centering
\renewcommand{\arraystretch}{1.3}
\begin{tabular}{llll} 
\toprule
Class     &  Violated & Conserved    & Interaction \\
\midrule
0 & & $C$, $P$, $T$, $CP$, $CT$, $PT$, $CPT$ & strong, electromagnetic\\
I & $C$, $P$, $CT$, $PT$ & $T$, $CP$, $CPT$ & (weak, with no KM phase or flavor-mixing) \\
II & $P$, $T$, $CP$, $CT$ & $C$, $PT$, $CPT$ & \\
III & $C$, $T$, $PT$, $CP$ & $P$, $CT$, $CPT$ & \\
IV & $C$, $P$, $T$, $CP$, $CT$, $PT$ & $CPT$ & weak \\
\bottomrule
\end{tabular}
\renewcommand{\arraystretch}{1.0}
\caption{Possible classes~I--IV of interactions that violate discrete spacetime symmetries, assuming $CPT$ invariance.
Electromagnetic and strong forces preserve $C$, $P$, and $T$ individually (class~0).
The weak force is ``maximally class~I'' since the $W$ boson couples only to left-handed particles; it is class~IV solely via the KM phase, i.e., only in processes involving flavor.
 \label{tab:symbreaking}}
\end{table}

We organize the remainder of our discussion based on symmetry.
Here it is customary to list the possible classes of discrete symmetry breaking that may occur assuming $CPT$ invariance,\footnote{$CPT$ violation in the $\eta,\eta^\prime$ sector has seen relatively little attention, though several channels have been proposed to test such effects.
These include searches for $CPT$-odd angular correlations in $\eta \to \pi^0 \ell^+ \ell^-$~\cite{Nefkens:2002sa} and a $CPT$-odd muon polarization asymmetry in $\eta \to \pi^\pm \mu^\mp \nu_\mu$~\cite{Gonzalez:2017fku, Gatto:2016rae}.
Additionally, $CPT$ violation from a departure from hermiticity (equivalent to a nonunitary $S$-matrix) can produce circularly polarized photons in $\eta \to \gamma\gamma$~\cite{Okun:2002pa} and $\eta \to \pi^+ \pi^- \gamma$~\cite{Geng:2005ff}.} shown in Table~\ref{tab:symbreaking}.
Section~\ref{sec:eta-2pi} focuses on $P,CP$-violating decays in class~II, the most famous of which is $\eta,\eta^\prime \to 2\pi$.
BSM scenarios along these lines have benefited from the most theoretical attention, but also receive strong constraints from EDM limits.
Section~\ref{sec:BSMCV} includes channels with $C$ violation.
These decays are mostly in class~III, but some channels such as $\eta \to 3\gamma$ may also be class~I.
Very little theoretical work has been done for these decays.
It is also sometimes argued that class~III models are ``EDM-safe'' since $P$ conservation prohibits a nonzero EDM.
However, since the weak interaction is $P$-violating, classes~II and III are mixed by electroweak radiative corrections, and both are effectively subsumed into class~IV~\cite{Khriplovich:1990ef}.
Within the context of $C$-violating $\eta,\eta^\prime$ decays, this connection with EDMs has not been explored.
Suffice to say, in the event of a positive experimental detection in any of these channels, it is not at all clear what the particle physics implications would be.
Lastly, we note that some of the symmetry-violating channels can be mimicked by decays to new light states and this discussion is deferred to Sect.~\ref{sec:BSM-lightparticles}.

\subsection{$P,CP$-violating decays and electric dipole moment constraints}
\label{sec:eta-2pi}

The decays $\eta,\eta^\prime \to 2\pi$ provide a test of $P$ and $CP$ violation since they are forbidden if either symmetry is preserved~\cite{Kobzarev}.
Angular momentum conservation requires the $2\pi$ final state to be $S$-wave and therefore $P$- and $CP$-even, while the initial $\eta$ is $P$- and $CP$-odd.
A related argument holds for $\eta \to 4 \pi^0$. Though not strictly forbidden, it is effectively so since $P$ and $CP$ invariance exclude all pions from being in relative $S$-wave states~\cite{Nefkens:1995dk}. $P$- and $CP$-conserving decays can occur only if the pions are in relative $D$-wave (or higher) states~\cite{Kupsc:2011ah}. 
This is suppressed by $\mathcal{O}(p^{10})$ in $\chi$PT power counting; coupled with the small available phase space energy $7.9 \MeV$, it yields an utterly negligible contribution to the branching ratio of $\mathcal{B}(\eta \to 4\pi^0) \sim 3 \times 10^{-30}$~\cite{Guo:2011ir}.
Other four-pion decays $\eta \to \pi^+\pi^- 2\pi^0$ and $2(\pi^+ \pi^-)$ are kinematically forbidden.
(See Sect.~\ref{sec:etap-4pi} for further discussion of $\eta,\eta^\prime \to 4\pi$.)

Experimental searches for the $2\pi$ channels have yielded only null results and the present constraints are 
\begin{align} \label{eq:twopionlimits}
{\mathcal B}(\eta \to \pi^+ \pi^- ) &< 4.4 \times 10^{-6} \, ,  & {\mathcal B}(\eta \to \pi^0 \pi^0 ) &< 3.5 \times 10^{-4} \,, \notag\\
{\mathcal B}(\eta^\prime \to \pi^+ \pi^- ) &< 1.8 \times 10^{-5} \, , & {\mathcal B}(\eta^\prime \to \pi^0 \pi^0 ) &< 4.5 \times 10^{-4} \, 
\end{align}
at 90\% C.L.~\cite{Ambrosino:2004ww,Babusci:2020jwb,Blik:2007ne,Ablikim:2011vg,Aaij:2016jaa}.
No decays have been observed in $\eta \to 4 \pi^0$ searches either, yielding an upper limit $\mathcal{B}(\eta \to 4\pi^0) < 6.9 \times 10^{-7}$ (90\% C.L.)~\cite{Prakhov:2000xm}.

Now let us consider the theoretical interpretation of these channels. 
Focusing on the $\eta \to 2\pi$ channel, we can write the $P,CP$-odd three-meson interactions as
\begin{equation} \label{eq:etapionint}
\mathcal{L}_{\eta \pi \pi} =\bar{ g}_{\pm} \eta \pi^+ \pi^- + \tfrac{1}{2} \bar{g}_{0} \eta \pi^0 \pi^0 \, ,
\end{equation}
where $\bar{g}_{\pm}, \bar{g}_{0}$ are real dimensionful parameters.
The $\eta \to \pi \pi$ branching fractions are 
\begin{eqnarray}  \label{eq:etatwopion}
\mathcal{B}(\eta \to \pi^+ \pi^-) 
= \frac{\bar{g}_{\pm}^2}{16\pi M_\eta \Gamma_\eta} \sqrt{1 - 4 M_{\pi^\pm}^2/M_\eta^2}
\, , \qquad
\mathcal{B}(\eta \to \pi^0 \pi^0) 
= \frac{\bar{g}_{0}^2}{32\pi M_\eta \Gamma_\eta} \sqrt{1 - 4 M_{\pi^0}^2/M_\eta^2} \, ,
\end{eqnarray}
where $\Gamma_\eta$ is the total width.  
The situation with the $\eta^\prime$ meson is identical to Eq.~\eqref{eq:etatwopion}, but with independent couplings $\bar g_\pm^\prime$, $\bar g_0^\prime$, and $M_\eta,\Gamma_\eta$ replaced by $M_{\eta^\prime},\Gamma_{\eta^\prime}$.

Originally $\eta \to 2\pi$ was proposed to test Purcell and Ramsey's notion of $P$ and $CP$ noninvariance in the strong interaction~\cite{Kobzarev}.
It was not until the arrival of QCD that this idea found some theoretical motivation.
There is a dimension-four $P$- and $CP$-odd operator for gluon fields
\beq \label{eq:thetaqcd}
\mathcal{L}_{\theta} = \frac{g_s^2 \theta}{32 \pi^2}
{G}_{\mu\nu}^{a}\widetilde{G}^{a \mu \nu} \, , 
\eeq
the so-called $\theta$ term of QCD~\cite{Callan:1976je}.
Even if $\theta$ were set to zero, it generically appears due to the noninvariance of the path integral measure from the anomaly when one performs a $U(1)_A$ rotation to put the quark mass matrix $\M$ in canonical form of Eq.~\eqref{eq:QCDLagrangian}.
The $\theta$ term thus shares a common origin with the mechanism that prevents the $\eta^\prime$ from being a pseudo-Nambu--Goldstone boson~\cite{Witten:1979vv,Veneziano:1979ec,Shifman:1979if}.
The fact that $\theta$ can be rotated into $\M$ and vice-versa implies that only the combination $\bar{\theta} = \theta + \arg \det(\M)$ is physical.

There is no reason to exclude the $\theta$ term from the Standard Model since it is renormalizable.
This term generates $\eta\pi\pi$ couplings 
\beq
\bar{g}_\pm = \bar{g}_0 = \frac{2 \bar{\theta} M_\pi^2 m_u m_d}{\sqrt{3} F_\pi (m_u + m_d)^2} 
\approx 0.05 \; \bar{\theta} \GeV
\eeq
in the limit $\bar\theta \ll 1$~\cite{Crewther:1979pi,Shifman:1979if,Pich:1991fq}, where $m_u/m_d \approx 0.45$ (see Sect.~\ref{sec:eta-3pi}).
The present constraint from $\eta \to \pi^+ \pi^-$ yields $|\bar\theta| < 4 \times 10^{-4}$.
Famously, the $\theta$ contribution to the neutron EDM yields a much stronger bound~\cite{Baluni:1978rf,Crewther:1979pi,Pich:1991fq}.
Studies using lattice QCD have further quantified this bound~\cite{Guo:2015tla,Dragos:2019oxn}, most recently yielding $d_n = - (1.5 \pm 0.7) \times 10^{-16} \, \bar\theta \, e \, {\rm cm}$.
According to present limits~\cite{Baker:2006ts,Afach:2015sja,Abel:2020gbr}, the $d_n$ constraint is therefore $|\bar\theta| \lesssim 10^{-10}$, far below the reach for $\ep \to \pi\pi$.
The experimental fact that $\bar\theta$ must be zero or extremely tiny is a puzzling issue known as the strong $CP$ problem~\cite{Peccei:2006as,Hook:2018dlk}.
This issue could in principle be solved in the Standard Model if the $u$ quark were massless.
Although disfavored, it is amusing to note parenthetically that in this case the $\eta^\prime$ would play the role of the axion to relax $\theta$ to zero dynamically (see Ref.~\cite{Hook:2018dlk} for discussion).

In the Standard Model, $CP$ violation arises solely through the KM phase~\cite{Kobayashi:1973fv}. Following Ref.~\cite{Jarlskog:1995ww}, the KM contributions to $\eta,\eta^\prime \to 2\pi$ are\footnote{We have updated Jarlskog and Shabalin's calculation~\cite{Jarlskog:1995ww} to account for the present experimental determination of the KM parameters.  
We take 
$s_2 s_3 \sin\delta \approx A^2 \lambda^4 \eta \approx 6 \times 10^{-4}$~\cite{Tanabashi:2018oca}, where $s_{2,3}$ and $\sin\delta$ refer to the mixing angles and phase of the original KM parameterization~\cite{Kobayashi:1973fv}, and $A,\lambda,\eta$ are from the Wolfenstein parameterization~\cite{Wolfenstein:1983yz}.}
\begin{align} \label{eq:eta2piKM}
{\mathcal B}(\eta \to \pi^+ \pi^- ) &\approx 3 \times 10^{-29} \, , & {\mathcal B}(\eta \to \pi^0 \pi^0 ) &\approx 1 \times 10^{-29} \,,\notag\\
{\mathcal B}(\eta^\prime \to \pi^+ \pi^- ) &\approx  8 \times 10^{-31} \, , & {\mathcal B}(\eta^\prime \to \pi^0 \pi^0 ) &\approx 3 \times 10^{-31} \,,
\end{align}
with $\mathcal{O}(1)$ uncertainties.  These values lie beyond any hope of experimental observation.

New $CP$-violating phases are prevalent in many BSM scenarios and need not be associated with flavor violation like the KM phase~\cite{Pospelov:2005pr,Engel:2013lsa,Chupp:2017rkp}.
Flavor-conserving $CP$ violation, e.g., in extended Higgs models~\cite{Jarlskog:1995gz,Jarlskog:2002zz}, can yield $\eta,\eta^\prime \to 2\pi$ branching fractions many orders of magnitude larger than Eq.~\eqref{eq:eta2piKM}.
However, EDMs are sensitive to these same $CP$-odd $\ep \pi\pi$ couplings~\cite{Gorchtein:2008pe,Gutsche:2016jap,Zhevlakov:2018rwo,Zhevlakov:2019ymi,Zhevlakov:2020bvr}.
Current limits require $\eta,\eta^\prime \to 2\pi$ branching fractions to be many orders of magnitude smaller than their direct limits in Eq.~\eqref{eq:twopionlimits} unless there occur extremely fine-tuned cancellations.
This can be shown on model-independent grounds, which we now discuss.

Our starting point is an effective $CP$-odd interaction between the pion $\pi^a$ and nucleon $N$, given by
\begin{equation} \label{eq:gbar}
\mathcal{L}_{\pi NN} = \bar{g}^{(0)}_{\pi NN} \bar{N} \tau^a N \pi^a + \bar{g}^{(1)}_{\pi NN} \bar{N}  N \pi^0 +  \bar{g}^{(2)}_{\pi NN} ( \bar{N} \tau^a N \pi^a - 3 \bar{N} \tau^3 N \pi^0 ) \, .
\end{equation}
Here $\bar{g}^{(0,1,2)}_{\pi NN}$ are isoscalar, isovector, and isotensor $CP$-odd couplings, respectively, generated by integrating out $CP$-violating physics at the higher scale, and $\tau^a$ is the Pauli matrix acting on isospin (see, e.g., Ref.~\cite{Pospelov:2005pr}).\footnote{Our sign convention for $\bar{g}_{\pi NN}^{(2)}$ is consistent with Ref.~\cite{Pospelov:2005pr}, but is opposite Refs.~\cite{Chupp:2014gka,Chupp:2017rkp}.} 
By Bose symmetry, $\eta, \eta^\prime \to 2\pi$ can only decay in an isoscalar or isotensor channel, and therefore only $\bar{g}^{(0)}_{\pi NN}$ and $\bar{g}^{(2)}_{\pi NN}$ are relevant for our discussion.  
The neutron EDM arising from the charged-pion loop is 
\begin{equation} \label{eq:nEDM}
d_n \approx \frac{e g_{\pi NN} }{4\pi^2 m_N} \big( \bar g_{\pi NN}^{(0)} + \bar g_{\pi NN}^{(2)} \big)  \log\left({m_N}/{M_\pi}\right)   \, , 
\end{equation}
retaining only the leading long-range contribution when the virtual $p$ and $\pi^-$ are separated by distance $\sim M_\pi^{-1}$~\cite{Crewther:1979pi}.
The $CP$-even pion--nucleon coupling is $g_{\pi NN} = 13.1(1)$~\cite{Baru:2010xn,Baru:2011bw}.
Short-distance effects also enter $d_n$~\cite{Engel:2013lsa,Chupp:2017rkp}, but since strong cancellations are not expected between high- and low-momentum contributions, Eq.~\eqref{eq:nEDM} can be treated as an approximate (calculable) lower limit.
Equation~\eqref{eq:gbar} also sources an atomic mercury EDM, $d_{\rm Hg}$, through the nuclear Schiff moment.
Nuclear and atomic many-body calculations yield
\begin{equation} \label{eq:HgEDM}
d_{\rm Hg} \approx - 4 \times 10^{-18} \; e \; {\rm cm} \times \bar{g}_{\pi NN}^{(0)} + 8 \times 10^{-18} \; e \; {\rm cm} \times \bar{g}_{\pi NN}^{(2)} \, ,
\end{equation}
taking preferred values from Ref.~\cite{Chupp:2014gka}, with large nuclear uncertainties in the coefficients~\cite{Ban:2010ea}.
$d_{\rm Hg}$ also receives short-distance contributions from four-nucleon contact interactions~\cite{Maekawa:2011vs}, but again no cancellation is expected between short and long distance scales.

\begin{figure}\centering
\includegraphics[width=0.3\linewidth]{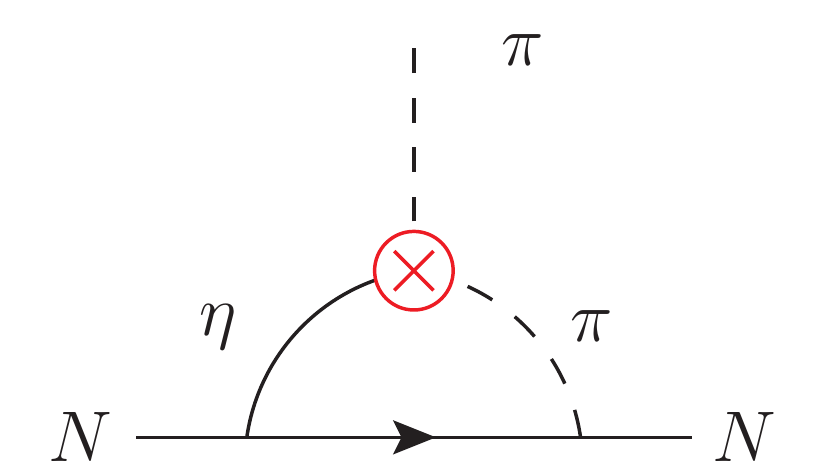}
\caption{$CP$-odd interactions for $\eta \to 2\pi$ in Eq.~\eqref{eq:etapionint}, shown by the red vertex ($\otimes$), source $CP$-odd pion--nucleon interactions in Eq.~\eqref{eq:gbar} that in turn contribute to EDMs.  }\label{fig:EDM_eta-pi-pi}
\end{figure}

Next, we consider how $\eta\pi\pi$ couplings in Eq.~\eqref{eq:etapionint} enter the EDMs.
Integrating out the $\eta$ meson through diagrams shown in Fig.~\ref{fig:EDM_eta-pi-pi}, we obtain $CP$-odd isoscalar and isotensor couplings
\begin{equation} \label{eq:cpoddpion}
\bar{g}^{(0)}_{\pi NN} =  \frac{ g_{\pi NN} g_{\eta NN} \mathcal{I}_\eta}{16 \pi^2 m_N} \left(\frac{ 2 \bar{g}_{\pm} + \bar{g}_{0} }{3} \right) \, , \qquad
\bar{g}^{(2)}_{\pi NN} =  \frac{ g_{\pi NN} g_{\eta NN} \mathcal{I}_\eta}{16 \pi^2 m_N} \left(\frac{ \bar{g}_{\pm} - \bar{g}_{0} }{3} \right) \, ,
\end{equation}
where $\mathcal{I}_\eta \approx 0.6$ results from the loop integration\footnote{The one-loop diagrams yield the Feynman parameter integral $\mathcal{I}_\eta = \int^1_0 dx \int^{1-x}_0 dy \, x/ \left(x^2+y (M_\eta^2/m_N^2) + (1-x-y)(M_\pi^2/m_N^2)\right)$.}
(see also Ref.~\cite{deVries:2015una}). 
The $CP$-even $\eta,\eta^\prime$-nucleon couplings are poorly known and we set $g_{\eta NN} = g_{\eta^\prime NN} =1$~\cite{Benmerrouche:1994uc,Chiang:2001as}.  
Combining these results, we have
\beq \label{eq:EDMresults}
d_n \approx 7 \times 10^{-16} \, \left( \frac{\bar g_\pm}{\rm GeV} \right) \; e \; {\rm cm} \, , \qquad 
d_{\rm Hg} \approx - 2 \times 10^{-20} \, \left( \frac{\bar g_0}{\rm GeV} \right) \; e \; {\rm cm}  \, .
\eeq
We note that the neutron and mercury EDMs are each sensitive predominantly to a single coupling corresponding to $\eta \to \pi^+ \pi^-$ or $\eta \to 2\pi^0$, respectively.\footnote{Subleading terms also arise, but are not retained here.
For the neutron, a contribution involving $\bar{g}_0$ enters through a $\pi^0$ loop involving the neutron magnetic dipole moment~\cite{Gorchtein:2008pe,Zhevlakov:2019ymi}, which is suppressed by $M_\pi^2/m_N^2$~\cite{Seng:2014pba}.
For mercury, a term involving $\bar{g}_\pm$ does not appear due to an accidental cancellation when isoscalar and isotensor terms in the Schiff moment differ by a factor of 2, as adopted in Ref.~\cite{Chupp:2014gka}. 
However, this cancellation is imperfect once nuclear uncertainties are taken into account~\cite{Ban:2010ea,Chupp:2014gka}.}
Given present constraints $|d_n| < 2.2 \times 10^{-26} \; e \, {\rm cm}$ (95\% C.L.)~\cite{Abel:2020gbr}
and $|d_{\rm Hg} | < 7.4 \times 10^{-30} \; e \, {\rm cm}$ (95\% C.L.)~\cite{Graner:2016ses}, we have $|\bar g_\pm| < 3 \times 10^{-11}\GeV$ and $|\bar g_0| < 4 \times 10^{-11}\GeV$.
The resulting constraints on the $\eta$ branching fractions are
\begin{equation} \label{eq:etalimits}
\mathcal{B}(\eta \to \pi^+ \pi^-) < 2 \times 10^{-17} \, , \qquad \mathcal{B}(\eta \to \pi^0 \pi^0) < 2 \times 10^{-17} \, .
\end{equation}
A similar discussion applies to the $\eta^\prime$ as well, with $\mathcal{I}_{\eta^\prime} \approx 0.4$.
Treating the $\eta^\prime \pi\pi$ couplings $\bar{g}_\pm^\prime$ and $\bar{g}_0^\prime$ separately, EDM bounds imply $|\bar g_\pm^\prime| < 5 \times 10^{-11}\GeV$ and $|\bar g_0^\prime| < 6 \times 10^{-11}\GeV$ and 
\begin{equation} \label{eq:etaplimits}
\mathcal{B}(\eta^\prime \to \pi^+ \pi^-) < 3 \times 10^{-19} \, , \qquad \mathcal{B}(\eta^\prime \to \pi^0 \pi^0) < 2 \times 10^{-19} \, .
\end{equation}
These results are in close agreement numerically with full two-loop $\chi$PT calculations for $d_n$~\cite{Zhevlakov:2018rwo,Zhevlakov:2019ymi}.
However, there is a qualitative difference in that chiral divergences $\sim \log M_\pi$, seen in Eq.~\eqref{eq:nEDM} from large virtual pion--proton separation, do not appear in the two-loop calculation~\cite{Zhevlakov:2018rwo}.
Additionally, previous studies have assumed the isospin-conserving limit $\bar g_\pm = \bar g_0$~\cite{Gorchtein:2008pe,Gutsche:2016jap,Zhevlakov:2018rwo,Zhevlakov:2019ymi}, whereas we allow for isospin violation and, moreover, highlight the complementarity of neutron and mercury EDMs in constraining $\bar{g}_\pm$ and $\bar{g}_0$ independently.
Isotensor interactions, which lead to $\bar{g}_\pm \ne \bar{g}_0$, are absent in $CP$-violating BSM operators up through mass dimension-six and only appear through higher-order electromagnetic or chiral corrections~\cite{deVries:2012ab}, but may be much larger in principle for $CP$-violating operators of higher dimension.

Our conclusion is quite pessimistic regarding the BSM physics reach for $\eta,\eta^\prime \to 2 \pi$. 
Even allowing for considerable uncertainty in our estimations, as well as some degree of cancellations, the branching ratio constraints \eqref{eq:etalimits} and \eqref{eq:etaplimits} place these channels far beyond foreseeable experimental reach.
It is plausible that EDM limits also constrain $\eta \to 4\pi^0$ as well, although this has not been explored.

Next, we turn to a further probe of $P$ and $CP$ violation through the decay $\eta \to \pi^+ \pi^- \gamma^{(*)}$~\cite{Herczeg:1974ik,Geng:2002ua}.
This effect can be parameterized by the amplitude
\beq \label{eq:etapipigammaCPV}
\mathcal{A}(\eta \to \pi^+ \pi^- \gamma^{(*)}) = M \, \epsilon_{\mu\nu\alpha\beta} p^\mu_+ p^\nu_- k^\alpha \epsilon^\beta
+ E \, \left( (\epsilon \cdot p_+) (k \cdot p_-) - (\epsilon \cdot p_- )( k \cdot p_+ ) \right) \, , 
\eeq
where $p_\pm$ are the $\pi^\pm$ momenta and $k,\epsilon$ are the $\gamma$ momentum and polarization vectors~\cite{Geng:2002ua,Petri:2010ea}.
Here $M$ and $E$ are coefficients representing magnetic and electric transitions, which are $P,CP$-even and $P,CP$-odd, respectively.\footnote{In general, there are four classes of transitions representing all combinations of $C=\pm 1$ and $P=\pm 1$~\cite{Herczeg:1974ik,Geng:2002ua}.
Equation~\eqref{eq:etapipigammaCPV} includes terms involving the fewest number of derivatives, which are $C$-even.
$C$-odd electric and magnetic terms may arise from BSM physics, involving more powers of momenta.
We do not consider these terms here except to note that the symmetry-breaking patterns are different in these cases, falling into classes~I and III, respectively, the former of which is the only truly EDM-safe option for this channel.}
The magnetic transition arises in the Standard Model through the chiral anomaly and is of order $M \sim e/F_\pi^3$ (see Sect.~\ref{sec:eta-pipigamma}), while the electric transition is sourced by BSM physics.
The interference between the two terms yields a $T$-odd correlation in the triple product $\boldsymbol{\epsilon} \cdot \mathbf k \times \mathbf{p}_+$, which is nonzero when the $\gamma$ has a linear polarization orthogonal to the decay plane in the $\eta$ rest frame~\cite{Geng:2002ua}. 
Alternatively, this effect may be measured for an off-shell photon process in the channel $\eta \to \pi^+ \pi^- e^+ e^-$.
The $P,CP$-odd interference term results in an asymmetry in the relative angle between the $\pi^+ \pi^-$ and $e^+ e^-$ decay planes~\cite{Gao:2002gq}.
This asymmetry is known to be large for the flavor-changing decay $K_L\to \pi^+\pi^-e^+e^-$, measured by the KTeV~\cite{AlaviHarati:1999ff} and NA48~\cite{Lai:2003ad} experiments, consistent with the Standard Model.
Extending this test to the $\eta,\eta^\prime$ sector allows for a test of flavor-conserving $CP$ violation beyond the KM paradigm.
The KLOE~\cite{Ambrosino:2008cp} and WASA-at-COSY~\cite{Adlarson:2015zta} experiments have measured the angular asymmetry in $\eta \to \pi^+ \pi^- e^+ e^-$ to be consistent with zero and no more than a few percent.

\begin{figure}\centering
\includegraphics[width=0.3\linewidth]{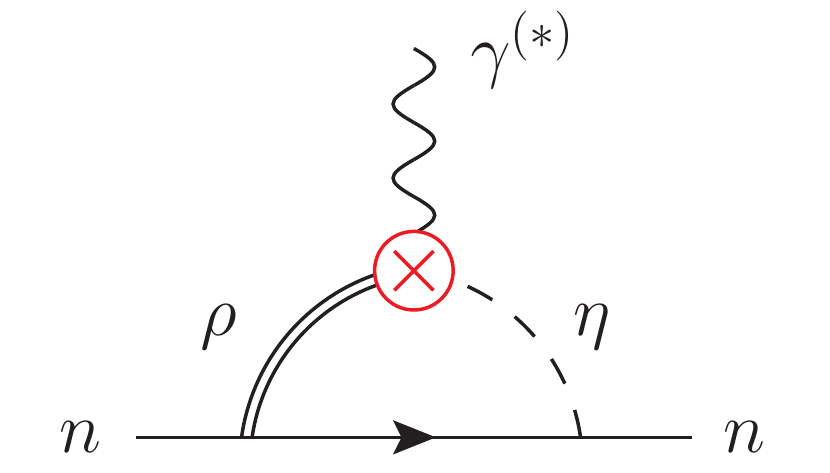}
\caption{One-loop contribution to neutron EDM from $CP$-odd interactions in Eq.~\eqref{eq:etaFrho0}, shown by the red vertex ($\otimes$). This provides strong constraints on $CP$-violating asymmetries in $\eta \to \pi^+ \pi^- \gamma^{(*)}$ since, in the $CP$-violating amplitude, the $\pi^+ \pi^-$ state has the same quantum numbers as the $\rho^0$ meson and the same BSM physics would source $\eta \to \rho^{0*} \gamma^{(*)}$ transitions.
}\label{fig:nEDM_eta-rho}
\end{figure}

Next, let us discuss the theoretical interpretation of this measurement. 
For kaons, the electric transition is dominated by $CP$-violating decay $K_L \to \pi^+ \pi^-$ with a bremsstrahlung $\gamma$~\cite{Heiliger:1993qt}.
For the $\eta$, this mechanism is precluded by the stringent EDM constraints on $\eta \to \pi^+ \pi^-$.
However, it is thought that some BSM four-fermion operators may source $P,CP$-odd transitions for $\eta \to \pi^+ \pi^- \gamma^{(*)}$ directly without contributing to $\eta \to 2\pi$ or EDMs~\cite{Geng:2002ua}.
We can give a qualitative argument why this is not the case.
To start, we recognize that the electric term in Eq.~\eqref{eq:etapipigammaCPV} corresponds to an effective operator 
\beq \label{eq:etaFpipi}
\mathcal{L}_{\rm eff} = - \frac{E}{2} \, \eta F^{\mu\nu} \left[ i \left(\partial_\mu \pi^+ \partial_\nu \pi^- - \partial_\nu \pi^+ \partial_\mu \pi^- \right) \right] \, .
\eeq
The term in brackets has the same quantum numbers as the field strength tensor $\rho^0_{\mu\nu} = \partial_\mu \rho_\nu^0 - \partial_\nu \rho_\mu^0$ for the $\rho^0$ meson.
Through strong dynamics, any BSM operator generating Eq.~\eqref{eq:etaFpipi} would certainly generate a lower-dimensional interaction of the form
\beq \label{eq:etaFrho0}
\mathcal{L}_{\rm eff} = \kappa \, \eta F^{\mu\nu} \rho^0_{\mu\nu} \, .
\eeq
The coefficient $\kappa$ can be estimated by dimensional analysis and is likely not much less than $\mathcal{O}(E \times F_\pi^2)$.
Equation~\eqref{eq:etaFrho0} contributes to the neutron EDM at one-loop order, shown in Fig.~\ref{fig:nEDM_eta-rho}.
Qualitatively, this yields $d_n \sim \kappa g_{\rho NN} g_{\eta NN}/(4\pi)^2$, where $g_{\rho NN} \sim 3$ is the $\rho$-meson--nucleon (vector) coupling~\cite{Brown:1994pq}.
It is not possible to be more precise since the loop integral is logarithmically divergent, which would be cut off around the QCD scale anyway.
Nevertheless, this yields an order-of-magnitude constraint at the level of $E/M \lesssim 10^{-11}$, far beyond present sensitivities for the angular asymmetry even allowing for the roughness of our estimates.\footnote{The foregoing argument excludes $CP$-violating $\eta \to \pi^+ \pi^- \gamma^*$ operators (as suggested in the literature) at an observable level, but strictly speaking does not preclude the contribution of local effective quark--lepton operators inducing $CP$-violating $\eta \to \pi^+ \pi^- \ell^+\ell^-$ decays.   
We expect significant constraints from EDM searches in paramagnetic atoms and molecules, which are sensitive to $CP$-odd electron--nucleon couplings~\cite{Chupp:2017rkp}. 
The consequences of similar quark--dimuon operators, as discussed in the following, have not been discussed for this specific decay yet.}

Recently, S{\'a}nchez-Puertas~\cite{Sanchez-Puertas:2018tnp,Sanchez-Puertas:2019qwm} proposed a new class of symmetry tests in $\eta$ decays to dimuon final states: 
\beq \label{eq:SPdecays}
\eta\to\mu^+\mu^- \, , \quad \eta\to\mu^+\mu^-\gamma \, , \quad
\eta\to\mu^+\mu^-e^+e^-  \, .
\eeq
$CP$-odd asymmetries in the first two channels involve the muon polarizations, which will be measurable at the planned REDTOP experiment~\cite{Gonzalez:2017fku, Gatto:2016rae}, while for the last decay, the simplest $CP$-odd observable is the angular asymmetry between the two dilepton planes (much like for $\eta\to\pi^+\pi^-e^+e^-$ discussed above).
In contrast to electronic final states, these decays avoid strong constraints from the electron EDM (and other $CP$-odd electron interactions, e.g.,~\cite{Chupp:2017rkp}); also, electron polarizations will not be measured in the proposed REDTOP detector~\cite{Gonzalez:2017fku, Gatto:2016rae}.

The minimal framework to generate $CP$-odd asymmetries in decays \eqref{eq:SPdecays} is to introduce $CP$ violation in the $\eta$--two-photon coupling.
The usual $\eta$ transition form factor~\eqref{eq:defTFF} is generalized to
\begin{align}
- \epsilon_{\mu\nu\alpha\beta} q_1^\alpha q_2^\beta F_{\eta\gamma^*\gamma^*}(q_1^2,q_2^2) 
& + \big[q_1\cdot q_2 \,g_{\mu\nu} - q_{2\mu}q_{1\nu} \big] F_{\eta\gamma^*\gamma^*}^{CP1}(q_1^2,q_2^2) \notag\\
& + \Big[q_1^2q_2^2 \,g_{\mu\nu} - q_1^2 \,q_{2\mu}q_{2\nu} - q_2^2 \,q_{1\mu}q_{1\nu} + q_1\cdot q_2\, q_{1\mu}q_{2\nu}\Big] 
  F_{\eta\gamma^*\gamma^*}^{CP2}(q_1^2,q_2^2) \,,
\end{align}
to include new $P$- and $CP$-violating TFFs $F_{\eta\gamma^*\gamma^*}^{CP1,2}$ that couple to scalar---instead of pseudoscalar---Lorentz structures (the contribution of $F_{\eta\gamma^*\gamma^*}^{CP2}$ vanishes as long as one of the photons is real).
\begin{figure}[tb!]
\centering
\includegraphics[width=0.3\linewidth]{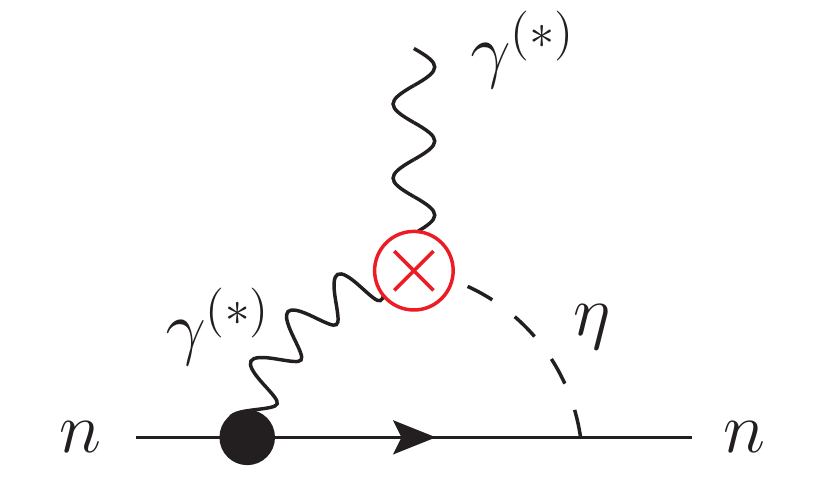} \hspace*{3cm}
\includegraphics[width=0.3\linewidth]{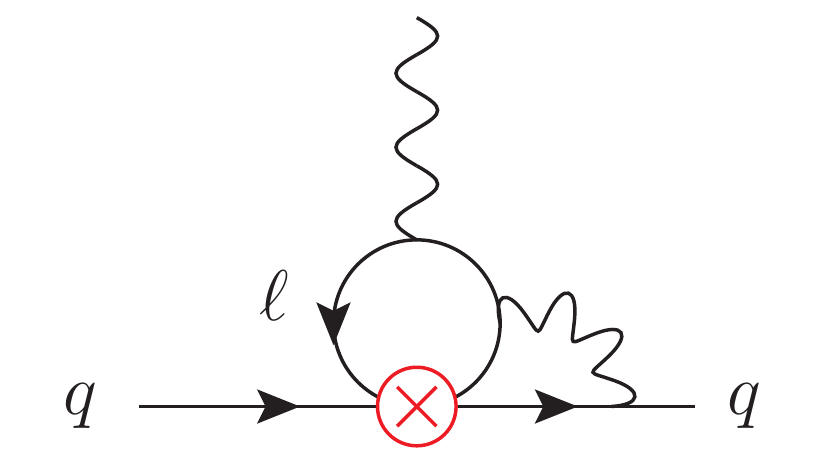}
\caption{\textit{Left:} One-loop contribution to the neutron EDM, induced by a $CP$-violating
$\eta\to\gamma^*\gamma^*$ TFF [denoted by the red vertex ($\otimes$)].
The photon--neutron coupling (black dot) involves the standard electromagnetic form factors.
A similar crossed diagram is not shown explicitly.
\textit{Right:} Two-loop contribution to the neutron EDM, induced by a $CP$-violating
quark--lepton four-fermion operator [denoted by the red vertex ($\otimes$)].
Neither crossed diagrams nor counterterm contributions at one-loop and tree level required for renormalization
are displayed explicitly.
}\label{fig:nEDM_eta-gamma}
\end{figure}
However, such new $\eta\gamma^*\gamma^*$ interactions induce contributions to the neutron EDM at the one-loop level via 
an $\eta\gamma$ intermediate state, shown in Fig.~\ref{fig:nEDM_eta-gamma} (left) (similar to the bounds on $\eta^{(\prime)}\to\pi\pi$ discussed above).
To evaluate the loop integral, the $q_i^2$-dependence of the $CP$-violating TFFs is modeled in a simple way motivated by the high-energy asymptotics of scalar TFFs in QCD~\cite{Kroll:2016mbt}.
The resulting indirect constraints on all dimuon asymmetries are several orders of magnitude stronger than projections for REDTOP based on a proposed statistics of $2\times 10^{12}$ produced $\eta$ mesons~\cite{Sanchez-Puertas:2018tnp}.

Alternatively, $CP$-violating dimuon asymmetries can also arise via $CP$-odd four-fermion operators between quarks and leptons in the Standard Model Effective Field Theory framework~\cite{Buchmuller:1985jz,Grzadkowski:2010es}.
In the notation of Ref.~\cite{Grzadkowski:2010es}, the relevant operators are
\begin{align}
\mathcal{L}_{\rm eff} = -\frac{1}{2v^2} \bigg\{ 
  \Im c_{\ell equ}^{(1)2211} \Big[(\bar\mu \mu) \Big(\bar u i\gamma^5 u\Big)+\Big(\bar\mu i\gamma^5\mu\Big)(\bar uu) \Big]
& - \Im c_{\ell edq}^{2211} \Big[(\bar\mu \mu) \Big(\bar d i\gamma^5 d\Big) -\Big(\bar\mu i\gamma^5\mu\Big)(\bar dd)\Big] \notag\\
& - \Im c_{\ell edq}^{2222} \Big[(\bar\mu \mu) \Big(\bar s i\gamma^5 s\Big) -\Big(\bar\mu i\gamma^5\mu\Big)(\bar ss)\Big]\bigg\} \,. 
\label{eq:CPV-ff}
\end{align}
The important observation here is that neutron EDM contributions from Eq.~\eqref{eq:CPV-ff} are comparatively suppressed since they begin at two-loop order, shown in Fig.~\ref{fig:nEDM_eta-gamma} (right)~\cite{Sanchez-Puertas:2018tnp}.
This is because the muon (pseudo)scalar bilinears cannot couple to only one photon via a single lepton loop, as the Green's functions $\langle 0 | T\{ V^\mu(x) S(P)(0)\}|0\rangle$ vanish in QCD+QED as a result of $C$ conservation. 
The resulting constraints were estimated to be
\beq
\big| \Im c_{\ell equ}^{(1)2211} \big| < 0.002 \,, \qquad
\big| \Im c_{\ell edq}^{2211} \big| < 0.003 \,, \qquad
\big| \Im c_{\ell edq}^{2222} \big| < 0.04 \,,
\eeq
assuming no cancellations between terms~\cite{Sanchez-Puertas:2018tnp}.
Taking advantage of $\SU(2)_L$ symmetry, a comparable limit $\big| \Im c_{\ell edq}^{2222} \big| < 0.02$ was determined from $D_s \to \mu \nu_\mu$ decays~\cite{Sanchez-Puertas:2019qwm}.

In light of these constraints, $CP$ violation between strange quarks and muons---sourced by $\Im c_{\ell edq}^{2222}$---remains an open possibility within REDTOP sensitivities that is not excluded by EDMs~\cite{Sanchez-Puertas:2018tnp}.
The most promising channel is $\eta \to \mu^+ \mu^-$.
From Eq.~\eqref{eq:CPV-ff}, the hadronic matrix elements of the pseudoscalar quark bilinears between $\eta$ states and the vacuum can be evaluated
in terms of rather well-known meson decay constants, as well as quark and meson masses.
The effective $CP$-odd interaction is represented as 
\beq
\mathcal{L} = - \mathcal{C} \, \eta \, \bar\mu \mu \, , \qquad
\mathcal{C} = \Big[ 1.57 \big(  \Im c_{\ell equ}^{(1)2211} + \Im c_{\ell edq}^{2211}\big) - 2.37 \, c_{\ell edq}^{2222} \Big] \times 10^{-6} \, , 
\eeq
which interferes with the $CP$-even Standard Model amplitude discussed in Sect.~\ref{sec:P-ll}.
The actual $CP$-odd asymmetries that are to be measured for this purpose rely on the polarized muon
decays; most promising is a forward--backward asymmetry of the electron/positron relative to their decaying parent
muons' line of flight (in the dimuon rest frame), which turns out to be more sensitive than an asymmetry in the relative
electron/positron azimuthal angles by about an order of magnitude. 
While the planned REDTOP statistics suggest a sensitivity for the various coefficients of $\Order(1)$ in the Dalitz decay
or worse in the four-lepton final states, the sensitivities of $\Order(10^{-2})$ in $\eta\to\mu^+\mu^-$ imply that 
in particular $\Im c_{\ell edq}^{2222}$ might induce $CP$ violation in this decay not yet excluded by EDMs and other constraints~\cite{Sanchez-Puertas:2018tnp,Sanchez-Puertas:2019qwm}.

\subsection{$C$ violation}
\label{sec:BSMCV}

\begin{table}
\centering
\renewcommand{\arraystretch}{1.3}
\begin{tabular}{llll} 
\toprule
Channel     & Branching ratio   & Note & Ref. 
\\ 
\midrule
$\eta \to 3\gamma$     
&  $<1.6\times 10^{-5}$     
&  
& \cite{Aloisio:2004di} 
\\ 
$\eta \to \pi^0\gamma$  
&  $<9\times 10^{-5}$ 
& Violates angular momentum conservation or gauge invariance
& \cite{Nefkens:2005ka} 
\\
$\eta \to \pi^0e^+e^-$       
&  $<7.5\times 10^{-6}$ 
&  $C$, $CP$-violating as single-$\gamma$ process 
& \cite{Adlarson:2018imw} 
\\ 
$\eta \to \pi^0 \mu^+ \mu^-$ 
& $< 5\times 10^{-6}$ 
&  $C$, $CP$-violating as single-$\gamma$ process
& \cite{Dzhelyadin:1980ti} 
\\ 
$\eta \to 2\pi^0\gamma$ 
&  $<5\times 10^{-4}$
&
& \cite{Nefkens:2005dp} 
\\ 
$\eta \to 3\pi^0\gamma$ 
& $<6\times 10^{-5}$ 
&
& \cite{Nefkens:2005dp} 
\\  
\midrule
$\eta^\prime \to 3\gamma$
&  $<1.0\times 10^{-4}$ 
&
& \cite{Alde:1987jt} 
\\  
$\eta^\prime \to \pi^0e^+e^-$
&  $<1.4\times 10^{-3}$
&  $C$, $CP$-violating as single-$\gamma$ process
& \cite{Briere:1999bp} 
\\  
$\eta^\prime \to \pi^0\mu^+\mu^-$
& $<6.0 \times 10^{-5}$   
&  $C$, $CP$-violating as single-$\gamma$ process
& \cite{Dzhelyadin:1980ti} 
\\  
$\eta^\prime \to \eta e^+e^-$
&  $<2.4\times 10^{-3}$
&  $C$, $CP$-violating as single-$\gamma$ process
& \cite{Briere:1999bp} 
\\  
$\eta^\prime \to \eta \mu^+\mu^-$
&  $<1.5\times 10^{-5}$ 
&  $C$, $CP$-violating as single-$\gamma$ process
& \cite{Dzhelyadin:1980ti} 
\\  
\bottomrule
\end{tabular}
\renewcommand{\arraystretch}{1.0}
\caption{All $C$-violating $\eta,\eta^\prime$ decay modes that have been searched for to date. Branching ratio limits are quoted at 90\% C.L. 
\label{tab:CVdecays} }
\end{table}

The $\eta,\eta^\prime$ mesons are $C=+1$ eigenstates 
and provide one of the few opportunities to test $C$ conservation in flavor-conserving strong and electromagnetic decays.
$C$ is violated for $\eta,\eta^\prime$ decaying into an odd number of photons (including off-shell photons) and any number of neutral pions.
For the $\eta^\prime$, $C$-violating channels may include an $\eta$ meson as well. 
Table~\ref{tab:CVdecays} lists all channels that have been tested experimentally and no evidence of $C$ violation has been found so far.
Other possibilities, such as  $\eta^\prime \to \eta \gamma$,  $\eta^\prime \to 2\pi^0 \gamma$, or $\eta^\prime \to \eta \pi^0 \gamma$ to name a few, have not been searched for.

Additionally, we discuss decays that are $C,CP$-violating only as single-photon processes at leading order in QED.
Some of these decays have been searched for experimentally, such as
\beq \label{eq:CV_scalar_decays}
\eta,\eta^\prime \to \pi^0 \ell^+ \ell^- \, , \quad
\eta^\prime \to \eta \ell^+ \ell^- \, ,
\eeq
where $\ell=e,\mu$, 
while no experimental data exists for other channels such as
\beq \label{eq:CV_ALP_decays}
\eta, \eta^\prime \to \pi^0 \pi^0 \ell^+ \ell^- \, , \quad
\eta^\prime \to \eta \pi^0 \ell^+ \ell^- \, .
\eeq
These decays are also relevant for searches for new ($C,CP$-conserving) light particles.
Decays in \eqref{eq:CV_scalar_decays} can arise for a new light scalar $S$, e.g., $\eta, \eta^\prime \to \pi^0 S \to \pi^0 \ell^+ \ell^-$~\cite{Ellis:1975ap}. 
Alternatively, decays in \eqref{eq:CV_ALP_decays} arise for a new light axion-like pseudoscalar $a$, e.g., $\eta, \eta^\prime \to 2 \pi^0 a \to 2 \pi^0 \ell^+ \ell^-$.
We discuss these models further in Sect.~\ref{sec:BSM-lightparticles}.

For the most part, these tests have been framed within the context of $T$-odd, $P$-even (TOPE) interactions (class~III in Table~\ref{tab:symbreaking}), which is equivalent to $C$ violation by the $CPT$ theorem.
This idea was proposed in the 1960s, following the discovery of $CP$ violation~\cite{Christenson:1964fg}, to be either a new electromagnetic-strength ``semi-strong'' force~\cite{Prentki:1965tt,Lee:1965zza,Lee:1965hi} or related to electromagnetic couplings for hadrons~\cite{Bernstein:1965hj}.
At second order in perturbation theory, the combination of the weak force (thought at the time to be class~I) and a flavor-conserving TOPE interaction could yield $CP$-violating decays for neutral kaons.
While nature did not realize this possibility for the Standard Model, the search for BSM physics and its connection with the matter--antimatter asymmetry has motivated ongoing experimental study.

Direct limits on flavor-conserving TOPE forces for hadronic interactions come from a variety of nuclear tests of $T$ invariance~\cite{Gudkov:1991qg}.
These limits are not very strong and one reason is that TOPE effects cannot be mediated through single-pion exchange and may only contribute via shorter-range nuclear forces~\cite{Simonius:1975ve}.
Therefore, it is very interesting that TOPE couplings may in principle be as large as $10^{-3}$ relative to strong couplings~\cite{RamseyMusolf:1999nk,Kurylov:2000ub} and, moreover, that considerable improvement can be made outside the nuclear arena through $\eta,\eta^\prime$ studies.\footnote{In contrast, constraints for flavor-conserving $T,P$-odd forces are much more stringent. For example, the neutron EDM must be at least $2 \times 10^{-12}$ times smaller than its ``natural'' size set by the nuclear magneton $\mu_N = e/(2m_N) \approx  10^{-14} \; e \; {\rm cm}$, which sets the scale of $d_n$ relative to strong dynamics.}
However, theoretical work on this front is severely lacking.
It remains unknown what BSM particle physics models would be tested and how $\eta,\eta^\prime$ measurements would compare relative to other constraints.
Very naively, we may estimate the latter by considering ratios of partial widths. 
For example, let us consider the recent limit on $\eta \to \pi^0 e^+ e^-$ from WASA-at-COSY~\cite{Adlarson:2018imw}. Treating the $\omega$ meson as a $C=-1$ proxy for the $\eta$, we have
\beq
\frac{\Gamma(\eta \to \pi^0 e^+ e^-)}{\Gamma(\omega \to \pi^0 e^+ e^-)} < 1.5 \times 10^{-6} \,,
\eeq
using $\Gamma(\omega \to \pi^0 e^+ e^-) = 6.5 \keV$~\cite{Tanabashi:2018oca}, which implies the relative TOPE amplitude to be no more than $10^{-3}$, comparable to limits from nuclear tests.
However, a futuristic constraint at the level $\BR(\eta \to \pi^0 e^+ e^-) < 10^{-10}$~\cite{Gatto:2016rae,Gonzalez:2017fku} would constrain the relative amplitude to be less than $10^{-5}$.

EDM limits place strong indirect bounds on flavor-conserving TOPE effects~\cite{Khriplovich:1990ef,Conti:1992xn,Haxton:1994bq,Engel:1995vv,RamseyMusolf:1999nk,Kurylov:2000ub}.
These constraints have been formulated in terms of nonrenormalizable operators in a low-energy effective theory for BSM physics.
The leading flavor-conserving TOPE operators involving only Standard Model fields arise at mass dimension seven~\cite{Khriplovich:1990ef}.
These include four-fermion operators~\cite{Conti:1992xn,Engel:1995vv}, e.g., 
\beq \label{eq:TOPEop1}
\frac{1}{\Lambda^3} \bar{\psi}_f \gamma_5 {D}_\mu  \psi_f \, \bar{\psi}_{f^\prime} \gamma^\mu \gamma_5 \psi_{f^\prime} + {\rm h.c.}\, ,
\eeq
where $f,f^\prime$ are distinct fermions, and fermion-gauge boson operators~\cite{RamseyMusolf:1999nk}, e.g.,
\beq \label{eq:TOPEop2}
\frac{1}{\Lambda^3} \bar{\psi}_f \sigma_{\mu\nu} \psi_f F^{\mu\lambda} Z^\nu_\lambda \, ,
\eeq
where $Z_{\mu\nu}$ is the $Z$ boson field strength tensor.
There is no BSM model in the literature for generating these operators, but they are simply represented by a characteristic energy scale $\Lambda$.\footnote{Since operators~\eqref{eq:TOPEop1} and \eqref{eq:TOPEop2} flip helicity, they arise only at mass dimension eight when expressed in terms of $\SU(2)_L \times U(1)_Y$-covariant fields, requiring at least one Higgs field insertion.
Therefore the coefficient $1/\Lambda^3$ can be interpreted as $v/\Lambda^4_{\rm BSM}$, where $\Lambda_{\rm BSM}$ is the mass scale for BSM physics and $v$ is the Higgs vacuum expectation value.}
Because electroweak radiative corrections mix TOPE operators with $P,T$-odd ones, these operators contribute to EDMs.
Following Ref.~\cite{RamseyMusolf:1999nk}, Eq.~\eqref{eq:TOPEop2} contributes to the EDM of an elementary fermion of at one-loop order, given approximately by
\beq \label{eq:TOPEdf}
d_f \sim \frac{e \, g^2}{(4\pi)^2} \frac{M_Z^2}{\Lambda^3} \approx 4 \times 10^{-25} \; e \; {\rm cm} \times \left( \frac{\Lambda}{10 \TeV} \right)^{-3} \, ,
\eeq
where $g$ is the $\SU(2)_L$ gauge coupling.
For light quarks $f=u,d$, this yields a comparable contribution to the neutron EDM~\cite{Pospelov:2005pr,Engel:2013lsa}, for which the present limit~\cite{Baker:2006ts,Afach:2015sja,Abel:2020gbr} implies $\Lambda \gtrsim 20 \TeV$.
For four-fermion operators \eqref{eq:TOPEop1}, $d_f$ arises at two-loop order, which is suppressed by an additional loop factor $1/(4\pi)^2$.
TOPE amplitudes must be proportional to $(p/\Lambda)^3$, where $p \sim 1 \GeV$ is a typical momentum scale.
According to the preceding argument, this factor must be around $10^{-11}$ or below, rendering the possibility of observing TOPE effects very remote.\footnote{A more detailed analysis in Ref.~\cite{Kurylov:2000ub} arrives at somewhat stronger limits for $\Lambda$. We also mention, as pointed out by Ref.~\cite{RamseyMusolf:1999nk}, that the proportionality between EDMs and TOPE operators holds only for models in which $P$ is restored at a scale below $\Lambda$. Otherwise, the contribution to $d_f$ in Eq.~\eqref{eq:TOPEdf} is likely only subleading compared to lower-dimensional operators that generically arise. In the absence of fine-tuned cancellations, this fact does not relax these constraints at all. Rather, it means that if a nonzero EDM is measured, it does not imply that TOPE effects are present.}
Nevertheless, these arguments have not been applied specifically to the $\eta,\eta^\prime$ sector and, at the very least, it would be desirable to know how these operators contribute to $\eta,\eta^\prime$ decays before making any definitive conclusions (see Ref.~\cite{M-RM-workshop}).

Now, let us consider specific $C$-violating channels in more detail.
The $\eta \to \pi^0 \gamma^{(*)}$ transition, for an on- or off-shell photon, violates $C$~\cite{Bernstein:1965hj}.
It is also $CP$-violating since angular momentum conservation requires a $P$-wave for the $\pi^0 \gamma^{(*)}$ state.
For an on-shell photon, the two-body decay $\eta \to \pi^0 \gamma$ additionally violates either angular momentum conservation or gauge invariance.
This can be shown easily from the decay amplitude 
\beq \label{eq:etapi0gamma}
\mathcal{A}(\eta \to \pi^0 \gamma) 
= \mathcal{A}^\mu(\eta \to \pi^0 \gamma) \varepsilon_\mu(q) \,,
\eeq
where we have factored out the $\gamma$ polarization vector $\varepsilon_\mu(q)$. 
By Lorentz invariance, the amplitude must take the form $\mathcal{A}^\mu =  f_1 (p^\mu + k^\mu) + f_2 q^\mu $ since the only available four-vectors are the momenta $p,k,q$ for the $\eta,\pi^0,\gamma$, respectively~\cite{Bernstein:1965hj}.
The coefficients $f_{1,2}$ are related by the Ward identity, $q_\mu \mathcal{A}^\mu = 0$, requiring $f_1 = - f_2 q^2/(M_\eta^2-M_{\pi^0}^2)$.
For a real, transverse photon in the final state, both terms in $\mathcal{A}$ must vanish since $q^2=0$ and $q^\mu \varepsilon_\mu(q) = 0$.
Perhaps due to a lack of theoretical motivation, this channel has seen only one experimental search to date~\cite{Nefkens:2005dp}.

The $C,CP$-violating transition $\eta \to \pi^0 \gamma^*$ can occur for an off-shell photon (without violating other, perhaps more cherished symmetries).
The preceding argument is evaded for $q^2 \ne 0$. (The $f_1$ term need not vanish, though the $f_2$ term will still not contribute since $q^\mu$ couples to a conserved current.)
Experiments have searched for this and related processes
\beq \label{eq:CVdilepton}
\eta,\eta^\prime \to \pi^0 \gamma^* \to \pi^0 \ell^+ \ell^-  \, , \qquad
\eta^\prime \to \eta \gamma^* \to \eta \ell^+ \ell^- \,,
\eeq
when the $\gamma^*$ decays into a lepton pair ($\ell = e,\mu$)~\cite{Dzhelyadin:1980ti,Briere:1999bp,Adlarson:2018imw}.
These channels are $C,CP$-violating only as a single-photon process.
However, a $C,CP$-conserving two-photon intermediate state is allowed in the Standard Model and gives a background to Eq.~\eqref{eq:CVdilepton}.
For $\eta \to \pi^0 \ell^+ \ell^-$, this rate has been calculated using several techniques~\cite{LlewellynSmith1967,Cheng:1967zza,Ng:1992yg,Ng:1993sc} and is estimated to be three or more orders of magnitude below present sensitivities (see Sect.~\ref{sec:etapi0ll}).

The $\eta,\eta^\prime \to 3 \gamma$ decays are another test of $C$ violation, arising either with or without $CP$ violation.
It was the related channel $\pi^0 \to 3 \gamma$ that was initially proposed~\cite{Prentki:1965tt} and studied~\cite{Berends:1965ftl,Tarasov,Galfi:1967rr} along these lines.
Gauge invariance and Bose symmetry require that the leading effective Hamiltonian for this transition contains at least seven derivatives, both for cases with~\cite{Berends:1965ftl,Galfi:1967rr} and without~\cite{Dicus:1975cz} $P$ invariance.
Since each power of momenta is proportional to the total allowed phase space $\mpiz$, this implies that $\Gamma(\pi^0 \to 3 \gamma)$ scales with a very large power of $\mpiz$.
The leading contribution from the Standard Model involves $CP$-conserving weak interactions (class~I in Table~\ref{tab:symbreaking}) at two-loop order.
Dicus has estimated
\beq \label{eq:pi03gamma}
\frac{\Gamma(\pi^0 \to 3\gamma)}{\Gamma(\pi^0 \to 2\gamma)}
\approx 1.2 \times 10^{-5} \, \frac{\alpha_{\rm em} G_F^2 \mpiz^4}{(2\pi)^5} \left( \frac{\mpiz}{\bar m} \right)^8 \, ,
\eeq
where $\bar{m}$ represents a hadronic scale of $\mathcal{O}(0.1\text{--}1 \GeV)$ appearing in the quark loop~\cite{Dicus:1975cz}. 
Varying within this range gives the order of magnitude for Eq.~\eqref{eq:pi03gamma} to be $10^{-32}$--$10^{-24}$, which even allowing for large uncertainties~\cite{Dicus:1975cz} is very far below the present experimental limit $\BR(\pi^0 \to 3\gamma) < 3.1\times 10^{-8}$ (90\% C.L.)~\cite{McDonough:1988nf}.
Due to the scaling with the pseudoscalar mass, it is thought that the ratios for $\eta,\eta^\prime \to 3\gamma$ can be dramatically enhanced compared to $\pi^0 \to 3 \gamma$~\cite{Herczeg:1988ep}.
Rescaling Eq.~\eqref{eq:pi03gamma} gives order-of-magnitude Standard Model predictions
\beq
\BR(\eta \to 3\gamma) \sim 10^{-25} \text{--} 10^{-17} \, , \qquad
\BR(\eta^\prime \to 3\gamma) \sim 10^{-23} \text{--} 10^{-15} \, ,
\eeq
also far below present limits.
The strongest limits to date on $\eta \to 3\gamma$ come from the KLOE~\cite{Aloisio:2004di} and
Crystal Ball~\cite{Nefkens:2005ka} experiments.
These searches were largely background limited. 
With improvements in $\gamma$ detection, the JEF experiment has the potential to improve sensitivities by an order of magnitude, reaching $\mathcal{O}(10^{-6})$ in the branching ratio (also for $\eta^\prime \to 3 \gamma$), due to a factor of 100 relative reduction in backgrounds~\cite{JEF-PAC42}.
By reducing ``missing photon'' backgrounds, JEF will also improve constraints for other $C$-violating channels with multi-$\gamma$ final states, such as $\eta \to 2 \pi^0 \gamma$ and $\eta \to 3 \pi^0 \gamma$~\cite{JEF-PAC42}.

The positive detection of $\pi^0, \eta, \eta^\prime \to 3 \gamma$ close to present sensitivities would certainly constitute BSM physics.
However, with little theoretical work, it is unclear what the interpretation for such a signal would be.
There are two general categories of models, depending on whether $CP$ is conserved or violated (classes~I or III, respectively, in Table~\ref{tab:symbreaking}).
In the latter case, EDM constraints are certainly relevant.
Herczeg estimated these indirect limits to be
\beq
\BR(\eta \to 3 \gamma) < 10^{-10} \, , \qquad \BR(\eta^\prime \to 3 \gamma) < 3 \times 10^{-9}
\eeq
from dimensional analysis and assuming the $C$-violating interactions source $d_n$ via electroweak radiative corrections~\cite{Herczeg:1988ep}.
(These upper limits should be reduced by two orders of magnitude since the $d_n$ bound~\cite{Abel:2020gbr} has improved by an order of magnitude compared to the value quoted in Ref.~\cite{Herczeg:1988ep}.)
On the other hand, the $CP$-conserving case is unconstrained by EDM limits.
One BSM example along these lines would be a $Z^\prime$ gauge boson with both vector and axial-vector couplings for quarks.
While $Z^\prime$ models are constrained by low-energy tests of $P$-violation~\cite{RamseyMusolf:1999qk}, these constraints are weakened for so-called leptophobic models with no tree-level $Z^\prime$ couplings to leptons (see, e.g., Refs.~\cite{Buckley:2012tc,Dobrescu:2014fca}).
Theoretical work is needed to determine how large the $\eta,\eta^\prime \to 3\gamma$ rates may be in such a model.

We also mention a third scenario for $\pi^0 \to 3 \gamma$ based on quantum electrodynamics extended to noncommutative spacetime~\cite{Grosse:2001ei}.
This model is Lorentz-violating and falls outside the classes of theories we have considered thus far.
In the $\pi^0$ rest frame, Ref.~\cite{Grosse:2001ei} obtained
\beq \label{eq:pi03gammaNC}
\frac{\Gamma(\pi^0 \to 3\gamma)}{\Gamma(\pi^0 \to 2\gamma)}
\approx \frac{\alpha_{\rm em} \theta^2 \mpiz^4}{120\pi}
\approx 6 \times 10^{-21} \,,
\eeq
where $|\theta| = 1 \TeV^{-2}$ was assumed for the noncommutivity scale.
Equation~\eqref{eq:pi03gammaNC} can be straightforwardly generalized to $\eta,\eta^\prime \to 3 \gamma$.
However, smaller values $|\theta| \ll 1 \TeV^{-2}$ are required by constraints on Lorentz noninvariance~\cite{Carroll:2001ws}, which implies that the branching ratios are unobservably small.

Lastly, we discuss $C,CP$-violating observables for $\eta\to\pi^+\pi^-\pi^0$~\cite{Prentki:1965tt,Lee:1965zza,Nauenberg:1965,Barrett:1965ia}.
Bose symmetry requires that the $3\pi$ final state with total isospin $I$ has $C = - (-1)^I$~\cite{Lee:1965zza}.
The interference of $I=0,2$ $C$-violating amplitudes with the $I=1$ $C$-conserving Standard Model amplitude, discussed in Sect.~\ref{sec:eta-3pi}, generates a charge asymmetry in $\pi^\pm$ distributions.
In particular, as an interference effect it scales linearly with the characteristic strength parameter of any $C$-violating BSM operator, while the rates of the $C$-forbidden processes discussed above are suppressed to quadratic order therein~\cite{Gardner:2019nid}.
The simplest possible observable is a left--right asymmetry
\beq
A_{LR} = \frac{N_+-N_-}{N_++N_-} 
\eeq
in the Dalitz distribution, where $N_\pm$ is the number of events with $T_+ \gtrless T_-$ and $T_\pm$ refers to the $\pi^\pm$ kinetic energy in the $\eta$ rest frame~\cite{Layter:1972aq}.
The study of more complicated quadrant and sextant asymmetries $A_{Q,S}$ in the Dalitz plot allows one to distinguish between isotensor and isoscalar amplitudes, respectively~\cite{Lee:1965zza,Nauenberg:1965,Layter:1972aq}.
Experimentally, these asymmetries are consistent with zero~\cite{Ambrosino:2008ht,Tanabashi:2018oca}
\beq
A_{LR} = 0.09^{+0.11}_{-0.12} \, , \qquad 
A_{Q} = 0.09(9) \, , \qquad 
A_{S} = 0.12^{+0.10}_{-0.11} \, .
\eeq
Alternatively, $C$ violation may be parameterized in the differential decay rate \eqref{eq:DalitzCharged} by terms that are odd in $T_+ - T_-$.
The $C$-odd parameters are $c$, $e$, $h$, $l$ and these may be fit from the Dalitz distribution simultaneously with the usual $C$-even parameters~\cite{Adlarson:2014aks,Ablikim:2015cmz,Anastasi:2016cdz}.
Only recently has KLOE-2 obtained the first joint fit allowing all four parameters to be nonzero, finding
\beq
c= 4.3(3.4) \times 10^{-3} \, , \qquad
e= 2.5(3.2) \times 10^{-3} \, , \qquad
h= 1.1(9) \times 10^{-2} \, , \qquad
l= 1.1(6.5) \times 10^{-3} \, , 
\eeq
again with no evidence for $C$ violation~\cite{Anastasi:2016cdz}.

Very recently, Ref.~\cite{Gardner:2019nid} proposed a new theoretical framework for $C$ violation in $\eta\to\pi^+\pi^-\pi^0$, which otherwise has been neglected by theoretical study since the 1960s.
In this work, the amplitude is decomposed as
\beq
\A_{\eta \to 3 \pi}(s,t,u) = -\frac{1}{Q^2} \frac{M_K^2 (M_K^2-M_\pi^2)}{3 \sqrt{3} M_\pi^2 F_\pi^2} \M(s,t,u) + \alpha \M_0^{\not C}(s,t,u) + \beta \M_2^{\not C}(s,t,u) \,,
\eeq
where the first term represents the $C$-conserving amplitude in Eq.~\eqref{eq:Q} and $\M_I^{\not C}(s,t,u)$ is the $C$-violating amplitude for isospin $I$.
The latter is parameterized by complex coefficients $\alpha$ and $\beta$ that in principle can be computed from the isoscalar or isotensor interactions, respectively, of the underlying BSM model.
The kinematic dependence of the amplitudes $\M_I^{\not C}(s,t,u)$ was calculated by a decomposition into single-variable functions up-to-and-including $P$-waves in analogy to Eq.~\eqref{eq:Mdecomp},
\begin{align}
\M^{\not C}_{0}(s,t,u) & = (s-t)M_1(u) + (u-s)M_1(t) + (t-u)M_1(s) \,, \notag\\
\M^{\not C}_{2}(s,t,u) & = (s-t)M_1(u) + (u-s)M_1(t) - 2(t-u)M_1(s) 
+ \sqrt{5} \big[M_2(u) - M_2(t) \big] \,, \label{eq:eta3pi-CV-decomp}
\end{align}
and assuming these functions are the same as for the Standard Model amplitude.
A global fit to the KLOE-2 data yielded
\beq
{\rm Re}(\alpha) = -0.65(80) \, , \qquad
{\rm Im}(\alpha) = 0.44(74) \, , \qquad
{\rm Re}(\beta) = -6.3(14.7) \times 10^{-4} \, , \qquad
{\rm Im}(\beta) = 2.2(2.0) \times 10^{-3} \, ,
\eeq
again consistent with no $C$ violation~\cite{Gardner:2019nid}.
This approach is physically more meaningful compared to the ``alphabetical'' parameterization above and moreover makes clear that $\eta \to 3\pi$ is far more sensitive to isotensor than isoscalar BSM physics.
This point was also made long ago: the isoscalar $\eta \to 3\pi$ transition is suppressed by a large angular momentum barrier~\cite{Prentki:1965tt}.
Bose symmetry requires that an effective $I=0$ interaction should involve at least six derivatives, e.g., 
\beq
\eta \epsilon_{abc} (\partial_\mu \partial_\nu \partial_\lambda \pi^a) (\partial^\mu \partial^\nu \pi^b) (\partial^\lambda \pi^c) \, , 
\eeq
implying a large suppression given the small available phase space.\footnote{As an alternative argument, we note that crossing symmetry requires $\M^{\not C}_0(s,t,u)$ to vanish along the three lines $s=t$, $s=u$, and $t=u$, while $\M^{\not C}_2(s,t,u)$ only vanishes along one $t=u$.
As long as both amplitudes are not varying rapidly across the Dalitz plot, this implies that 
$\M^{\not C}_0$ is generally more suppressed kinematically compared to $\M^{\not C}_2$.} 

\subsection{Lepton flavor violation}\label{sec:LFV}

Another class of fundamental symmetry tests in $\pi^0$, $\eta$, and $\eta'$ decays concerns processes that violate charged-lepton flavor.
It is well-known that massive neutrinos are lepton flavor nonconserving, but their impact on flavor violation for charged leptons is negligible.
A positive detection of the latter would be unambiguous evidence for BSM physics (see Ref.~\cite{Calibbi:2017uvl} for a review).
Examples are
\beq
\pi^0,\eta,\eta'\to\mu^+e^-+\text{c.c.} \,, \qquad \pi^0,\eta,\eta'\to\gamma\mu^+e^-+\text{c.c.} \,, \qquad \eta,\eta'\to\pi^0\mu^+e^-+\text{c.c.} \,, \qquad \eta'\to\eta\mu^+e^-+\text{c.c.} \,.
\eeq
The PDG cites the following upper limits (90\% C.L.)~\cite{Abouzaid:2007aa,Appel:2000tc,White:1995jc,Appel:2000wg,Briere:1999bp}:
\begin{align}
\BR(\pi^0\to\mu^+e^-+\mu^-e^+) &< 3.6\times 10^{-10}  \,, &
\BR(\pi^0\to\mu^-e^+) &< 3.4\times 10^{-9}  \,, \notag\\
\BR(\eta\to\mu^+e^-+\mu^-e^+) &< 6\times 10^{-6}  \,, &
\BR(\pi^0\to\mu^+e^-) &< 3.8\times 10^{-10}  \,, \notag\\
\BR(\eta'\to\mu^+e^-+\mu^-e^+) &< 4.7\times 10^{-4}  \, .
\label{eq:LFV-exp}
\end{align}
Other processes (e.g., involving additional photons or pseudoscalars in the final state) could be envisioned, but no upper limits seem to have been obtained so far.
Similar matrix elements for other lepton generations have been discussed in the context of 
lepton-flavor-violating $\tau$ decays~\cite{Arganda:2008jj,Celis:2014asa}, where the most rigorous bounds 
have been obtained by the Belle collaboration~\cite{Miyazaki:2007jp}.
Ref.~\cite{Hazard:2016fnc} specifically investigates $\eta$ and $\eta'$ decays (along with those of other quarkonia) and lists the effective BSM operators whose Wilson coefficients can be constrained via limits as in Eq.~\eqref{eq:LFV-exp}.

It seems not unlikely that stricter upper limits on the $\eta$ and $\eta'$ decays might be inferred,
at least with some moderate model dependence, from the rather rigorous bounds on $\mu\to e$ conversion
on nuclei, invoking $\eta^{(\prime)}$ exchange mechanisms.  While the pseudoscalar nuclear matrix elements,
in which $\eta^{(\prime)}$ exchange can appear, do not contribute to coherent conversion, i.e., they involve the
nucleons' spins and hence do not add up in large nuclei (see, e.g., Ref.~\cite{Cirigliano:2009bz} and references therein),
they might still influence spin-dependent conversion~\cite{Cirigliano:2017azj,Davidson:2017nrp}.
The corresponding matrix elements have been discussed in the context of dark matter direct detection~\cite{Hoferichter:2015ipa}, but no limits on the $\eta^{(\prime)}\to\mu e$ decays seem to have been worked
out so far. 

REDTOP even plans to study upper bounds on double lepton flavor violation in decays 
such as $\eta\to\mu^+\mu^+e^-e^-+\text{c.c.}$~\cite{Gatto:2016rae,Gonzalez:2017fku}.  These would likely not be directly constrained
from $\mu\to e$ conversion, and therefore pave the way towards an investigation of another set of BSM operators and processes.

\section{Searches for new light hidden particles}
\label{sec:BSM-lightparticles}

New symmetries and forces have long been theorized to exist beyond the Standard Model.
These interactions are mediated by scalar or vector particles that can be searched for experimentally and a discovery would have profound implications for our understanding of fundamental physics.
Traditional motivations for extended scalar and gauge sectors derive from electroweak symmetry breaking or unification, in which case new states would lie around or (far) above the TeV scale~\cite{Hewett:1988xc,Leike:1998wr,Langacker:2008yv}.
However, other top-down models predict much lighter mediators, below the GeV scale (see, e.g., Refs.~\cite{Fayet:1980rr,Goodsell:2009xc}). 
Such light hidden states would have escaped detection thus far if they are very weakly coupled to the Standard Model.

Meson studies have a unique opportunity to discover new vector or scalar mediators in the MeV--GeV mass range~\cite{Nelson:1989fx,Fayet:2006sp}.
To have a sizeable new physics rate, new particles must be light enough to be produced on-shell in meson decays; they are detected when decaying into $e^+ e^-$ or other visible final states if kinematically allowed.
Below the MeV scale, astrophysical constraints become very stringent~\cite{An:2013yfc,Redondo:2013lna} and production rates at colliders must be very suppressed.
Consequently, meson factories have little sensitivity to ultralight particles (far) below the MeV scale, such as QCD axions or other weakly interacting sub-eV particles (WISPs), which must be tested elsewhere~\cite{Jaeckel:2010ni}.
In addition to production via meson decays, light-meson facilities also have the capacity to search for new light states produced directly in primary collisions~\cite{Babusci:2014sta,Fanelli:2016utb,Aloni:2019ruo}.

Dark matter provides one motivation for light mediator particles. 
Dominating the matter density of the Universe, its cosmological abundance and stability may point toward the existence of a dark sector with its own forces, symmetries, and spectrum perhaps as rich as the Standard Model.
Annihilation into light scalar or vector mediators provides an efficient channel to set the dark matter relic density~\cite{Boehm:2003hm,Pospelov:2007mp}.
A conserved gauge charge for dark matter may also explain why it is stable (in analogy with stability of the electron)~\cite{Boehm:2003hm,Pospelov:2007mp,Ackerman:mha,Feng:2009mn}.
In these theories, both dark matter and baryons may share a unified baryogenesis mechanism that can account for the similar cosmic densities of both forms of matter~\cite{Davoudiasl:2012uw,Petraki:2013wwa,Zurek:2013wia}.

Recently, MeV--GeV-scale mediators have gained a lot of attention in the phenomenology community, driven partly by several observational anomalies.
These include several purported excesses in high-energy cosmic rays, which could be explained by dark matter annihilation~\cite{ArkaniHamed:2008qn,Pospelov:2008jd}.
Scalar- or vector-mediated dark forces can also explain long-standing issues with galactic rotation curves and other small scale structure observations within the framework of self-interacting dark matter (see Ref.~\cite{Tulin:2017ara} for a review).
Lastly, there are the muon $g-2$ anomaly~\cite{Gninenko:2001hx,Fayet:2007ua,Pospelov:2008zw} and an anomalous $e^+ e^-$ resonance observed in $^8$Be decay~\cite{Krasznahorkay:2015iga}.
While none of these 
hints are definitive, if interpreted in terms of new physics all point toward mediator particles in the MeV--GeV mass range.
This in turn has inspired new approaches for discovering new physics at high-luminosity experiments that are complementary to high-energy colliders~\cite{Fayet:2007ua,Reece:2009un,Bjorken:2009mm,Batell:2009yf} (see Refs.~\cite{Essig:2013lka,Alekhin:2015byh,Alexander:2016aln,Battaglieri:2017aum} for more recent reviews).
If light mediators are discovered, this could provide the first glimpse into the unknown physics of dark matter.

Within the broader landscape of searches for light hidden particles, precision studies of $\eta,\eta^\prime$ mesons are noteworthy for several reasons.
First, they predominantly probe the couplings of hidden particles to light quarks (and gluons), complementary to tests from the leptonic and heavy flavor sectors.
Second, $\eta$ meson decays are a leading production channel for inclusive hidden particle searches at LHCb~\cite{Ilten:2016tkc,Ilten:2019xey} and various experiments relying on proton beam dumps~\cite{Batell:2009di,deNiverville:2011it,Gninenko:2012eq,Alekhin:2015byh,Berlin:2018pwi,Tsai:2019mtm}.
Exclusive searches within the $\eta$ (and $\eta^\prime$) system are therefore essential for disambiguating any positive signals.
Third, while many models can yield $\ell^+ \ell^-$ resonances, tagged $\eta,\eta^\prime$ decays allow for easy discrimination between different models through other particles in the final state.
Below, we discuss the different signatures from vector, scalar, and pseudoscalar mediators.
Fourth, while proton beam dump experiments will far exceed the typical luminosity for dedicated $\eta,\eta^\prime$ factories, both will explore complementary regimes for light mediators; the former targets long-lived mediators decaying far from the interaction point~\cite{Bergsma:1985is,Bernardi:1985ny,Blumlein:1990ay,Blumlein:1991xh,Astier:2001ck,Gninenko:2012eq,Alekhin:2015byh,Berlin:2018pwi,Tsai:2019mtm} or invisible decays to dark matter~\cite{Batell:2009di,deNiverville:2011it}, while the latter probes prompt mediator decays.

The main focus of this section is on fully visible signatures from BSM particles produced on-shell in $\eta,\eta^\prime$ decays.\footnote{We limit our attention to BSM interactions that are $CP$-conserving, which otherwise are subject to stringent EDM constraints~\cite{Kirpichnikov:2020tcf,Kirpichnikov:2020lws}.}
There are three (nonexclusive) search strategies: {\it (i) resonance searches,} i.e.~bump-hunting, {\it (ii) rare decays,} channels that are highly suppressed in the Standard Model but can be mimicked by BSM physics (without symmetry breaking), and {\it (iii) displaced vertices}, potentially down to $\mathcal{O}(100\, \mu m)$~\cite{t-REDTOP}.
Lastly, we also discuss invisible final states in $\eta,\eta^\prime$ decays.

\subsection{Vector bosons from new gauge symmetries}\label{sec:BSM-vector}

Light $U(1)^\prime$ vector bosons (generically dubbed $Z^\prime$) can be searched for in $\eta,\eta^\prime$ decays.
Signatures and rates depend on the gauge coupling $g^\prime$ and mass $m_{Z^\prime}$, as well the gauge charges $Q^\prime$ assigned to Standard Model particles.
The $Z^\prime$ interactions with Standard Model fermions $f$ can be written as
\begin{equation} \label{eq:generalZprime}
\mathcal{L}_{\rm int} = - g^\prime j_{Z^\prime}^\mu Z^\prime_\mu \, , 
\end{equation}
where $j_{Z^\prime}^\mu = \sum_i Q_i^\prime \bar{f}_i \gamma^\mu f_i$ is the corresponding current.
The minimal framework along these lines is that of a ``hidden'' or ``secluded'' $U(1)^\prime$ with $Q^\prime = 0$ for all Standard Model particles.
However, the physical $Z^\prime$ field can still inherit couplings to quarks and leptons through hypercharge kinetic mixing~\cite{Okun:1982xi,Holdom:1985ag} or mass-mixing with the $Z$ boson~\cite{Babu:1997st}.
The former is the well-known ``dark photon'' model, with the $Z^\prime$ (often denoted $A^\prime$) coupling to the $P$-conserving electromagnetic current $j_{\rm em}^\mu$.
We discuss this model further in Sect.~\ref{sec:dark_photons}.
The latter has been dubbed a ``dark $Z$'' and couples the $Z^\prime$ to the $P$-violating weak neutral current $j_{\rm NC}^\mu$~\cite{Davoudiasl:2012ag}.
The mass of the $Z^\prime$ can arise through the Higgs mechanism, in which case additional scalar mediators may be relevant for phenomenological studies~\cite{Schabinger:2005ei}, or through the Stueckelberg mechanism with no extra physical degrees of freedom~\cite{Kors:2004dx,Feldman:2007wj}.

More general possibilities arise if Standard Model fermions carry $U(1)^\prime$ charges.
Anomaly cancellation puts a constraint on the charges $Q^\prime$ to preserve a consistent quantum theory.
If couplings are universal for all generations, there are only two possibilities without introducing exotic fermions: $Q^\prime$ must be proportional to either $Y$ (hypercharge) or $B-L$, with the latter requiring three right-handed neutrino fields.
Other possibilities, such as $U(1)_B$, are allowed only by introducing additional exotic fermions that carry electroweak quantum numbers.
Collider searches for the anomaly-canceling fermions can provide strong indirect bounds on the light $Z^\prime$~\cite{Dobrescu:2014fca}.
If the anomaly-canceling fermions are chiral under $U(1)^\prime$, there are also stringent constraints from rare decays (e.g., $K \to \pi Z^\prime$, $B \to K Z^\prime$, $Z \to Z^\prime \gamma$) due to enhanced longitudinal production of a light $Z^\prime$~\cite{Dror:2017nsg,Dror:2017ehi}.

In this section, we focus on the purely vector case where couplings are independent of chirality ($Q^\prime_{f_L} = Q^\prime_{f_R} \equiv Q^\prime_{f}$) and the $Z^\prime$ can be assigned $J^P = 1^-$.\footnote{Light axial (or mixed) vectors have received much less attention in the literature (see, e.g.,~\cite{Dobrescu:2014fca,Kahn:2016vjr}).
The well-known connection with $\eta,\eta^\prime$ physics is $\eta,\eta^\prime \to \ell^+ \ell^-$, mediated by an off-shell $Z^\prime$ in the $s$-channel (discussed in Sect.~\ref{sec:weak}).
Production of an on-shell $Z^\prime$ through its axial vector couplings in $\eta,\eta^\prime$ decays has not been studied in the literature to our knowledge.
}
One leading production process for $\eta,\eta^\prime$ decays is $\eta,\eta^\prime \to Z^\prime \gamma$. 
This is related to the usual triangle diagram for $\eta, \eta^\prime \to \gamma \gamma$ with one $\gamma$ replaced by a $Z^\prime$.
Assuming ideal mixing for the vector mesons, we present a general formula for these decays following VMD:
\begin{align}
\frac{\Gamma(\eta \to Z^\prime \gamma)}{\Gamma(\eta \to \gamma \gamma)} &=
\frac{2 \alpha^\prime}{\alpha_{\rm em}} \left(1 - \frac{m_{Z^\prime}^2}{\meta^2} \right)^3 \left| \frac{ 9 (Q^\prime_u - Q^\prime_d) F_{\rho_0}(m_{Z^\prime}^2) + 3 (Q^\prime_u + Q^\prime_d) F_\omega(m_{Z^\prime}^2) + 6 \sqrt{2} Q^\prime_s (F_q/F_s) \tan\phi \, F_\phi(m_{Z^\prime}^2) }{10 - 2 \sqrt{2} (F_q/F_s) \tan\phi  } \right|^2 \,, \notag \\
\frac{\Gamma(\eta^\prime \to Z^\prime \gamma)}{\Gamma(\eta^\prime \to \gamma \gamma)} 
&= \frac{2 \alpha^\prime}{\alpha_{\rm em}} \left(1 - \frac{m_{Z^\prime}^2}{\metap^2} \right)^3 \left| \frac{ 9 (Q^\prime_u - Q^\prime_d) F_{\rho_0}(m_{Z^\prime}^2) + 3 (Q^\prime_u + Q^\prime_d) F_\omega(m_{Z^\prime}^2) - 6 \sqrt{2} Q^\prime_s (F_q/F_s) \cot\phi \, F_\phi(m_{Z^\prime}^2) }{10 + 2 \sqrt{2} (F_q/F_s) \cot\phi  } \right|^2 \, , \label{eq:etaZpg}
\end{align}
where $\alpha^\prime = g^{\prime 2}/(4\pi)$ and $F_V(q^2)$ are form factors.
For $\eta$--$\eta^\prime$ mixing, we assume a common mixing angle $\phi$ in the quark-flavor basis and $F_{q,s}$ are the relevant decay constants (see Sect.~\ref{sec:large-Nc}).\footnote{This mixing angle scheme is compatible with present experimental constraints for $\eta$--$\eta^\prime$ mixing. 
Previous works~\cite{Tulin:2014tya,Batell:2014yra,Ilten:2018crw} have adopted the single-angle mixing scheme in the octet--singlet basis, which although known to be broken both experimentally and theoretically, typically suffices for phenomenological estimates.}
For the $\eta^\prime$, the fact that $\BR(\eta^\prime \to \pi^+ \pi^- \gamma)$ is an order of magnitude larger than $\BR(\eta^\prime \to \gamma\gamma)$ suggests that $\BR(\eta^\prime \to \pi^+ \pi^- Z^\prime)$ may be a promising channel for BSM searches (in particular, for a $Z^\prime$ that mixes with the $\rho^0$ meson), though further study is required to predict the branching rates.
For completeness, we also note a similar decay formula for the $\pi^0$
\beq \label{eq:pi0Zpg}
\frac{\Gamma(\pi^0 \to Z^\prime \gamma)}{\Gamma(\pi^0 \to \gamma \gamma)} = \frac{2 \alpha^\prime}{\alpha_{\rm em}} \left(1 - \frac{m_{Z^\prime}^2}{\mpiz^2} \right)^3 \left| \frac{3}{2} (Q^\prime_u + Q^\prime_d) F_{\rho_0}(m_{Z^\prime}^2) + \frac{1}{2} (Q^\prime_u - Q^\prime_d) F_\omega(m_{Z^\prime}^2) \right|^2 \, .
\eeq
Under VMD, the form factors take the standard Breit--Wigner form $F_V(q^2)= (1 - q^2/m_V^2 - i \Gamma_V/m_V)^{-1}$.
For $Q^\prime \propto Q$ (dark photon model), the term under the absolute value is the (normalized) singly-virtual transition form factor
$\bar F_{\eta\gamma^*\gamma^*}\big(q^2,0\big) = F_{\eta\gamma^*\gamma^*}\big(q^2,0\big)/F_{\eta\gamma\gamma}$, discussed in Sect.~\ref{sec:DRetaTFF}, evaluated at $q^2 = m_{Z^\prime}^2$.
In the general case, the terms become a linear combination of form factors and the methods of Sect.~\ref{sec:DRetaTFF} could be used to refine theoretical predictions beyond VMD.

\subsubsection{Dark photons}
\label{sec:dark_photons}

The dark photon is the benchmark model for gauge mediators accessible at low energies~\cite{Fayet:2007ua,Reece:2009un,Bjorken:2009mm,Batell:2009yf} (see \cite{Raggi:2015yfk,Fabbrichesi:2020wbt,Filippi:2020kii} for recent reviews).
We have a $U(1)^\prime$ gauge boson $A^\prime$ that couples to electric charge by kinetic mixing with the photon~\cite{Okun:1982xi,Holdom:1985ag}.
The kinetic mixing term is
\beq \label{eq:kinmix}
\mathcal{L}_{\rm kin.mix.} = - \frac{\varepsilon}{2 \cos\theta_W} F^\prime_{\mu\nu} B^{\mu \nu} \, ,
\eeq
where $F^\prime_{\mu\nu}$ ($B_{\mu\nu}$) is the $U(1)^\prime$ (hypercharge $U(1)_Y$) field strength tensor and $\theta_W$ is the weak mixing angle.
Putting the gauge Lagrangian in canonical form requires a redefinition of the neutral gauge fields.
If $m_{A^\prime}$ is nonzero, this amounts to a shift in the photon field $A_\mu \to A_\mu + \varepsilon A^\prime_\mu$, while the $A_\mu^\prime$ and $Z_\mu$ fields are left unchanged (at leading order in $\varepsilon, m_{A^\prime}^2/m_Z^2 \ll 1$).
Therefore, even if Standard Model fields carry no $U(1)^\prime$ charges to begin with, they inherit couplings to the dark photon of the form
\begin{equation} \label{eq:Lint_darkphoton}
\mathcal{L}_{\rm int} = - e \varepsilon j_{\rm em}^\mu A^\prime_\mu \,.
\end{equation}
Since the kinetic mixing parameter $\varepsilon$ is constrained to be small, these couplings are far weaker than electromagnetism.

Dark photons are a feature of many dark sector models where dark matter is charged under a $U(1)^\prime$ gauge symmetry~\cite{Boehm:2003hm,Feldman:2006wd,Pospelov:2007mp,Pospelov:2008jd,Feng:2008mu}.
Its many guises include inelastic~\cite{TuckerSmith:2001hy}, light (sub-GeV)~\cite{Boehm:2003hm}, mirror~\cite{Foot:2014mia}, asymmetric~\cite{Davoudiasl:2012uw,Petraki:2013wwa,Zurek:2013wia}, and self-interacting~\cite{Tulin:2017ara} dark matter models, to name a few.
Assuming that dark matter is neutral under the Standard Model gauge group and vice-versa, the so-called ``vector portal'' of Eq.~\eqref{eq:kinmix} is one of the few renormalizable operators for interactions between the Standard Model and dark sector.
The couplings are suppressed by $\varepsilon$, which may arise by integrating out heavy states charged under both $U(1)^\prime$ and $U(1)_Y$.
This framework came to prominence due to several astrophysical electron/positron excesses reported for indirect detection searches~\cite{Boehm:2003hm,Boehm:2003bt,Pospelov:2007mp,Pospelov:2008jd,ArkaniHamed:2008qn}.
Provided the $A^\prime$ is in the MeV--GeV range, dark matter annihilation into dark photons can yield a flux of $e^+ e^-$ pairs when they decay (without an excess flux of antiprotons). 
Another motivation for the $A^\prime$ has been the apparent discrepancy for $(g-2)_\mu$ measurements~\cite{Bennett:2006fi,Abi:2021gix,Albahri:2021ixb} compared to the Standard Model prediction~\cite{Aoyama:2020ynm}. 
The $A^\prime$ provides one BSM explanation for reconciling this discrepancy, provided $\varepsilon \sim 10^{-2} \text{--} 10^{-3}$~\cite{Fayet:2007ua,Pospelov:2008zw}.

The branching ratio to produce one dark photon per $\eta$ decay is
\begin{equation}
\mathcal{B}(\eta \to A^\prime \gamma) = 2 \varepsilon^2 \mathcal{B}(\eta \to \gamma\gamma) \left|\bar{F}_{\eta \gamma^* \gamma^*}(m_{A^\prime}^2,0)\right|^2 
\left(1-m_{A^\prime}^2/M_\eta^2 \right)^3 \, .
\end{equation}
A similar formula can be written for the $\eta^\prime$.
Branching ratios to two dark photons are suppressed, proportional to $\varepsilon^4$.
Experimental search strategies depend on how the $A^\prime$ decays.
Presently we consider the case where the $A^\prime$ decays into Standard Model particles via Eq.~\eqref{eq:Lint_darkphoton}, predominantly $A^\prime \to e^+ e^-$, $\mu^+ \mu^-$, or $\pi^+ \pi^-$ in the MeV--GeV mass range.\footnote{Hadronic partial widths are calculated from the experimentally measured ratios $\frac{\sigma(e^+ e^- \to \mathcal{F})}{\sigma(e^+ e^- \to \mu^+ \mu^-)}$, where $\mathcal{F}$ is a given hadronic final state~\cite{Ilten:2018crw}.}
The primary strategy is bump-hunting in $\eta, \eta^{\prime} \to \ell^+ \ell^- \gamma$ or $\pi^+ \pi^- \gamma$~\cite{Reece:2009un}. 
The $q^2$-dependence of these channels is already of interest for studies of transition form factors and the $A^\prime$ would appear as a resonance at $q^2=m_{A^\prime}^2$.
For detectors with vertexing capability, another strategy is searching for displaced $\ell^+ \ell^-$ decay vertices in the case that $A^\prime$ decays nonpromptly (but not so long-lived that it escapes the detector invisibly before decaying).
Alternatively, the $A^\prime$ can decay invisibly, e.g., into light dark matter particles.
We defer invisible signatures to Sect.~\ref{sec:invis}.

\begin{figure}[t]
\centering
\includegraphics[width=10cm]{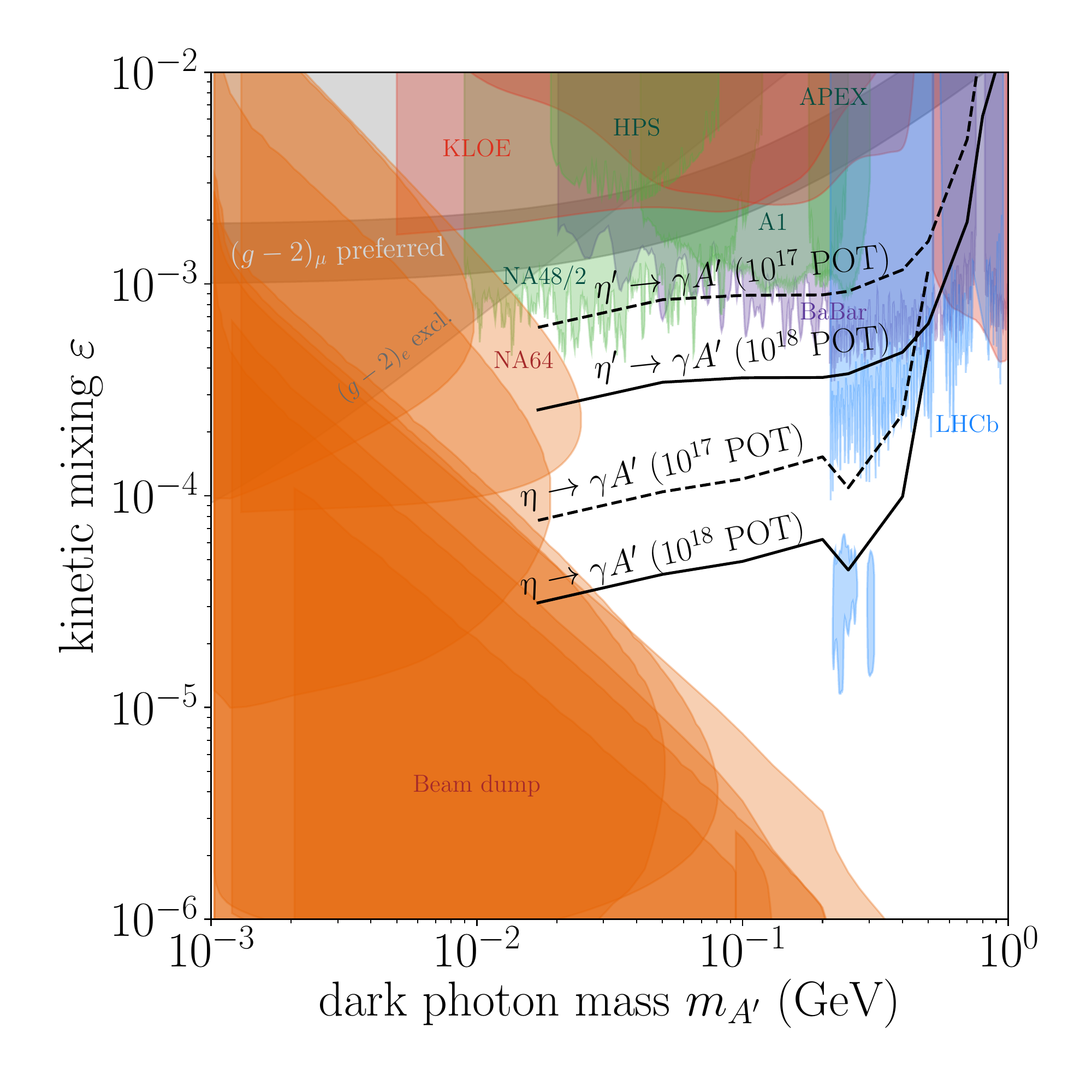}
\caption{Projected sensitivities for visibly-decaying $A^\prime$ from $\eta,\eta^\prime$ decays at REDTOP, for expected flux $10^{17}$ (dashed) or $10^{18}$ (solid) protons on target (POT), depending on experimental facility. 
Dark shaded band is preferred to explain $(g-2)_\mu$ anomaly, while other shaded regions are exclusions.
Figure made in part using the {\tt Darkcast} package~\cite{Ilten:2018crw}.}
\label{fig:darkphoton} 
\end{figure}

We summarize current constraints on the dark photon model in Fig.~\ref{fig:darkphoton}.
The orange exclusion regions show constraints on displaced $A^\prime$ decays from old beam dump experiments~\cite{Bergsma:1985is,Konaka:1986cb,Riordan:1987aw,Bjorken:1988as,Bross:1989mp,Davier:1989wz,Bernardi:1985ny,Blumlein:1990ay,Blumlein:1991xh,Astier:2001ck} that have been recast for the $A^\prime$~\cite{Bjorken:2009mm,Blumlein:2011mv,Blumlein:2013cua,Andreas:2012mt,Gninenko:2012eq,Tsai:2019mtm}, as well recent results from the NA64 experiment~\cite{Banerjee:2019hmi}.
Several among these are proton beam dump experiments (CHARM~\cite{Bergsma:1985is}, PS191~\cite{Bernardi:1985ny}, $\nu$-CAL~\cite{Blumlein:1990ay,Blumlein:1991xh}, and NOMAD~\cite{Astier:2001ck}) that have produced large samples of $\eta$ mesons and limit their possible decays to long-lived dark photons~\cite{Gninenko:2012eq,Tsai:2019mtm}.
Green exclusions show constraints on prompt $A^\prime$ decay from recent fixed target experiments APEX~\cite{Abrahamyan:2011gv}, A1~\cite{Merkel:2014avp}, NA48/2~\cite{Batley:2015lha}, and HPS~\cite{Adrian:2018scb}.
Other collider constraints from KLOE~\cite{Babusci:2012cr,Anastasi:2015qla,Anastasi:2018azp} (red), BaBar~\cite{Lees:2014xha} (purple), and LHCb~\cite{Aaij:2019bvg} (blue) are shown as well.
At LHCb, $\eta$ decays are one of the leading $A^\prime$ production channels for the limits shown~\cite{Ilten:2016tkc,Ilten:2019xey}.
Lastly, following Ref.~\cite{Endo:2012hp}, the gray region is excluded from $(g-2)_e$ at $5\sigma$ based on a recent measurement of $\alpha_{\rm em}$~\cite{Parker:2018vye}.\footnote{The experimental situation concerning $\alpha_{\rm em}$ is not entirely settled.
Recent determinations of $\alpha_{\rm em}$ with atom interferometry for $^{133}$Cs~\cite{Parker:2018vye} and $^{87}$Rb~\cite{Morel:2020dww} differ by $5.4\sigma$ from one another.
The former leads to a value for $(g-2)_e$ that is $-2.4\sigma$ below the Standard Model prediction (i.e., with opposite sign to the $(g-2)_\mu$ anomaly), while the latter yields a value $+1.6\sigma$ above.}
The preferred region to explain the $(g-2)_\mu$ anomaly within $2\sigma$~\cite{Tanabashi:2018oca} (dark band) is excluded.  

Despite null searches to date, the dark photon model remains an active target for many future experiments at the intensity frontier, including future $\eta,\eta^\prime$ factories such as the REDTOP experiment~\cite{Beacham:2019nyx}.
The black contours show the expected REDTOP sensitivities to $\eta, \eta^\prime \to A^\prime \gamma \to \ell^+ \ell^- \gamma$.
The dashed (solid) limits have been forecasted for $10^{17}$ ($10^{18}$) protons on target, corresponding to a possible beam flux for running at CERN (Fermilab or Brookhaven)~\cite{Gatto:2019dhj}.
This reach is complementary to other proposed experiments operating on similar timescales~\cite{Beacham:2019nyx,Tsai:2019mtm}.

\subsubsection{Other hidden photons}

Beyond the dark photon, other light mediator models have been proposed by gauging various global $U(1)$ symmetries for the Standard Model.
Many of these yield dilepton resonances and fall under the purview of dark photon searches.
One candidate is the $Z^\prime_{B-L}$ associated with $U(1)_{B-L}$, which is anomaly-free with three generations of right-handed neutrinos~\cite{Marshak:1979fm}.
Previous works have considered this model in the context of MeV--GeV-scale physics~\cite{Harnik:2012ni,Heeck:2014zfa,Bilmis:2015lja,Ilten:2018crw,Bauer:2018onh}.
For $\eta,\eta^\prime$ mesons, the phenomenology is very similar to the dark photon, appearing as a resonance in 
\beq \label{eq:ZBL}
\eta,\eta^\prime \to Z^\prime_{B-L} \gamma \to \ell^+ \ell^- \gamma \, .
\eeq
However, an important difference between the $A^\prime$ and $Z^\prime_{B-L}$ is that the latter couples to neutrinos.
The dilepton signal is reduced by the large invisible $Z^\prime_{B-L}$ width to neutrinos, the value of which depends on whether they are Dirac~\cite{Heeck:2014zfa} or Majorana~\cite{Bauer:2018onh}.
Also, there are strong experimental constraints from neutrino scattering data~\cite{Harnik:2012ni,Bilmis:2015lja} and presently the gauge coupling limited to be $g_{B-L} \lesssim 10^{-4}$~\cite{Ilten:2018crw,Bauer:2018onh}.
Taking $Q_u^\prime = Q_d^\prime = Q_s^\prime = \tfrac{1}{3}$ in Eq.~\eqref{eq:ZBL}, the branching ratio for the $\eta$ is constrained to be less than $10^{-8}$.
For the $\eta^\prime$, the branching ratio may be as large as $10^{-8}$ near the $\omega$ mass, but otherwise is a few orders of magnitude smaller.
Other anomaly-free $Z^\prime$ candidates can arise by gauging various lepton flavor combinations $L_e - L_\mu$, $L_e - L_\tau$, or $L_\mu - L_\tau$ (see, e.g., Ref.~\cite{Bauer:2018onh}). 
These are less interesting for $\eta,\eta^\prime$ physics without any direct coupling to quarks.

Another important case is inspired by the $^8$Be anomaly reported by Krasznahorkay et al.~\cite{Krasznahorkay:2015iga}. 
They measured the decay of an excited $^8$Be state to the ground state via internal pair creation, \mbox{$^8\textrm{Be}^* \to \,^8\textrm{Be} \; e^+ e^-$}.
By measuring the $e^+ e^-$ opening angle spectrum, they found an anomalous resonance that could be fit by a $16.7\MeV$ boson at $>5\sigma$.
The same authors recently reported a similar signal found in $^4\textrm{He}$ transitions~\cite{Krasznahorkay:2019lyl}.
With no compelling nuclear physics explanation thus far~\cite{Zhang:2017zap}, various light mediators have been proposed.
One candidate is a so-called ``protophobic'' vector boson, dubbed the $X$ boson~\cite{Feng:2016jff,Feng:2016ysn}.
Starting from a generic vector boson model in Eq.~\eqref{eq:generalZprime}, the authors found that the signal process 
\beq
\;^8\textrm{Be}^* \to \,^8\textrm{Be} \; X \to \,^8\textrm{Be} \; e^+ e^-
\eeq
could be explained by $X$ boson couplings satisfying
\beq \label{eq:8Becouplings}
|\varepsilon_p + \varepsilon_n| \approx \frac{0.01}{\sqrt{ {\mathcal B}(X \to e^+ e^-)}} \, , \qquad |\varepsilon_p| < \frac{0.0012}{\sqrt{ {\mathcal B}(X \to e^+ e^-)}}  \, ,
\eeq
where $\varepsilon_{p,n}$ are the $X$ charges of the proton and neutron in units of $e$.\footnote{Following the notation of Refs.~\cite{Feng:2016jff,Feng:2016ysn}, fermion couplings to the $X$ boson are expressed in units of $e$. To connect with our notation in Eq.~\eqref{eq:generalZprime}, we have $g^\prime Q_f^\prime = e \varepsilon_f$. 
The proton and neutron couplings are $\varepsilon_{p} = 2 \varepsilon_{u} + \varepsilon_{d}$ and $\varepsilon_{n} = 2 \varepsilon_{d} + \varepsilon_{u}$, respectively.}
The latter constraint on $\varepsilon_p$ comes from null searches for $\pi^0 \to X \gamma \to e^+ e^- \gamma$ at NA48/2~\cite{Batley:2015lha} and is conveniently evaded if the $X$ boson couples predominantly to neutrons over protons (e.g., $X$ may couple proportional to $B-Q$)~\cite{Feng:2016jff}.
The $X$ boson is also consistent with a $^4$He decay signal, not only in mass but the decay rate as well~\cite{Feng:2020mbt}. 

\begin{figure}[t]
\centering
\includegraphics[width=10cm]{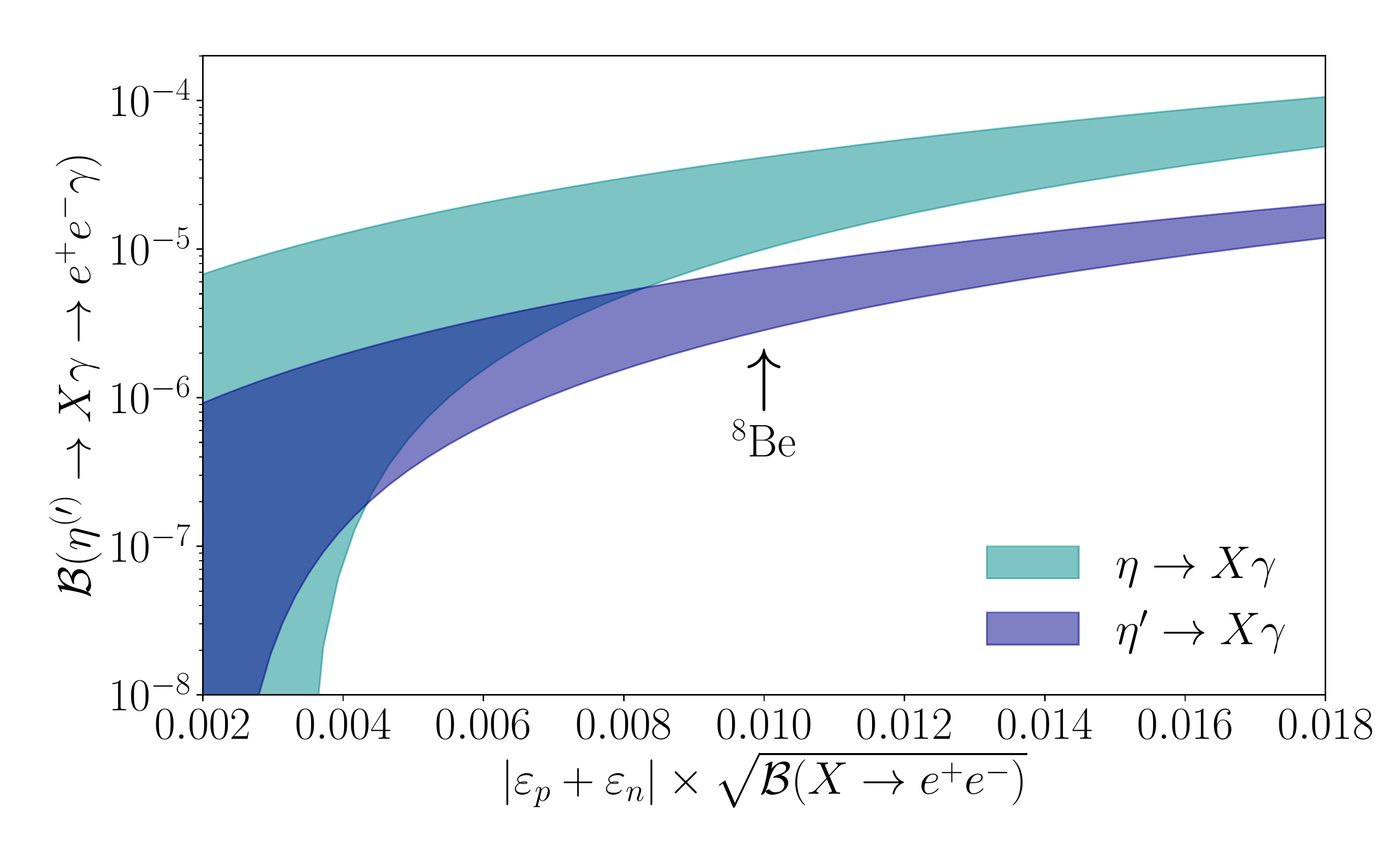}
\caption{Shaded bands show predicted range from the $^8\mathrm{Be}$ anomaly for $\eta,\eta^\prime \to X \gamma \to e^+ e^- \gamma$, detectable as an $e^+ e^-$ resonance at $m_X = 17\MeV$.
Per Eq.~\eqref{eq:8Becouplings}, abscissa should be $\approx 0.01$ to fit $^8\mathrm{Be}$ signal, while band width corresponds to allowed range in $|\varepsilon_p|$ from $\pi^0 \to e^+ e^- \gamma$ constraints.}
\label{fig:Xboson} 
\end{figure}

The protophobic $X$ boson can be searched for as a narrow resonance in
\beq
\eta,\eta^\prime \to X \gamma \to e^+ e^- \gamma \, .
\eeq
According to Eqs.~\eqref{eq:etaZpg} and \eqref{eq:pi0Zpg}, each of the $\eta$, $\eta^\prime$, and $\pi^0$ decay channels depend on different linear combinations of charges.
While $\pi^0$ decay is small (by construction), the situation is favorable for $\eta,\eta^\prime$ studies.
Since the couplings are relatively large, the $\eta^{(\prime)}$ decay rate can be larger compared to other hidden photon models.
Moreover, while ongoing searches at NA64 have constrained the $X$ boson solely through its coupling to the electron~\cite{Banerjee:2019hmi}, $\eta^{(\prime)}$ studies more directly test the quark couplings governing the $^8\mathrm{Be}$ signal.
The expected branching ratios are shown in Fig.~\ref{fig:Xboson}.\footnote{These rates depend on the $X$ charge of the strange quark, $\varepsilon_s$, which is in principle an independent parameter.
Our estimates assume flavor universality, such that $\varepsilon_d = \varepsilon_s$, to avoid possible contributions to flavor-violating observables (see Appendix~C of Ref.~\cite{Dror:2017nsg}).
A direct limit on $\varepsilon_s$ follows from resonance searches in $\phi \to \eta e^+ e^-$ at KLOE~\cite{Babusci:2012cr}, yielding $|\varepsilon_s| < 0.007/\sqrt{\mathcal{B}(X \to e^+ e^-)}$ for $m_X \approx 17\MeV$.}
The abscissa is the combination of parameters entering the $^8\mathrm{Be}$ signal and must be around 0.01 (vertical arrow) to fit the data for $m_X = 17\MeV$~\cite{Feng:2020mbt}.\footnote{Smaller couplings may also be allowed if $m_X$ is larger~\cite{Feng:2016ysn}. This correspondingly reduces the predicted $\eta,\eta^\prime$ branching ratios.}
The $X$ boson can be searched for in parallel with $\eta, \eta^\prime$ transition form factor studies (see Sect.~\ref{sec:DRetaTFF}), though reaching small invariant mass $m_{e^+ e^-} \equiv \sqrt{q^2}$ may be challenging.
For example, data for $\eta \to e^+ e^- \gamma$ from A2 at MAMI includes only $m_{e^+ e^-} > 30\MeV$ in their analysis due to $e^+ e^-$ shower overlap~\cite{Adlarson:2016hpp}. 
New data on $\eta^{(\prime)} \to e^+ e^- \gamma$ will be expected from the JEF experiment in the coming years.

\subsubsection{Leptophobic $U(1)_B$ boson}

New vector mediators interacting primarily with quarks are a blindspot for dark photon searches.
The usual strategy of searching for $e^+ e^-$ or $\mu^+ \mu^-$ resonances can miss a mediator decaying predominantly to hadrons.
The minimal model along these lines is that of a $U(1)_B$ gauge symmetry for baryon number.
This idea was first proposed by Lee and Yang in 1955~\cite{Lee:1955vk} and subsequently discussed extensively in the literature~\cite{Pais:1973mi,Nelson:1989fx,Foot:1989ts,Rajpoot:1989jb,He:1989mi,Carone:1994aa,Bailey:1994qv,Carone:1995pu,Aranda:1998fr,FileviezPerez:2010gw}.
Baryon number symmetry may be related to dark matter as well~\cite{Agashe:2004ci,Farrar:2005zd,Davoudiasl:2010am,FileviezPerez:2010gw,Graesser:2011vj,Duerr:2013lka}.
In these models, dark matter is stabilized since it carries a conserved baryon number charge.
Dark matter can share a unified baryogenesis mechanism with regular baryons, explaining the apparent coincidence that both forms of matter have similar cosmic densities~\cite{Nussinov:1985xr}.
A $U(1)_B$ gauge symmetry has also been proposed as a natural framework for the Peccei--Quinn solution to the strong $CP$ problem~\cite{Foot:1989ts,Duerr:2017amf}.

How one searches for the $U(1)_B$ gauge boson depends on its mass.
For masses below the MeV scale, new baryonic interactions are strongly constrained by low-energy neutron scattering~\cite{Barbieri:1975xy,Leeb:1992qf,Nesvizhevsky:2007by} and other tests for long-range forces~\cite{Adelberger:2003zx}.
For masses above the GeV scale, gauge bosons have been searched for directly at high-energy colliders, e.g., in dijet resonance searches~\cite{Barger:1996kr,Dobrescu:2013coa}.
The intermediate MeV--GeV mass range remains less explored~\cite{Nelson:1989fx,Tulin:2014tya}. 
Constraints from heavy quarkonia and $Z$ decays are relatively weak~\cite{Carone:1994aa,Bailey:1994qv,Carone:1995pu,Aranda:1998fr} and stronger limits can be obtained through studies of meson decays~\cite{Tulin:2014tya}.
Alternatively, $U(1)_B$ gauge bosons may decay invisibly to dark matter.
This case can be tested at neutrino factories by producing a beam of dark matter particles that scatters in the downstream detector~\cite{Batell:2014yra}, or as semi-visible jets at the LHC (for much higher mass around the TeV scale)~\cite{Cohen:2015toa}.
Here we limit our attention to visible decays relevant for $\eta,\eta^\prime$ studies.

The $U(1)_B$ model has an interaction Lagrangian
\beq \label{eq:Lint_Bboson}
\mathcal{L}_{\rm int} =  \left( \tfrac{1}{3} g_B  + \varepsilon e Q_q \right) \bar q \gamma^\mu q  B_\mu - \varepsilon e \bar{\ell} \gamma^\mu \ell B_\mu \, ,
\eeq
where $B$ is the new gauge boson~\cite{Nelson:1989fx} and $g_B$ is its gauge coupling.
The $U(1)_B$ analog of the fine structure constant is $\alpha_B = g_B^2/(4\pi)$.
Since quarks carry both electromagnetic and $U(1)_B$ charges, radiative corrections induce kinetic mixing with the photon~\cite{Carone:1995pu,Aranda:1998fr}.
Hence, Eq.~\eqref{eq:Lint_Bboson} includes subleading photon-like couplings to fermions, where the kinetic mixing parameter $\varepsilon$ is expected to be $\mathcal{O}(g_B e/(4\pi)^2)$.
$B$ boson interactions preserve the low-energy symmetries of QCD, namely $C$-, $P$-, and $T$-invariance, as well as isospin (and $\SU(3)$ flavor symmetry) at zeroth order in $\varepsilon$.

\begin{figure}[t]
\centering
\includegraphics[width=8cm]{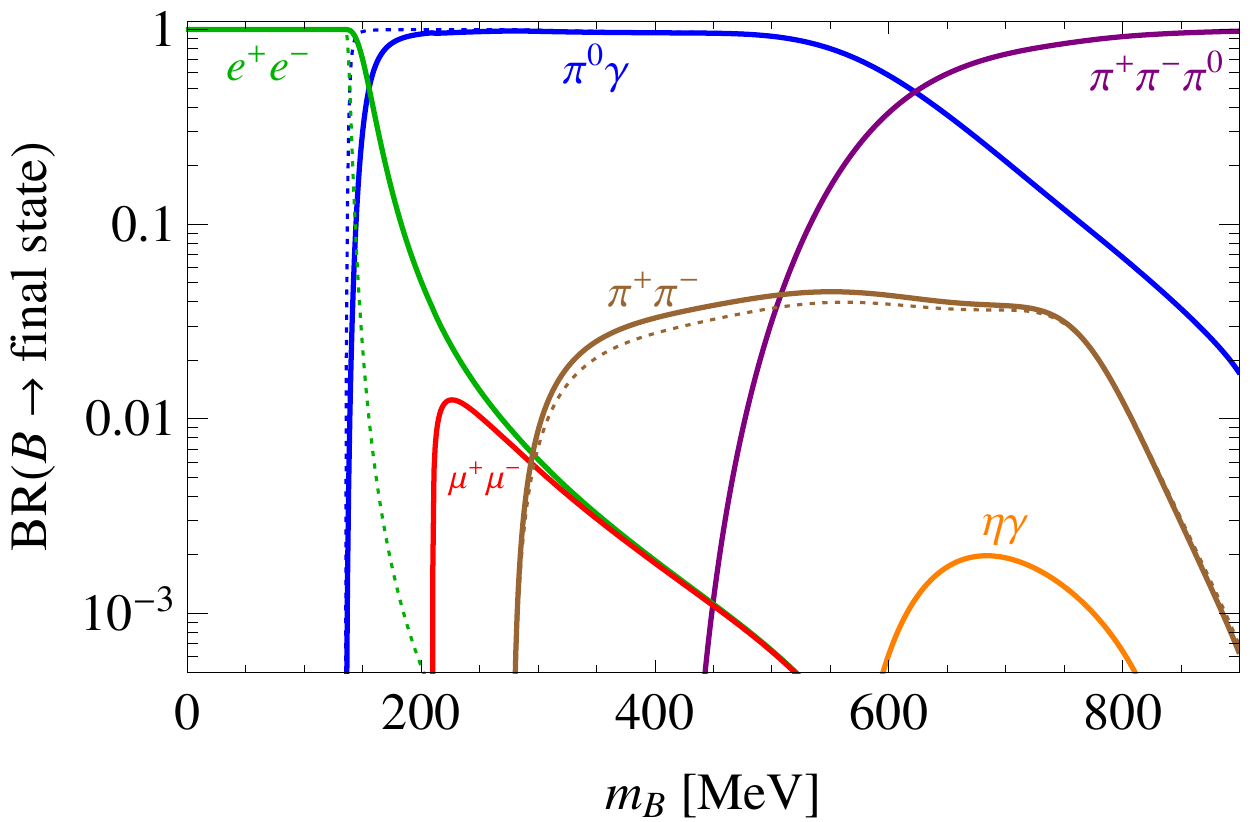}
\caption{$B$ boson branching ratios for kinetic mixing parameter $\varepsilon = g_B e/(4\pi)^2$ (solid lines) and $0.1 \times g_B e/(4\pi)^2$ (dotted lines). $\mathcal{B}(B \to \mu^+ \mu^-)$ for the latter case is not shown and lies two orders of magnitude below the solid red curve. Reprinted from Ref.~\cite{Tulin:2014tya}.}
\label{B-branchingratios} 
\end{figure}

Branching ratios for $B$ boson decay in the MeV--GeV mass range are shown in Fig.~\ref{B-branchingratios}.
These were calculated using VMD in Ref.~\cite{Tulin:2014tya} (see also \cite{Ilten:2018crw}, which extends to larger $m_B$).
Above single-pion threshold, the leading channels are 
\beq
B \to \pi^0 \gamma \quad (M_{\pi^0} \lesssim m_B \lesssim 620 \MeV) \, , \qquad B \to \pi^+ \pi^- \pi^0 \quad (620 \MeV \lesssim m_B \lesssim 1\GeV) \, ,
\eeq
which are not covered in dark photon searches.
These modes are reminiscent of $\omega$ decays since the $B$ boson can be assigned the same quantum numbers $I^G(J^{PC}) = 0^-(1^{++})$ as the $\omega$ meson.
Other subleading decays $B \to \eta \gamma$ and $B \to \pi^+ \pi^-$ are also possible.
The latter is suppressed, being forbidden by $G$-parity, and arises predominantly via $\rho$--$\omega$ mixing.
Since this mixing is poorly constrained away from the $\omega$ peak, these estimates for $B \to \pi^+ \pi^-$ should be viewed with skepticism and are likely an overestimation (discussed in Ref.~\cite{Tulin:2014tya}).
Below single pion threshold, hadronic decays are forbidden and the leading channel is $B \to e^+ e^-$ through kinetic mixing.
The decays $B \to e^+ e^-, \; \mu^+ \mu^-, \; \pi^+ \pi^-$ fall under the purview of dark photon searches, albeit with important differences in production rates between hadronic and leptonic processes.
The {\tt Darkcast} package is useful for mapping dark photon constraints onto the parameter space of the $B$ boson (and other vector mediators)~\cite{Ilten:2018crw}.

\begin{figure}[t]
\centering
\includegraphics[width=10cm]{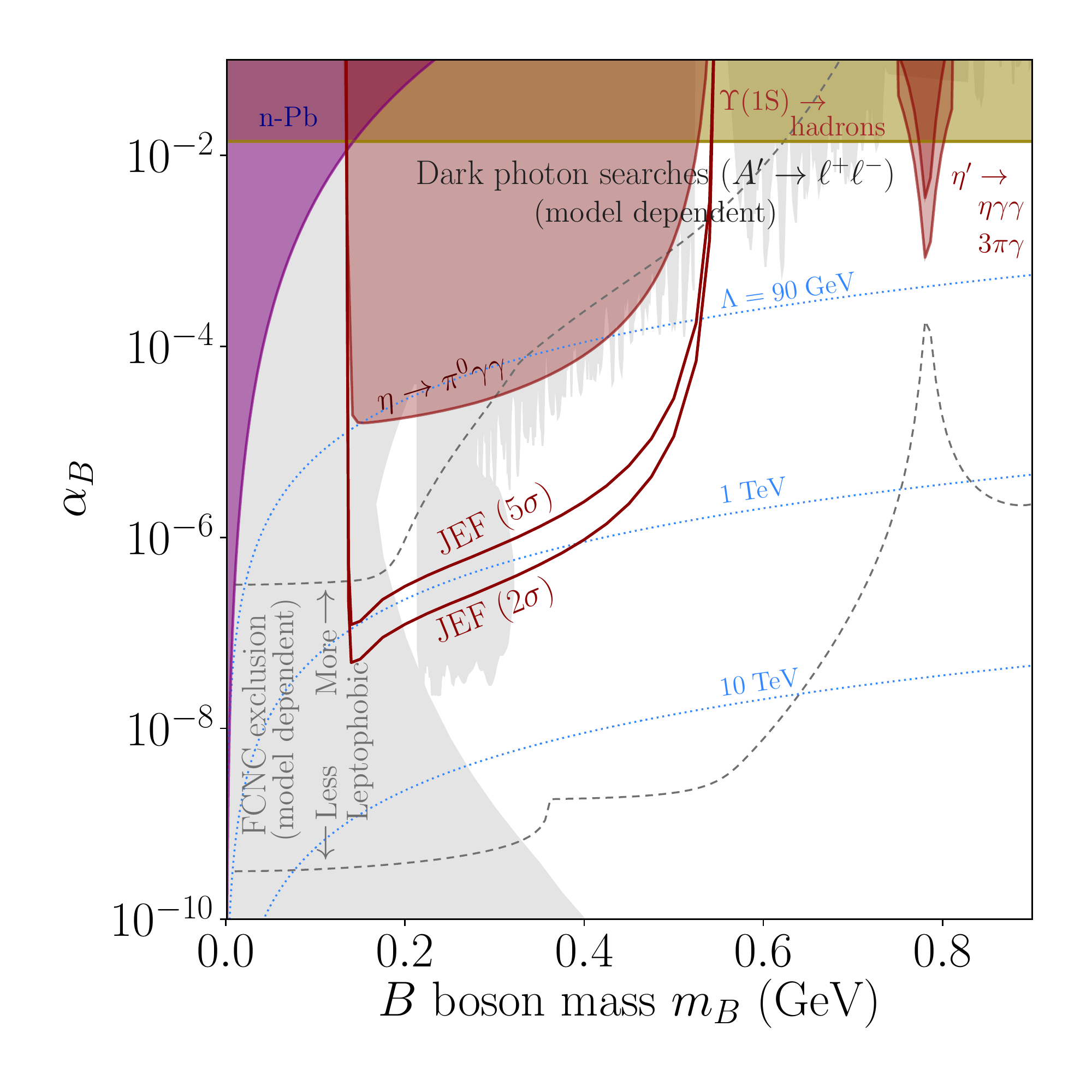}
\caption{
Color-shaded regions are model-independent constraints on leptophobic $B$ boson mass and coupling $\alpha_B$ from rare $\eta, \eta^\prime$ decays (red), hadronic $\Upsilon (1S)$ decays (yellow), and low-energy $n$-Pb scattering (purple).
Solid red contours show the projected $2\sigma/5\sigma$ sensitivity reach for the JEF experiment.
Gray shaded regions and contours are model-dependent and involve leptonic couplings via kinetic mixing $\varepsilon$.
Gray shaded regions are excluded by dark photon searches for dilepton resonances, $A^\prime \to \ell^+ \ell^-$, for $\varepsilon_{0.1}$ (made using \texttt{Darkcast} package~\cite{Ilten:2018crw}).
Gray dashed contours are upper limits on $\alpha_B$ from FCNC $b \to s \ell^+ \ell^-$ and $s \to d \ell^+ \ell^-$, for $\varepsilon_{0.001}$ (upper line) and $\varepsilon_1$ (lower line).
Dotted blue contours denote the
upper bound on the mass scale $\Lambda$ for new 
electroweak fermions needed for anomaly cancellation.
The contour $\Lambda=90\GeV$ is an approximate upper limit on $\alpha_B$ given null direct searches from LEP. 
}
\label{B-reach-map} 
\end{figure}

The $B$ boson mimics rare decay modes for the $\eta$, $\eta^\prime$, and other mesons~\cite{Nelson:1989fx,Tulin:2014tya}.
These include 
\beq \label{eq:eta_B_decays}
\eta \to B \gamma  \to \pi^0 \gamma \gamma \, , 
\quad 
\eta^\prime \to B \gamma \to \pi^0 \gamma \gamma \, ,
\qquad
\eta^\prime \to B \gamma \to  \pi^+ \pi^- \pi^0 \gamma \, ,
\qquad 
\eta^\prime \to B \gamma \to  \eta \gamma \gamma
\, .
\eeq
At present, the only constraints on these modes are based on total rates, which are shown in Fig.~\ref{B-reach-map} (excluded regions shaded red).
For the first three modes in \eqref{eq:eta_B_decays} we assume the $B$ boson contribution does not oversaturate the observed branching ratios~\cite{Tanabashi:2018oca} at $2\sigma$ (the second of these is rather weak and is not shown in Fig.~\ref{B-reach-map}), while for the last mode we take the 90\% upper limit recently reported from BESIII~\cite{Ablikim:2019wsb}.

A more promising strategy is bump-hunting in invariant mass distributions, with the $B$ boson appearing as a narrow resonance.\footnote{Standard Model contributions to \eqref{eq:eta_B_decays} are an irreducible background to $B$ boson searches.
These are discussed in Sects.~\ref{sec:eta-omegag} and \ref{sec:eta-pi0gg}.}
The Belle experiment published the first $B$ boson constraint along these lines, searching for a $\pi^+ \pi^-$ resonance associated with the decay $\eta \to B \gamma \to \pi^+ \pi^- \gamma$~\cite{Won:2016pjz}.
However, since $B \to \pi^+ \pi^-$ is suppressed, the resulting constraint is not as stringent as the total rate limit from $\eta \to \pi^0 \gamma \gamma$.
The golden mode for the $B$ boson is a $\pi^0 \gamma$ resonance~\cite{Tulin:2014tya}.
Since this channel is subdominant for the $A^\prime$ (and other hidden photons), it is a unique signature for the $B$ boson.
The JEF experiment plans to search for a $\pi^0 \gamma$ resonance in $\eta \to \pi^0 \gamma \gamma$, covering the mass range $0.14 < m_B < 0.54 \GeV$~\cite{JEF-PAC42}.
Fig.~\ref{B-reach-map} shows the projected sensitivities at the $2\sigma/5\sigma$ levels (red solid lines) based on 100~days of beam time.
This search will be improving over the total rate limit by two orders of magnitude, reaching $\alpha_B \sim 10^{-7}$.
Similar studies are also in the works by the KLOE-2 collaboration~\cite{PerezdelRio:2019liy}.
They will search for a $\pi^0 \gamma$ resonance via $\phi \to \eta B \to \eta \pi^0 \gamma$ and $\eta \to B \gamma \to \pi^0 \gamma \gamma$.
The $\eta^\prime$ can be used to explore larger $B$ boson masses.
JEF searches for new $3\pi$ resonances in $\eta^{\prime}\to\pi^+\pi^-\pi^0\gamma$ (beyond the $\omega$ meson) 
or in direct photoproduction $\gamma p\to Bp\to\pi^+\pi^-\pi^0 p$
would further extend the experimental reach up to $m_B \sim 1\GeV$. 

In addition to $\eta,\eta^\prime$ physics, Fig.~\ref{B-reach-map} shows several other constraints on the $B$ boson model.
The color-shaded regions denote model-independent exclusions for $\alpha_B$.
These include hadronic $\Upsilon (1S)$ decays~\cite{Aranda:1998fr} (yellow) and 
low-energy $n$-Pb scattering~\cite{Barbieri:1975xy} (purple).
The gray-shaded regions denote exclusions from dark photon searches for $A^\prime \to \ell^+ \ell^-$~\cite{Babusci:2012cr,Konaka:1986cb,Riordan:1987aw,Bjorken:1988as,Bross:1989mp,Davier:1989wz,Banerjee:2019hmi,Tsai:2019mtm,Astier:2001ck,Bernardi:1985ny,Batley:2015lha,Aaij:2019bvg} that have been recast for the $B$ boson~\cite{Ilten:2018crw}.
These constraints are more model-dependent, depending on the value of the kinetic mixing parameter $\varepsilon$ governing $B \to \ell^+ \ell^-$.
Here we consider different values denoted by $\varepsilon_x$, defined by 
\beq \label{eq:eps_x}
\varepsilon_x = x \frac{e g_B}{(4\pi)^2} \, .
\eeq
The ``natural'' value from one-loop $B$--$\gamma$ mixing is $\mathcal{O}(\varepsilon_1)$, but we consider smaller value $\varepsilon_{0.1}$ in Fig.~\ref{B-reach-map} to allow for more decoupling from leptons.

It is well-known that $U(1)_B$ is anomalous in the Standard Model, requiring new baryonic fermions with electroweak charges~\cite{Nelson:1989fx,Foot:1989ts}.
These fermions are necessarily electrically charged and direct searches at LEP place a lower limit $\sim 90 \GeV$~\cite{Dobrescu:2014fca}.
If the fermions are vector-like under $U(1)_B$, they must be chiral under $\SU(2)_L \times U(1)_Y$.
This leads to observable deviations in Higgs decays $h \to \gamma\gamma, Z \gamma$~\cite{Bizot:2015zaa}.
Precision Higgs studies and direct searches at the LHC constrain this case to be only marginally viable~\cite{Dror:2017nsg}.
Alternatively, if the fermions are vector-like under $\SU(2)_L \times U(1)_Y$, they must be chiral under $U(1)_B$.
In this case, perturbativity implies a maximum mass scale $\Lambda$ where these fermions can appear, $\Lambda \sim 4\pi m_B/g_B$~\cite{Williams:2011qb,Dobrescu:2014fca}, which is shown in Fig.~\ref{B-reach-map} by the dotted lines.
If the $B$ boson is discovered within the sensitivity range of JEF, these states would be within the reach of high-energy colliders. 

The case where anomaly-canceling fermions are chiral under $U(1)_B$ is subject to additional constraints from flavor-changing neutral current (FCNC) transitions $b \to sB$ and $s\to dB$ induced the emission of a longitudinal $B$ boson~\cite{Dror:2017nsg,Dror:2017ehi}.
For leptonic $B$ boson decay, this yields
\beq \label{eq:FCNC_leptons}
B_q^\pm \to K^{(*)\pm} \ell^+ \ell^- \, , \quad
B_q^0 \to K^{(*)0} \ell^+ \ell^- \, , \quad
K^\pm \to \pi^\pm \ell^+ \ell^- \, , \quad
K^0_L \to \pi^0 \ell^+ \ell^- \, .
\eeq
Even though the leptonic decay $B \to \ell^+ \ell^-$ is typically suppressed above single pion threshold, these processes have a longitudinal $1/m_B^2$ enhancement that results in very stringent constraints.
Following Ref.~\cite{Dror:2017nsg} and assuming the rates for \eqref{eq:FCNC_leptons} do not oversaturate the observed values at $2\sigma$ (and taking the 90\% upper limit for $K^0_L \to \pi^0 \ell^+ \ell^-$)~\cite{Tanabashi:2018oca}, we find the exclusion contours shown in Fig.~\ref{B-reach-map} (dashed gray lines), which depend on the value of $\varepsilon$ assumed.
For a natural value $\varepsilon_1$
much of the parameter space accessible to $\eta,\eta^\prime$ studies is excluded~\cite{Dror:2017nsg}, but for smaller (more strongly leptophobic) values these constraints are weakened.
In principle, other FCNC decays such as 
\beq
B_q^{\pm} \to K^{(*)\pm} \pi^+ \pi^- \, , \quad
B_q^{0} \to K^{(*)0} \pi^+ \pi^- \, , \quad
B_q^{\pm,0} \to K^{\pm,0} \eta \gamma \, , \quad
K_L^0 \to \pi^0 \pi^0 \gamma 
\eeq
can provide a more model-independent constraint on hadronic $B$ boson decays that are largely independent of $\varepsilon$ (e.g., $B \to \pi^0 \gamma, \pi^+ \pi^-, \eta \gamma$). 
However, one must remember that $B \to \pi^+ \pi^-$ has its own uncertainties due to the modeling of $\rho$--$\omega$ mixing, discussed above, so conclusions from this channel would need to be carefully assessed.

\subsection{Scalar bosons from extended Higgs sectors}\label{sec:scalars}

The existence of the Higgs boson has fueled the possibility of other fundamental spin-zero particles beyond the Standard Model, perhaps much lighter than the electroweak scale.
Such states can been proposed within the contexts of inflation~\cite{Shaposhnikov:2006xi,Bezrukov:2009yw}, supersymmetry~\cite{Dermisek:2006py,He:2008zw}, dark sectors~\cite{Pospelov:2008jd,Schmidt-Hoberg:2013hba,Krnjaic:2015mbs}, the relaxion~\cite{Graham:2015cka}, and anomalous measurements for $^8\textrm{Be}$~\cite{Ellwanger:2016wfe} and $g-2$ for the muon~\cite{TuckerSmith:2010ra}, among others (see \cite{Essig:2013lka,Alekhin:2015byh,Beacham:2019nyx} for further references).
Here we consider the possibilities for $\eta,\eta^\prime$ studies to discover such states if they fall within the MeV--GeV mass range.

We consider a new fundamental $J^P = 0^+$ scalar boson $S$ coupling to quarks and gluons.
If kinematically allowed, $S$ can be produced in $\eta,\eta^\prime$ decays and then it may decay $S \to \ell^+ \ell^-$ if it couples to leptons ($\ell = e,\mu$).
The signal channels are
\beq \label{eq:scalar_channels}
\eta, \eta^\prime \to \pi^0 S \to \pi^0 \ell^+ \ell^- 
\, , \quad
\eta^\prime \to \eta S 
\to \eta \ell^+ \ell^- 
\, ,
\eeq
with $\ell^+ \ell^-$ invariant mass equal to $m_S$.
These channels have been discussed previously in the context of $C$ violation (Sect.~\ref{sec:BSMCV}).
While $C$ invariance forbids Eq.~\eqref{eq:scalar_channels} as single-photon processes, they arise in the Standard Model via $C$-conserving two-photon exchange (see Sect.~\ref{sec:etapi0ll}).
This yields, for example, $\BR(\eta\to\pi^0\ell^+\ell^-) \sim 2\ldots3 \times 10^{-9}$, which is more than three orders of magnitude below present sensitivities (see Table~\ref{tab:CVdecays}).
This channel was considered long ago in the context of searching for the (possibly very light) Higgs boson in the Standard Model~\cite{Ellis:1975ap}.

On the other hand, if the new scalar $S$ has no lepton couplings, its decays are dominated by $S \to 2\pi$ if kinematically allowed or $S \to \gamma\gamma$ otherwise~\cite{Batell:2018fqo}.
The former may be observed as a resonance in the Dalitz distribution for $\eta \to 3\pi$, while the latter can be searched for as a $2\gamma$ resonance in $\eta \to \pi^0 \gamma \gamma$ (and analogous channels for the $\eta^\prime$).
Alternatively, $S$ may be invisible at $\eta,\eta^\prime$ facilities if it decays to neutrinos, dark matter, or is otherwise too long-lived to be detected.
This case is constrained by searches for FCNCs with invisible final states~\cite{Bird:2004ts,Artamonov:2009sz}, as well as searches for long-lived new states with displaced decays~\cite{Beacham:2019nyx}.

To discuss production in $\eta,\eta^\prime$ decays, we begin with the general interaction Lagrangian for $S$ with quarks and gluons
\beq \label{eq:Lint_scalar}
\L_{\text{int}} = - \sum_{q=u,d,s,c,b,t} \kappa_q \frac{m_q}{v} \bar{q} q S + \kappa_G \frac{\alpha_s}{12\pi v} S G^{a}_{\mu\nu} G^{a\mu\nu} \, ,
\eeq
following Ref.~\cite{Batell:2018fqo}.
Here $\kappa_q$ and $\kappa_G$ are model-dependent coefficients and $v=( \sqrt{2} G_F)^{-1/2} \approx 246\GeV$ is the Higgs vacuum expectation value.
The minimal and most widely-studied framework along these lines is the Higgs mixing model~\cite{Bird:2004ts,Bezrukov:2009yw,Clarke:2013aya,Winkler:2018qyg}. 
In this case, $S$ couples solely via the ``Higgs portal'' operator $|H|^2$~\cite{Patt:2006fw}, which induces mixing between $S$ and the Standard Model Higgs boson. 
$S$ couples to fermions proportional to mass, with proportionality $\kappa_q = \sin\theta_S$, where $\theta_S$ denotes the scalar mixing angle.
On the other hand, more general constructions are viable as well, such as $S$ coupling preferentially to light quarks~\cite{Batell:2017kty}.
In this case, $\eta,\eta^\prime$ decays provide a key probe to test this model~\cite{Batell:2018fqo}.
For the gluon coupling, $\kappa_G$ is nonzero if there exist new BSM states carrying color, but otherwise $\kappa_G=0$ in minimal models.

Next, we derive the $\eta,\eta^\prime \to \pi^0 S$ and $\eta^\prime \to \eta S$ transition amplitudes.
To start, one integrates out the heavy quarks $(c,b,t)$ and expresses Eq.~\eqref{eq:Lint_scalar} as
\beq \label{eq:Lint_scalar_2}
\L_{\text{int}} = - \left( \sum_{q=u,d,s} \lambda_q \bar qq + \frac{2\lambda_G }{27 v} \Theta_\mu^\mu \right) S\, ,
\eeq
where the gluon coupling is eliminated in favor of the trace of the energy-momentum tensor~\cite{Shifman:1978zn,Donoghue:1990xh}
\beq
\Theta_\mu^\mu = -\frac{9\alpha_s}{8\pi}G_{\mu\nu}^a G^{a\mu\nu}
+\sum_{q=u,d,s} m_q \bar qq \, ,
\eeq
with $\lambda_G = \kappa_G + \kappa_c + \kappa_b + \kappa_t$ including heavy-quark loops and 
$\lambda_q = (\kappa_q - \tfrac{2}{27} \lambda_G) m_q/v$.
We remark that the couplings $\lambda_q$ depend on the QCD renormalization scale since only the product $m_q\bar qq$ is an invariant.
Within the large-$N_c$ $\chi$PT framework, the low-energy effective theory is obtained by a suitable modification of Eq.~\eqref{L0}~\cite{Batell:2018fqo}.
Namely, only the replacement $\M \to \M + \lambda S$, where $\lambda = {\rm diag}(\lambda_u, \lambda_d, \lambda_s)$, is necessary for transitions at leading order.
The decay amplitudes are
\beq \label{eq:Amp_to_scalar}
\mathcal{A}(\eta^{(\prime)} \to \pi^0 S) =  - \left\{ (\lambda_u - \lambda_d) B_0 \Gamma^{u-d}_{\pi \eta^{(\prime)}}(m_S^2) + (\lambda_u + \lambda_d) B_0 \Gamma^{u+d}_{\pi \eta^{(\prime)}}(m_S^2) + \lambda_s B_0 \Gamma^{s}_{\pi \eta^{(\prime)}}(m_S^2) \right\} \, .
\eeq
Here the isovector transition form factor is 
\beq \label{eq:isovec}
\big\langle \pi^0(p) \big| \tfrac{1}{2} (\bar{u}u - \bar d d) \big| \eta^{(\prime)}(k) \big\rangle = B_0 \Gamma^{u-d}_{\pi\eta^{(\prime)}}(t) \, ,
\eeq
with $t = (k-p)^2 = m_S^2$, and the strange and nonstrange isoscalar transition form factors are, respectively,
\beq \label{eq:isoscal}
\big\langle \pi^0(p) \big|  \bar s s \big| \eta^{(\prime)}(k) \big\rangle = B_0 \Gamma^{s}_{\pi\eta^{(\prime)}}(t) \, ,
\qquad
\big\langle \pi^0(p) \big| \tfrac{1}{2} (\bar{u}u + \bar d d) \big| \eta^{(\prime)}(k) \big\rangle = B_0 \Gamma^{u+d}_{\pi\eta^{(\prime)}}(t) \, ,\eeq
which have been considered in the context of $a_0$--$f_0$ mixing~\cite{Hanhart:2007bd}.
$B_0$ is proportional to the quark condensate in the chiral limit, defined in Eq.~\eqref{eq:defs}.
At leading order in the chiral and large-$N_c$ expansions, these form factors take constant ($t$-independent) values 
\begin{subequations} \label{eq:Gamma_scalar}
\begin{align}
\Gamma_{\pi\eta}^{u-d}(t) &= \frac{1}{\sqrt 3}\left(\cos\theta_P - \sqrt 2 \sin\theta_P \right) 
\approx \sqrt{ \frac{2}{3} } \, , &
\Gamma_{\pi\eta}^{s}(t) &= -2 \Gamma_{\pi\eta}^{u+d}(t) = \frac{2\sqrt 2}{3} \epsilon \left(\sqrt 2 \cos\theta_P + \sin\theta_P \right) 
\approx \frac{2\sqrt 2}{3} \epsilon \,, \label{eq:Gamma_scalar_eta}
\\
\Gamma_{\pi\eta^\prime}^{u-d}(t) &= \frac{1}{\sqrt 3}\left(\sqrt{2} \cos\theta_P +  \sin\theta_P \right)
\approx  \frac{1}{\sqrt 3} 
\, , &
\Gamma_{\pi\eta^\prime}^{s}(t) &= -2 \Gamma_{\pi\eta^\prime}^{u+d}(t) = \frac{2\sqrt 2}{3} \epsilon\left( \sqrt{2} \sin\theta_P - \cos\theta_P \right) 
\approx - \frac{4}{3}  \epsilon \,,
\label{eq:Gamma_scalar_etap}
\end{align}
\end{subequations}
where $\theta_P \approx \arcsin(-1/3)$ and $\epsilon \approx 0.012$ are the $\eta$--$\eta^\prime$ and (isospin-breaking) $\pi^0$--$\eta$ mixing angles defined in Eqs.~\eqref{LOmixing} and \eqref{pi0-eta-mixing-eps}, respectively.
While isoscalar form factors are suppressed by $\epsilon$ to isovector ones, isospin symmetry is not necessarily a good guide to prioritize the different form factors if $\lambda_s \gg \lambda_u,\lambda_d$.
For the Higgs mixing model, e.g., the $\Gamma^{u-d}_{\pi \eta^{(\prime)}}$ and $\Gamma^{s}_{\pi \eta^{(\prime)}}$ terms in Eq.~\eqref{eq:Amp_to_scalar} have similar size.
Transition form factors from $\Theta_\mu^\mu$ and the $\SU(3)$-singlet operator $\sum \bar q q$ vanish at leading order (in addition to breaking isospin) and will not be considered further. 
(The latter implies the leading-order relation $\Gamma_{\pi\eta^{(\prime)}}^{u+d} + 2 \Gamma_{\pi\eta^{(\prime)}}^{s}= 0$.)

For the $\eta^\prime \to \eta S$ channel, the decay amplitude is
\beq \label{eq:Amp_to_scalar_2}
\mathcal{A}(\eta^\prime \to \eta S) =  - \left\{ (\lambda_u + \lambda_d) B_0 \Gamma^{u+d}_{\eta \eta^{(\prime)}}(m_S^2) + \lambda_s B_0 \Gamma^{s}_{\eta \eta^{(\prime)}}(m_S^2) \right\} \, ,
\eeq
where the form factors are defined analogously to Eq.~\eqref{eq:isoscal}.
These take constant values
\beq
\Gamma_{\eta\eta^\prime}^{s}(t) = -2 \Gamma_{\eta\eta^\prime}^{u+d}(t) = -\frac{1}{3} \left(2 \sqrt 2 \cos2 \theta_P - \sin2\theta_P \right) \approx - \frac{2\sqrt 2}{3}
\eeq
at leading order in the chiral and large-$N_c$ expansions.
In contrast to $\eta^{(\prime)} \to \pi^0 S$, here the isoscalar form factors are not suppressed by the isospin-breaking parameter $\epsilon$, while the isovector form factor $\Gamma_{\eta\eta^\prime}^{u-d}$ is proportional to $\epsilon$ and has been neglected from Eq.~\eqref{eq:Amp_to_scalar_2}.\footnote{We also disregard the $\eta'\to\eta S$ form factor of the trace of the energy-momentum tensor $\Theta_{\eta\eta'}(t)$, which 
vanishes at leading order in the chiral expansion and should thus be suppressed
relative to those of the isoscalar quark densities.
Nontransition form factors of $\Theta_\mu^\mu$, in particular for pions and kaons, 
have been studied on various occasions in the 
literature~\cite{Donoghue:1990xh,Donoghue:1991qv,Celis:2013xja,Winkler:2018qyg}.}

For the Higgs mixing model, the result for $\eta \to \pi^0 S$ has been quoted in the literature~\cite{Bezrukov:2009yw} based on rescaling the old Standard Model prediction for a light Higgs boson~\cite{Leutwyler:1989xj}, computed at $\mathcal{O}(p^4)$ in $\chi$PT, by the scalar mixing angle.
Transition matrix elements vanish at $\mathcal{O}(p^2)$, reflecting the fact that Higgs interactions are diagonalized in the mass eigenstate basis~\cite{Ellis:1975ap,Vainshtein:1980ea}.
Numerically evaluating the formula from Ref.~\cite{Leutwyler:1989xj}, we find for the branching ratio 
\beq \label{eq:LSHiggs}
\BR(\eta \to \pi^0 S) \approx 1.8 \times 10^{-6} \sin^2\theta_S \times \lambda^{1/2}\left(1, \frac{\mpiz^2}{\meta^2}, \frac{m_S^2}{\meta^2} \right)  \, ,
\eeq
with an $\mathcal{O}(20\%)$ uncertainty.
The factor of $\lambda^{1/2}$, defined in Eq.~\eqref{eq:lambda}, is a phase space factor that is unity in the limit $\meta^2 \gg \mpiz^2, m_S^2$.
Eq.~\eqref{eq:LSHiggs} depends on the isospin-breaking mass difference $\Delta m = m_d - m_u$ and we have taken $B_0 \Delta m \approx 0.006 \GeV^2$.
This result can be understood within the large-$N_c$ $\chi$PT framework as well, which gives a nonzero transition amplitude at $\mathcal{O}(p^2)$ due to $\eta$--$\eta^\prime$ mixing.
From Eqs.~\eqref{eq:Amp_to_scalar} and \eqref{eq:Gamma_scalar}, and setting $\lambda_q = \tfrac{7}{9} m_q \sin\theta_S/v$, the predicted branching ratios are 
\begin{subequations}
\begin{align}
\BR(\eta \to \pi^0 S) &= \frac{3 B_0^2 \Delta m^2  (7/9)^2 \sin^2 \theta_P \sin^2\theta_S}{32 \pi \meta v^2 \Gamma_\eta}  \lambda^{1/2}\left(1,\frac{\mpiz^2}{\meta^2}, \frac{m_S^2}{\meta^2} \right) \, , \label{eq:BRetascalar}
\\
\BR(\eta^\prime \to \pi^0 S) &= \frac{3B_0^2 \Delta m^2  (7/9)^2 \cos^2 \theta_P \sin^2\theta_S}{32 \pi \metap v^2 \Gamma_{\eta^\prime}} \lambda^{1/2}\left(1,\frac{\mpiz^2}{\metap^2}, \frac{m_S^2}{\metap^2}\right) \, ,
\label{eq:BRetapscalar} 
\\
\BR(\eta^\prime \to \eta S) &= \frac{B_0^2 (m_s - \hat m)^2  (7/9)^2 \sin^2\theta_S}{144 \pi \metap v^2 \Gamma_{\eta^\prime}} \left(2\sqrt 2 \cos 2\theta_P - \sin2\theta_P\right)^2 \lambda^{1/2}\left(1,\frac{\meta^2}{\metap^2}, \frac{m_S^2}{\metap^2}\right) \, .
\label{eq:BRetapscalar2} 
\end{align}
\end{subequations}
Numerically, Eq.~\eqref{eq:BRetascalar} is identical to the result~\eqref{eq:LSHiggs} from Ref.~\cite{Leutwyler:1989xj}.
For the $\eta^\prime$, Eqs.~\eqref{eq:BRetapscalar} and \eqref{eq:BRetapscalar2} predict
\begin{subequations}
\begin{align}
\BR(\eta^\prime \to \pi^0 S) &\approx 5.4 \times 10^{-8} \sin^2\theta_S \times \lambda^{1/2}\left(1,\frac{\mpiz^2}{\metap^2}, \frac{m_S^2}{\metap^2}\right) \, , \\
\BR(\eta^\prime \to \eta S) & \approx 4.7 \times 10^{-5}  \sin^2\theta_S \times \lambda^{1/2}\left(1,\frac{\meta^2}{\metap^2}, \frac{m_S^2}{\metap^2}\right) \, .
\end{align}
\end{subequations}
Present constraints from CHARM~\cite{Bergsma:1985qz}, E949~\cite{Artamonov:2008qb}, and LHCb~\cite{Aaij:2016qsm} on light scalar production in $K \to \pi S$ and $B \to K S$ require $\sin^2\theta_S \lesssim 10^{-6}$ over the entire range of $m_S$ accessible in $\eta,\eta^\prime$ decays~\cite{Winkler:2018qyg,Beacham:2019nyx}.
This limits $\eta,\eta^\prime$ decays in the Higgs mixing model to be unobservable in the near future, e.g., with projected sensitivities at REDTOP only down to $\sin^2\theta_S \sim 10^{-4}$~\cite{Beacham:2019nyx,Gatto:2019dhj}.

Nevertheless, $\eta,\eta^\prime$ studies play a complementary role in light scalar models beyond the minimal Higgs mixing framework~\cite{Gatto:2019dhj}.
FCNC tests are predominantly sensitive to the top coupling $\kappa_t$, while $\eta, \eta^\prime$ decays probe light quark couplings. 
If a signal were observed in the latter but not the former, it would favor a scalar model with enhanced couplings to light quarks~\cite{Batell:2017kty}.
On the other hand, if the Higgs mixing model is realized in nature, signals would first appear in FCNC tests and $\eta,\eta^\prime$ decays would still give a complementary (albeit challenging) corroboration.

Recent studies have explored light scalars coupling preferentially to light quarks~\cite{Batell:2017kty,Liu:2018qgl,Batell:2018fqo}.
Ref.~\cite{Batell:2018fqo} proposed a ``hadrophilic'' scalar model in which $S$ couples exclusively to the $u$ quark (only $\kappa_u \ne 0$), while Ref.~\cite{Liu:2018qgl} considered more general couplings to both $u$ and $d$ quarks as well as leptons $e,\mu$ (to address the $(g-2)_\mu$ and proton radius puzzles).
In these models, production is generically dominated by the isovector form factor. 
The $\eta,\eta^\prime \to \pi^0 S$ partial widths are
\beq \label{eq:eta_to_scalar_u1}
\Gamma\big(\eta^{(\prime)}\to\pi^0 S\big) = \frac{(\lambda_u-\lambda_d)^2 B_0^2}{16\pi M_{\eta^{(\prime)}}}
 \big|\Gamma_{\pi\eta^{(\prime)}}^{u-d}\big(m_S^2\big)\big|^2  \lambda^{1/2}\left(1,\frac{\mpiz^2}{M_{\eta^{(\prime)}}^2}, \frac{m_S^2}{M_{\eta^{(\prime)}}^2}\right)
 \,.
\eeq
Lattice results yield $B_0 = 2.39(18)\GeV$~\cite{Bazavov:2009fk,Aoki:2019cca} at a scale of $2\GeV$.
(Since $B_0$ is a scale-dependent quantity, the couplings $\lambda_{u,d}$ should be evaluated at the same scale.)
From Eq.~\eqref{eq:Gamma_scalar}, the predicted branching ratios for the hadrophilic model are~\cite{Batell:2018fqo}
\begin{subequations} \label{eq:eta_to_scalar_u3}
\begin{align}
\BR(\eta \to \pi^0 S) &\approx 5.2 \times10^{-2} \, \left(\frac{ \lambda_u}{ 7 \times 10^{-4} }\right)^2 \times \lambda^{1/2}\left(1,\frac{\mpiz^2}{\meta^2}, \frac{m_S^2}{\meta^2} \right)\\
\BR(\eta^\prime \to \pi^0 S) &\approx 1.0 \times10^{-4} \, \left(\frac{ \lambda_u}{ 7 \times 10^{-4} }\right)^2 \times \lambda^{1/2}\left(1,\frac{\mpiz^2}{\metap^2}, \frac{m_S^2}{\metap^2} \right)  \, .
\end{align}
\end{subequations}
Similarly, the prediction for $\eta^\prime \to \eta S$ is
\beq \label{eq:eta_to_scalar_u2}
\BR(\eta^\prime \to \eta S) \approx 6.5 \times 10^{-5}\, \left(\frac{ \lambda_u}{ 7 \times 10^{-4} }\right)^2 \times \lambda^{1/2}\left(1,\frac{\meta^2}{\metap^2}, \frac{m_S^2}{\metap^2} \right) \, .
\eeq
It is assumed $\lambda_u$ is induced by a nonrenormalizable operator with mass scale $M$ and fine-tuning arguments favor $|\lambda_u| \lesssim 7 \times 10^{-4} \, (m_S/100\MeV) \, (M/2 \TeV)^{-2}$~\cite{Batell:2018fqo}.

Theoretical predictions for Eqs.~\eqref{eq:eta_to_scalar_u1}--\eqref{eq:eta_to_scalar_u2} can be assessed (and in principle improved) by looking at higher-order $\chi$PT corrections to the scalar form factors.
For $\eta$ decay, it suffices to consider a linear approximation of the form
\beq \label{eq:Gamma_ud_p4}
\Gamma_{\pi\eta}^{u-d}(t) = \Gamma_{\pi\eta}^{u-d}(0) \bigg\{ 1 + \frac{1}{6}\langle r^2\rangle_{\pi\eta}^{u-d} \,t + \Order\big(t^2\big) \bigg\} \, ,
\eeq
given the small available phase space and large distance to the first resonance $a_0(980)$.
Assuming isospin conservation, Eq.~\eqref{eq:Gamma_ud_p4} is given in terms of two real parameters\footnote{There are no hadronic on-shell intermediate states in the accessible kinematic range $0 \leq t = m_S^2 \leq (\meta-\mpiz)^2$. The lowest-lying isovector state is simply $\eta \pi^0$.} that have been calculated in Ref.~\cite{Albaladejo:2015aca}.
The overall normalization $\Gamma_{\pi\eta}^{u-d}(0)$ is not very precisely fixed, depending on two low-energy constants that are poorly known, and Ref.~\cite{Albaladejo:2015aca} quotes values in the range $\Gamma_{\pi\eta}^{u-d}(0) \approx  0.6\ldots1.0$.
We note that the value $\Gamma_{\pi\eta}^{u-d}(0) \approx \sqrt{2/3}$ in Eq.~\eqref{eq:Gamma_scalar_eta} is in the middle of this range and therefore we estimate $\Gamma_{\pi\eta}^{u-d}(0) = 0.8(2)$.
On the other hand, the scalar radius $\langle r^2\rangle_{\pi\eta}^{u-d}$ depends on a single low-energy constant at one-loop order.  With different values employed for the latter~\cite{Bijnens:2014lea}, the radius has been evaluated to
$\langle r^2\rangle_{\pi\eta}^{u-d} = 0.092(7)\,\text{fm}^2$~\cite{Albaladejo:2015aca} or
$\langle r^2\rangle_{\pi\eta}^{u-d} = 0.067(7)\,\text{fm}^2$~\cite{Lu:2020qeo},
which is smaller by at least a factor of 2 compared to the corresponding $K\to\pi$ scalar radius. 
The authors of Ref.~\cite{Albaladejo:2015aca} expected the first value to be enhanced at higher orders, but probably by no more than 30\%, while the fit of the $\pi\eta$ $T$-matrix to $\gamma\gamma\to\pi^0\eta$ data~\cite{Lu:2020qeo} rather yields a further suppression to 
$\langle r^2\rangle_{\pi\eta}^{u-d} = 0.027\big({}^{+41}_{-26}\big)\,\text{fm}^2$.

For $\eta^\prime$ decays, a wider range of $m_S$ is accessible, but further study of the $\pi\eta'$ and $\eta\eta^\prime$ scalar form factors is necessary to put these channels on more secure theoretical footing.
Little is known about the isovector $\pi\eta^\prime$ form factor, in particular at low energies.  
While Eq.~\eqref{eq:Gamma_scalar_etap} predicts $\Gamma_{\pi\eta'}^{u-d}(0) \approx 1/\sqrt{3}$, R$\chi$T seems to suggest a significant suppression below this value~\cite{Escribano:2016ntp}.
The $t$-dependence of the normalized form factor would be the same as $\Gamma_{\pi\eta}^{u-d}(t)$ in a naive single-meson-dominance picture~\cite{Paver:2011md}, though obviously  a linear approximation for 
$\Gamma_{\pi\eta^\prime}^{u-d}(t)$ is insufficient at higher invariant masses.  
Higher-order corrections to the $\eta\eta^\prime$ scalar form factors have not been studied to the best of our knowledge. 
At leading order in $1/N_c$, the scalar radii for $\Gamma_{\eta\eta^\prime}^{u+d}$ and $\Gamma_{\eta\eta^\prime}^{s}$ are determined by the same low-energy constant entering the corresponding $\pi K$ and $\pi\eta$ scalar radii, but loop corrections clearly break this equality since the latter differ by about a factor of 2.
Moreover, the energy dependence of the $\eta\eta'$ scalar form factors may be significantly affected by isoscalar $\pi\pi$ intermediate states, whose coupling to $\eta\eta'$ is badly controlled in the chiral expansion (see Sect.~\ref{sec:etap-etapipi}).\footnote{A sum rule for the $\eta\eta'$
scalar radii, based on the contribution of $\pi\pi$ intermediate states only, reads 
\[
\langle r^2 \rangle^{u+d/s}_{\eta\eta'} \approx \frac{1}{\Gamma_{\eta\eta'}^{u+d/s}(0)}
\frac{6}{\pi}\int_{4M_\pi^2}^{4M_K^2} \diff x \frac{\Im \Gamma_{\eta\eta'}^{u+d/s}(x)}{x^2} \,, \qquad 
\Im \Gamma_{\eta\eta'}^{u+d/s}(t) = \frac{3\sigma(t)}{32\pi} \Big[f_{\eta'\to\eta\pi\pi}^0(t)\Big]^* \Gamma_{\pi\pi}^{u+d/s}(t)\,,
\]
where $f_{\eta'\to\eta\pi\pi}^0(t) = \M_0^0(t)+\hat\M_0^0(t)$ in the notation of Sect.~\ref{sec:etap-etapipi-dispersive}
denotes the $\pi\pi$ $S$-wave projection of the $\eta'\to\eta\pi\pi$ decay amplitude~\cite{Isken:2017dkw}, 
and $\Gamma_{\pi\pi}^{u+d/s}(t)$ are the nonstrange and strange scalar form factors of the pion, respectively.
This sum rule, however, does not lead to useful constraints on the radii due to its very strong dependence 
on the input near the $K\bar K$ threshold.}

Next, we consider $S$ decay modes in the mass range relevant for $\eta,\eta^\prime$ decays ($1 \lesssim m_S \lesssim 800\MeV$).
These are
\beq \label{eq:Sdecays}
S \to e^+ e^-, \, \mu^+ \mu^-, \, \pi^+ \pi^-, \, \gamma\gamma  \, .
\eeq
While the first three are dominant in the Higgs mixing model~\cite{Bezrukov:2009yw}, only $\pi^+ \pi^-$ and $\gamma\gamma$ are present in the hadrophilic model in which lepton couplings are explicitly set to zero~\cite{Batell:2018fqo}.
Here we consider all possibilities in \eqref{eq:Sdecays}, e.g., allowing for subleading leptonic couplings within the hadrophilic model.

For the lepton and photon modes, we generalize the low-energy Lagrangian~\eqref{eq:Lint_scalar_2} to include scalar couplings to these final states
\beq
\L_{\rm int} = - \left( \sum_{\ell = e,\mu} \lambda_\ell \bar{\ell}\ell - \lambda_\gamma \frac{\alpha_{\rm em}}{6\pi v} F^{\mu\nu} F_{\mu\nu} \right) S \, .
\eeq
The electron coupling $\lambda_e$ is strongly constrained by many of the same experimental and astrophysical results as the dark photon~\cite{Liu:2016qwd,Knapen:2017xzo}.
On the other hand, Refs.~\cite{TuckerSmith:2010ra,Liu:2016qwd} have motivated nonzero values for the muon coupling $\lambda_{\mu} \sim 10^{-3}$ in connection with the $(g-2)_\mu$ anomaly~\cite{Bennett:2006fi,Abi:2021gix,Albahri:2021ixb} and for the combination of proton and muon couplings to explain the observed Lamb shift in muonic hydrogen~\cite{Pohl:2010zza,Antognini:1900ns}.
While Ref.~\cite{Liu:2018qgl} investigated resulting implications for $\eta$ decays, we find that further study is warranted before any definitive predictions can be made.\footnote{Ref.~\cite{Liu:2018qgl} presents limits on the scalar--nucleon couplings $\lambda_{p,n}$, defined here as $\L_{\rm int} = - S(\lambda_p \bar p p + \lambda_n \bar n n)$, from muonic hydrogen and other nuclear physics constraints.
The implications for $\eta$ decays, in turn, depend on the quark-level couplings $\lambda_{u,d}$.
Ref.~\cite{Liu:2018qgl} translates these couplings using the relations $\lambda_p=2\lambda_u+\lambda_d$ and $\lambda_n=\lambda_u+2\lambda_d$, which instead should be~\cite{Crivellin:2013ipa,Hoferichter:2015dsa}
\[
\lambda_p = 8.6(7) \lambda_u + 8.3(6) \lambda_d \,,\qquad
\lambda_n = 7.8(7) \lambda_u + 9.1(6) \lambda_d \, .
\]
In this convention, $\lambda_{p,n}$ inherit QCD scale dependence and have been converted from 
Ref.~\cite{Hoferichter:2015dsa} using 
light quark masses from lattice QCD~\cite{Fodor:2016bgu}, evaluated at $\mu=2\GeV$.
Ref.~\cite{Liu:2018qgl} also derived a value for $B_0 \Gamma_{\pi\eta}^{u-d}(0) \approx 825\MeV$ smaller than the value ($\approx\! 1.9\GeV$) we have advocated above.
}
The decay width into leptons is
\beq
\Gamma(S\to \ell\bar \ell) = \frac{\lambda_\ell^2 m_S}{8\pi}\Bigg(1-\frac{4m_\ell^2}{m_S^2}\Bigg)^{3/2} \,. \qquad
\eeq
The photon coupling $\lambda_\gamma$ arises at one loop by integrating out heavy states with electric charge, including possible BSM states, that couple to $S$.
In the Higgs mixing model this includes $c$, $b$, $t$, $\tau$, and $W^\pm$ and yields $\lambda_\gamma \approx -(5/4) \sin\theta_S$, while in the hadrophilic model $\lambda_\gamma=0$.
The two-photon decay width is 
\beq
\Gamma_f(S\to\gamma\gamma) = \frac{ \alpha_{\text{em}}^2 \left|{\lambda}^{\rm eff}_\gamma\right|^2 m_S^3}{144\pi^3 v^2} \, ,
\label{eq:Gamma_S-ff-gg}
\eeq 
where the total effective $S \to \gamma\gamma$ coupling $\lambda_\gamma^{\rm eff} = \lambda_\gamma + \lambda_\gamma^{(\ell)} + \lambda_\gamma^{(q)}$ includes contributions from light quarks and leptons.
The lepton contribution is straightforward to write down
\beq \label{eq:lambda_lepton}
\lambda_\gamma^{(\ell)} = \sum_{\ell=e,\mu} \frac{\lambda_\ell  v}{m_\ell} \mathcal{F}_{1/2}\left(m_S^2/m_\ell^2\right) \, ,
\eeq
where the standard fermion loop function is (e.g., \cite{Gunion:1989we})
\beq \label{eq:photon_loop_function}
\mathcal{F}_{1/2}(x) = 
\frac{6}{x^2} \Big( x+(x-4) f(x) \Big) \,, 
\qquad
f(x) = \left\{ \begin{array}{cl} \arcsin^2(\sqrt x/2) & x < 1 \\
-\frac{1}{4}\bigg(\log\frac{1+\sqrt{1-4/x}}{1-\sqrt{1-4/x}}-i\pi\bigg)^2 & x > 1
\end{array} \right. \, ,
\eeq 
normalized to $\mathcal{F}_{1/2}(0) = 1$.
The light-quark contribution has been evaluated in the literature~\cite{Liu:2018qgl,Batell:2018fqo} using a similar formula, e.g., for $u,d$ quarks,
\beq \label{eq:lambda_cons_quark}
\lambda_\gamma^{(q)} = \sum_{q=u,d} \frac{N_c Q_q^2 \lambda_q  v}{m_q} \mathcal{F}_{1/2}\left(m_S^2/m_q^2\right) \, 
\eeq
by taking constituent quarks of mass $200 \ldots 350\MeV$ in the loop.
However, assuming charged-pion and -kaon loops dominate the decay (cf.~analogous calculations of $K_S\to\gamma\gamma$~\cite{DAmbrosio:1986zin,Goity:1986sr}), the light-quark contribution can be computed in $\chi$PT and we obtain
\beq \label{eq:lambda_chipt}
\lambda_\gamma^{(q)} = \frac{(\lambda_u + \lambda_d) B_0 v}{8\mpipm^2} \mathcal{F}_0\left(m_S^2/\mpipm^2\right)
+ 
\frac{(\lambda_u + \lambda_s) B_0 v}{8\mKpm^2} \mathcal{F}_0\left(m_S^2/\mKpm^2\right)
\eeq
at leading order, where the scalar loop function is (e.g., \cite{Gunion:1989we})
\beq
\mathcal{F}_0(x) = 
-\frac{12}{x^2} \Big( x - 4 f(x) \Big) \,,
\eeq
again with $\mathcal{F}_0(0) = 1$.
For the hadrophilic model, Eq.~\eqref{eq:lambda_chipt} leads to an order-of-magnitude enhancement compared to Eq.~\eqref{eq:lambda_cons_quark}, although higher-order contributions may be numerically relevant. 

Decays into two pions are the leading channels for a Higgs-mixed scalar~\cite{Bezrukov:2009yw} and hadrophilic scalar~\cite{Batell:2018fqo} in the mass range $2\mpi < m_S < \metap - \mpi$.
The partial widths are 
\beq \label{eq:S_to_pipi}
\Gamma(S\to \pi^+\pi^-) = 2\Gamma(S\to \pi^0\pi^0) = \frac{1}{16\pi m_S}\sqrt{1-\frac{4M_\pi^2}{m_S^2}} 
\bigg|(\lambda_u+\lambda_d)B_0\Gamma^{u+d}_{\pi\pi}\big(m_S^2\big)
+ \lambda_s B_0\Gamma^s_{\pi\pi}\big(m_S^2\big)
+ \frac{2}{27v}\lambda_G \Theta_{\pi\pi}\big(m_S^2\big) \bigg|^2 \, .
\eeq
These channels depend on the same parameters that govern production in $\eta,\eta^\prime$ decays and therefore necessarily occur when kinematically allowed.
The $\pi\pi$ scalar form factors are defined in analogy to 
Eq.~\eqref{eq:isoscal}, while the additional form factor from the trace of the energy-momentum tensor
is
\beq
\langle \pi^i(p) \big| \Theta_\mu^\mu \big| \pi^j(k) \rangle
= \delta^{ij} \Theta_{\pi\pi}(t) \, .
\eeq
In Eq.~\eqref{eq:S_to_pipi} we retain only isoscalar form factors, neglecting the isovector (isospin-breaking) one, $\Gamma_{\pi\pi}^{u-d}$.
At leading order in the chiral expansion, these are given by
\beq
\Gamma^{u+d}_{\pi\pi}(t) = 1 \,, \qquad
\Gamma^s_{\pi\pi}(t) = 0 \,, \qquad
\Theta_{\pi\pi}(t) = 2M_\pi^2+t \,.
\eeq
As we have discussed many times throughout this review, however, rescattering or unitarity corrections to these form factors, which produce pion pairs in isospin $I=0$ $S$-waves, are large and need to be resummed dispersively, treating 
pions and kaon--antikaon pairs in coupled channels (see Sect.~\ref{sec:DispTheory}).
Such corrections have been considered previously in the literature for $2\pi$ decays of a light Higgs boson or Higgs-mixed scalar~\cite{Truong:1989my,Donoghue:1990xh,Monin:2018lee,Winkler:2018qyg}, as well as other contexts~(e.g.,~\cite{Moussallam:1999aq,DescotesGenon:2000ct,Daub:2012mu,Celis:2013xja,Daub:2015xja,Ropertz:2018stk}).
All form factors are complex but share the same phase $\delta_0^0(t)$ provided $S$ is below $K\bar{K}$ threshold~\cite{Winkler:2018qyg}.
For the range $4M_\pi^2\leq t\leq(M_\eta-M_\pi)^2$ relevant for $\eta \to \pi^0 S \to 3\pi$ decay, the form factor moduli are in the ranges
\begin{align}
1.42 \lesssim \Big| \Gamma^{u+d}_{\pi\pi}(t)\Big|  \lesssim 1.68 \,, \quad  
0.04 \lesssim \Big| \Gamma^s_{\pi\pi}(t)\Big|  \lesssim 0.11 \,, \quad
9.5 M_\pi^2 \lesssim \Big| \Theta_{\pi\pi}(t) \Big|  \lesssim 20.7 M_\pi^2 \, .
\end{align}
For a larger range of $t$ relevant for $\eta^\prime$ decays, we refer the reader to Ref.~\cite{Winkler:2018qyg}.

Within the range of scalar mass and couplings relevant for $\eta,\eta^\prime$ decays, we can summarize the situation as follows.
Low-mass scalars (below $10\MeV$) are completely excluded, while higher-mass scalars (above $2\mpi$) are viable and will generically decay into $2\pi$ and can be tested in $\eta,\eta^\prime \to 3\pi$ and $\eta^\prime \to \eta \pi \pi$.
The intermediate range is also viable, but more model-dependent and may include photon or lepton final states.
Let us expand these statements in more detail.
\begin{itemize}
\item {\it Low mass} ($m_S \lesssim 10 \MeV$): This range is not interesting for $\eta,\eta^\prime$ decay studies.
The only visible kinematically allowed channels are $S \to \gamma\gamma$ and $e^+ e^-$.
Big bang nucleosynthesis arguments exclude a sizeable branching fraction $S \to \gamma \gamma$ for $m_S < 20\MeV$, which would otherwise spoil the successful prediction for the deuterium abundance~\cite{Batell:2018fqo}.
On the other hand, $S \to e^+ e^-$ is constrained by beam dump experiments and $(g-2)_e$ such that $|\lambda_e| \lesssim 3 \times 10^{-8}$ for $m_S \lesssim 10\MeV$~\cite{Liu:2016qwd}.
This implies a long decay length
\beq
\gamma c \tau \gtrsim 10^6 \, {\rm cm} \times \left( \frac{m_S}{10 \MeV} \right)^{-2}  
\eeq
that is larger than typical detector sizes for $\eta,\eta^\prime$ studies.
For example, the recent WASA-at-COSY analysis, which yielded $\BR(\eta\to\pi^0e^+e^-) < 7.5\times 10^{-6}$~\cite{Adlarson:2018imw}, vetoed $e^+ e^-$  vertices displaced more than $\mathcal{O}(10\, {\rm cm})$ transverse to the beamline~\cite{Adlarson:2015zta}.
Therefore, this limit does not provide a strong constraint on $\eta \to \pi^0 S \to \pi^0 e^+ e^-$ since the BSM signal is penalized by a small acceptance fraction    (cf.~Ref.~\cite{Liu:2018qgl}).

\item {\it Intermediate mass} ($10 \MeV \lesssim m_S < 2\mpi$): Current exclusions already rule out visible signals for future $\eta,\eta^\prime$ studies for the hadrophilic model (where only $\lambda_u$ is nonzero)~\cite{Batell:2018fqo}.
In this case, the leading decay mode is $S \to \gamma\gamma$ via a $u$ quark loop, which yields, e.g., $\eta \to \pi^0 S \to \pi^0 \gamma \gamma$.
The entire mass range is already excluded from the total $\eta \to \pi^0 \gamma\gamma$ rate measured by MAMI~\cite{Nefkens:2014zlt} (sensitivity down to $|\lambda_u| \sim 2 \times 10^{-4}$), as well as beam dump and astrophysical constraints on long-lived $S \to \gamma\gamma$ decays (for smaller $\lambda_u$ values)~\cite{Batell:2018fqo}.
However, in a generalized model, these latter constraints may be weakened if $S$ decays promptly. 
Therefore, it is worthwhile to improve sensitivities in $\eta \to \pi^0 \gamma\gamma$ beyond the total rate limit by searching for resonances in the $\gamma\gamma$ invariant mass distribution.
This may occur if $S \to \gamma\gamma$ is enhanced via a large direct photon coupling $\lambda_\gamma$ or if there is an additional decay channel, e.g., $S \to \ell^+ \ell^-$, to ensure a short decay length.
In this mass range, the electron coupling may be as large as $|\lambda_e| \sim 10^{-4}$~\cite{Liu:2016qwd,Knapen:2017xzo}, while a muon coupling of $|\lambda_\mu| \sim 5 \times 10^{-3}$ can accommodate the $(g-2)_\mu$ measurement.
Hence, $S$ can be searched for in all three channels
\beq \label{eq:eta_S_decays}
\eta \to \pi^0 S \to \pi^0 \gamma \gamma , \: \pi^0 e^+ e^-, \: \pi^0 \mu^+ \mu^- \, ,
\eeq
the latter only if $S \to \mu^+ \mu^-$ is kinematically allowed.
Analogous channels can in principle be searched for in $\eta^\prime$ decays, but the estimated rates \eqref{eq:eta_to_scalar_u3} and \eqref{eq:eta_to_scalar_u2} are much smaller. With ongoing construction of an upgraded forward electromagnetic calorimeter and a muon detector by the GlueX collaboration, searches for $S$ in $\eta^{(\prime)}$ decays into those three final states will be anticipated from the JEF experiment in the coming years.

\item  {\it High mass} ($m_S > 2\mpi$): In the hadrophilic model, this remains the only viable mass range for future $\eta,\eta^\prime$ studies~\cite{Batell:2018fqo}.
In this case, the leading decays are $S \to \pi^+\pi^-, \pi^0\pi^0$ and can be detected as a resonance in the Dalitz plots for 
\beq
\eta, \eta^\prime \to \pi^0 S \to 3\pi , \quad
\eta^\prime \to \eta S \to \eta\pi\pi \, .
\eeq
Bump hunting in $\eta \to 3\pi$ data projected for REDTOP~\cite{Gatto:2016rae} can push the limit on $\lambda_u$ beyond that from current KLOE data~\cite{Anastasi:2016cdz} to below the level of $10^{-6}$~\cite{Batell:2018fqo}.
Moreover, $\eta^\prime$ decays can probe the scalar mass range $\meta - \mpiz < m_S < \metap - \mpiz$ that is inaccessible in $\eta$ decays.
Ref.~\cite{Batell:2018fqo} quoted $|\lambda_u| \lesssim 10^{-2}$ based on the total $\eta^\prime \to 3\pi$ rate from BESIII~\cite{Ablikim:2016frj}.
In a more general scalar model, other decay channels \eqref{eq:eta_S_decays} can arise and can be searched for as well.
\end{itemize}
A more detailed roadmap for light scalars in $\eta,\eta^\prime$ decays is desirable, but more work is needed to fully map out the existing constraints and allowed parameter space for quark, lepton, and photon couplings along the lines of Refs.~\cite{Liu:2016qwd,Batell:2017kty,Liu:2018qgl,Batell:2018fqo}.

\subsection{Axion-like particles}\label{sec:axions}

Light pseudoscalars arise in many new physics scenarios, for example, as pseudo-Nambu--Goldstone bosons from spontaneously broken global symmetries.
Since their masses are protected by an approximate shift symmetry, such states are naturally light and may in fact be the only detectable remnant of BSM physics at high mass scales beyond experimental reach.
The most famous example is the QCD axion from a broken Peccei--Quinn symmetry, proposed to solve the strong $CP$ problem in QCD (reviewed in Refs.~\cite{Peccei:2006as,Hook:2018dlk}).
In recent years, the study of axions has expanded to encompass a broader landscape of axion-like particles (ALPs), denoted $a$, where the usual QCD-scale relation between axion mass $m_a$ and decay constant $f_a$ is relaxed.
This is expressed as
\beq \label{eq:ALP_mass}
m_a^2 = m_0^2 + \frac{M_\pi^2 F^2_\pi}{2f_a^2} \left(\frac{m_u m_d}{(m_u + m_d)^2} \right) \, .
\eeq
The second term is the well-known mass for the QCD axion~\cite{Weinberg:1977ma,Wilczek:1977pj,Shifman:1979if}, which must be much lighter than the QCD scale if $f_a \gg F_\pi$. 
For ALPs, $m_a$ and $f_a$ are decoupled by introducing an explicit symmetry-breaking mass term $m_0^2$.
This broader parameter landscape has attracted much attention in the phenomenology community, motivating a number of current and future searches~\cite{Alekhin:2015byh,Dobrich:2015jyk,Berlin:2018pwi,Curtin:2018mvb,Banerjee:2020fue} (see also~\cite{Essig:2013lka,Beacham:2019nyx} and references therein).

While the QCD axion is a well-known candidate for dark matter, the masses and couplings accessible in $\eta,\eta^\prime$ decays make ALPs short-lived and therefore unsuitable to be dark matter.
However, ALPs have been proposed within the dark sector framework to mediate interactions between dark matter and the Standard Model~\cite{Freytsis:2010ne,Boehm:2014hva,Dolan:2014ska}.
From a phenomenological point of view, one appealing feature is that direct detection constraints are severely weakened since pseudoscalar interactions are suppressed in nonrelativistic dark-matter--nucleon scattering.
Consequently, collider-based studies (and other measurements) have greater importance in exploring these models.

Searches for ALPs span from the low-energy frontier, far below the eV-scale~\cite{Jaeckel:2010ni,Essig:2013lka}, to the LHC~\cite{Mimasu:2014nea,Jaeckel:2015jla,Brivio:2017ije,Bauer:2017ris}.
At $\eta,\eta^\prime$ facilities, production channels include $\eta, \eta^\prime \to \pi\pi a$, analogous to $\eta,\eta^\prime \to 3\pi$, where the ALP appears via mixing with pseudoscalar mesons~\cite{Aloni:2018vki,Landini:2019eck}.
Therefore, the accessible mass ranges for $\eta,\eta^\prime$ decays, respectively, are
\beq
2m_e < m_a < \meta - 2\mpi \approx 270\MeV \, , \qquad 
2m_e < m_a < \metap - 2\mpi \approx 680\MeV \, . \qquad 
\eeq
For $\eta$ decay, the only allowed channels are $a\to \gamma\gamma$, $e^+e^-$, $\mu^+ \mu^-$.
Since typically only visible decays are searched for in $\eta,\eta^\prime$ studies, ALP couplings must be large enough to yield a prompt decay.
The $a \to \gamma\gamma$ case is strongly constrained by a combination of beam dump experiments~\cite{Hewett:2012ns,Jaeckel:2015jla}, pion decays~\cite{Altmannshofer:2019yji}, astrophysics~\cite{Jaeckel:2010ni}, and cosmology~\cite{Millea:2015qra,Depta:2020wmr}.
Only the range $m_a \gtrsim 10\MeV$ is allowed~\cite{Beacham:2019nyx}.
On the other hand, the $a\to e^+ e^-$ case remains viable down to $e^+e^-$ threshold~\cite{Bauer:2017ris,Beacham:2019nyx}, though partly excluded by beam dump experiments~\cite{Liu:2017htz}.
Another possibility is $a \to \mu^+ \mu^-$ for $m_a > 2m_\mu$.
Direct ALP couplings to leptons are motivated to explain the $(g-2)_\mu$ anomaly.
While the one-loop Schwinger contribution for a pseudoscalar mediator famously has the wrong sign to fix the discrepancy, there is also a one-loop Barr--Zee contribution that depends on the product of couplings $c_\gamma c_\mu$ that can have either sign~\cite{Marciano:2016yhf}.
The sum of the terms gives a viable BSM explanation for $(g-2)_\mu$ for the ALP mass range $\sim 50\MeV$--$5\GeV$~\cite{Bauer:2017ris} that is accessible partly in $\eta,\eta^\prime$ decays.
For $\eta^\prime$ decays, larger ALP masses are allowed and other final states are possible, namely $a \to \pi^+ \pi^- \gamma$ and $3\pi$~\cite{Dobrescu:2000jt,Aloni:2018vki}.\footnote{Analogous to the discussion in Sect.~\ref{sec:BSMCV}, we neglect $a \to \pi^0 \gamma$ ($C$-violating and angular momentum nonconserving), $\pi^0\pi^0 \gamma$ ($C$-violating), $\pi\pi$ ($CP$-violating), $\pi^0 \gamma\gamma$ (order $\alpha_{\textrm{em}}^2$), and $\pi^0 e^+ e^-$ (order $\alpha_{\textrm{em}}^2$ or $C$-violating).}

The total $\eta$ decay chains would be
\beq \label{eq:eta_ALP_decay}
\eta \to \pi\pi a \to \pi\pi \gamma\gamma \,, \;
\pi\pi e^+ e^-  \,, \; \pi\pi \mu^+\mu^- \, .
\eeq
For the $\eta^\prime$, we have these same channels plus
\beq \label{eq:etap_ALP_decay}
\eta^\prime \to \pi\pi a \to \pi\pi \, \pi^+ \pi^- \gamma \,, \;  5\pi \, , \qquad
\eta^\prime \to \eta\pi^0 a \to \eta\pi^0 \gamma\gamma \,, \; \eta\pi^0 e^+ e^- \,, \; \eta\pi^0 \mu^+ \mu^- \, .
\eeq
These rather complex decay modes have seen very little attention and only a few have been searched for to date~\cite{Tanabashi:2018oca}.
We discuss what is known about them at the end of this section.

Next, we turn to the theoretical description of ALPs.
For $m_a \lesssim 1\GeV$, ALP--hadron interactions are described by chiral effective theory~\cite{Georgi:1986df}.
This framework was recently reviewed in several works~\cite{Bauer:2017ris,Aloni:2018vki,Landini:2019eck,Ertas:2020xcc}, but for completeness we briefly recap the setup here as well, following the original arguments of Ref.~\cite{Georgi:1986df}.
We start with the Lagrangian for ALPs coupled to quarks $q=u,d,s$, gluons, leptons $\ell = e,\mu$, and photons
\beq \label{eq:L_ALP}
\L_{\text{ALP}} =  \L_{\rm QCD} +  \frac{1}{2} (\partial_\mu a)(\partial^\mu a) - \frac{1}{2} m_0^2 a^2 - \frac{\alpha_s}{8\pi f_a} a\, G^a_{\mu\nu} \tilde G^{a\mu\nu} - \frac{\alpha_{\text{em}} c_{\gamma}}{8\pi f_a} a\, F_{\mu\nu} \tilde F^{\mu\nu}
- \frac{\partial^\mu a}{2f_a} \bar q c_q \gamma_\mu\gamma_5 q 
- \frac{\partial^\mu a}{2f_a} \bar \ell c_\ell \gamma_\mu\gamma_5 \ell 
\eeq
where, e.g., $\tilde{F}^{\mu\nu} = \frac{1}{2} \epsilon^{\mu\nu\alpha\beta} F_{\alpha\beta}$ and $\epsilon^{0123} = +1$.
We assume the direct ALP couplings to quarks $c_q$ and leptons $c_\ell$ are flavor-diagonal.\footnote{ALPs are expected to couple to heavier Standard Model particles as well (see, e.g., Refs.~\cite{Dolan:2014ska,Jaeckel:2015jla,Brivio:2017ije,Bauer:2017ris}).
Since we only consider flavor-conserving decays, it suffices to integrate out these states and absorb their ALP couplings into the Wilson coefficients in Eq.~\eqref{eq:L_ALP}. 
Such contributions are typically small anyways for the ALP mass range of interest here~\cite{Bauer:2017ris}. We also consider only $CP$-conserving ALP couplings, which otherwise are subject to EDM constraints~\cite{Kirpichnikov:2020lws}.}
Next, it is customary to eliminate the ALP--gluon coupling by a chiral rotation of the quark fields 
\beq \label{eq:chiral_rot}
q \to \exp\left( i \frac{a\kappa_q}{2f_a} \gamma_5 \right) q \, ,
\eeq
where $\kappa_q$ is a diagonal matrix that must satisfy $\langle \kappa_q \rangle = 1$.
Since there remains some freedom in $\kappa_q$, for convenience one takes $\kappa_q =\M^{-1}/\langle\M^{-1}\rangle$ to eliminate ALP--octet mass mixing ($\M$ is the quark mass matrix).
At the hadron level, the theory is described by the leading-order chiral effective Lagrangian, generalizing Eq.~\eqref{L0},
\begin{align} 
\L_{\text{eff}}& = 
\frac{F^2}{4} \left\langle\partial_\mu U^\dagger\partial^\mu U +  \hat\chi(a) U^\dagger + \hat\chi(a)^\dagger U \right\rangle -
\frac{1}{2}M_0^2\eta_0^2 +
\frac{1}{2} (\partial_\mu a)(\partial^\mu a) -\frac{1}{2} m_a^2 a^2 - \frac{\alpha_{\rm em} \bar{c}_{\gamma}}{8\pi f_a} a\, F_{\mu\nu} \tilde F^{\mu\nu} 
\notag \\
& \quad - i \frac{F^2 \partial^\mu a}{4f_a}  
\left\langle \bar{c}_q \left(U^\dagger D_\mu U-U D_\mu U^\dagger\right) \right\rangle 
- \frac{\partial^\mu a}{2f_a} \bar \ell c_\ell\gamma_\mu\gamma_5 \ell \,,
\label{eq:axion-Leff}
\end{align}
where $\bar{c}_q = c_q + \kappa_q$.
Due to the chiral anomaly, the $a$-dependence from the gluon term reappears in the quark mass term via
\beq \label{eq:chi_hat_a}
\hat \chi(a) = 2B_0 \exp \bigg(i \frac{a\kappa_q}{2f_a} \bigg) \M \exp \bigg(i \frac{a\kappa_q}{2f_a}\bigg)
= 2B_0 \M + \frac{i\,B_0}{f_a} \left\{\kappa,\M\right\} a
- \frac{B_0^2 a^2}{4f_a^2} \left\{ \kappa_q \left\{ \kappa_q, \M \right\} \right\} + \ldots  \, ,
\eeq
the last term of which yields the QCD contribution to the ALP mass in Eq.~\eqref{eq:ALP_mass}.
The photon coupling is also shifted to be
\beq \label{eq:c_bar_ALP}
\bar c_\gamma = c_\gamma - 2N_c \langle \kappa Q^2 \rangle = c_\gamma - 1.92(4) \, ,
\eeq
where the numerical value is from Ref.~\cite{diCortona:2015ldu} and includes NLO corrections. 
Improved input on the NLO LECs~\cite{Kampf:2009tk} slightly changes the latter to $-2.05(3)$~\cite{Lu:2020rhp}.
In the case of an ALP coupled only to gluons~\cite{Aloni:2018vki,Landini:2019eck}, the direct couplings $c_q = (c_u, c_d, c_s)$, $c_\ell=(c_e, c_\mu)$, and $c_\gamma$ all vanish and all processes are functions of two parameters, $m_a$ and $f_a$.
It is worth emphasizing that a purely gluon-coupled ALP still couples to photons: there is a part from the chiral anomaly---the $-1.92$ term in Eq.~\eqref{eq:c_bar_ALP}---as well as a part from ALP--pseudoscalar mixing, discussed below.

The physical states of the theory are obtained by diagonalizing the full four-pseudoscalar system $(a,\pi^0, \eta, \eta^\prime)$~\cite{Choi:1986zw}.
Eq.~\eqref{eq:axion-Leff} includes both kinetic and mass mixing between the ALP and neutral mesons $P = \pi^0, \eta, \eta^\prime$, parameterized as
$\mathcal{L}_{\rm eff} \supset K_{aP} (\partial_\mu a)( \partial^\mu P) - M_{aP}^2 a P$~\cite{Aloni:2018vki}.
This requires a shift in the meson fields
\beq \label{eq:ALP_shift}
\pi^0 \to \pi^0 + \frac{F}{f_a} \theta_{a\pi} a \, , \;\;
\eta \to \eta + \frac{F}{f_a} \theta_{a\eta} a \, , \;\;
\eta^\prime \to \eta^\prime + \frac{F}{f_a} \theta_{a\eta^\prime} a \, ,
\eeq
where the mixing angles have $F/f_a$ explicitly factored out and are assumed to be small, $F\theta_{aP}/f_a \ll 1$.
These are
\beq
\theta_{aP} = \frac{ M^2_{aP} + m_a^2 K_{aP}}{M_P^2 - m_a^2} \,,
\eeq
provided the ALP is not degenerate with one of the mesons.
The kinetic and mass-mixing terms are compactly expressed, respectively, as
\beq
K_{aP} = \langle \bar{c} T_P \rangle \, , \qquad
M_{aP}^2 = - k \mpi^2 \langle T_P \rangle \, ,
\eeq
where we define
\beq
k = \frac{ B_0 }{ \langle \M^{-1} \rangle \mpi^2 } = \frac{m_u m_d m_s}{(m_u + m_d)(m_u m_d + m_u m_s + m_d m_s)} \approx 0.2
\eeq
and we take the $U(3)$ generators to be
\begin{align}
T_{\pi^0} &= {\rm diag}\left( \frac{1+\sqrt{3} \epsilon}{2}, -\frac{1-\sqrt{3} \epsilon}{2}, - \frac{\epsilon}{\sqrt 3} \right) \,, \qquad
T_{\eta} = {\rm diag}\left( \frac{1}{\sqrt{6}} - \frac{2\sqrt 2 \epsilon}{3}, \frac{1}{\sqrt{6}} + \frac{2\sqrt 2 \epsilon}{3}, - \frac{1}{\sqrt{6}} \right) \,,  \notag \\
T_{\eta^\prime} &= {\rm diag}\left( \frac{\sqrt 3 - \epsilon}{6} , \frac{\sqrt 3 + \epsilon}{6},  \frac{1}{\sqrt{3}} \right) \, .
\end{align}
These generators are obtained from the original $U(3)$ basis $(\pi^0, \eta_8, \eta_0)$ in Eq.~\eqref{eq:goldstones-prime} and rotating to the basis of would-be physical meson eigenstates in the Standard Model (i.e., without ALP mixing).
We have adopted the convenient single-angle $\eta$--$\eta^\prime$ mixing scheme~\eqref{LOmixing}, with $\theta_P = \arcsin(-1/3)$, and we work to linear order in the isospin-breaking $\pi^0$--$\eta$ mixing angle $\epsilon \approx 0.012$, defined in Eq.~\eqref{pi0-eta-mixing-eps}.
Lastly, we note that the ALP field must also be shifted, but this is irrelevant for processes considered here at linear order in $1/f_a$.

ALP-producing $\eta,\eta^\prime$ decays \eqref{eq:eta_ALP_decay} and \eqref{eq:etap_ALP_decay} are three-body decays of the form $\eta^{(\prime)} \to P_1 P_2 a$, where $P_{1,2}$ are mesons (with masses $M_{1,2}$).
The general formula for the decay rate is
\beq
\Gamma(\eta^{(\prime)} \to P_1 P_2 a) = 
\frac{1}{256\pi^3 M_{\eta^{(\prime)}} \,S}
\int_{(M_1 + M_2)^2}^{(M_{\eta^{(\prime)}}-m_a)^2} \diff s \, \lambda^{1/2}\left( 1 , \frac{M_1^2}{s}, \frac{M_2^2}{s} \right)
\lambda^{1/2}\left(1,\frac{s}{M_{\eta^{(\prime)}}^2},\frac{ m_a^2}{M_{\eta^{(\prime)}}^2}\right)
\big| \A(\eta^{(\prime)} \to P_1 P_2 a)\big|^2 \, ,
\eeq
where the phase space factor $\lambda$ is defined in Eq.~\eqref{eq:lambda}, and the symmetry factor is $S=2$ for identical particles ($P_1 P_2 = 2\pi^0$) and $S=1$ otherwise.
The decay amplitudes follow from Eq.~\eqref{eq:axion-Leff}, after rotation~\eqref{eq:ALP_shift}, and are given by
\allowdisplaybreaks{
\begin{subequations} \label{eq:eta-pipiaxion}
\begin{align}
\A(\eta \to 2\pi^0 a) &= 
\frac{1}{F f_a} \left( -\frac{2\sqrt 2}{\sqrt 3}  k \mpi^2 
- \frac{\sqrt 2 (3 \meta^2 - 8 \mpi^2)}{3} \epsilon \theta_{a\pi} 
+ \frac{2\mpi^2}{3} \theta_{a\eta} 
+ \frac{\sqrt 2 \mpi^2}{3} \theta_{a\eta^\prime} \right) \,,\\
\A(\eta^\prime \to 2\pi^0 a) &=
\frac{1}{F f_a}\left( -\frac{2}{\sqrt 3}  k \mpi^2
- \frac{3 \meta^2 - 11 \mpi^2}{3} \epsilon \theta_{a\pi} 
+ \frac{\sqrt 2 \mpi^2}{3} \theta_{a\eta} 
+ \frac{\mpi^2}{3} \theta_{a\eta^\prime} \right) \,,\\
\A(\eta \to \pi^+ \pi^- a) &= \frac{1}{F f_a} \left( - \frac{2 \sqrt 2}{\sqrt 3} k \mpi^2 + 
\frac{2\sqrt 2(\bar{c}_u - \bar{c}_d)}{9}  \left( 3 \meta^2 + m_a^2 - 3 s\right)  \epsilon \right. \notag \\
& \qquad\quad \left. + \frac{\sqrt 2}{9} \left(4 m_a^2 - 12 s + \meta^2 + 16 \mpi^2\right) \epsilon \theta_{a\pi} 
+ \frac{2\mpi^2}{3} \theta_{a\eta}
+ \frac{\sqrt 2 \mpi^2}{3} \theta_{a\eta^\prime}
\right)
\,, \\
\A(\eta^\prime \to \pi^+ \pi^- a) &= \frac{1}{F f_a} \left( - \frac{2}{\sqrt 3} k \mpi^2 + 
\frac{\bar{c}_u - \bar{c}_d}{18} \left( 3 \metap^2 + m_a^2 - 3 s \right)  \epsilon \right. \notag \\
& \qquad\quad \left. + \frac{1}{9} \left(\metap^2 + m_a^2 - 3 s - 3 \meta^2 +  13 \mpi^2\right) \epsilon \theta_{a\pi} 
+ \frac{\sqrt 2\mpi^2}{3} \theta_{a\eta}
+ \frac{\mpi^2}{3} \theta_{a\eta^\prime} 
\right) \,, \\
\A(\eta^\prime \to \eta \pi^0 a) &=
\frac{1}{F f_a}\left(  \frac{\sqrt 2 \mpi^2}{3} \theta_{a\pi}
- \frac{4(3 \meta^2 + \mpi^2)}{9} \epsilon \theta_{a\eta} 
+ \frac{\sqrt 2 (3 \meta^2 -2 \mpi^2)}{9} \epsilon \theta_{a\eta^\prime} \right) \, .
\end{align}
\end{subequations}}
These amplitudes depend only on the Mandelstam variable $s = M_{12}^2 = (p_1 + p_2)^2$, which is the $P_1 P_2$ invariant mass.

\begin{figure}[t]
\centering
\includegraphics[width=12cm]{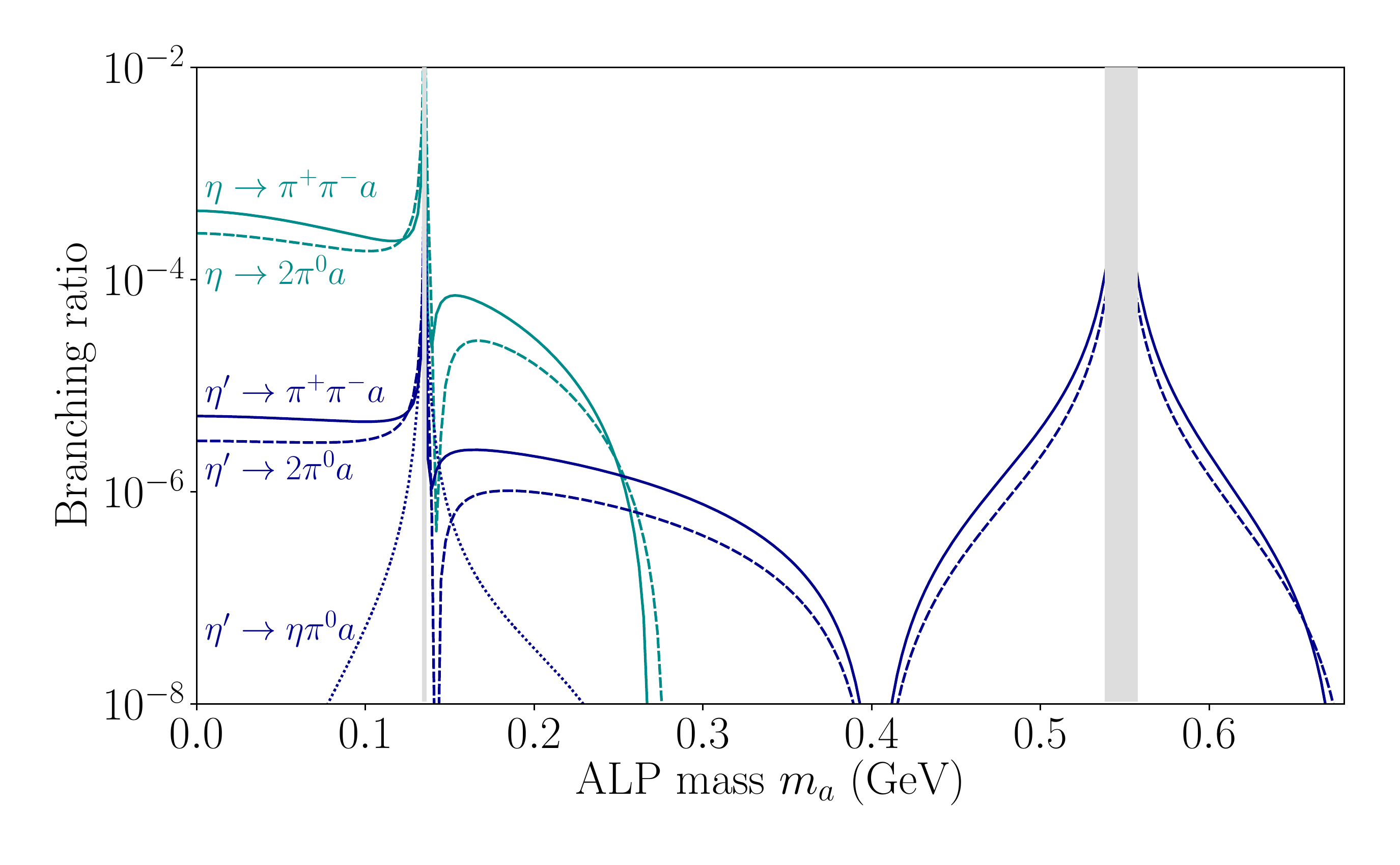}
\caption{
Branching ratios for ALP production in $\eta$ decays (light blue) and $\eta^\prime$ decays (dark blue), assuming no direct quark-ALP coupling ($c_q=0$).
Channels include \mbox{$\eta,\eta^\prime \to \pi^+ \pi^- a$~(solid)}, \mbox{$\eta,\eta^\prime \to 2\pi^0 a$ (dashed)}, and $\eta^\prime\to\eta \pi^0 a$ (dotted).
Branching ratios scale as $1/f_a^2$ and here ALP decay constant is fixed to $f_a = 10\GeV$, equivalent to an effective mass scale $\approx 3\TeV$.
Regions around $m_a = \mpiz$ and $m_a = \meta$ are shaded out where ALP--meson mixing angles become larger that 0.1.
Final states depend on how ALP decays, which include $a \to \gamma \gamma, e^+e^-, \mu^+ \mu^-, 2\pi \gamma, 3\pi$ depending on mass $m_a$ and couplings.
}
\label{eta_ALP_decays} 
\end{figure}

The $\eta,\eta^\prime$ branching ratios are shown in Fig.~\ref{eta_ALP_decays}.
For simplicity, we assume these channels are dominated by ALP--gluon couplings (setting $c_q=0$), as in Ref.~\cite{Aloni:2018vki}, so that the only free parameters are $m_a$ and $f_a$.
We fix the latter to be $f_a = 10\GeV$, but since branching ratios scale with $1/f_a^2$, they are easily rescaled to other values. 
The decay constant corresponds to an effective BSM mass scale $\Lambda/|C_{GG}| = 32\pi^2 f_a \approx 3\TeV$, where $C_{GG}$ is the ALP--gluon Wilson coefficient~\cite{Bauer:2017ris,Aloni:2018vki}.

It would be worthwhile to improve these predictions beyond the LO calculations presented here.\footnote{See also
Ref.~\cite{Landini:2019eck} for a different calculation of these decays and
Refs.~\cite{Bauer:2017ris,Aloni:2018vki}
for similar calculations of the isospin-violating amplitude $a\to3\pi$ using $\SU(2)$ and $\SU(3)$ chiral Lagrangians,
respectively.}
Final-state rescattering effects can be taken into account by multiplying, e.g., $\A(\eta\to\pi\pi a)$ by $\Omega_0^0(s)$, the isospin $I=0$ $\pi\pi$ $S$-wave Omn\`es function.
Additionally, there are indications that NLO corrections may be significant for these channels.
Since the ALP couples (in part) by mixing with the $\eta,\eta^\prime$, the decay amplitudes \eqref{eq:eta-pipiaxion} are related to the chiral amplitudes
for $\pi\eta$ scattering~\cite{Bernard:1991xb} or the decay amplitude for $\eta'\to\eta\pi\pi$ (cf.\ Sect.~\ref{sec:etap-etapipi}), both of which are constant and proportional to $\mpc^2$
at leading order, but receive large corrections at $\Order(p^4)$.
NLO corrections for the QCD axion mass and photon coupling have been studied in Ref.~\cite{diCortona:2015ldu}, but remain unknown for the processes considered here.

Next, we discuss ALP decays. 
For the mass range $m_a < \meta - 2\mpi$ accessible in $\eta \to \pi \pi a$ decays, the only kinematically allowed channels are $a \to \gamma\gamma, \ell^+\ell^-$.
The partial widths are
\beq \label{eq:light_alp_widths}
\Gamma(a \to \gamma\gamma) = \frac{\alpha_\text{em}^2 m_a^3 |c_\gamma^{\rm eff}|^2}{256\pi^3 f_a^2} \, ,
\qquad
\Gamma(a \to \ell^+ \ell^-) = \frac{m_\ell^2 m_a |c_\ell|^2}{8\pi f_a^2} \sqrt{ 1 - 4m_\ell^2/m_a^2} \, ,
\eeq
where the effective photon coupling is
\beq
c_\gamma^{\rm eff} = \bar{c}_\gamma +  c_\gamma^{(q)} + c_\gamma^{(\ell)} \, .
\eeq
The one-loop lepton contribution is~\cite{Bauer:2017ris}
\beq
c_{\gamma}^{(\ell)} = \frac{c_\ell}{16\pi^2} \left( 1- \frac{4 m_\ell^2}{m_a^2} f\big(m^2_\ell/m_a^2\big) \right) \,,
\eeq
where $f(x)$ is defined in Eq.~\eqref{eq:photon_loop_function}.
The light-quark contribution is described by mixing with  pseudoscalar mesons, which gives
\beq
c_{\gamma}^{(q)} = 2\left( \theta_{a\pi} + \frac{4 \sqrt 2}{3 \sqrt 3} \theta_{a\eta} + \frac{7}{3\sqrt 3} \theta_{a\eta^\prime} \right) 
\eeq
in the single-angle $\eta$--$\eta^\prime$ mixing scheme with $\theta_P = \arcsin(-1/3)$ and neglecting isospin breaking.
Ref.~\cite{Aloni:2018vki} considered a purely gluon-coupled ALP ($c_q = c_\ell = c_\gamma = 0$), in which case $a \to \gamma\gamma$ is the dominant decay.
However, in light of the $(g-2)_\mu$ discrepancy, it is worthwhile to search for dilepton decays as well.
Whether $a \to \gamma\gamma$ or $a \to \ell^+ \ell^-$ is dominant depends nontrivially on ALP couplings and mass since the partial widths scale differently with $m_a$~\cite{Bauer:2017ris}.

For larger ALP masses accessible in $\eta^\prime$ decays, the ALP may also decay $a \to \pi\pi\gamma, 3\pi$.
For the purely gluon-coupled ALP, Ref.~\cite{Aloni:2018vki} calculated both channels and showed that they dominate the decay rate when kinematically allowed.
Again, however, the $(g-2)_\mu$ motivates a more general ALP model and it is worthwhile searching for dilepton and diphoton final states even for higher-mass ALPs.

Here we summarize possible signal channels and what is known about them:
\begin{itemize}
\item {\it $\eta,\eta^\prime \to 2\pi a \to 2 \pi \gamma \gamma$:}
This channel can in principle explore the full range of ALP masses accessible in $\eta,\eta^\prime$ decays, although the region $m_a \lesssim 10\MeV$ is already excluded~\cite{Beacham:2019nyx}.
A bump-hunt in the diphoton invariant mass ($M_{\gamma\gamma}$) spectrum is desirable for separating the ALP signal from the large background from $\eta,\eta^\prime \to 3\pi$.
Standard Model predictions have been made for the $M_{\gamma\gamma}$ distributions for $\eta \to \pi^0 \pi^0 \gamma \gamma$ and $\eta \to \pi^+ \pi^- \gamma \gamma$, in particular, for the nonresonant part away from the $\pi^0$ pole~\cite{Knochlein:1996ah,Bellucci:1996fm,Ametller:1997qy}.
The charged-pion mode is not interesting from a $\chi$PT perspective, being dominated by bremsstrahlung from $\eta\to\pi^+\pi^-\gamma$.
The only experimental studies, dating back to 1967~\cite{Price:1967zz,Baltay:1967zzb}, put a limit $\BR(\eta \to \pi^+\pi^-\gamma\gamma) < 2.1\times 10^{-3}$ for $M_{\gamma\gamma} > 195\MeV$. 
The neutral-pion mode has seen greater interest since the nonresonant part is sensitive to chiral loop corrections~\cite{Bellucci:1996fm,Ametller:1997qy}.
Limits come from the Crystal Ball detector at BNL-AGS~\cite{Nefkens:2005ka} and the GAMS-$4\pi$ experiment~\cite{Binon:2006bs}, the former giving a stronger limit, $\BR(\eta \to \pi^0\pi^0\gamma\gamma) <1.2\times 10^{-3}$ with a cut on $M_{\gamma\gamma}$ values near $\mpiz$. 
To our knowledge, no similar theoretical or experimental work has been done for $\eta^\prime \to 2\pi \gamma\gamma$ decay.
Since $\eta^\prime$ decays access a much larger range of ALP masses, such studies are strongly encouraged.

\item {\it $\eta^\prime \to \eta \pi^0 a \to \eta \pi^0 \gamma \gamma$:} 
This channel is not a good probe for ALPs.
According to Fig.~\ref{eta_ALP_decays}, $\BR(\eta^\prime \to \eta \pi^0 a)$ is negligible outside a limited mass range near $m_a \approx \mpiz$ (driven by $\pi^0$--ALP mixing).
It may be challenging to distinguish this BSM decay from the Standard Model decay $\eta^\prime \to \eta \pi^0 \pi^0$.

\item {\it $\eta,\eta^\prime \to \pi^+\pi^- a \to \pi^+\pi^- \ell^+ \ell^-$:} 
The connection with $(g-2)_\mu$---along with fact that ALP couplings to leptons are less constrained than to photons~\cite{Bauer:2017ris}---motivates the search for ALPs in these channels.
Presently, the partial widths for $\ep \to \pi^+ \pi^- e^+ e^-$ have been measured~\cite{Bargholtz:2006gz,Ambrosino:2008cp,Ablikim:2013wfg} and only upper limits exist for $\ep \to \pi^+ \pi^- \mu^+ \mu^-$~\cite{Berlowski:2007aa,Ablikim:2013wfg} (see Tables~\ref{tab:eta} and \ref{tab:etaprime}).
However, there is impetus from the Standard Model side for improving these measurements further to obtain dilepton invariant mass distributions, which access the doubly-virtual transition form factors entering the light-by-light contribution to $(g-2)_\mu$ (see Sect.~\ref{sec:eta-pipill}).
Bump-hunt searches for ALPs could be made in parallel.
More detailed differential distributions would further aid in discriminating ALPs from Standard Model processes.
In Standard Model decays, the $\pi^+\pi^-$ emerge in a relative $P$-wave (i.e., via the $\rho$ resonance), while in BSM decays \eqref{eq:eta-pipiaxion} they will be dominantly produced in a relative $S$-wave.  Also, the distribution of the leptons from an ALP is different than from a virtual photon.

\item {\it $\eta,\eta^\prime \to 2\pi^0 a \to 2 \pi^0 \ell^+ \ell^-$:} 
Like the preceding channel, experimental searches would be worthwhile to access ALPs in connection with $(g-2)_\mu$.
However, this final state is $C,CP$-violating as a single-photon process at leading order in QED and has not been studied in any context.
As a probe of $C$ violation, there is little motivation to search for these decays given that the simpler channels $\eta,\eta^\prime \to \pi^0 \ell^+ \ell^-$ access the same physics and are all consistent with zero~\cite{Dzhelyadin:1980ti,Briere:1999bp,Adlarson:2018imw}.
However, in ALP models (where $C,CP$ are conserved), ALP-mediated decays $\eta,\eta^\prime \to 2\pi^0 \ell^+ \ell^-$ are allowed, while $\pi^0 \ell^+ \ell^-$ final states are forbidden.
The Standard Model background arises at one-loop order in QED, which is an advantage over the preceding channel.

\item {\it $\eta^\prime \to \eta \pi^0 a \to \eta \pi^0 \ell^+ \ell^-$:} 
Similar to the preceding channel, this decay is $C,CP$-violating as a single-photon process, but is $C,CP$-conserving as an ALP-mediated decay.
It also has not been studied to date, though it is limited as probe for ALPs since $\BR(\eta^\prime \to \eta \pi^0 a)$ is negligible unless $m_a \approx \mpiz$.
In this case, only $\eta^\prime \to \eta \pi^0 e^+ e^-$ is allowed.
If the ALP and $\pi^0$ are sufficiently close in mass, there is an irreducible background of $\BR(\eta^\prime \to \eta \pi^0 e^+ e^-) \sim 3 \times 10^{-8}$ from $\eta^\prime \to \eta \pi^0 \pi^0$ where one pion decays $\pi^0 \to e^+ e^-$, as well as a background at one-loop order in QED.

\item {\it $\eta^\prime \to  2\pi a \to 2\pi \, \pi^+ \pi^- \gamma $:} Possible final states are $2(\pi^+\pi^-)\gamma$ and $2\pi^0 \pi^+ \pi^- \gamma$, for the kinematically-allowed ALP mass range $279 < m_a < 688\MeV$.
No experimental searches have been made for these decays, nor has this process been studied in the Standard Model. Ref.~\cite{Aloni:2018vki} has placed limits on the related process $\phi \to a \gamma \to \pi^+ \pi^- \gamma\gamma$ based on null searches for this final state at CMD-2~\cite{Aulchenko:2008zz}.

\item $\eta^\prime \to  2\pi a \to 5 \pi$:
The only experimental constraint is from CLEO~\cite{Naik:2008aa}, $\BR(\eta^\prime \to 2(\pi^+\pi^-)\pi^0) < 1.8\times 10^{-3}$~\cite{Tanabashi:2018oca}, while $3\pi^0 \pi^+\pi^-$ and $5\pi^0$ final states have not been studied to date.
We are not aware of any predictions for the Standard Model rates, which are likely very small since they violate isospin and are suppressed by five-body phase space.
In contrast, the BSM decay $\eta^\prime \to  2\pi a$ is not isospin-violating and is only a three-body decay, followed by $a \to 3\pi$ which has an $\mathcal{O}(1)$ branching fraction when kinematically allowed.
The CLEO result gives the strongest limit on a purely gluon-coupled ALP within the kinematically-allowed mass range, $414 <  m_a < 679 \MeV$~\cite{Aloni:2018vki}.
\end{itemize}
Other search strategies may be more practical for low mass ALPs decaying $a \to e^+ e^-$.
If the ALP is sufficiently boosted and the $e^+ e^-$ opening angle is sufficiently small, the $e^+ e^-$ pair could appear as a fake (converted) photon in the detector.
The most promising channel to search for this would be \beq
\eta \to 2\pi^0 a \to 2\pi^0 \, \gamma_{\rm fake} \, ,
\eeq
given the fact that $\eta \to 2\pi^0 \gamma$ is $C,CP$-violating and has a negligible Standard Model rate. This can be explored by JEF with its highly boosted $\eta$  and strongly suppressed backgrounds of the multi-photon final states.

\subsection{Invisible decays}
\label{sec:invis}

Light dark matter (below the GeV scale) has attracted significant attention in the literature, motivated by the absence of definitive signals at dark matter direct detection experiments which are most sensitive to larger masses~\cite{Essig:2013lka,Alekhin:2015byh,Alexander:2016aln,Battaglieri:2017aum,Knapen:2017xzo}.
Therefore, it is possible that $\eta,\eta^\prime$ decays can produce light dark matter particles (or other long-lived hidden states) if they are within kinematic range and couple to light quarks~\cite{McElrath:2005bp,Fayet:2006sp}.
Since dark matter is typically invisible to detectors, it yields ``missing energy'' signatures at $\eta,\eta^\prime$ factories.

Experimental limits on fully invisible decays, $\ep \to {\rm invisible}$, were first obtained by BESII~\cite{Ablikim:2006eg} and later by CLEO~\cite{Naik:2008aa} and BESIII~\cite{Ablikim:2012gf}.
In these searches, since the $\ep$ is assumed to decay fully invisibly and is not reconstructed directly, $\ep$ tagging relies on visible associated particles that arise in the production decay chain with well-defined kinematics.
BESII/III searched for $J/\psi \to \phi \ep \to K^+K^- + {\rm invisible}$, where $\phi \to K^+K^-$ is used to tag these events~\cite{Ablikim:2006eg,Ablikim:2012gf}, while CLEO searched for $\psi(2S) \to \pi^+ \pi^- J/\psi$, with $J/\psi \to \gamma \eta^\prime$, using the $\pi^+ \pi^-$ and monochromatic $\gamma$ as a tag.
Present constraints are~\cite{Ablikim:2012gf,Tanabashi:2018oca}
\beq \label{eq:inv_limits}
\BR(\eta \to {\rm invisible}) < 1 \times 10^{-4} \, , \qquad 
\BR(\eta^\prime \to {\rm invisible}) < 5 \times 10^{-4} \;\; (90\%~{\rm C.L}) \, .
\eeq
Since a positive detection would be evidence for BSM physics, further proposals have been made to search for these channels at NA64~\cite{Gninenko:2014sxa,Beacham:2019nyx}.
The SM rates are essentially negligible, dominated by $\ep \to \nu\bar\nu \nu \bar\nu$ with the branching ratio of $\mathcal{O}(10^{-18})$~\cite{Gao:2018seg}; the $\ep \to \nu \bar\nu$ process is even smaller since it is helicity-suppressed by small neutrino masses~\cite{Arnellos:1981bk}.
Limits on partially-invisible decays were obtained by the Crystal Barrel experiment~\cite{Amsler:1994gt,Amsler:1996hb}
\beq
\BR(\eta \to \gamma\!+\!{\rm invisible}) < 6 \times 10^{-5} \, , \qquad 
\BR(\eta^\prime \to \gamma\!+\!{\rm invisible}) < 4 \times 10^{-5}  \;\; (90\%~{\rm C.L}) \, .
\eeq
Ref.~\cite{Arnellos:1981bk} calculated the  branching fractions in the Standard Model to be $\BR(\ep \to \gamma \nu \bar \nu) \sim 2 \times 10^{-15} \; (2 \times 10^{-14})$.

Invisible signatures can be realized for any of the light mediator models discussed previously.
Since dark sector mediators naturally couple more strongly to dark matter than Standard Model particles, invisible decays to the former would dominate the branching ratio if kinematically allowed.
Alternatively, the mediator particles may be so long-lived that they are effectively invisible at $\eta,\eta^\prime$ factories.
This possibility has recently come to the fore in connection with kaon decays~\cite{Kitahara:2019lws,Fabbrichesi:2019bmo,Egana-Ugrinovic:2019wzj,Dev:2019hho,Liu:2020qgx,Jho:2020jsa,Hostert:2020gou} as a possible explanation for anomalous $K_L \to \pi^0 \nu\bar\nu$ events reported preliminarily from the KOTO experiment~\cite{KOTOslides}.

The aforementioned constraints are model-independent and apply to any scenario where the invisible state is undetected.
However, for specific models, much more stringent limits exist from high-intensity beam dumps and other experiments where $\eta,\eta^\prime$ mesons are copiously produced.
If $\eta,\eta^\prime$ decays produce dark matter particles, the latter can scatter and be detected in a downstream detector~\cite{Batell:2009di,deNiverville:2011it,Izaguirre:2013uxa,NA64:2019imj}.
Alternatively, if $\eta,\eta^\prime$ decays produce long-lived mediators, the highly displaced vertex can be detected~\cite{Gninenko:2011uv,Gninenko:2012eq,Tsai:2019mtm,Aaij:2019bvg} or the mediator can Primakoff scatter in the detector to yield a meson~\cite{Gninenko:1998pm}.
Mono-photon searches at $B$-factories provide additional constraints on light invisible mediators~\cite{Lees:2017lec}.

The channels given above are only a few of the possibilities for $\eta,\eta^\prime$ decays with invisible final states.
Here we describe the different classes of signals and their theoretical motivations.
\begin{itemize}
\item {\it Fully-invisible decays:} $\ep \to {\rm invisible}$ can arise from decays to pairs of light mediators, each decaying invisibly to dark matter~\cite{Fayet:2006sp}, or by $\ep$ decays to pairs of dark matter particles directly~\cite{McElrath:2005bp}.
Ref.~\cite{Darme:2020ral} recently explored the latter case for several models within an effective theory framework, which covers models with heavy (electroweak-scale) mediators that have been integrated out.
Given existing collider constraints, the branching ratios cannot be much larger than $\BR(\ep \to {\rm invisible}) \sim 10^{-11} \; (10^{-13})$~\cite{Darme:2020ral}, which is a challenging prospect for tagged $\eta,\eta^\prime$ factories.
\item {\it Partially-invisible decays:} 
$\eta,\eta^\prime$ final states can include both visible and invisible particles.
Possible scenarios are the same vector, scalar, and ALP mediator models discussed above but where the mediator decays invisibly into light dark matter or is too long-lived to be detected.
Signatures include
\beq
\ep \to \gamma \!+\! {\rm invisible}, \;\; \pi^0 \!+\! {\rm invisible}, \;\; \pi\pi \!+\! {\rm invisible} \, , \label{eq:part_invis}
\eeq
for the vector, scalar, and ALP models, respectively. 
The first channel mimics $\ep \to \gamma\gamma$ with an undetected $\gamma$, except the detected $\gamma$ has a smaller (monochromatic) energy $E_\gamma < M_{\ep}/2$ in the $\ep$ rest frame.
For the second channel, the $\pi^0$ is also monochromatic in the $\ep$ rest frame.
Alternatively, $\ep \to \gamma + {\rm invisible}$ can arise as a three-body decay directly into a pair of dark matter particles (plus a photon).
Within an effective theory framework, the branching ratio cannot be much larger than $\mathcal{O}(10^{-16})$~\cite{Darme:2020ral}.

\item {\it Invisible decays with a displaced vertex:} 
Dark sector states produced in $\eta,\eta^\prime$ decays can include a higher state that de-excites down to the lightest hidden state, e.g., by emitting an $\ell^+ \ell^-$ pair.
This situation is motivated by the dark photon parameter window for $(g-2)_\mu$, which has been excluded for both visible and invisible $A^\prime$ decays but remains viable for a more complicated inelastic dark matter model~\cite{Mohlabeng:2019vrz}.
In this setup, the dark photon decays $A^\prime \to \chi_1 \chi_2$, where $\chi_1$ is dark matter and $\chi_2$ is a next-to-lightest state that de-excites $\chi_2 \to \chi_1 \ell^+ \ell^-$~\cite{Mohlabeng:2019vrz,Tsai:2019mtm}.
The total decay chain is
\beq
\ep \to \gamma A^\prime \to \gamma \, \chi_2 \chi_1
\to \gamma \, \ell^+ \ell^- \, \chi_1 \chi_1 \,,
\eeq
where $\chi_1$ particles are invisible.
The final state signal is $\gamma \, \ell^+ \ell^- \!+ \!{\rm invisible}$, where the $\ell^+ \ell^-$ pair may be displaced from the primary vertex. 
In general, displaced states can be searched for in addition to any of the invisible channels discussed above (see also~\cite{Hostert:2020gou}).
\end{itemize}
We also note other related work on invisible signatures for flavored mesons~\cite{Badin:2010uh,Bhattacharya:2018msv,Gninenko:2015mea,Barducci:2018rlx,Tan:2020gpd}.

Among proposed $\eta,\eta^\prime$ factories, the REDTOP experiment will have the capability to target many of the signals discussed above.
The upgraded t-REDTOP detector for REDTOP run-II (and beyond) will be able to fully measure the scattering kinematics to reconstruct the $\ep$ four-momentum and will therefore be sensitive to invisible final states~\cite{t-REDTOP}.
Moreover, the detector will include vertexing capability to detect displaced decays.
However, more work is needed motivate these studies within the landscape of theories and other existing and proposed experiments.
The latter include high-intensity beam dump experiments~\cite{Alekhin:2015byh,Berlin:2018pwi} and other long-lived particle searches at the LHC~\cite{Curtin:2018mvb,Ariga:2018uku} that can explore the invisible $\eta,\eta^\prime$ frontier down to very small branching ratios~\cite{Darme:2020ral}.

\section{Summary and outlook}\label{sec:summary}

\begin{table}
\centering
\renewcommand{\arraystretch}{1.3}
\begin{tabular}{lccc}
\toprule
Decay channel & Standard Model & Discrete symmetries & Light BSM particles \\
\midrule
$\eta\to\pi^+\pi^-\pi^0$ & light quark masses & $C$/$CP$ violation & scalar bosons (also $\eta^\prime$) \\
$\eta^{(\prime)}\to\gamma\gamma$ & $\eta$--$\eta'$ mixing, precision partial widths & & \\
$\eta^{(\prime)}\to\ell^+\ell^-\gamma$  & $(g-2)_\mu$ & & $Z'$ bosons, dark photon \\
$\eta\to\pi^0\gamma\gamma$ & higher-order $\chi$PT, scalar dynamics & &  $U(1)_B$ boson, scalar bosons\\
$\eta^{(\prime)}\to\mu^+\mu^-$ & $(g-2)_\mu$, precision tests & $CP$ violation & \\
$\eta\to\pi^0 \ell^+\ell^-$ & & $C$ violation & scalar bosons \\
$\ep\to\pi^+\pi^- \ell^+\ell^-$ & $(g-2)_\mu$ &  & ALPs, dark photon \\
$\ep\to\pi^0\pi^0 \ell^+\ell^-$ & & $C$ violation & ALPs \\
\bottomrule
\end{tabular}
\renewcommand{\arraystretch}{1.0}
\caption{Summary of high-priority $\eta^{(\prime)}$ decays with emphasis on synergies across 
Standard Model and BSM investigations.
\label{tab:summary} }
\end{table}

Decays of $\eta$ and $\eta'$ mesons offer a vast multitude of opportunities to investigate fundamental physics under the broad theme of symmetry and symmetry-breaking, both within the Standard Model and beyond.  
Instead of an all-encompassing summary of this review, we highlight here what we perceive as the most important challenges and issues in this field for the coming years, for experiment and theory alike.
Table~\ref{tab:summary} presents our recommended list of highest-priority $\eta$ and $\eta^\prime$ decay channels, emphasizing those that allow for simultaneous investigations of high-precision Standard Model tests and 
searches for physics beyond, looking either for interesting discrete symmetry violations or potential new light particles.

\subsection*{Standard Model decays}
\begin{itemize}
\item $\eta^{(\prime)}\to3\pi$.  
Several laboratories plan to measure the $\eta\to3\pi$ Dalitz plots with yet improved precision. 
Combined with a more accurate determination of the $\eta$ radiative width to sharpen the $\eta\to3\pi$
partial widths, this will allow for a yet-more-precise extraction of the quark mass double ratio $Q$. 
To this end, theoretical issues to be addressed concern a more precise matching to the $\chi$PT representation, which currently constitutes the dominating theoretical uncertainty in the determination of $Q$~\cite{Colangelo:2018jxw}, 
but also an improved implementation of radiative corrections and other higher-order isospin-breaking 
effects~\cite{Ditsche:2008cq,Schneider:2010hs}.

In contrast, a comparably fundamental interpretation of measurements of $\eta'\to3\pi$ requires first of all 
serious theoretical advances: while a dispersive representation to describe the Dalitz plot distributions is 
within reach, its matching to a systematic effective field theory comparable to the program for $\eta\to3\pi$ ---
$\chi$PT, even if supplemented with large-$N_c$ arguments---seems to be working marginally at best, and clearly
requires further developments to extract quark mass ratios also from $\eta'$ decays.

\item $\eta^{(\prime)}\to\gamma\gamma$.
High-precision determinations of the two-photon decay widths of $\eta$ and $\eta'$ are of utmost importance 
for a variety of reasons: they serve as reference channels for many other decays and therefore indirectly impact
the physics goals of those; they yield the normalization of the corresponding transition form factors and thus
contribute significantly to our understanding of hadronic light-by-light scattering in the anomalous magnetic moment
of the muon; and their pattern allows us to extract $\eta$--$\eta'$ mixing parameters to sharpen
our understanding of the $U(1)_A$ anomaly and its interplay with chiral symmetry.

\item $\eta^{(\prime)}$ transition form factors both have a heavy impact on hadronic contributions 
to $(g-2)_\mu$, and allow us to learn about hadronic structure.  Data in different kinematic regimes, from 
all possible decay and production processes, ought to be analyzed simultaneously, with theoretical representations
apt for such analytic continuation.  Besides direct measurements, the statistical leverage of using hadronic
decay channels to reconstruct the transition form factors dispersively should be taken advantage of, 
in particular via data on decays such as $\eta'\to2(\pi^+\pi^-)$,
$\eta'\to\pi^+\pi^-e^+e^-$, or $\eta'\to\omega e^+e^-$ (or, in a production reaction, $e^+e^-\to\eta\pi^+\pi^-$).
In all these, detailed differential information has to be the primary goal.

\item $\eta\to\pi^0\gamma\gamma$ and similar $\eta'$ decays.
JEF follows ambitious plans to significantly improve on the precision of differential decay information
of these doubly-radiative $\eta$ decays.  While the suppression of the $\eta\to\pi^0\gamma\gamma$ decay
amplitude in the chiral expansion and the intricate interplay between vector exchange and scalar $S$-wave
rescattering dynamics make for a highly interesting theoretical challenge to predict its properties 
with high accuracy, we believe that the interpretation in terms of fundamental insights still ought to 
be sharpened: are there general properties to be learnt from simultaneous analyses of 
$\eta\to\pi^0\gamma\gamma$ and $\eta'\to\eta\gamma\gamma,\,\pi^0\gamma\gamma$?  Do these processes also 
impact hadronic light-by-light scattering significantly?  ---  Experimentally, again these decays
offer high synergy effects through the simultaneous search for new light particles and a high-precision
investigation of QCD dynamics.
\end{itemize}

\subsection*{Discrete symmetry tests and lepton flavor violation}
\begin{itemize}
\item $P$ and $CP$ violation.
From a theoretical point of view, a large number of $P,CP$-violating $\eta^{(\prime)}$ decays are indirectly
excluded through extremely stringent neutron EDM bounds.
The only exception currently known to us is the investigation of the muon polarization asymmetries in $\eta\to\mu^+\mu^-$, probing flavor-conserving $CP$-violation in the second generation.
Measurements would probe effective dimension-6 strange-quark--muon operators that evade EDM limits so far and should be within reach of planned REDTOP statistics~\cite{Sanchez-Puertas:2018tnp}.
Within this class of discrete symmetry tests, we therefore assign a high-precision experimental test of 
$\eta\to\mu^+\mu^-$ the highest priority.
Improved experimental bounds on classic channels such as $\eta^{(\prime)}\to\pi\pi$ are certainly welcome, but 
seem very unlikely to find evidence of BSM physics in the foreseeable future.

\item $C$ and $CP$ violation.
On the theoretical side, most of the literature on $C,CP$-violating $\ep$ decays predates the Standard Model and very little has been done within a modern context.
The main task, therefore, lies with the theory community to motivate searches for these channels.
At the very least, the relevant BSM operators need to be identified and their contributions to $\ep$ decays, as well as indirect limits from EDMs, need to be quantified.
We also point out that the traditional $C,CP$-violating channel $\ep \to 3\gamma$ is also sensitive to BSM physics that is $C,P$-violating but $CP$-conserving.
It is worthwhile investigating further how large the $\ep \to 3\gamma$ rate could be in such a scenario since it would be safe from EDMs.
On the experimental side, the study of $C$- and $CP$-violating asymmetries in the $\eta\to\pi^+\pi^-\pi^0$
Dalitz plot offer excellent synergies with the QCD/Standard Model motivation for high-precision investigations
of this channel.  Similarly, many channels originally proposed to test $C,CP$-violation can also be used to search for new light particles.
These include, e.g., $\ep\to\pi^0\ell^+\ell^-$ and $\ep\to 2\pi^0\ell^+\ell^-$ which are $C,CP$-violating as single-$\gamma$ processes in QED, but can arise from new light scalar and pseudoscalar bosons, respectively.

\item Lepton flavor violation.
In light of strong constraints on $\mu \to e$ conversion on nuclei, further theoretical study is needed to motivate searching for $\ep \to e^\pm \mu^\mp$.
On the other hand, decays that violate charged lepton flavor by two units, $\ep \to e^\pm e^\pm \mu^\mp \mu^\mp$, would not be similarly constrained.

\end{itemize}

\subsection*{New light particles (MeV--GeV) beyond the Standard Model}

\begin{itemize}

\item Dark photons and other hidden vector bosons. 
Particle models and motivations are well-established on the theory side.
With many other experimental efforts underway, searches at precision $\ep$ factories can have an impact in this field and, moreover, have synergies with key Standard Model studies.
Dark photons (and variants) appear as resonances in the dilepton invariant mass spectrum for $\ep \to \ell^+ \ell^- \gamma$, which can be studied in conjunction with transition form factor measurements.
REDTOP is projected to make substantial improvements in covering dark photon parameter space, complementary to other proposed experiments~\cite{Beacham:2019nyx,Gatto:2019dhj}.
The protophobic $X$ boson, related to the $^8$Be anomaly~\cite{Feng:2016jff}, is another target, with a predicted branching ratio $\BR(\eta \to X \gamma \to \gamma e^+ e^-) \approx 10^{-5}$.
Leptophobic vector bosons, e.g., the $B$ boson from gauged $U(1)_B$ symmetry, yield nonleptonic signatures, e.g., a $\pi^0 \gamma$ resonance in the rare decay $\eta \to B \gamma \to \pi^0 \gamma\gamma$~\cite{Nelson:1989fx,Tulin:2014tya}.
This decay is targeted by JEF~\cite{JEF-PAC42} and REDTOP~\cite{Gonzalez:2017fku} for its implications for both BSM physics and QCD physics.

\item Scalar particles. 
While most theoretical and experimental attention has focused on the Higgs-mixed scalar (Higgs portal model), $\ep$ studies are unlikely to push the frontier for this model.
On the other hand, $\ep$ decays have a complementary role in setting among the strongest limits on a scalar $S$ coupled preferentially to light quarks instead of heavy quarks~\cite{Batell:2018fqo,Liu:2018qgl}.
For a leptophobic scalar, signal channels are $\ep \to \pi^0 \gamma\gamma$ or $3\pi$ (both modes important also for QCD studies) with $S$ appearing as a $\gamma\gamma$ or $2\pi$ resonance, respectively.
In more general models, $S$ can be discovered as a dilepton resonance in $\ep \to \pi^0 \ell^+ \ell^-$, a channel that has largely been motivated only within a $C,CP$-violation context. 
However, work remains for phenomenologists to map out the more general parameter space for these decays, as well as for $\chi$PT theorists to pin down the scalar transition form factors for $\eta^\prime$ decays needed to access a wider range of $m_S$ (these are under better control for the $\eta$).

\item Axion-like particles (ALPs). 
ALP searches in $\ep$ decays is a new and potentially rich avenue that has not been widely studied in the literature.
Collaboration between BSM phenomenologists, $\chi$PT theorists, and experimentalists would be fruitful toward making progress.
Important questions to address: 
Which among the at least four $\eta$ and eleven $\eta^\prime$ signal channels---all with four- and five-body final states, several of which have never been studied before in any context---are most promising to measure from both theoretical and experimental points of view?
What are the predicted branching ratios in light of existing ALP constraints?
While we have given LO formulae for these decay rates, NLO corrections are likely large and would need to be included for robust predictions.

\item Invisible final states. 
Motivated by light dark matter and related dark sector models, possible channels include $\ep \to {\rm invisible}$, $\gamma+{\rm inv.}$, $\pi^0+{\rm inv.}$, and $\pi\pi+{\rm inv.}$, with only the first two having been searched for to date.
Additional displaced vertices (e.g., $\ell^+ \ell^-$) can arise on top of these final states for multi-state dark sectors~\cite{Mohlabeng:2019vrz,Tsai:2019mtm}.
Upcoming searches at NA64 will target invisible meson decays~\cite{Beacham:2019nyx} and further progress could be made at tagged $\eta,\eta^\prime$ factories such as REDTOP (run-II and beyond)~\cite{t-REDTOP}.
However, it is important to situate these measurements within the broader landscape of searches for light dark sectors, since final states that are invisible at $\eta,\eta^\prime$ factories can be visible for other types of experiments. 
In particular, several experiments using proton beam dumps produce large numbers of $\eta,\eta^\prime$ mesons and can detect their would-be invisible by-products as highly-displaced vertices from long-lived mediators~\cite{Gninenko:2012eq,Tsai:2019mtm} or by scattering events in downstream neutrino detectors~\cite{Batell:2009di,deNiverville:2011it}.

\end{itemize}

\subsection*{Experimental perspectives}

The future success of nuclear and particle physics will rely on high-precision and high-statistics data. The ongoing scientific programs at JLab will offer an exciting opportunity for $\eta$ and $\eta^{\prime}$ physics at the precision frontier. 

\begin{itemize}
\item
The comprehensive \textit{Primakoff program} on the measurements of two-photon decay widths and the transition form factors of $\pi^0$, $\eta$, and $\eta^{\prime}$ has been in progress.  Two experiments performed for the $\pi^0$ in the JLab $6\GeV$ era achieved an unprecedented precision of 1.5\% on $\Gamma(\pi^0\to\gamma\gamma)$.  Two data sets for a measurement of $\Gamma(\eta\to\gamma\gamma)$ on $^4$He, one of two physics targets ($^4$He and hydrogen), were collected in spring 2019 and in fall 2021 using a $\sim 11\GeV$ photon beam.  The third run will be expected in 2022.  The final goal of 3\% precision for $\Gamma(\eta\to\gamma\gamma)$  will help resolving a longstanding puzzle of the systematic discrepancy between the previous Primakoff and the $e^+e^-$ collider measurements.  Looking forward, the first Primakoff measurement of $\Gamma(\eta^{\prime}\to\gamma\gamma)$ at an accuracy of 4\% is planned; and the space-like $Q^2$ region of the transition form factors of $\pi^0$ and $\eta^{(\prime)}$ will be mapped out as low as $10^{-3}\GeV^2$. These anticipated results will provide sensitive probes to the symmetry structure of QCD at low energies.

\item
The ongoing \textit{JLab Eta Factory experiment} 
will open a new avenue for precision measurements of various decays of $\eta$ and $\eta^{\prime}$ in one setting, with unprecedented low backgrounds in rare decays, particularly in neutral modes.  Highly boosted $\eta$ and $\eta^{\prime}$ by a $\sim 12\GeV$ photon beam will help reducing the experimental systematics, offering complementary cross checks on the results from A2, BESIII, KLOE-II, WASA-at-COSY, and future REDTOP experiments, where the produced mesons have relatively small kinetic energies in the lab frame. 
The data collection for nonrare decays has been in progress since fall 2016.  A significant improvement on the light quark mass ratio will be achieved in the next 3--4~years by a combination of a new Primakoff measurement of the $\eta$ radiative decay width and the improvements in the Dalitz distributions of $\eta\to 3\pi$ for both charged and neutral channels. 
The second phase of JEF will run with an upgraded forward calorimeter.  Within 100~days of beam time for the phase~II, JEF will have sufficient precision to explore the role of scalar meson dynamics in chiral perturbation theory for the first time, to search for sub-GeV dark gauge bosons (vector, scalar, and axion-like particles) by improving the existing bounds by two orders of magnitude that is complementary to the ongoing worldwide efforts on invisible decays or decays only involving leptons.

\item
The proposed future \textit{REDTOP experiment}, on the other hand, will be the champion to push the limit on the high-intensity frontier with the projected $\eta$ production rate at the level of $2\times 10^{12}$ (a factor of ten more for phase~II) per year. 
Although the backgrounds in REDTOP will be expected about several orders of magnitude higher than in the JEF experiment, this will be compensated for by an enormous $\eta$ yield. The recoil detection technique considered for  phase II will help further reducing the backgrounds. The proposed muon polarimeter (and an optional photon polarimeter) for the REDTOP apparatus will offer additional capability to measure the longitudinal polarization of final-state muons (and possibly photons), which are not available in most other experiments, including JEF. In the foreseeable future, REDTOP will offer the most sensitive probes for the rare charged decay channels, while the JEF experiment will remain leading in the rare neutral decays because of lower backgrounds and higher experimental resolutions.  The JEF and REDTOP experiments are complementary to each other, promising a new exciting era for $\eta^{(\prime)}$ physics.

Another proposal to build a REDTOP type of $\eta$-factory at the High-Intensity Heavy-Ion Accelerator Facility (HIAF), Huizhou, China, has been under development recently~\cite{HIAF-eta}. A $\sim 2 \GeV$ proton beam will be used to produce $\eta$ mesons at the projected rate of $3\times 10^{13}$ per year.   The HIAF is a new facility that is currently under construction.

\end{itemize}

Continuing programs of $\eta^{(\prime)}$ mesons at different kinematics and via different production mechanisms will be important for controlling the experimental systematics.  Although the KLOE-II and the WASA-at-COSY collaborations completed data collections, only a part of their data has been analyzed.  More new results will be anticipated from these two groups.  The A2 and the BESIII collaborations have been active in the recent years to produce publications based on their existing data.  They still have potential to collect more data in the coming decade~\cite{Ablikim:2019hff}.
The upgraded Belle-II experiment at SuperKEKB will cover $Q^2$ up to $40\GeV^2$ in the two-photon fusion reaction.  These global experimental efforts at different facilities will offer opportunities for discoveries in the $\eta^{(\prime)}$ sector.

\section*{Acknowledgements}
We would like to thank Brian Batell, Gilberto Colangelo, Igor Danilkin, Luc Darme, Jonathan Feng, Alexander Friedland, Susan Gardner, Corrado Gatto, Sergei Gninenko, Jos\'e Goity, Sergi Gonz\`alez-Sol\'is, Feng-Kun Guo, Christoph Hanhart, Martin Hoferichter, Bai-Long Hoid, Simon Holz, Philip Ilten, Tobias Isken, Felix Kahlhoefer, Yannis Korte, Andrzej Kup\'s\'c, Stefan Lanz, Heiri Leutwyler, Jan L\"udtke, Bachir Moussallam, Malwin Niehus, Pablo S{\'a}nchez-Puertas, Yotam Soreq, Peter Stoffer, Simon Taylor, Andreas Wirzba, Tevong You, and Yongchao Zheng for useful discussions on topics covered in this review.  
We gratefully acknowledge financial support by the Deutsche Forschungsgemeinschaft (DFG, German Research Foundation) (CRC 110,``Symmetries and the Emergence of Structure in QCD''), the Natural Sciences and Engineering Research Council of Canada, the U.S.\ Department of Energy (contract DE-AC05-06OR23177), and the U.S.\ National Science Foundation (PHY-1714253, PHY-1506303, PHY-1812396, and PHY-2111181). This research was also supported by the Munich Institute for Astro- and Particle Physics (MIAPP), which is funded by the DFG under Germany's Excellence Strategy -- EXC-2094 -- 390783311.

\vfill\eject

\end{document}